# The Multiverse:
# a Philosophical Introduction

Jeremy Butterfield

*To Mari*

# Preface

This is an introduction to the idea that our universe is just one of many universes: it is part of a multiverse. This idea is very topical. A multiverse of one kind or another is seriously advocated by many philosophers. And similarly for physics: many physicists advocate a multiverse---usually of a different kind than that of the philosophers. So the time seems ripe to assess the various versions of this idea. In this book, I will assess three versions of it. One is from philosophy: more specifically, from logic's treatment of possibility. The other two are from physics: more specifically, the Everettian interpretation of quantum theory, and inflationary cosmology. I will discuss these in order; and then in a final Chapter, compare them and relate them to each other.

I should declare at the outset that the book has three main limitations. First: I set aside completely the treatment of the multiverse in countless novels, plays and films. Agreed, the parallel worlds (so called) in, for example, the films 'Sliding Doors' and 'Everything Everywhere All at Once', raise plenty of philosophical issues. But in this book, I will pursue the issues in themselves, in an academic manner. I have no aptitude for literary matters; so I set aside (albeit with some regret) these entertaining examples.

Second: within physics and philosophy, there are multiverse proposals other than the three I will assess. But among these, there are two which I will discuss during my assessments of my main three. The first of these is the multiverse of string theory; (string theory is a speculative physical theory). This proposed multiverse, often called 'the landscape', gives the main rationale for the cosmological multiverse; and so I will discuss it in that context. The second is the 'Pythagorean' proposal that all of reality is mathematical, and that for this reason our universe is one of many universes. I will discuss this towards the end of my assessment of the philosophical multiverse, based on logic's treatment of possibility.

Third: this book is a philosopher's introduction. I mean this both in the sense that I am a philosopher, not a physicist; and that my intended readership is people who are interested in philosophy and the philosophical aspects of physics---but who are newcomers to it. So I endeavour to explain every term which is "in-house" for the disciplines of either physics or philosophy. For example, I will explain the terms 'truth-table' (in philosophy) and 'wave-function' (in physics). But the tone is brisk and breezy. References are few, and almost entirely confined to suggested readings as the end of each Chapter. Thus it will be evident throughout that my discussion is inconclusive, and my conclusions tentative.

Of course, being inconclusive is hardly surprising. For a subject as large in every sense (and more importantly: multi-faceted) as the multiverse, even a learned and exhaustive investigation would surely be inconclusive. No human mind could know all the relevant strands of evidence and of argumentation, from all of physics and philosophy. And even if one did know all that, the process of weighing them against each other---for there are bound to be conflicts---would involve judgments which could no doubt be disputed.

So while I admit that my discussion is inconclusive, I console myself with the thought that a brief and non-exhaustive investigation can serve a purpose. Indeed, I hope it can satisfy the curiosity of the newcomer reader, and prompt their own thinking, about as well as a long and learned book could. (In Chapter 1, I will say more in defence of this philosophical outlook: namely, that one should "dive in" and speculate, even when one's evidence is small and perhaps defective.)

I am grateful to many people for comments and corrections on drafts of this material and for discussion. For very helpful comments and corrections on drafts of various Chapters, I am especially grateful to: Feraz Azhar, Mark Burgman, Frank Cudek, Richard Dawid, Alex Oliver, Ruward Mulder, Bryan Roberts, Simon Saunders and David Wallace. I am also grateful to participants in the 2022 Cambridge-LSE Philosophy of Physics "Bootcamp" online discussion group, especially: Frank Cudek, Henrique Gomes, Josh Hunt, Brendan Kolb, Klaas Landsman, Joanna Luc, Tushar Menon, Ruward Mulder, Neil Dewar---and especially Bryan Roberts. I am also grateful for the comments and corrections by two anonymous readers for the British Society for the Philosophy of Science. My thanks also to the audiences at talks about various parts of this material, held in the mathematics, philosophy and physics departments and societies at the Universities of Amsterdam, Cambridge, Kings College London, Oxford, Salzburg and Vienna. Finally, it is a pleasure to thank some physicist colleagues for discussion of the Everettian and cosmological multiverses, and for encouragement: John Barrow, Bernard Carr, George Ellis, Jonathan Halliwell, Jim Hartle and Martin Rees.

I am also grateful to Helen Beebee, Oxford University Press, and the Firestone Library, Princeton University, for generous advice and for permission to reprint (in Chapter 3, Section 10) a letter by David Lewis: which is reprinted in Volume 1 of *Philosophical Letters of David K. Lewis*, edited by H. Beebee and A.R.J. Fisher, Oxford University Press (2020).

This book is dedicated to my wife Mari, for all her love and support.

JNB
Trinity College,
Cambridge

# Table of Contents with Chapter Summaries

First, I introduce the idea that our universe is just one of many universes. Then I announce the book's plan. I will present and assess three versions of the idea that have been proposed: one from philosophy, and two from physics. In short, the proposals are: all the logically possible worlds; all the branches of the quantum state, in an Everettian interpretation of quantum theory; and all the bubbles of inflationary cosmology. For each proposal, I choose one main philosophical question to discuss in depth. They are, respectively: what is a possible world; what is chance; and what is explanation. My treatment of these proposals and their associated questions is in my main central Chapters: 3, 4 and 5. Before that, Chapter 2 will set the stage by reviewing physics and philosophy from about 1600 to about 1900; and the final Chapter 6 will compare and contrast the proposals. The rest of Chapter 1 is a statement of my philosophical method and temperament, in addressing these proposals. I also admit in advance that I do not outright believe these proposals. But I argue that this is a merely autobiographical fact, and that the reader must of course make their own assessment of them.

This Chapter reviews the development of physics, and of philosophy, and their increasing separation from each other, from the mid-seventeenth century onwards. I stress three themes. The first is how counter-intuitive, and opposed to the preceding mechanical philosophy, Newtonian mechanics is: especially as regards action-at-a-distance. My second theme is the effect on philosophy of the success of Newtonian mechanics: especially in prompting Hume's modest conceptions of inductive inference and of empirical knowledge---which I endorse. My third theme is that within philosophy, logic was downplayed from the seventeenth century to the mid-nineteenth century: after which it revived, in part due to the crisis in the foundations of mathematics. Both the revival and the crisis were crucial in placing logic centre-stage in twentieth-century philosophy---and so in preparing us for Chapter 3's philosophical multiverse.

The philosophical multiverse is, in short, the set of all the logically possible worlds, i.e. all the ways the universe could be. My discussion has four stages. First, I review logic's role in twentieth-century philosophy. Second, I stress that our thought and language, both everyday and scientific, constantly deals with possibilities, i.e. ways that some fragment of reality could be other than the way it actually is. Thus I emphasise logic's treatment of modality, i.e. of necessity and possibility. Thirdly, I report how postulating a framework of possible worlds, i.e. maximally specific possibilities, has the benefit of providing clear and persuasive accounts of various notions that are important for philosophy. For example: laws of nature, counterfactual conditionals, supervenience and determinism. Fourthly, I discuss the philosophical question: what exactly is a possible world? My discussion is inconclusive. I consider four answers. I criticize the first three as wrong; and although the fourth---David Lewis' famous modal realism---is coherent, I admit that I cannot believe it.

The first of my two multiverse proposals from physics is the Everettian interpretation of quantum theory. So I first sketch the ideas of quantum theory. I emphasise how different quantum states are from states in classical physics. Indeed, they are problematic: for they are assignments of (square roots of) probabilities to each of a class of classical states; and these probabilities are understood in an instrumentalist way, as probabilities for measurement results. Hence we arrive at the measurement problem, epitomised by Schroedinger's cat. In short, the problem is that the quantum state is indefinite for (in the jargon: is a superposition for) the quantities, such as the position of macroscopic objects, that appear definite. I briefly list some main responses to it, but expound in detail only the Everettian view. The Everettian's key idea is that the quantum state is a sum of states that are definite for such quantities---and we should take all these states as physically real, i.e. actual. This implies that in the Schroedinger's cat example, there are indeed two cats, one alive and one dead. Nowadays, Everettians defend this proposal by invoking the rapid and ubiquitous process of decoherence. The main idea is that decoherence prevents one from seeing the other cat, in the other world. Finally, I discuss the philosophical question how we should understand probability in the context of the Everettian interpretation.

**Chapter 5: All the worlds from the primordial bubbles** 121

The second of my two multiverse proposals from physics is the bubbles, or domains, that inflationary cosmology postulates in the very early universe. Since inflationary cosmology of course uses quantum theory, my first job is to relate this proposal to the previous Chapter's Everettian multiverse. (I also relate it to the landscape of string theory: though this landscape is a multiverse proposal which I consider only for cosmology.) I then describe how in about 1980 cosmology, despite its great successes, faced some pressing why-questions about the values of some of its parameters: and how postulating a very short and very early period of accelerating expansion (called 'inflation') promised to answer these questions. This leads in to my chosen philosophical topic: explanation. About this topic, I distinguish between two main strategies: (i) explaining the value of a parameter by showing the value to be generic or typical, and (ii) explaining it as likely to be observed, even though not generic or typical. The second strategy leads to discussions of: (a) selection effects and the anthropic principle; and (b) how we might confirm a cosmological multiverse.

In this concluding Chapter, I gather the threads in various ways. First, I say what I myself believe about the three proposals. Then I address the question, for each proposal, why we do not experience the other worlds; which of course bears on the question whether we could have evidence for them. Then I turn to the question whether one of the multiverse proposals could encompass one or both of the other two. I focus on the proposal that the Everettian multiverse encompasses the other two, especially the philosophical multiverse---as advocated by Alastair Wilson). This proposal has advantages---but I do not endorse it. Finally, I end with some closing quotations.

# Chapter 1: Introduction

I write these words, believing that I have always lived in one world: and that it is the same world in which my family, friends, historical figures, all humankind, including you, dear reader, live or have lived or will live. But our topic is the speculation, made by many authors past and present, that this world is by no means all there is: that it is one of a vast collection of worlds---dubbed a <u>multiverse</u>.

Here and throughout this book, I will use the word 'world', not for a planet, but for a cosmos, a universe, extended throughout all of space and all of time. So the actual world---as I will use that phrase---contains all objects and events that are at some distance, no matter how great, from us here on Earth. And it contains all objects and events that are in the distant past or distant future, as well as those that are now. As one might put it: the actual world contains all objects and events at any 'temporal distance', no matter how great, from us now, just as it contains all objects and events at any spatial distance.

So our topic is the idea that the actual world, though very inclusive, is not all of reality. Indeed, it is only a tiny part of reality. Let me say this a bit more precisely. The idea is that there is a vast collection of worlds, i.e. universes, differing in myriad ways one from another. The actual world---the universe, as we usually use that word---is just one member, one element, of this vast collection. And it is no more real than all the other members. Agreed: it seems "especially real" to us. But, so the idea goes, that is only because we are in it, rather than in another world.

This is supposed to be rather like how in the actual world, any place---such as Westminster Abbey, London, or the Sydney Opera House---can seem especially real to a person at it. But such a person will readily agree that other places are equally real. They are just spatially distant: and therefore they are usually hard to know about since, for example, they are not visible. Similarly, the idea is that the proposed other worlds are just as real as the actual world: though they are in general harder to know about than the actual world. This vast collection, or multitude, of universes is <u>the multiverse</u>. (Some people say 'pluriverse', but I will always say 'multiverse'.)

The multiverse is a timely topic. For in the last thirty years, 'multiverse' has become a buzz-word in both physics and philosophy. In both disciplines, it has been proposed that our universe is just a tiny part of a multiverse.

Of course, what is meant by a multiverse, and therefore the reasons given for it, differ between the disciplines. We can glimpse this variety already in my introductory words above: when I said that according to proponents of a multiverse, the actual world, comprising everything at some spatial or temporal distance from us-here-now, is just one of a vast collection. For we shall see that on some multiverse proposals, the objects in the other worlds are indeed related in space and time to us-here-now. It is just that these relations are very different from the distances across space and time that we are familiar with (even very large ones).

Besides, even <u>within</u> a discipline, authors differ about their reasons for believing in a multiverse. Broadly speaking, these differences are as one might expect. Physicists who propose a multiverse tend to take their reasons for it to be empirical. Here, 'empirical' does not just mean 'derived from immediate experience'. It also includes data from experiments, maybe very advanced or delicate ones. Thus physicists tend to argue for a multiverse on the grounds that postulating it explains---or explains better than rival suggestions do---some significant physical facts, that would be otherwise puzzling, even mysterious. But in propounding these arguments, physicists tend to down-play the ways in which conceptual, i.e. <u>non</u>-empirical, considerations can also contribute to the explanation.

On the other hand, philosophers who propose a multiverse tend to take their reasons for it to be conceptual, i.e. non-empirical, even with 'empirical' understood in a wide or liberal sense that includes data from arcane experiments. Thus philosophers proposing a multiverse tend to argue that it provides the best account of some problematic concept; not that it explains, or explains better than rival ideas, some empirical evidence. (As we shall see in the next Section, the main problematic concept at issue is that of possibility.)

So philosophers tend not to consider whether any empirical considerations---in particular, physicists' reasons for <u>their</u> notion of a multiverse---bear on the account of the concepts, like possibility, that the philosophers focus on.

But in fact, the various proposals share important common themes. Besides, the reasons for (and against) each proposal combine empirical and conceptual considerations. So the topic calls for an interdisciplinary treatment. Thus my aim will be to assess these proposals, by comparing them with each other and by articulating common themes. Among these themes, there will be major open philosophical problems.

Chapter 1 Section 1: The plan: three multiverse proposals

My plan will be to discuss, in order, three different multiverse proposals. There will be one from philosophy, and two from physics. The philosophical proposal is about <u>logically possible worlds</u>. The first physical proposal is about the <u>many worlds</u> of <u>the Everettian interpretation</u> of quantum mechanics; and the second is about the <u>bubble universes</u> proposed by inflationary cosmology and string theory. For each proposal, I will explain it, and the reasons why people advocate it.

As one would expect, each proposal comes in various versions that have, in relation to each other, various advantages and disadvantages: in other words, features that are agreed, by at least some parties to the debate, to make the version in question, respectively, more plausible or less plausible. To keep things simple (as befits an Introduction), I will by and large not try to formulate different versions, but instead focus on one version that is mainstream: at least in the uncontentious sense of having got plenty of attention, for and against, in the literature. But I should admit at the outset that each of these mainstream versions has been met in some quarters with incredulity, and even indignation. As we will see, here physics joins philosophy in being very controversial.

And as befits a philosophical Introduction, I will for each proposal, emphasize a philosophical question that it raises, which then moulds my discussion of the later proposals. Each question will be a major philosophical problem---a problem which is unsolved: indeed, a problem about which I of course hope this book will prompt further effort.

So there will be six Chapters (including this one): as follows. The three main Chapters, discussing my three multiverse proposals, are Chapters 3, 4 and 5. Chapter 3 is about the philosophical multiverse, the logically possible worlds. Chapter 4 is about Everettian quantum mechanics; and Chapter 5 is about the cosmological multiverse. But before these central Chapters, I need to do some stage-setting. In this Chapter, I will describe how one assesses such proposals---or at least, how I propose to assess them. This will largely be a matter of being wary of pitfalls, and being self-conscious about one's assumptions. In Chapter 2, I will review those aspects of physics and philosophy from 1600 to 1900 that we will need, in order to understand how in both physics and philosophy, the ground was fertile, by about 1970, for multiverse proposals. Then follow the three central Chapters. In the final Chapter, Chapter 6, I review the relations between the multiverse proposals, and conclude.

As a result of this plan, Chapters 1, 2 and especially 3 are more about philosophy, while Chapters 4 and 5 are more about physics. To a large extent, all these five Chapters can be read independently, so that a reader can follow their interests. The main exception is that Chapter 5's discussion of the cosmological multiverse will make a slight appeal to Chapter 4's explanation of Everettian quantum mechanics.

To give a glimpse of what follows, here is some more detail about the three proposals, and the three questions they raise.

The first proposal I consider is from philosophy (Chapter 3). It says that all the logically possible worlds---all the myriadly many ways that the universe could be, without contradiction---are equally real. We 'just happen' to be in one of these worlds. (As I announced in this Chapter's preamble: here and throughout, 'world' means, not a planet, but a cosmos or universe, throughout all space and time.) The rationale for this proposal is that it gives the best account of a concept that is problematic but apparently indispensable: namely the concept of possibility. For example, an everyday statement like 'it might have rained today' (said on a dry day) is surely about an alternative way the world could be. So to make sense of such statements being true, we surely need to countenance, i.e. accept as existing, the alternative ways the world could be: other <u>possible worlds</u>.

Besides, there are several other concepts that are closely related to that of possibility (and equally problematic but apparently indispensable), which can be readily understood in terms of possible worlds. So much so that the arch-advocate of possible worlds, David Lewis, dubbed the possible worlds 'a philosopher's paradise'. (Although Lewis' name is unknown to the general public, unlike e.g. Bertrand Russell and Ludwig Wittgenstein, he is generally agreed by philosophers to be one of the greatest philosophers of the twentieth century. He died in 2001, at the age of 60.) Details of this paradise are in Chapter 3. For now, I just note that one such related concept is that of a statement or proposition, i.e. the meaning of a sentence.

I call Lewis 'the arch-advocate' although the idea of a set of all the possible worlds of course goes back much further than the recent decades of philosophy (at least to Leibniz). I do so for two reasons. First, he defended his own version of the proposal with great clarity, imagination and resourcefulness.

Second, his own version is, despite his fine defence, almost universally rejected, i.e. not believed by other philosophers. For Lewis proposes that the other worlds are, in their nature, just like the actual world. Thus we all believe that the actual world consists (at least partly) of material objects that are, in the philosophical jargon, <u>concrete</u> rather than <u>abstract</u>: like a table, a rock, a molecule or an animal, rather than, say, a number or a proposition or an idea. Lewis argues that the other worlds are just as concrete as the actual world; they also consist (at least partly) of concrete material objects. This is a doctrine that almost all other philosophers, even those who endorse using possible worlds to address philosophical problems, find impossible to believe. They say: surely every non-actual possible world is in some way abstract, rather like a number or a proposition? So in Chapter 3, I will report this debate between Lewis (whose version is called 'modal realism') and other "more abstract" versions of the philosopher's paradise.

This yields the philosophical question associated to this proposed multiverse. Though philosophy is of course a subject in which any question leads rapidly to several others, I choose to emphasize the obvious one: what exactly are these different 'ways' that the universe could be? That is: what exactly is a <u>possible world</u>, or a <u>possibility</u>?

In fact, we will see at the end of Chapter 3 that the concrete/abstract distinction is not in good order. For it can be made precise in several different ways that cut across one another. This undermines the traditional idea that numbers, triangles and the other entities mentioned in mathematics, are abstract. And as a result, some have advocated the 'Pythagorean" view that all of reality---the apparently concrete objects like a table, no less than the apparently abstract ones like a number---is mathematical. This leads to <u>another</u> multiverse proposal. For since there are, presumably, vastly many possible mathematical structures, reality being wholly mathematical would make for a <u>mathematical multiverse</u>. So at the end of Chapter 3, I will also discuss this proposal: albeit more briefly, as an epilogue to my assessment of the multiverse of possible worlds.

I turn to the two proposals from physics. They are from quantum theory and from cosmology.

Quantum theory is famous---one might say: notorious---for the conundrums about how to interpret it. Above all, there is the measurement problem. It arises from the fact that quantum theory attributes to the objects it successfully describes, e.g. atoms, a lack of definite properties; or in other words, indefinite properties. For example, an object such as a particle that one expects to have a definite position gets attributed no definite position. (The buzz-word here will be 'superposition'.) One might accept this indefiniteness of properties for the unfamiliar and unvisualizable realm of microscopic objects like atoms. But the problem is that this indefiniteness of properties can be transmitted from the atomic realm to the everyday macroscopic realm---where such indefiniteness seems ludicrous. This is the measurement problem. It is summarized iconically by Schroedinger's cat. In a famous paper in 1935, Schroedinger described how in principle, an indefiniteness about a radioactive atom (viz. about whether it has decayed, or not) could be transmitted by a quantum measurement apparatus, so as to render a cat indefinite as to the property of being alive. That is: at the end of the process that Schroedinger describes, the cat is neither alive nor dead. And this indefiniteness is not due to the vagueness of our words 'alive' and 'dead', but concerns the cat in itself. It is somehow "in limbo".

The second multiverse proposal (Chapter 4) is a proposal for how best to interpret quantum theory, and especially how to solve the measurement problem. It was first suggested in 1957 by Hugh Everett. (Like the philosophers' proposal, there are various versions in the subsequent literature; I will focus on one mainstream version developed in the last thirty years.) So it is sometimes called the 'Everettian interpretation' of quantum theory. But it is also often called the 'many worlds interpretation'. For the idea is that there is a quantum state of the multiverse as a whole; and this state encodes myriadly many different macroscopic <u>worlds</u> (sometimes called <u>branches</u> or <u>realms</u>), including the various macroscopic worlds that include (one each of) the various possible results of a quantum measurement process. So in particular, for Schroedinger's thought-experiment, the quantum state at the end of the process encodes (at least) two macroscopic worlds. There is one, i.e. at least one, with a cat that is alive; and also at least one with a cat that is dead. Besides, and here is the punch-line: all these macroscopic worlds are equally real.

As we will see, this proposal links back to the philosophers' multiverse proposal; and it raises many conceptual, indeed philosophical, questions. But I will emphasize one question: namely, what is <u>chance</u>? Here, 'chance' means <u>objective probability</u>, i.e. a probability that is made true by the subject-matter concerned. Nowadays, a standard example is radioactivity (quite apart from Schroedinger's mention of it in his thought-experiment). For example, there is the chance of this Uranium atom decaying in the next hour; and it has a particular numerical value. So chance is contrasted with <u>subjective probability</u>, which are degrees of belief about a subject-matter, e.g. my degree of belief that this horse will win the race. Such degrees of belief are made true by my state, not the horse's. (For they are shown in my behaviour, for example by what odds I would be willing to accept in a bet on the race.) As we will see, the nature of chance is a central aspect of the statement, and the assessment, of the many worlds interpretation.

The third proposal (Chapter 5) is from cosmology. The last sixty years have been a golden age for the science of cosmology. Our understanding of the structure and evolution of the universe has grown immeasurably. So it is now an established fact that the universe we see, and see to be expanding, originated some 13.8 billion years ago in a very hot dense fireball---which itself originated, perhaps, in a singularity of infinite density, dubbed 'the Big Bang'. However, since the early 1980s, cosmologists have also speculated (prompted by good empirical reasons) that very early on, there was a very brief period of rapidly accelerating expansion, called 'inflation'.

It is this conjectured inflationary epoch that leads to the multiverse proposal. For the conjectured mechanism driving inflation also yields countless bubble, or pocket, domains (so called) that branch off from all the material that then expanded to become the observable universe which we now see. And each of these domains would itself expand and become a universe. So each of the countless bubbles (pockets, domains) is a universe, and the whole collection is a multiverse.

As we will see in Chapter 5, there are two main sources of this multiverse proposal. The first source, in the early 1980s, was what came to be called 'fine-tuning'. For the cosmological theory developed in the 1980s to be adequate, the value of certain physical quantities (such as a measure of how dense is the matter in the universe) had to be "just-so". That is, their exact numerical value was theoretically constrained to many decimal places. For example, the quantity about matter-density was constrained to sixteen decimal places: for the theory to be adequate, its value could differ from 1 by at most $10^{-16}$. This is a minuscule difference, i.e. a very tight constraint. For $10^{-16}$ is approximately the ratio between the width of a human hair (viz. a tenth of a millimeter) and the average distance between Earth and Mars (viz. 225 million kilometres)! In the early 1980s, cosmologists realized that this glaring conundrum---why should there be such fine-tuning?---could be resolved if there was, very soon after the Big Bang, a very brief period of rapidly accelerating expansion. Roughly speaking, such a period made these quantities' values generic rather than exquisitely fine-tuned. But this speculation, for all its merits as an explanation of the quantities' values, prompts the question: what could be the mechanism of this brief accelerating expansion?

Here enters the second source. Namely, string theory: this is a speculative physical theory (developed from the mid-1980s onwards) that aims to reconcile quantum theory with Einstein's general relativity, which is our best theory of gravity. It suggests not only a mechanism, but also one that generates countless bubble (pocket, domain) universes. Besides, these bubble universes will in general differ from one another as regards the value of physical quantities, such as the speed of light or the electric charge on an electron, that we normally call 'constants of nature', since we find them to be constant across the whole observable universe. So the cosmological multiverse, as elaborated using string theory, envisages a variety in the values of these so-called constants.

Again, this proposal (of which, again, there are many versions) links back to the previous ones, and raises many questions, including philosophical ones. Of these, I will emphasize the question that is most directly suggested by the conundrum of exquisite fine-tuning: what counts as an explanation?

So that is the overall plan of the book. In short: there will be three multiverse proposals, and three associated questions. But throughout the book, I will also discuss how various themes which the book downplays---such as (i) the proposal for a Pythagorean or mathematical multiverse, and (ii) the philosophical justification of induction---fit in.

Chapter 1 Section 2: What do I believe?
So much by way of a prospectus. You will want to know—maybe so as to gauge my sanity, before deciding whether to read what follows---where I stand about these speculations.

I began this Chapter by saying I believe that I have always lived in one world. Of course, that could be so, even while there are many other worlds. It could even be so, while there are many other worlds and also, we can have some knowledge of them; or at least, we have some warranted beliefs about them. So you will want to press the question: do I believe there are such worlds?

To cut a long story very short: my answer is 'Yes, No and Maybe'. (Needless to say, my reasons, about all three proposals, will be less than conclusive.) That is: I believe in the philosophical multiverse, though not in Lewis' modal realist version with its "concrete" worlds.

But I do not believe in quantum theory's Everettian multiverse. And for the cosmological multiverse, I say, as the film producer Sam Goldwyn is meant to have done: 'a definite Maybe'.

In Chapter 6, I will discuss these verdicts in more detail, in the light of the evidence and arguments we will by then have in hand. But for the moment, I want to emphasize two points. The first is about what I mean by 'belief'. The second is about what my beliefs imply for you, the reader. This will take a longer discussion, which is in the next Section.

First: here and throughout the book, I use the word 'believe' in an everyday sense: a belief is a conviction, on which I am willing to bet a great deal, even my life. We all make these kinds of bet all the time. I believe the plane will fly safely, so I get on it without worrying. Agreed, some people have their doubts about planes. But the same point is made, even more vividly, by even more humdrum examples. As I walk across the room, I believe the floor will continue to support me; as I eat the bread, I believe it will not poison me. And so on. Agreed: we all have, for many propositions, degrees of belief that fall short of conviction. Recall the example above, of my degree of belief (subjective probability) that this horse will win the race. But for most of this book, we can set subjective probabilities aside, and so take belief to involve a subjective probability so close to 1 (100%) that the difference is negligible. Hence my word above, 'conviction'.

So I intend my beliefs as reported above, 'Yes, No and Maybe', in this everyday sense. But stated so briefly, they are also "merely autobiographical". They just report that after surveying the evidence and arguments as well as I can, I cannot believe this, while I could believe that.

Chapter 1 Section 3: What should you believe?
This leads to the second point. Being "merely autobiographical", these verdicts should have little weight with you, the reader. While I am happy to tell you straight-up what I believe, you should, and of course will, make up your own mind. You may well reach more positive conclusions about these multiverse proposals than I have. For as we will see: much deeper thinkers than I have believed in a multiverse, and made an extended case for the multiverse they believe in. I only hope that my survey of the evidence and arguments is open-minded and clear-headed enough to give you good material for reaching your own conclusions.

The reason you and I may differ is that in the current state of knowledge, and combining the insights of both physics and philosophy, it is impossible to now know for sure about any of the three multiverse proposals. Thus I do not give my Yes and No, respectively, to the first two multiverse proposals, on the basis of some evidence or argument that I take to be irrefutable, or 'knock-down'. Indeed, I doubt that we could get such irrefutable evidences or arguments, either for or against these proposals.

Within philosophy, this situation---of admitting that while one finds some evidence or argument cogent, and even persuasive, it is certainly not conclusive or irrefutable---is of course familiar. It is also to be expected. For philosophy is by its nature controversial. Since the problems it addresses are abstract and general, it is hard to pinpoint what evidence, or considerations, would definitively solve them. (Or if one prefers to think of philosophy as asking questions: it is hard to pinpoint what evidence, or considerations, would provide a definitive answer.)

Of course, to say that philosophy's problems are abstract and general makes it sound like mathematics, or perhaps physics. But there is a difference. Philosophy's problems are also about concepts that are either not completely precise, and-or are contested, i.e. rejected as bad concepts by some people. (Here, 'bad' means, roughly speaking, 'useless and even misleading', e.g. because the concept has a mistaken presupposition.) This is obvious for concepts that are the focus of moral and political philosophy: concepts like freedom, responsibility, equality, justice, class, and just war. For example, someone might reject the concept of political equality, in the sense that they maintain its only role is to mislead: namely, the ruling elite uses it as a slogan to deceive the ruled that they have a say in how politics and government are run.

But as we will discuss, imprecision and being contested are similarly features of concepts that are a focus of philosophy of science, and so of this book---even within its discussion of the multiverse proposals from the less controversial discipline of physics.

For example, think of the concepts in the three questions that I listed as being raised by the three multiverse proposals: namely, possibility, chance and explanation. Each of these concepts is not precise; and even when it is made precise, philosophers disagree over claims involving it; and some philosophers reject the concept (even once made precise) as bad. Besides, we shall see that these concepts are linked to several other such concepts: that is, concepts that are apparently valuable for making philosophical sense of science, but are not completely precise, and-or are contested. Two examples are the concept of a law of nature, and the concept of causation: to both of which we will return.

So for any problem whose formulation uses an imprecise and perhaps even contested concept, it is inevitably controversial how we should assess proposed solutions to the problem. For there will be no prior agreement on the kinds of evidence, either empirical or conceptual, to which a solution should be answerable. And as I said: this is as expected---all in a day's work---within philosophy.

Agreed: physics is, by and large, far less controversial than philosophy. Its concepts are more precise, and less contested; and so its problems are better defined. And so also are the kinds of evidence to which a proposed solution is answerable.

But obviously, for physics' two multiverse proposals, these contrasts with philosophy fall away. The Everettian multiverse is one proposal of several in the debate about the best way to interpret quantum theory. That debate still rages, with a mixture of empirical and conceptual reasons for and against the various proposals. So assessing the Everettian multiverse is as controversial as most of philosophy.

The cosmological multiverse is also controversial, though for rather different reasons. It is not in the first instance an interpretative proposal, in the way that the Everettian multiverse is. So one would expect the empirical evidence, or indeed conceptual considerations, that would count for or against the proposal to be easier to state, i.e. to be better defined---easier for different people to agree on. But any relevant empirical evidence is fearfully hard to get: as of course one would expect, since the proposal is precisely that there are universes other than---beyond---all that we can directly observe. And it is also hard to agree on the relevant conceptual considerations that would weigh one way or the other. For as we will see, the proposal raises interpretative, and therefore controversial, issues; even though it is not primarily an interpretative proposal. One main way that it raises such issues is that it turns out to be closely related to the debate about the best way to interpret quantum theory. (Hence it is clearest to postpone discussing it until after the Everettian multiverse.)

So in short: my personal assessments of the three proposals, 'Yes, No and Maybe', will be tentative. I do not urge them as definitive. For we must recognize that different people are very likely to disagree about how to weigh the various pieces of evidence and lines of argument; not least because they may disagree about the usefulness of the concepts in which the evidence or argument is formulated.

Besides, it seems too much to hope that all such disagreements could be resolved in principle, by arranging some resolutely open-minded exchange of opinions that was allowed to last "as long as it takes". If in order to resolve such a disagreement, you were to lock "the jury" in a room with coffee and refreshments, and let them out only when they agree---argh, the jury might never come out. Agreed, some such disagreements might be thus resolved. For example, a conflicting assessment of some specific line of argument might be shown to turn on people using different versions of some controversial concept such as explanation. But I doubt that all can be thus resolved.

But this is not a bland or indifferent agnosticism. I agree that it would be good---indeed, wonderful---to <u>know</u> whether any of the three proposals is true. Or if knowledge is not to be had: at least to have conclusive reasons for belief, one way or the other. But we must recognize that enquiry about these proposals is both inconclusive and fallible; (as it is also, no doubt, for many topics). For these multiverse proposals, the best we can do is to marshal the available evidence, both empirical and conceptual, and to try to be open-minded and clear-headed in assessing the proposals.

Chapter 1 Section 4: What would you risk? Confidence vs. caution

I have just delivered a right-minded sermon about what one might call 'cognitive modesty'. The summary is: we should accept that our views are tentative, not conclusive; that even the concepts with which we formulate our views may be contested by other people; and that maybe some of our disagreements with them could not be resolved by an open-minded exchange of opinions, no matter how long we allowed the exchange to go on.

But there is another aspect of cognitive modesty that I should also register---and extoll. It arises from the fact that people differ in their attitude to risk, i.e. in how willing they are to take a risk. We will see that these attitudes influence people's views, especially about the topics of this book.

We are all familiar with the fact that what is an unacceptably large risk to one person can be a tolerably low, or even negligible, risk to another. We are also all familiar with the fact that, although to some extent one can urge reasons on someone to change their attitude to risk: beyond a certain point, such attitudes are a matter of basic temperament, and one cannot expect reasons to change the person's attitude.

The same goes, say I, for enquiry. People differ in their attitude to risk in enquiry, just as much as in action. It is just that in enquiry, the risk is of error, of false belief, rather than of some traumatic event. (Of course, false belief can engender traumatic events.) So each of us, when pursuing abstract and general questions, that cannot be easily settled by some well-defined body of evidence, takes a stance about how tolerant, or how averse, we are to ending up with a false belief. (Of course, the falsity may seem harmless as regards our personal safety and well-being, just because the topic of the belief lies so far from practical matters.)

Of course, this stance is almost never a matter of a decision being made consciously. Did anyone, even a philosopher, ever say to themselves: 'I hereby decide that I am too cautious, too averse to having a false belief, to either endorse or reject this philosophical proposition (about, say, a multiverse proposal)---I must remain entirely agnostic'? I doubt it. Nevertheless, each of us, when we engage with philosophical debates, in particular the debates in this book, thereby endorses or rejects, or at least assesses, various philosophical propositions. And we thereby adopt some stance, in the spectrum from tolerance to aversion, about the risk of false belief.

I believe that like the everyday examples of attitude to risks about actions, rather than about beliefs, this stance is ultimately a matter of temperament. And this is even so, for beliefs whose topic is far-removed from practical matters; such as philosophical beliefs.

Like the everyday examples, your stance can be changed, to some extent, by reasons. Other people engaged in the same philosophical debate, i.e. assessing the same philosophical propositions as those you are focussed on, can offer you reasons to change your stance. Thus they might say to you: 'You should be more willing to endorse this proposition about the multiverse, because your background philosophical beliefs about possibility (or about explanation, or what-not) make it more plausible.' Or they might say: 'your background beliefs are such that, even if it is false, this would spell little damage to---force only a minor revision of--- your other philosophical beliefs.' (These examples show that the reasons urged for being less risk-averse can concern either a specific proposition, or the coherence of the pattern of one's beliefs. Similarly of course for reasons for being more risk-averse.)

But I maintain that beyond a certain point, such reasons cannot persuade. Your stance cannot be wholly determined by discursive reasoning, i.e. by reasons that can be put in a discourse of words and arguments. It is ultimately a matter of what I would call 'intellectual temperament'. So when the words give out in this way, the most that can be reasonably asked of you is that you should be self-conscious about this matter of temperament. And *a fortiori*, you should not be dogmatic about it: you should not proclaim that it is the only stance that is defensible, or rational.

Because a person's attitude to the risk of false belief, as moulded both by reasons and by their individual temperament, will be a factor in---will play a role in---the position they take in various debates throughout this book, it will be convenient for us to have a label for the range of attitudes. That is: I recommend adopting a label for this spectrum of risk-tolerance through to risk-aversion. So I will say: 'confident' vs. 'cautious'. (Another possible label is: 'ambitious' vs. 'modest'.)

Besides, this discussion of one's attitude to the risk of false belief---confidence vs. caution---can be generalized. Hitherto, I discussed the topic simply in terms of whether a belief is true or false, without distinguishing whether the proposition believed is: (i) "mildly" or "merely" false in that, though it is false, all the concepts it involves are correct, or at least are concepts that the agent herself does not reject or contest; or (ii) "more sickly" false, or wrong-headed, in that some of the concepts it involves are rejected or contested, at least by the agent herself.

But we must allow for (ii). That is: we must allow that someone might be cautious about using a concept, whatever claim is then made using it; (and so they will be tempted to reject it). So one can be cautious about a concept, as well as about a claim or proposition. In subsequent Chapters, we will see several examples of philosophers and physicists (including myself, and maybe you the reader) being cautious about, or definitely rejecting, some concept or other.

For the moment, let me give an example of this distinction between (i) and (ii), by considering the broad enterprise of "making sense" of physical science. Think of how physicists go about their business. They invent general theories; they specialize them in various ways with models and approximations; and they do experiments, to help improve the theories, the models and the approximations. Now let us ask: does making sense of this overall enterprise *need* the concept of a *law of nature*? Of course, 'law of nature' is vague, and different advocates will make it precise in different ways. But the main idea is that a law of nature is an especially informative proposition about how the natural world "works": a proposition that is true, but which can be unknown, even un-formulated, by us humans. Some philosophers accept this concept (making it precise in one way or another). And some even say that it is a central goal of physics, or of all of science, to discover laws of nature.

But the point here is: a person might reject the very concept of a law of nature. That is: a person might reject the idea of a true and especially informative proposition about nature; especially such a proposition that is not yet formulated, but nevertheless said to be the goal of enquiry. They might say it is an illusion, a will of the wisp. So according to this view, we can, and should, make sense of the overall enterprise of physics---the theorizing, modelling, approximating and experimenting---without ever invoking the idea of a law of nature, in any precise version. Using my jargon of 'confident' vs. 'cautious': such a person, such a view, is cautious. (Chapter 3 will return to this example.)

Again, I should come clean about my own attitudes, my own position in the spectrum from confidence to caution. In later Chapters, I will give details in the context of each discussion. But to try and be honest and clear-headed about my intellectual temperament: let me say in advance that broadly speaking, about the dozen or so contested (usually philosophical) concepts that arise in multiverse proposals, I am inclined to be:

(i): confident, i.e. accepting, of concepts that are proposed within physics, or in logic or in metaphysics; examples of such concepts include: the quantum state of the universe, logical necessity, possible world, supervenience; but:

(ii): cautious, i.e. rejecting, about concepts proposed within epistemology and methodology; examples include: the idea of a law of nature, the idea of explanation.

I should also say in advance that for several philosophical issues, my views are close to David Hume's. This will be most evident in the stage-setting Chapter 2, especially its Sections 4 and 5; and, to a lesser extent, in Chapter 3. But I think my assessments of the Everettian and cosmological multiverses, in Chapters 4 and 5, are largely independent of my Humean sympathies.

So as the book proceeds, it may be useful to you, the reader, to know that these are my tendencies. But again, this report of my intellectual temperament is "merely autobiographical". So do not let them have undue weight. As I said above: each of us must, in the end, decide their position for themselves.

Chapter 1 Section 5: Beware the beguiling power of words
I have just followed my sermon about 'cognitive modesty' with an admission of the role of intellectual temperament, and a confession of my own temperament. I turn to giving a warning about how confusing words can be.

The warning is this. Once one has a word to use, one readily falls in to thinking that it represents a concept in good order: that one understands, or can explain, something---though often one doesn't understand, and cannot explain anything. This warning is of course of a piece with my previous point that a person may reject a concept as bad, because misleading: recall the example of rejecting the concept of political equality, on the grounds that it is only the elite's tool for duping those they rule.

This is a time-honoured warning. Sometimes, it is expressed as a joke. In Moliere's play The Hypochondriac, the target of the joke is doctors who give a learned label, suggestive of understanding, to something they do not understand at all. When asked to explain why opium induces sleep, they answer in a learned tone of voice---as if they knew something---that opium has a 'dormitive virtue'. (Here, derived from Latin: 'virtue' means 'causal power', so that 'dormitive virtue' means 'tendency to induce sleep', and the doctors' answer merely repeats the question.)

This warning also occurs in some great philosophical texts. Since the next Chapter will discuss the natural philosophers, i.e. philosophers-cum-physicists, of the seventeenth century, let us enjoy the prose of one such author, John Locke, in a famous passage.

Locke, in the `Epistle to the Reader' at the start of his An Essay Concerning Human Understanding (1690) praises the contemporary great physicists (as we would now call them), Huygens and Newton; for whom Locke sees himself as an under-labourer, who can help by doing what one might call 'conceptual house-keeping'---and in particular by seeing through beguiling words. Thus he writes:

'The commonwealth of learning is not at this time without master-builders, whose mighty designs, in advancing the sciences, will leave lasting monuments to the admiration of posterity: but every one must not hope to be a Boyle or a Sydenham; and in an age that produces such masters as the great Huygenius and the incomparable Mr. Newton, with some others of that strain, it is ambition enough to be employed as an under-labourer in clearing the ground a little, and removing some of the rubbish that lies in the way to knowledge; which certainly had been very much more advanced in the world, if the endeavours of ingenious and industrious men had not been much cumbered with the learned but frivolous use of uncouth, affected, or unintelligible terms, introduced into the sciences, and there made an art of ... Vague and insignificant forms of speech, and abuse of language, have so long passed for mysteries of science ... that it will not be easy to persuade either those who speak or those who hear them, that they are but the covers of ignorance, and hindrance of true knowledge. ... Few are apt to think they are deceived in the use of words; or that the language of the sect they are of has any faults in it ...'

By the way: similar sentiments, also famous, can be found in Francis Bacon, who warns against the danger of being misled by what he calls the 'idols of the market-place': i.e. false ideas engendered by human communication and abuse of language. He also warns against three other idols, i.e. sources of false ideas. Roughly speaking, they are: (i) universal human tendencies, such as relying uncritically on perception, and jumping to conclusions (called 'idols of the tribe', where 'tribe' means humankind); (ii) idiosyncratic or communal prejudices and other deficiencies of judgment (called 'idols of the den', where 'den' refers to a benighted community, as in the metaphor of the cave in Plato's <u>Republic</u>); (iii) being misled by abstract, general and high-falutin' theories (called 'idols of the theatre', where 'theatre' connotes a fantastical representation).

In short: we have been warned …

Chapter 1 Section 6: Can we be sure that we are in the same universe?
Finally, let me broach a teasing question: if there is a multiverse, how can we be sure that we are in the same universe? In particular, how can I, as I write this book, be sure that you the reader are in the same universe as me?

Of course, this question is more pressing for advocates of a multiverse, than for agnostics. But it needs to be addressed. For if the answer is 'in a multiverse, the reader of my book might be in a different universe from me', then writing a book becomes a curious enterprise: especially if most (the vast majority?) of the readers are not in the author's universe. And the worry is of course not just about writing and reading, or speaking and hearing, or communication in general. It seems that in a multiverse, there could be all sorts of "disconcerting" causal processes from one universe to another: for example, with objects disappearing from one universe and appearing in another. (Of course, science fiction makes great play with these ideas; with the channels between universes called 'portals' and 'wormholes.)

But rest assured. We will see (especially in Chapter 6.2) that on the three multiverse proposals I will consider, causal processes between objects, or events or states of affairs in different universes, are either downright impossible or very rare and arcane---because they require very special circumstances.

So the overall situation about whether you and I, as reader and writer, are in the same universe is as follows. Agreed: any or all of these multiverse proposals may be very weird, and-or very hard to believe, and-or plain false. But even if such a proposal is true: at least there is nothing problematic about an advocate of such a proposal believing that you, dear reader, are in the same universe as them (and as me). And accordingly, they write their books . . .

Chapter 1: Notes and Further Reading
The subsequent Chapters give more detailed suggestions for further reading. Here, I will do just four things. (1): I advertise some internet resources. (2): I list three other books about the idea of a multiverse in general. (3): I list two academic but introductory books about topics at the interface of philosophy and physics, <u>other than</u> the multiverse. (4): I add a little to the "sermon" about philosophical method that I gave in Sections 3, 4 and 5.

(1): For all of philosophy, the Stanford Encyclopedia of Philosophy is an excellent resource; with many excellent entries in philosophy of science, and philosophy of physics in particular. It is Open Access.

Also excellent is the Routledge Encyclopedia of Philosophy (requiring a subscription). At: https://www.rep.routledge.com

There are electronic archives for individual articles (both research and expository) in both physics and philosophy.

For physics, the main archive is the 'arxiv', at: https://arxiv.org. For this book, the main sections of it are: in the Section called 'Physics', the sub-section 'History and philosophy of physics'; the Section called 'General relativity and quantum cosmology' and the Section called 'Quantum Physics'.

For philosophy of science, the main archive is maintained by the University of Pittsburgh, at: http://philsci-archive.pitt.edu.

In all of these archives, one can search by author, title, subject-area etc.

(2): A fine academic book that philosophically assesses the cosmological multiverse (the topic of my Chapter 5), and gives briefer discussions of the philosophical and Everettian multiverses, is:
S. Friederich, Multiverse Theories: a philosophical perspective, Cambridge University Press, 2021: https://www.cambridge.org/core/books/multiverse-theories/68CE18BE78DE31550C67855107A57942

There are several good popular books by physicists that discuss the multiverse in detail. Two main ones are Max Tegmark's The Mathematical Universe (2014), which advocates the multiverse (in several senses, both physical and philosophical); and Sabine Hossenfelder's Lost in Math (2018) which is very critical of multiverse proposals in physics, and of several other trends in recent physics.

I reviewed these books, and stand by what I wrote. (Although this book does not repeat the content of those long reviews, there will be some overlap in later Chapters.) The reviews are most conveniently available at the Pittsburgh archive: respectively at: https://philsci-archive.pitt.edu/10760/ ; and: https://philsci-archive.pitt.edu/15724/ .

(3): There are many topics apart from the multiverse, lying at the interface of philosophy and physics; and many academic but introductory books about them. Two excellent ones are:
D. Wallace, Philosophy of Physics; which is in Oxford University Press' Very Short Introduction series; the series has many related books.
N. Huggett, Everywhere and Everywhen, Oxford University Press, 2010; also available online at: https://academic.oup.com/book/9556?searchresult=1

(4): I think my view, in Sections 3, 4 and 5, of philosophy and its method is very widespread in academic philosophy. After all, hardly anyone would disagree that in debates involving concepts that are not precisely defined, and are perhaps contentious, one cannot expect conclusive arguments; and so the best we can do is try to assess all the evidence and arguments in an open-minded way---and beware of the beguiling power of words. But three further comments may interest philosopher readers.

First: I wrote throughout in an "objectivist" manner. I presupposed that there are truths about the concepts of interest, even though the concepts are not precisely defined. (More precisely: there are truths about those concepts one does not reject as bad; cf. Section 3.) Or in other words: I presupposed that philosophy addresses genuine questions or problems that have a correct answer or solution. By this, I stand. I agree that even if the concepts were made precise (perhaps in an arbitrary stipulative way), the truths would remain fearfully hard to know, just because the concepts are multiply connected with other concepts, in an open sea of both empirical and conceptual considerations. But in the face of this, we must not get disheartened about the validity of our enquiry. We must not let the difficulty of the problems prompt a failure of nerve.

Second: In Sections 3, 4 and 5 (and just above) I for the most part combined together the sorts of evidence one can get from science, and from philosophy. I combined in a single phrase, such as 'empirical and conceptual considerations' or 'evidence and arguments', the contingent facts we might learn from the sciences, and the apparently necessary (non-contingent or analytic) facts we might learn by analysing our concepts (by 'armchair reflection', in the traditional image of philosophy). So let me here be more explicit. I do indeed think that empirical science and philosophy are continuous with one another, in that both are about---and should attend assiduously to---both sorts of consideration. In philosophy, this view is much associated with Quine. But one can endorse it without also joining him (in his essay 'Two dogmas of empiricism': 1951) in denying the validity of the analytic-synthetic distinction. (In fact, I do not join him in the denial. I endorse the distinction---and Putnam's diagnosis (in his essay 'The analytic and the synthetic': 1962) of the grain of truth in Quine's view. Incidentally, this diagnosis is prefigured in Kemeny's insightful review of Quine's essay, immediately after its publication: in the Journal of Symbolic Logic for 1952.)

My third comment is more programmatic. Namely: my view of philosophy and its method (and indeed, my acceptance of the analytic-synthetic distinction) is reminiscent of that great influence on analytic philosophy, David Hume. Thus recall my comment at the end of Section 4. In particular: I should announce now that in Chapters 2 and 3, I will endorse his scepticism about the idea of necessity in nature; and thereby, his modest and low-key conception of the sort of explanation and understanding of nature that empirical science, indeed all human enquiry, can provide us with.

# Chapter 2: Physics and Philosophy from 1600 to 1900

The main aim of this Chapter is to review some aspects of philosophy and of physics in the three centuries from 1600 to about 1900. That may seem a tall order. But we will only need those aspects that will help me explain how by about 1970, both philosophy and physics were ripe for formulating the three multiverse proposals, on which Chapters 3, 4 and 5 will focus. So those Chapters will also review the developments in the twentieth century that prepared the ground for their multiverse proposal. (The time-frames will vary between the three cases. For example, the relevant twentieth-century developments in logic and philosophy cover a century, from 1870 to 1970; while for cosmology, they happened in the sixty-year period, from 1910 to 1970.)

The historical aspects up to about 1900, reviewed in this Chapter, and the twentieth-century developments in the early Sections of Chapters 3, 4 and 5, will not just be stage-setting. They will also help us to assess the multiverse proposals.

For philosophy, this Chapter will mostly be about how a modest conception of scientific enquiry, separated from the idea of necessity and from the framework of logic, emerged in the eighteenth century. In this development, the main name will be David Hume. For physics, the Chapter will mostly be about the rise of mechanics; for which the main name will be, of course, Isaac Newton.

I begin with the tradition of <u>natural philosophy</u>: a tradition from which both physics and philosophy, as we now conceive those disciplines, sprang. This will lead in to the rise of mechanics, especially Newton's mechanics. I will emphasize how accepting action-at-a-distance in Newton's theory of gravity paved the way to the modern, Humean, modest---one might even say: pessimistic---conception of what it is to understand the natural world. At the end of this Chapter, we will see that this conception, and the emergence in the nineteenth century of the distinction between applied mathematics and pure mathematics, contributed to the rise of logic as central to philosophy. These factors also prompt a philosophical question, 'What is pure mathematics about?'. This question has been at the centre of twentieth-century philosophy, and we will return to it in Chapter 3.

<u>Chapter 2, Section 1: The tradition of natural philosophy</u>
'Natural philosophy' is a venerable phrase. It refers to enquiry into the natural world. It encompasses enquiry that is empirical, including experiments as well as everyday observation; and also enquiry that is conceptual, including using quantitative e.g. mathematical methods. It is especially associated with the seventeenth century. Figures such as Galileo, Descartes and Newton all saw themselves as engaged in natural philosophy. Indeed, Newton's masterpiece that propounds his theory of mechanics and gravitation (published in 1687) is entitled: '<u>The Mathematical Principles of Natural Philosophy</u>'. But this field of enquiry goes much further back, to ancient times: as the seventeenth century thinkers of course recognized. Thus when figures such as Aristotle, Lucretius, Aquinas, Galileo and Newton asked what was the nature of space, or of time, or of matter, or of causation, they shared a common field of enquiry---however great the disagreements between their resulting answers.

Various developments from the eighteenth century onwards broke up this intellectual unity between philosophy and what we now call 'the sciences', especially physics. The parting of the ways is symbolized by the invention of the word 'scientist' (by Whewell in 1834), and the phrase 'natural philosophy' falling completely out of use by the late nineteenth century. (But the phrase remains, even today, in some British universities' title of one of their Professorships of physics. This is, for anyone enthusiastic about the connections between physics and philosophy, an evocative reminder of yesteryear's synergy between the disciplines.)

The broadest of these developments that broke up the unity might be summed up as: the growth of knowledge. For our purposes, the main development within physics was the establishment of Newtonian mechanics as describing with increasing detail, and quantitative accuracy, not just astronomical observations but also many terrestrial phenomena. (I will give some more details below.) As the eighteenth century went on, this success increasingly used technical notions and advanced mathematics: which of course made for intellectual specialization.

Agreed: great figures in physics, such as Euler, continued to write natural philosophy: as did some great figures of nineteenth century physics, such as Helmholtz, Maxwell and Mach---despite the explosion of knowledge within physics, during the 1800s. But broadly speaking, the increasing specialization of physics over the course of these two centuries, 1700 to 1900, meant that such writings became less central to physicists' detailed researches.

So much by way of a lightning summary of how physics grew away from philosophy between 1700 and 1900. During the same period, philosophy also grew away from physics, and more generally from science. This was not only due to the obvious point that philosophy is about so many topics other than the natural world and our knowledge of it (such as moral and political philosophy). Also there was, even within metaphysics (i.e. the theory addressing the general nature of all entities) and epistemology (the theory addressing what knowledge is), a self-conscious turning away from the details of physics, and of science. This occurred as part of the legacy of Kant (and so it occurred especially in German philosophy). The reason in short is that Kant announced a new and ambitious conception of metaphysics and epistemology that rendered them autonomous from other disciplines, and in particular independent of the details of the sciences. Though Kant himself wrote a lot of natural philosophy, much of German philosophy got more and more separated from science after his death, in the work of figures like Fichte and Hegel. (In my opinion, it also got more and more obscure and high-falutin'.)

Then in the early twentieth century the quantum and relativity revolutions erupted, and brought tumult to physics; (more details in Chapter 4). The controversies about how to develop, and even how to "just" understand, these new theories threw physicists back to addressing basic conceptual questions, like those I listed above: what is space, what is time, what is matter, what is causation? The ensuing debates in which philosophers, especially in Vienna and Berlin (which were centres of the new physics), took part, had an enormous influence on philosophy. Indeed, they moulded the logical positivist movement; and this led to the idea of philosophy of science as a sub-discipline of philosophy. In this way, natural philosophy---under the new name 'the metaphysics and epistemology of the sciences'--- became again, by the mid-twentieth century, a research subject.

We will see later (in Chapters 4 and 5) how since 1970, the metaphysics and epistemology of physics has really taken off: being nowadays called 'the philosophy of physics'. But in this Chapter, I will develop three themes from natural philosophy in the earlier period, between 1650 and 1900. We will need these themes in order to understand our multiverse proposals, especially the first two proposals (from philosophy, and from quantum physics).

The first theme is about what the natural philosophers believed. The second is about their optimism that, by adopting their views, humans could achieve an understanding of the natural world that was completely clear and satisfying. Here my point will be that this optimism was dashed by the success of Newton's physics, and indeed, by Hume's philosophy. The third theme is the status of logic during these three centuries. In short, logic was for most of this period in the doldrums; but from the mid-nineteenth century onwards, it became vigorous---which will lead in to Chapter 3.

Chapter 2, Section 2: The mechanical philosophy

Several of the greatest natural philosophers of the seventeenth century believed that all the processes in the natural world would ultimately be explained in terms of objects' parts, including their tiniest parts, interacting with one another by causal "pushes and pulls". Here, the phrase 'the natural world' was to include biological processes, such as the growth of plants. For some of these authors, it was to include also psychological processes, such as perception, both in humans and in animals. And for most of these authors, the causal "pushes and pulls" required that the two objects (usually called 'bodies') touch one another, i.e. be in contact, like the gear-wheels in a machine such as a windmill.

Hence the phrase 'mechanical philosophy'. Other jargon: the principle that there could be no interaction without contact came to be called 'the principle of contact action'. Another slogan for the same idea was that there should be no 'action-at-a-distance'. (So here, the main meaning of 'action-at-a-distance' is causation between spatially separated bodies being unmediated by something in between them, rather than the causation being instantaneous—that is secondary.)

Advocates of this view included Galileo, Hobbes and Descartes. They knew, of course, that their view had ancient precursors: in particular, the ancient atomists, Democritus and Lucretius, who maintained that matter ultimately consisted of small indivisible lumps in a void (vacuum). Thus Democritus (ca. 460-370 BC) said: 'The first principles of the universe are atoms and empty space; everything else is merely thought to exist . . . Sweet exists by convention, bitter by convention, colour by convention; atoms and Void [alone] exist in reality.' (Here of course, 'by convention' means something like 'as a result of how humans' sensory organs happen to work'; rather than the modern meaning, viz. 'by a human decision that could have been made otherwise'.)

Of course, the mechanical philosophers disagreed amongst themselves. Some, such as Descartes, denied that there is a void (vacuum). They said instead that matter fills space completely (even on the tiniest length-scales), and what seems to be empty space, e.g. between the planets, is filled with a "very thin" fluid through which solid objects pass easily, rather like a boat through water. And some (including again Descartes) denied that all biological and-or psychological processes could be thus explained. So they limited the scope and ambition of their mechanical philosophy to explaining, in terms of contact-action, all processes in the inanimate world: or all processes in either the inanimate world or within organisms that are not sentient.

Nowadays, the conception of matter as lumps in the void is the everyday conception. It is so-called 'common sense'. For we all learn in primary school that matter is made of atoms, which are separated from each other by empty space i.e. a vacuum. (Agreed: our atoms differ from Democritus', in that they can be divided.) But it is important to recognize that this vision was hard-won over the centuries. At the time, the mechanical philosophers' vision seemed not just radical, but thoroughly implausible.

Indeed, there are four points here.

First: to the extent that the lumps-in-the-void conception of matter is true, it is not at all obvious that it is true. This is so even for inanimate objects, nevermind the objects involved in biological and psychological processes. It is far from obvious that air and other gases are mostly empty space; or indeed that liquids and solids consist of tiny particles jostling each other, with interstices between them. On the contrary, the naïve appearance of all the gases, liquids and solids we see around us is that they consist of matter that fills space completely (even on the tiniest length-scales): such matter is called 'continuous'.

Second: even if we accept that all matter consists of tiny lumps in a void, these philosophers' other claim---that all processes can be explained as accumulations of microscopic interactions, "pushes and pulls", occurring when the lumps are in contact---by no means follows. Indeed, it seems a radical, even thoroughly implausible, speculation. One naturally asks: how could all the variety and complexity of the processes we see, including biological and even psychological processes, be explained in such simple terms? It is really only in the last hundred years or so that the claim has become believable---indeed well-confirmed by countless pieces of detailed evidence. Think of how the existence of atoms, and the way they compose molecules, was understood only in the early twentieth century, together with how they explain chemical processes. And the understanding of biological processes, such as how muscles contract, how nerves send signals, or how genes are passed from parent to offspring, came only with the rise, in mid-twentieth century, of physiology and molecular biology.

There is, of course, another aspect to these achievements. Namely, they require much more subtle interactions between the tiny lumps in the void than the phrase "pushes and pulls", i.e. collisions or impacts, suggests. In fact, they require quantum physics with its strange features. A classical physics of "tiny billiard balls bouncing off each other" certainly <u>cannot</u> explain all these processes. But I postpone quantum physics till Chapter 4. At the moment, I want to stress two other ways in which the claims of the mechanical philosophers were radical: and so, at the time, hard to believe. Again, the moral will be that we should recognize that their later acceptance was not really common sense, but was instead hard-won. So these will be my third and fourth points.

Third: the mechanical philosophers fashioned concepts with which to describe with quantitative accuracy the contact, and collisions, of bodies. I mean concepts such as velocity, acceleration, mass, momentum, and energy. Nowadays these concepts are everyday notions: they are broadly familiar from e.g. car travel, and are taught in detail in school. But we should remember how non-obvious they are. For example: it took decades for physics to settle on each of the following. (i): The idea that acceleration is change in velocity during a given interval of time; rather than change in velocity during traversal of a given distance. (ii): The idea that mass is an intrinsic property of a body, different from its weight, that is a measure of its resistance to being accelerated. (iii): The idea that although in most cases the sum of bodies' momenta is evidently not conserved (i.e. constant over time) before and after a collision---think of a car-crash, or more safely, of throwing two marshmallows together---nevertheless, it is useful to consider the special, simple cases where the sum of the bodies' momenta is indeed the same before and after the collision.

Fourth: So far, I have summarized mechanical philosophy's claims that matter consists simply of lumps, and that mere contact-action among such lumps (especially tiny ones), theorized in terms of momentum etc., can explain the great variety we see around us. And I have praised these claims as bold and implausible, at the time---but vindicated by history. But I confess: I have over-simplified. There is an elephant in the room. Namely, Newton and his theory of gravity.

Chapter 2, Section 3: Newton's theory of gravity: unbelievable?
For Newton's theory <u>denies</u> the principle of contact action, that bodies cannot interact unless they are in contact. And in its description of gravity, the theory's replacement for this principle is utterly precise. What came to be called 'Newton's universal law of gravitation' says that at any time any two bodies attract one another with a force along the geometrical line <u>between them at that time</u>; and that this mutual attraction is unmediated---it occurs even when there is a vacuum between them. (The law also says how the force decreases with the distance between them, and how it depends on the bodies' masses. I will discuss these other features in a moment.)

So the law explicitly proclaims <u>action-at-a-distance</u>: exactly what Galileo, Descartes and the other mechanical philosophers denied. It is indeed very hard to believe. Take as the two bodies, the Sun and the Earth; and think of how light takes eight minutes to travel through the vacuum from the Sun to the Earth. Newton's law says that if somehow you could shift the entire Sun during the course of, say, a minute, by some distance, say a thousand miles, then the direction along which the Sun pulls the Earth would be different at the end of the minute--- before the light arrived. Indeed, the direction of the pull would change instantaneously during the course of the minute, sweeping through the sky like a lighthouse-beam. Of course, the angle through which it sweeps would be tiny, because the Sun-Earth distance is so much greater than a thousand miles. But that is only because I imagined shifting the Sun by a "modest" astronomical distance. If instead I had imagined shifting the Sun by, say, half the radius of the Earth's orbit, then the direction of the Sun's pull would have changed by a much larger amount---but again, instantaneously during the course of the shift. So the important point is that the attraction is unmediated, and that the change in direction is instantaneous.

This law is yet harder to believe when one notices that it claims this instantaneous attraction-at-a-distance occurs between <u>any</u> two bodies; for example, between an apple and a planet.

Besides, the law says that the attractive forces are <u>equal in size</u> (though opposite in direction). So the apple pulls on the planet with exactly the same "strength" that the planet pulls on the apple. That is well-nigh incredible. In particular, notice that it goes far beyond the familiar anecdote about Newton seeing the apple fall and thinking of the Earth as pulling the apple down. In that anecdote, as it is usually told, there is no suggestion that the apple pulls the Earth upward---let alone that it does so with a force equal in size to the force exerted by the Earth.

Newton's explanation of our not noticing the Earth's upward acceleration is that the Earth's mass is so vastly greater than the apple's mass, that its acceleration is correspondingly smaller. For according to Newton, force = mass times acceleration. So the equal and opposite forces make the small mass of the apple accelerate fast enough for us to observe and measure it; but they make the vast mass of the Earth accelerate only a minuscule, and unobservable, amount.

What are we to make of all this? As regards physics and its history, the verdict has been, in short: gradual acceptance. Indeed, the acceptance was eventually so complete that by about 1800 or 1850, physicists were sanguine, even placid, about the idea of action-at-a-distance, and about the idea that the gravitational forces between any two bodies are equal in size (though oppositely directed).

Let me spell this out in a bit more detail. Newton himself favoured the principle of contact action. For example, in a now-famous letter to Bentley, he wrote: 'That gravity should be innate inherent and essential to matter so that one body may act upon another at a distance through a vacuum without the mediation of any thing else by and through which their action or force may be conveyed from one to another is to me so great an absurdity that I believe no man who has in philosophical matters any competent faculty of thinking can ever fall into it.' Besides, he wrote in a Scholium (i.e. comment) added to his <u>Principia</u> that he had endeavoured to find a means, a mechanism, by which gravity acted, and had been unsuccessful. (I will return to this admission at the end of Section 6.) In short: it is no wonder that initially, Newton's readers were incredulous about his theory.

But Newton also argued, with considerable justice, that as regards the gravitational force between each planet and the Sun, the detailed astronomical observations which he had in hand (especially those encapsulated in what became known as 'Kepler's laws') implied that he could <u>deduce</u> that the gravitational force acted along the instantaneous line between the planet and the Sun. Furthermore, he argued that he could <u>deduce</u> that the force decreased with distance in the precise way his law of gravitation said. (Namely, by what is called 'the inverse-square'. This means: doubling the distance reduces the force by a factor of 4, i.e. multiplies it by a quarter; tripling the distance reduces the force by a factor of 9, i.e. multiplies it by a ninth; and so on.)

This quantitative precision of Newton's theory meant that, once combined with astronomical observations, it made precise predictions about the planets' movements. Many such predictions were tested in the decades following his theory's publication (i.e. after 1687)--- and they turned out to be true. In time, this impressive quantitative success trumped people's doubts; in particular, their incredulity about action-at-a-distance.

(These successes depended of course on developing the calculus that Newton and Leibniz had invented, and applying it in ever greater detail to mechanics and astronomy. These successes were mostly achieved, not in Britain, but in continental Europe by figures such as Euler and Lagrange.)

Thus by about 1800, most natural philosophers accepted the claims that gravity involved action-at-a-distance, and equal and opposite forces that decreased with distance according to the inverse-square. And so it went. The theory garnered more and more successes; so that by 1850, physicists---by then, professionally identified as such---were sanguine, even placid, about these claims.

Agreed: there were dissenting voices, such as the physicist-philosopher Ernst Mach. And calm comes before a storm. In the early twentieth century, Einstein (inspired in part by Mach's misgivings) created an amazing new theory of gravity, which he called 'general relativity' (1915). According to this theory, there is no action-at-a-distance. Gravitational influence <u>does</u> take time to propagate across space: namely, it travels at the same speed as light. So for my imaginary example above, in which the entire Sun is shifted by a thousand miles during the course of a minute, Einstein's general relativity says that eight minutes must elapse (from the beginning of the Sun's shift) before the direction along which the Sun pulls the Earth begins to change. For eight minutes is the travel-time from Sun to Earth, for gravity as well as for light

In Chapters 4 and 5, I will briefly return to general relativity. But for this book's purposes, it is the <u>philosophical</u> consequences of Newton's postulate of action-at-a-distance that will matter---more than its two centuries of success, followed by its demise at the hands of Einstein.

For as I shall explain in the next Section, the success of Newton's theory also contributed to the decline of the mechanical philosophers' extraordinary "cognitive optimism" about our ability to understand nature's innermost workings. Another factor in that decline was David Hume's philosophy: which, together with Newton's theory, paved the way for what is now the mainstream "modest", or "pessimistic", picture of how much humans can understand nature.

Chapter 2, Section 4: Optimism about understanding nature: 'we will soon deduce the cause from the effect'

We have seen that the mechanical philosophers had a bold and ambitious vision, about understanding all the processes of nature as mechanical. Some of them, including Galileo and Descartes, were also accomplished proselytizers---one might say, propagandists---for the movement. They confidently proclaimed that detailed and successful mechanical explanations would soon be achieved---"if not tomorrow, then the next day". (And of course, they promised that the explanations would conform to their own principles, rather than some rival's favoured principles.)

But there was also another strand to the mechanical philosophers' confidence. It is about the quality of understanding that such prospective explanations were expected to provide. Crudely and metaphorically: the quality was going to be the very best. To go beyond metaphor, I need to invoke a topic that my account has so far suppressed. It is the ultimate "elephant in the room": namely, God and God's understanding of nature.

Thus the mechanical philosophers believed that God had complete insight into the innermost workings of nature; any natural process was completely "intellectually transparent" to God. So far, so unsurprising. After all, He is meant to have created the natural world. But they believed also that we humans, being made 'in the image of God' (Genesis 1:27), can hope to emulate this complete insight and understanding. Of course, we are finite creatures, and God is infinite. So we cannot hope for such understanding all at once, and for all of nature: to hope for that would be grossly hubristic. But for individual "patches" of nature, perhaps "small" ones--- for example: the collisions of solid bodies moving in straight lines---we can attain an insight and understanding, as complete and intellectually transparent as God's own. (Or rather, most of the mechanical philosophers believed these claims. Of course, theological controversies abounded as much as philosophical ones; even to the extent that some of them, for example Hobbes, were accused of atheism.)

One main way in which this idea of complete understanding was made more precise was in terms of <u>deduction</u>. So here at last we broach the discipline of <u>logic</u>. In the Western tradition, Aristotle had founded the subject, mainly by classifying <u>valid patterns</u> of argument.

Thus recall what it means to say that an argument with premises and conclusion is <u>valid</u>: in other jargon, that one can <u>deduce</u> the conclusion from the premises, or that the premises <u>imply</u>, or <u>entail</u>, the conclusion. All these different jargons are synonymous. Namely: if all the premises are true, or (supposing them to be in fact false) if they <u>were</u> true, then the conclusion <u>must</u> be true. That is: any way in which all the premises are made true must also make the conclusion true.

In many cases, the validity of an argument turns on the placing within the premises and conclusion of words like 'all', 'some', 'none', 'and', 'or' and 'not', irrespective of the other words. In such cases, we say there is a <u>valid pattern</u>. An elementary Aristotelian example turns on the behaviour of the word 'all'. Consider the argument-pattern: 'Premise: all As are Bs. Premise: all Bs are Cs. Therefore, Conclusion: All As are Cs'. This pattern is evidently valid, whatever 'A', 'B' and 'C' stand for, i.e. whatever plural nouns or noun phrases ('horses', 'red things' etc.) one puts in for them.

Since medieval times, logic (based on Aristotle's work) had been a basic part of university studies. (Along with grammar and rhetoric, the three disciplines together comprised the 'trivium', the 'three ways'). So it was natural for the mechanical philosophers to conceive the complete understanding, that they were proclaiming to be imminent in their description of nature, in terms of deduction.

Besides, there is a tempting metaphor that yields an analogy between, on the one hand, the relation between premises and conclusion, and on the other, the relation between cause and effect. Namely, the metaphor of <u>containment</u>. Since the premises being true forces the conclusion to be true, it is natural to say that the conclusion is contained in the premises. Or rather, since the premises and conclusion are sentences, i.e. pieces of language, we should express this as: the proposition expressed by the conclusion, i.e. the <u>content</u> of the conclusion, is contained in the conjunctive proposition expressed by all premises taken together. And analogously for causation. It is natural to say that the effect is contained in the cause: at least, provided that the cause is described in sufficient detail so that all relevant factors are included.

This analogy suggests that from a sufficiently detailed description of the cause, one should be able to deduce a description of the effect: rendering the effect completely comprehensible "to the light of reason". Indeed, Descartes says exactly this in his famous <u>Meditations</u> (the Third Meditation). He writes: 'Now it is already clear by the light of nature that the complete efficient cause must contain at least as much as the effect of that cause. For where, pray, could the effect get its reality if not from the cause? And how could the cause supply it, without possessing it itself?'

This argument (with, of course, variations in its exact formulation) occurs frequently in the writings of Descartes and his contemporaries. As an illustrative example, one variation appeals to the idea that there can be no creation <u>ex nihilo</u>, i.e. creation out of nothing; (except of course by God, as in the creation of the world). So the effect with 'its reality' (as Descartes puts it) must somehow be latent in what occurred before: which prompts the argument above.

Thus the common theme is that over the next hill---"if not tomorrow, then the next day"---there will be a science ("mine, not that of my rivals") whose concepts and claims will be so clear to the light of reason that they do not merely command our assent, but also provide complete understanding and certain knowledge. In particular, this science will provide deductions of effects from their causes: the premises describing the causes will entail the conclusion describing the effect.

Chapter 2, Section 5: Lowering our sights: Hume
In the eighteenth century, this optimistic view withered away. Two main reasons for this were Hume's critique of the view, and the success of Newton's theories. These reasons are the topic of this Section and the next one.

Hume argued (in his <u>Treatise of Human Nature</u> (1739) and his <u>Enquiry concerning Human Understanding</u> (1748)) that, whatever the concepts and claims of a successful science might turn out to be, there is no hope at all of a genuine <u>deduction</u> of effect from cause. For a deduction of a proposition E stating the effect from a proposition C stating the cause would require that <u>if</u> C is true, then it is logically impossible for E to be false. This implies that no matter how detailed one's description of the cause, it is hopeless to aim for a deduction. For it is logically possible that the cause as described occurs, with the effect being absent. In support of his claim of logical possibility, Hume takes two examples: that an impact of a body, say of a billiard ball, causes another body, another ball, to move; and that bread causes nourishment. For these examples, he points out that one can conceive, i.e. imagine, that the impact occurs without the second ball moving away; and that I eat the bread (just as it is, in look, smell and chemical composition) without getting nourished.

To make a genuine deduction of the effect, i.e. a valid argument whose conclusion states that the effect occurs, one obviously needs an extra premise. This could be a general premise. It could be along the lines: all impacts of such-and-such a kind (including the one in question) are followed by the second ball moving. And similarly for the bread example: anyone eating a loaf of such-and-such a kind, is later nourished. Here, I say 'followed by' and 'is later', since for the aim of validly implying that the effect occurs, the extra premise need not claim a causal relation. It is enough---but essential---that it claims the effect's occurrence. Or the extra premise could be a specific one, along the lines: this impact is followed by the second ball moving; my eating this bread is followed by my being nourished.

Hume agrees that once such an extra premise, either general or specific, is supplied, there is undoubtedly a valid argument to the effect. Indeed, these arguments will illustrate the simple and familiar pattern, <u>modus ponens</u>, viz: 'Premise: P; Premise: if P then Q. Therefore, Conclusion: Q'. For we have here the pattern: 'Premise: The cause occurs; Premise: If the cause occurs then the effect occurs. Therefore, Conclusion: The effect occurs'. Here, the second premise, the if-then premise, can be either general or specific; as we have seen. And in both premises 'the cause' is to be understood, i.e. described, in sufficient detail as to justify the second premise. The impact must be sufficiently forceful, the first ball rigid enough (not made of jelly) etc.; the bread must be made from wheat or barley or … but not from cyanide.

All this is nowadays so obvious to us that it is tempting to criticize Hume as flogging a dead horse. One thinks: 'Of course, the later effect---the second ball moving away, the person being nourished---does not follow with sheer logical necessity from the earlier state, no matter how detailed our specification of that earlier state. Only on the assumption of a suitable linkage, like the extra premises above, will there be a deduction.' Agreed: that is so. But it being so does not mean that Hume's critique was misdirected, i.e. that all Hume's predecessors acknowledged the point. As I urged in the last Section: they did not.

This point is often put in terms of the idea of rationality. For the topic of whether or not there is, in cases like the billiard balls or the bread, a deduction can be put in terms of the question: why is it rational to believe that, given a sufficiently detailed specification of the impact or of the bread, the second ball will move, or that the person will be nourished? Of course we all do believe this. But why? And is it rational to do so?

When we put the question in this way, the temptation to criticize Hume is very likely to be expressed as follows. 'Yes it is rational to believe these propositions: though it is not a matter of deduction, as Hume emphasizes. But Hume's emphasizing this shows that he is using an unduly narrow notion of rationality. He recognizes only deductive rationality: that is, the obligation on rational beings to believe the deductive (sheer logical) consequences of what they believe. But not everything that it is rational for us to believe follows by sheer logic (a deductively valid argument) from other propositions we already believe. In short: Hume should loosen up about what rationality requires of us.'

To which I reply, on behalf of Hume---at least, the historical Hume---as follows. Once we set aside deductive rationality---which is a notion, and an obligation, that we can surely all agree on---rationality is a contested concept, in the sense I discussed in Chapter 1. For let us ask: given some collection of propositions we already believe, what else should we believe (additional of course to the deductive consequences of the collection)? That is: what principles, additional to deduction, should govern the formation of beliefs from other beliefs, taken as evidence? That is a very difficult and multi-faceted question which philosophers, and of course scientists and statisticians, have addressed in many different ways since Hume's time. For example, some believe probability must be the central idea for answering the question, while others abjure it. Large bodies of theory, with specialist names like 'inductive logic', 'statistical inference' and 'causal inference', have been developed---and debated. For the jury is still out concerning how best to make this question precise (perhaps as several sub-questions); and accordingly, about what the answers are.

To which I say: 'More power to your elbow: tough work, and we should all look forward to, and value, the answers'. But what is relevant here is that Hume also, not just we moderns, can say this. He need not deny---he has no reason to deny---that there are principles about belief-formation that go well beyond deductive rationality.

For his point remains: that his predecessors thought they did <u>not</u> need to formulate and assess any such principles. They thought their new science would need <u>only</u> deductive rationality. Thus the objection that Hume is using an unduly narrow notion of rationality, and should loosen up about what rationality demands, is doubly wrong. For first: Hume is merely examining the same deductive notion that his predecessors touted as both sufficient for science, and as promising a complete understanding of cause-effect relations. Hume shows that it is not sufficient, and it does not usher in any such complete understanding. And second: Hume can and should accept (along with the rest of us) that there are principles about the non-deductive formation of beliefs. And he, like us, can investigate what they are.

Chapter 2, Section 6: Newton again
So much by way of expounding, and defending, Hume. I turn to my second reason why the mechanical philosophers' optimism died way: why, to use the jargon above, eighteenth century natural philosophers stopped claiming that their description of an effect, or of how it came about, was 'clear to the light of reason'. This second reason is: the success of Newton's theory of gravity with its action-at-a-distance.

We saw above how radical, indeed unbelievable, Newton's theory was. The point now is that during the course of the eighteenth century, it gathered ever more empirical successes, so that the conclusion became inescapable: our most successful framework for quantitative empirical knowledge explicitly abjures there being any <u>intelligible</u> ('clear to the light of reason') causes of gravitation.

For indeed, it is not intelligible that a change in the position of the Sun would instantaneously alter the direction of pull felt by the Earth; since any process propagating from the change of position would take time to arrive at Earth. For example, a process at the speed of light would take eight minutes. But though unintelligible, this conditional proposition, 'If the Sun were to move …, the Earth's direction of pull would instantaneously alter', is a consequence of our well-confirmed theory---and we should accept it. Thus the idea that the effect is necessarily connected to the cause, in particular by its being somehow contained in the cause, withers away. We must accept that the effect is just an event that invariably succeeds the cause.

This situation shows that intelligibility in the above sense is not a <u>sine qua non</u>, a necessary condition, of exact empirical knowledge. And it shows that, although you might want intelligibility in your scientific theory, seeking intelligibility is sometimes (i.e. at some stages in enquiry, and for some aspects of nature---here, mechanics) not fruitful.

This lowering of one's sights about what our understanding of nature should involve was formulated already by Newton himself in a famous passage in the General Scholium that he added to the second edition (1713) of his 1687 masterpiece <u>The Mathematical Principles of Natural Philosophy</u>. The passage includes his reporting that (as I mentioned above) he had tried, but not succeeded, to understand gravity other than as action-at-a-distance. Thus he writes:

'Thus far I have explained the phenomena of the heavens and of our sea by the force of gravity, but I have not yet assigned the cause of gravity. Indeed, this force arises from some cause that penetrates as far as the centres of the sun and planets, without any diminution of its power to act . . . I have not as yet been able to deduce from phenomena the reason for these properties of gravity, and I do not feign hypotheses. . . . And it is enough that gravity really exists, and acts according to the laws that we have set forth and is sufficient to explain all the motions of the heavenly bodies and of our sea.'

Newton's final words neatly sum up this discussion. His magisterial 'it is enough' (in Latin: '<u>satis est</u>') lowers our sights about what to require in scientific theories: we cannot require intelligibility in the strong sense above. But it also offers the solace that even without such intelligibility, we can achieve amazing quantitative accuracy. Thus Newton resolutely sets the path of the future of physics.

Besides, the subsequent history of physics has vindicated him, in the sense that amazing quantitative accuracy has been achieved in various fields of physics while giving up, not just intelligibility in the specific sense above viz. requiring contact-action, but also in other senses. In Chapter 4, we will see examples of this. The first will be the treatment of matter by classical physics (from about 1750) as made of <u>point-particles</u>, i.e. masses that are not just tiny but extensionless---located at just one point of space. (So their density, understood in the usual way as the mass per unit volume---in other words: as mass divided by volume---is infinite.) But the main example will be quantum theory, with its paradoxical combination of amazing quantitative accuracy and notorious interpretative difficulties.

<u>Chapter 2, Section 7: Logic in the doldrums---and its revival</u>
For this Chapter, one task remains: to describe the changing fortunes from 1600 to 1900 of the discipline of logic. In short, it went from bad times (1600 to 1850) to good (from 1850 onwards). But the revival of logic in the late nineteenth century can only be understood as part of a wider change within mathematics: namely, the emergence of the distinction between applied mathematics and pure mathematics. So the next (and final) Section will be about that distinction, and its impact on the landscape of logic and philosophy.

But let us start with the earlier period, from 1600 to about 1850. I have said that the mechanical philosophers knew their logic, and envisaged a science in which one could validly deduce the effect from the cause. Nevertheless, it is fair to say that in their time, and more generally in the period 1600 to 1850, logic was in the doldrums. Indeed, it was in the doldrums in two senses.

First, it was generally regarded as a completed subject, in which the last word had been said. Logic was of course to be respected in any discourse. But there was no recognition that valid patterns of argument additional to those codified by Aristotle and his successors (including medieval successors) might yet be discovered and codified. (For the mechanical philosophers, this indifference was part of their rebellion against the Aristotelian tradition.) Thus in philosophical texts, the teachings of logic were sometimes summed up as the principle of 'excluded middle', i.e. 'P or not-P'; or as the principle of non-contradiction, i.e. 'not both P and not-P'. Venerable principles indeed: but there is so much more to the subject.

It was also in the doldrums in a second, and more subtle or controversial, sense (which is connected with the first). In short, philosophers in this period tend not to address what for us are natural philosophical questions about logic: questions such as what exactly is the nature of logical necessity, and exactly which propositions are indeed necessary. Their discussions of such questions often presuppose that Euclidean geometry and arithmetic are both, indeed, necessary. But what exactly makes them necessary is not a question that they really engage with. (I shall return to this sort of question shortly.)

I admitted that this second sense is subtle or controversial, because the work of each of two great philosophers, Leibniz and Kant, prompts a qualification. Leibniz aimed to reconcile the insights of the new learning of the mechanical philosophers with the doctrines of the Aristotelian tradition. And famously, his philosophy engaged with the nature of necessity. He proposed that there is a realm of all possible worlds, and a proposition's being necessary is a matter of it being true in any, and so in all, of them. Similarly, a contingent proposition is true in some but not all worlds; and an impossible proposition is true in none. He also said that God in his omnipotence created just one world; and in his benevolence, created the best possible world---a claim later satirized devastatingly in Voltaire's <u>Candide</u>. Setting these theological aspects aside, Chapter 3 will of course return to assess the idea that necessity consists of truth in all possible worlds.

The second qualification is that although Kant said little about necessity as such, he said a great deal about how a proposition could have that apparently similar and associated feature, of being <u>a priori</u>: meaning, roughly, 'knowable (and so true) independently of any experience'. For he believed that the propositions of both Euclidean geometry and of arithmetic are <u>a priori</u>. But he also believed they are informative about the world. (His label for this was 'synthetic', as against 'analytic': which he construed as 'the predicate being contained in the subject', though his usage corresponds well to the common modern formulation of 'being true in virtue of the meanings of the words'.) How a proposition could be both <u>a priori</u> and synthetic thus became his central question: which his masterpiece, <u>The Critique of Pure Reason</u> (1781), answered by saying that they result from the way our human cognitive constitution moulds, or imposes structure on, raw experience.

About this answer, the jury is still out. Agreed: we all accept, in everyday thought as much as technical science, that our concepts, and so what we perceive, believe and even know, reflect contributions from the side of the subject, us, as well as the from the side of object thought about or perceived. We also accept that such contributions can be common to the species (and so presumably of biological origin), or to a population (and so of historical or cultural or linguistic origin); or specific to an individual (so idiosyncratic, in a non-perjorative sense). But Kant's conception of these contributions was less straightforward, and less empirical, than this. He maintained that by philosophical reflection he could formulate the contributions, as being a necessary condition of objective experience---and that he could thus justify, not just that there are some synthetic <u>a priori</u> propositions, but also that Euclidean geometry and arithmetic consist of such.

I myself think that even his first claim is wrong: there are no synthetic <u>a priori</u> propositions. (This of course goes with my admiration for Hume, evident in the previous Sections.) But even if his first claim is right, his second claim fails. For as I will report in the next

Section: in the nineteenth century, mathematicians developed various non-Euclidean geometries and showed their consistency: the world, and our experience of it, <u>could</u> be described by each such geometry. (Kant also claimed that the so-called 'law of causality', that every effect has a cause, and several principles of Newtonian mechanics were synthetic <u>a priori</u>: claims that failed with the advent of twentieth-century physics.)

But this is not the place to belabour Kant. Instead, I wanted just to record his work as prompting a qualification of my summary statement that logic was in the doldrums.

As to <u>why</u> logic was thus side-lined between 1600 and 1850, one obvious reason was its being associated with the Aristotelian tradition, so that it became a target of the mechanical philosophers' rebellion. But another reason arises from the cognitive optimism of those philosophers. As I discussed in Section 4, they believed their forthcoming science would be <u>certain</u>, and "intellectually transparent": clear to the light of reason. Being convinced of achieving such certainty naturally prompted them to ignore questions about whether their science's doctrines are necessary. Besides, if such questions had been pressed, the doctrine that the science would indeed deduce an effect from a cause would prompt the confident reply, that Yes, the doctrines are necessary. For after all: to every valid argument---say: Premise: P1, Premise: P2; Therefore, Conclusion: C---there corresponds a necessary proposition, viz: 'If P1 and P2, then C'. In short: if these philosophers' optimistic programme had succeeded, the proposition saying that the effect is deducible would indeed be necessary.

The broader context of mathematical, indeed scientific, thought is that from ancient times until the mid-nineteenth century, mathematics was taken to consist of, on the one hand, the study of numbers (arithmetic, and algebra in the sense of equations about numbers with variables 'x' and so on; and later, the calculus); and on the other hand, the study of space, viz. Euclid's geometry. And both these were universally regarded as providing an absolutely certain body of knowledge, that could never be overturned.

Here, I use the single word 'mathematics' very deliberately. For no distinction was made, as we now do, between: (i) applied mathematics, which describes the physical world, objects in space and time, using mathematical concepts (numbers and geometrical concepts); and (ii) pure mathematics, which is about numbers, triangles etc. "in themselves", regardless of what is in the physical world.

From a modern philosophical viewpoint, the first thing to say about this distinction is that pure mathematics, so understood, is obviously problematic from a philosophical viewpoint. For no matter how certain we may agree its claims to be, resulting as they do from rigorous mathematical proofs, there is the question: how do we come to have such knowledge? For we are physical organisms, embodied in space and time. So presumably our ideas, beliefs and knowledge originate in our experience of the physical world. But since the subject-matter of pure mathematics (numbers, triangles etc.) is not in the physical world, i.e. not located in space and time, it is then a pressing question how we come to have any ideas about that subject-matter. Besides, assuming we have such ideas: how can we come to believe, even know, propositions about this subject-matter, by following mathematical proofs?

This is a central, perhaps the central, question of the philosophy of (pure) mathematics since about 1850, i.e. after the applied/pure distinction gets articulated. I will briefly discuss this question in the next Section and the next Chapter. But to philosophers and mathematicians of the eighteenth century, this question was simply invisible. One main reason for this was that they conceived numbers in terms of lines within physical space. This conception, we will see, led to trouble.

<u>Chapter 2, Section 8: Houses built on sand---and how to repair them</u>
To introduce this, let us recall how we learn in school that besides the integers, positive and negative, there are, firstly, the rational numbers, where 'rational' stands for ratio or proportion.

These numbers are a ratio of integers like 1/3, 2/5, 10/2 (= 5), or - 42/9. Expressed as a decimal, they either terminate, e.g. 2/5 = 4/10 = 0.4, or recur, e.g. 1/3 = 0.333… with the 3s going on forever. Here, the idea of recurring includes eventually settling down to a finite sequence of digits that then repeats forever, e.g. 137.95421372137213721372137…, where '2137' repeats forever. But as the ancient Greeks discovered, there are also <u>irrational</u> numbers, e.g. the square root of 2, that cannot be expressed as a ratio of integers. When expressed as a decimal, these numbers neither terminate nor recur. The decimal expression goes on forever. But it never settles down into a repeating digit, nor even into a repeating sequence of digits. For example, the square root of 2 begins as 1.41421… but it never settles down. Another example is $\pi$ (the Greek letter 'pi'), defined as the ratio of the length of a circle's circumference to its diameter. It begins as 3.14159… but it never settles down. The set including all rational numbers (taken as including the integers, as in 10/2 = 5), and also all irrational numbers, is called the set of <u>real</u> numbers.

So for our purposes, the point here is that until the mid-nineteenth century, philosophers and mathematicians conceived of real numbers in an intuitive way, as segments of physical space. For example, they took the square root of 2 to be the diagonal of a square whose sides are of length 1. With this intuitive treatment, mathematicians produced amazing developments in the theory of real numbers; and from 1690 onwards, in the calculus that Newton and Leibniz had invented, and in the applications of these theories within mechanics and astronomy.

But it was a house built on sand. For the intuitive treatment I have just sketched led to paradoxes. One could construct apparently valid arguments within the calculus whose conclusions were contradictions. Agreed: with talent and care, mathematicians could insulate their work from these paradoxical arguments. But they remained as discomforts, so to speak; and they prompted efforts in the nineteenth-century to make the calculus more rigorous, and thereby expunge the paradoxes.

These efforts went along with a more general movement towards rigour, and especially rigorous proof, and therefore towards <u>formalizing</u> and <u>axiomatizing</u> mathematical theories. By the end of the nineteenth century, 'formalizing' came to mean writing the theory, not in a natural language such as English or Latin (augmented of course with technical terms like 'isosceles triangle' or 'limit of a sequence'), but in an artificial language with: a precisely specified vocabulary (usually a very small one); and precise rules of grammar dictating exactly which sequences of vocabulary items count as grammatical sentences; and precise <u>rules of inference</u> dictating exactly which passages from finite sequences of such sentences (thought of as premises) to another sentence (thought of as conclusion) count as an allowed inference. Accordingly, 'axiomatizing' came to mean that, having written all the claims of a theory in a formal language of this sort, one finds a small subset of these claims with the feature that any claim of the theory can be inferred from some choice of finitely many elements of the subset as premises, using only the proclaimed rules of inference. Thus the small subset are the <u>axioms</u>, and all the other claims are <u>theorems</u>. (In almost all cases, the set of axioms is not unique: even assuming a fixed formal language, there are several equally good ways to axiomatize the theory.)

This movement towards formalization and axiomatization was also prompted by two other developments, additional to the efforts at rigorizing the calculus. The first was the rise of what one might call 'heterodox' theories of numbers and geometrical figures; the second was the rise of the theory of <u>sets</u>.

First: new mathematical theories were proposed that surprised, even shocked, mathematicians by the fact that their subject-matters (numbers and geometrical figures) explicitly disobeyed the familiar postulates and rules, that had traditionally been considered necessary for numbers and figures. Yet these new theories seemed consistent: scrutinizing the arguments within the theories revealed no contradictions. The best-known examples of such theories are the non-Euclidean geometries, which were developed from the 1830s. As I mentioned in the last Section, the consistency of these geometries, and the fact that the world and our experience of it could be described by them, spelled trouble for Kant's view that the propositions of Euclidean geometry are synthetic _a priori._

In addition to the new geometries, there were also new notions of quantity different from the familiar real numbers, which obeyed strange rules. For example, both Hamilton and Grassmann introduced (in two different ways) theories in which multiplication of numbers was not commutative, i.e. did not obey the rule that x times y = y times x. So these new theories were, to put it mildly, unintuitive. The new ideas of quantity underlying the strange rules were hard to understand; and indeed, hard to accept as correct mathematics. To do so, mathematicians needed to adopt, and did adopt, an abstract and formal approach. The idea was similar to that for geometry: "Just follow the postulated rules, and you will see they they lead to a novel, but consistent and even elegant, algebra". And the claim of consistency, the reassurance, could be secured more easily if the theory was written in a formal language, with its precise rules of grammar, and of inference.

Here I should stress that until these new theories were proposed, the only mathematical theory that had been conceived as an axiomatized theory was Euclidean geometry. And if one adopted these new nineteenth-century standards of 'formal' and 'rigorous', then the venerable textbook that had been used for two millennia, Euclid's _Elements,_ was very informal and unrigorous---whether written in English or in Latin. Accordingly, mathematicians developed axiomatisations of Euclidean geometry in the modern style. Thus their formal languages had very small vocabularies. For example, the language might have just four basic predicates, such as '… is a point', '… is a line', '… lies on …' (ascribed to a point and a line), and '… is between … and …' (ascribed to three points). From a few axioms making claims, using only this tiny set of predicates, about points and lines, all the hundreds of theorems of Euclidean geometry would follow.

The second development that prompted formalization and axiomatization arose from another new theory from the late nineteenth-century, that led to paradox: to another house built on sand. Namely, mathematicians developed the theory of _sets_. Its basic ideas are simple, and nowadays familiar from school mathematics; e.g. the intersection of two sets of objects is the set of those objects that are elements of both the given sets. But considering infinite sets, i.e. set with infinitely many elements led, like the calculus had done earlier, to paradoxes. Again, one could construct apparently valid arguments, with plausible premises about infinite sets, whose conclusions were contradictions.

Taken together, these various problems---about calculus, about the new theories of geometry and algebra, about sets---amounted to a crisis in the foundations of mathematics. In response, over the years 1870 to 1930, various different repairs were proposed. That is, several mathematical research programmes were launched, with distinctive proposals about how to rigorize, and thereby vindicate as free of paradoxes, all these mathematical fields: the real numbers, the new geometries and algebras, the theory of sets. Thus ensued a vigorous multi-faceted debate, that lasted some sixty years.

And in all this, the role of logic was second to none. This was not just because diagnosing the errors in paradoxical arguments is obviously a job for logic. Also, paradoxes apart, the effort to rigorize proofs was a matter of breaking them down in to simpler steps that can be explicitly checked as conforming to some announced rule of inference: clearly, a matter of logic.

Furthermore, there were deep similarities between logic and set theory. Since arguments can be about anything, and sets can be made up of any objects, both fields seem to have no specific subject-matter. In philosophical jargon, they are topic-neutral. And the truths of the two fields seem similar, or even the same. For example, recall the valid argument pattern I mentioned: 'All As are Bs, and all Bs are Cs; therefore all As are Cs'. That corresponds exactly to the truth of set theory that if a set A is a subset of a set B, and B is a subset of another set C, then A is a subset of C.

Indeed, these similarities inspired one of the research programmes mentioned above. The great German logician Frege created the logicism programme. He proposed that all of pure mathematics was really logic. This proposal promises a ready explanation of why pure mathematics is necessary. For Frege took logic to be a body of necessarily true propositions. We saw the idea here near the end of Section 7: it is that to every valid argument---say: Premise: P1, Premise: P2; Therefore, Conclusion: C---there corresponds a necessary proposition, viz: 'If P1 and P2, then C'. Thus where Aristotle and countless logicians after him saw the study of logic as the investigation of valid arguments, Frege took it as the production of necessary truths. So if mathematics is "just" a part of logic, so understood, it is guaranteed to be necessary.

So Frege endeavoured to show that the necessary propositions of logic are the axioms and theorems of a formal axiomatic system; and then that from this system, all of pure mathematics, e.g. arithmetic, the calculus, geometry etc., could be derived merely by adding appropriate definitions of the various mathematical symbols (like numerals). This came to be called the reduction of mathematics to logic (or reducing mathematics to logic).

Logicism was enormously influential in philosophy from 1900 to 1930, partly through the writings of Russell and Whitehead, and later, the logical positivists. To cut a long story short: the details of this turned out to depend on writing pure mathematics in terms of a (paradox-free) theory of sets, and then arguing that this theory of sets is really logic in disguise. Broadly speaking, the first aim was achieved: all of pure mathematics could indeed be presented in terms of sets, and nowadays most textbooks proceed in this way. But Frege and the other logicists failed in their second, philosophical, aim, i.e. showing that theory of sets is really logic in disguise. Nowadays, the consensus is that after all, the theory of sets is not really logic; and so logicism failed.

But for our purposes, what matters is not so much the failure of logicism, as its historical role and its legacy. In this Section, we have sketched how it arose from the applied/pure mathematics distinction, and the concurrent crisis in the foundations of mathematics. In the next Chapter, we will see its legacy: placing logic, and so the nature of logical necessity, at the centre of philosophy.

Chapter 2: Notes and Further Reading
For the topics of this Chapter, the best reading list is of course the original masterpieces themselves. Though daunting, one should at least dip into them. For Newton, Hume, Kant and Frege, I suggest the following.

I. Newton, The Principia: Mathematical Principles of Natural Philosophy; translated by I.B. Cohen and A Whitman; with a Guide by I.B. Cohen; University of California Press 1999. (In Section 6 above, I quoted this translation of Newton's General Scholium; which is discussed in Chapter 9 of Cohen's Guide.)

D. Hume's Treatise of Human Nature (1739) and Enquiry concerning Human Understanding (1748) are available in many editions. For example, the Treatise is published by Penguin (1969), edited by E. Mossner; and the Enquiry by Oxford University Press (1894 onwards), edited by L. Selby-Bigge. The pre-eminent passages for Hume's discussion of causation and inductive inference (Section 5 above) are: the Treatise, Book 1: Part I, Sections 1, 4 and 5; and Part III, Sections 12 and 14; the Enquiry, Sections 3 to 7.

I. Kant's Critique of Pure Reason (1781; second edition 1787) is available in English in many editions. The pre-eminent passages for his claim that geometry and arithmetic consist of synthetic a priori propositions are in: The Preface, Introduction, and the Transcendental Aesthetic.

G. Frege's logicism is best approached through his Foundations of Arithmetic (1884), which is available in many editions, and translated into English by J.L. Austin (Blackwell, Oxford UK: 1950 onwards). Frege's proposed definition of the numbers 0, 1, 2 and all the positive whole numbers, in terms of logic is in the final Part IV; (after a critique of previous authors' accounts of arithmetic, including Kant).

On the internet, there are of course editions, sometimes definitive, of these works and many others by these Maestri. The first three are especially well served; Frege less so. I recommend:

For Newton: The Newton Project, which is at: https://www.newtonproject.ox.ac.uk
For Hume: Hume Texts Online, which is at: https://davidhume.org
For Kant: The Gutenberg Project, which is at:
https://www.gutenberg.org/files/59023/59023-h/59023-h.htm
For Frege: The Foundations of Arithmetic is online, in German, at the Gutenberg Project, namely at: https://www.gutenberg.org/ebooks/48312. Also, a selection of his philosophical writings (ed. P. Geach and M. Black: 1960) is on the Internet Archive at https://archive.org/details/the-philosophical-writings-of-gottlob-frege. But a very complete selection of his writings, including extracts from The Foundations of Arithmetic, is The Frege Reader, ed. M. Beaney, Wiley-Blackwell 1997.

For secondary reading, there are excellent entries about all the topics of this Chapter in the internet resources, such as The Stanford Encyclopedia of Philosophy, suggested in the Notes and Further Reading for Chapter 1.

Among these excellent entries are many about natural philosophy, and even the history of physics, in its philosophical aspects. For example, the Newton scholar G. Smith has an entry on Newton's Principia. It is at: https://plato.stanford.edu/entries/newton-principia/

A superb survey of the role of philosophical ideas in the historical development of physics (which also covers twentieth century physics, especially relativity and quantum theory) is: J. Cushing, Philosophical Concepts in Physics (Cambridge University Press, 1998; online 2012). Available at: https://www.cambridge.org/core/books/philosophical-concepts-in-physics/F285F13FE71F225BD8BE01F754F8C2E5

I also recommend the British Academy's Dawes Hicks series in the history of philosophy (both Lectures and Symposia). They are all available online at:
https://www.thebritishacademy.ac.uk/events/lectures/listings/dawes-hicks-lectures-philosophy/
For this Chapter, I especially recommend:
(i) the lecture by I. Hacking, Leibniz and Descartes: Proof and Eternal Truths (1973); which is at: https://www.thebritishacademy.ac.uk/documents/2191/59p175.pdf ; and
(ii) the Symposium Mathematics and Necessity: Essays in the History of Philosophy, which is three essays by J. Bennett, M. Burnyeat and I. Hacking (ed. T. Smiley): which is at: https://www.thebritishacademy.ac.uk/publishing/proceedings-british-academy/103/

History apart, the justification of induction and the nature of inductive inference, remain central topics in philosophy of science. And about these, J. Norton's two excellent books, The Material Theory of Induction, and The Large-Scale Structure of Inductive Inference, represent the current state-of-the-art. They are both available in the British Society for Philosophy of Science Open Access series at: https://press.ucalgary.ca/series/bsps-open/

Beyond this, I will here just give a bit more detail about three of this Chapter's themes, as follows. (1): My interpretation of Hume (my Sections 4 and 5). (2): The development of physics from 1600 to 1900 (my Sections 1, 2, 3 and 6). (3): The nineteenth-century development of logic (Section 7 and 8).

(1): In Sections 4 and 5, I reported Hume's critique of the cognitive optimism of his predecessors, such as Descartes. I should add here that although my account is part of an interpretation of Hume that is widely endorsed, there is a rival interpretation.

The difference, in short, is between: (i) arguing that we can know about some problematic concept X, because once we analyse X carefully, we see that there is less to know than we first thought---and so no problem; and (ii) arguing that indeed we cannot know about X, while still saying we do have the original concept X.

In reading Hume, the main example of X one needs to consider is the concept of causation. Thus (i) becomes his view that once we analyse our concept of causation, we see that it is really the concept of the cause and effect invariably accompanying one another (simultaneously or one soon after the other). That is: the property by which we specify the cause, e.g. this object being bread, is invariably accompanied by the property by which we specify the effect, e.g. this object being nourishing. In Hume's famous phrase, causation is <u>constant conjunction</u>.

There is no doubt that much of Hume's writing supports this interpretation, (i). For he maintains in general that the analysis of a concept requires tracing its origin in our experience; (in his jargon: the analysis of an <u>idea</u> requires tracing the <u>impressions</u> from which it originated). And then he argues at length that our idea of causation can be traced back to our experience of constant conjunction. There is no more to it than that. We have no experience of, nor insight into, a necessary---in particular, a deductive---link from cause to effect.

So in the bread example: not only do we believe bread causes nourishment---in ordinary language: 'this piece of bread will be nourishing'---because of our previous experiences. This 'because' claim is naturally understood as a surely uncontentious causal claim about our psychology. Hume is also arguing that the <u>content</u> of our belief is only that being bread and being nourishing accompany one another. There is no more to the content of the belief than that. (Again, the example can be varied without affecting the issues. The accompaniment can be either simultaneous or soon-thereafter; the belief can be either about this new piece of bread, not yet tasted, or about bread in general.) In other words, we have no concept of "necessitation-in-nature": a sort of ontological "oomph", by which the cause "forces" the effect to occur---as alleged by Hume's predecessors, such as Descartes.

As I said, this interpretation of Hume is widely endorsed. One persuasive statement of it which makes the connection with Newton (my Sections 3 and 6), and is connected with my later theme (Chapter 4) of probability, is: I. Hacking, <u>The Emergence of Probability</u> (Cambridge University Press, 1975), Chapter 5. For more detailed support of this interpretation of Hume, I especially recommend: E. Craig, <u>The Mind of God and the Works of Man</u> (Oxford University Press, 1987), Chapters 2 and 3. This book is a superb overview of philosophy from 1600 to the present day. As its title hints, it articulates two dominant philosophical themes: the first about knowledge, and the second about action. The first, from 1600 till about 1800, is that we finite creatures can, and should, aspire to know nature, or rather parts of it, with the full understanding that God enjoys for all of nature (cf. my Section 4). The second, from 1800 till now, is that we ('Man') are, not passive knowers of nature, but active agents in it, imposing our will on it. The transition between these themes is, very neatly, Kant. For his doctrine that our cognitive constitution imposes structure on raw experience (cf. my Section 7) keeps the first theme's stress on knowledge but adds the idea of the human mind as active.

But agreed, there are passages in Hume's texts that suggest the interpretation which I labelled (ii), again with X taken as the concept of causation. That is: some passages say---or seem to say---that we <u>do</u> have a concept of necessitation-in-nature, of ontological "oomph"; but that nevertheless, we cannot know anything about how it "works". A recent full defence of this sort of interpretation is: G. Strawson, <u>The Secret Connexion: causation, realism and David Hume</u> (Oxford University Press, 1989, revised edition 2014.)

(2): In treating physics from 1600 to 1900, this Chapter has focused solely on mechanics, especially as applied to astronomy (Sections 1, 2, 3 and 6). I shall now: (a) add some details about mechanics; (b) mention other fields of physics; and (c) return to the topic at the end of Section 6, about physics after 1700 having forsaken intelligibility in senses additional to forsaking contact-action (i.e. to requiring action-at-a-distance, as Newton's gravity did).

Although the references in (2a) to (2c) are about the history of physics, I choose them for their emphasis on philosophical issues.

(2a): A magisterial study of theories of motion, from Aristotle to Newton and beyond, and focused on the contrast between "absolute" and "relative" conceptions of motion, and is: J. Barbour, <u>Absolute or Relative Motion?</u> (Cambridge University Press, 1989). Barbour favours the relative conception, following in the spirit of Ernst Mach (mentioned at the end of Section 3). The book was reprinted by Oxford University Press in 2001, entitled '<u>The Discovery of Dynamics</u>'. It is available at; https://academic.oup.com/book/54639.

An excellent popular book describing physics' changing conceptions of the vacuum, not just from 1600 to 1900, but also in contemporary physics, is: J. Weatherall, <u>Void: the strange physics of nothing</u> (Yale University Press, 2016). It gives many philosophically important references, and is on the Internet Archive at: https://archive.org/details/voidstrangephysi0000weat_u9b4.

For details of Newton himself, two excellent biographies are: R. Westfall <u>Never at Rest</u> (Cambridge University Press, 1980), and A.R. Hall, <u>Isaac Newton: adventurer in thought</u> (Cambridge University Press, 1992). In this Press' <u>Companion</u> series (usually for philosophers), <u>The Cambridge Companion to Newton</u> (ed. I.B. Cohen and G. Smith, 2002), which is available at: https://www.cambridge.org/core/books/cambridge-companion-to-newton/B92293E01C97D041CA42B30396E2EA22 ; while Oxford University Press has <u>The Oxford Handbook of Newton</u> (ed. E. Schliesser and C. Smeenk, 2017), which is available at: https://academic.oup.com/edited-volume/34749.

(2b): M. Hesse, <u>Forces and Fields: the concept of action-at-a-distance in the history of physics</u> (Philosophical Library, London, 1961) is a fine overview of the struggles from 1700 onwards to accept action-at-a-distance, as in Newton's theory of gravity, and its gradual replacement, from 1850 onwards, by the concept of an all-pervasive field, as in Maxwell's electromagnetism concept (discussed in Chapter 4). It is on the Internet Archive at: https://archive.org/details/forcesfieldsconc0000hess.
P. Harman, <u>Energy, Force and Matter</u> (Cambridge University Press, 1982) is a general history of nineteenth-century physics. It is available at: https://www.cambridge.org/core/books/energy-force-and-matter/00A35E995E821EEF4A20A7AE1D37202F.
<u>The Oxford Handbook of the History of Physics</u> (ed. J Buchwald and R. Fox: Oxford University Press, 2013) is a fine anthology whose Parts I, II and III cover the period from 1600 to 1900. It is available at: https://academic.oup.com/edited-volume/38638

(2c): Section 6 ended by saying that physics after 1700 had forsaken intelligibility in senses additional to forsaking contact-action (i.e. additional to requiring action-at-a-distance, as Newton's gravity did). I gave the example of point-particles (introduced ca. 1750), and said that quantum theory (Chapter 4) will add more. There are two further points worth making here.

First: in the early days of quantum theory, it was much debated whether a physical theory should be visualizable (in German: <u>anschaulich</u>), in the way that Schroedinger's version of quantum theory (wave mechanics) seemed to be and Heisenberg's version (matrix mechanics) was not.

Second, recently philosophers of science have sought a general account of intelligibility of a theory (usually, under the label 'understanding'): accounts that usually do not require visualizability, nor subsume understanding as just as aspect or result of having a scientific explanation (which is Chapter 5's topic). In this trend, a good and influential book is H. De Regt <u>Understanding Scientific Understanding</u> (Oxford University Press, 2017): which discusses nineteenth-century mechanical models, as well as Newton's action-at-a-distance, and the failure of <u>Anschaulichkeit</u> in quantum theory. It is available at: https://academic.oup.com/book/36363.

(3): About the nineteenth-century development of logic, and of rigour and formalization in pure mathematics,(Section 7 and 8), there is an enormous literature; and excellent coverage in Philosophy curricula and internet resources like <u>The Stanford Encyclopedia of Philosophy</u>. So I will be brief (also because the next Chapter will give ample references to logic). Thus apart from Frege (above) I recommend: (a) two superb overviews, and (b) two superb specialist books.

(3a): M. Kline's 1200-page book, <u>Mathematical Thought from Ancient to Modern Times</u> (Oxford University Press, 1972) is now available very conveniently as three paperbacks. For our topics, Chapters 36, 37, 40-43, and 51 are relevant. It is available at: https://global.oup.com/academic/product/mathematical-thought-from-ancient-to-modern-times-9780195061376?lang=en&cc=gb

Much more recent (and emphasising logic, rather than mathematics) is: M. Potter, <u>The Rise of Analytic Philosophy, 1879-1930</u> (Routledge, 2020). As the dates (and the sub-title, 'From Frege to Ramsey') hint, this book also discusses not just Frege, but also Russell Wittgenstein and Ramsey (in roughly equal measure). It is available at: https://www.taylorfrancis.com/books/mono/10.4324/9781315776187/rise-analytic-philosophy-1879–1930-michael-potter

(3b): J. Alberto Coffa, <u>The Semantic Tradition from Kant to Carnap: to the Vienna station</u> (Cambridge University Press 1991) is a deep study of the origins of logical positivism. (So its details about German and Austrian philosophers apart from Frege make a good complement to the books in (3a).) It is available at:
https://www.cambridge.org/core/books/semantic-tradition-from-kant-to-carnap/E448B2413A076ED2275A87C87811D419

For philosophers of physics, geometry provides a central "sub-plot" in the story of increasing rigour and formalization, from 1850 onwards. A deep study of this is: R. Torretti, <u>The Philosophy of Geometry from Riemann to Poincare</u> (Springer, 1978). It is on the Internet Archive at:
https://archive.org/details/philosophy-of-geometry-from-riemann-to-poincare-roberto-torretti#:~:text=It%20is%20a%20technical%20but,physics%20%26%20mathematics%20and%20its%20applications

# Chapter 3: All the logically possible worlds

In this Chapter, I will proceed in four stages. At the end of the last Chapter, I said that although logicism failed, it was a major reason why early twentieth-century philosophy placed logic centre-stage. Since then, logic has remained central, and this Chapter's first stage (Sections 1 and 2) will be to present some details about this. We will then be ready to discuss the philosophers' multiverse, in three further stages.

In the second stage (Section 3), I will urge that in everyday life, and technical science, and philosophy, we are up to our necks in <u>modality</u>. This word is philosophers' jargon for the topic of necessity, possibility and impossibility. That is: in order to state what we believe to be true, whether in everyday life or in technical science, we need to accept non-actual possibilities. Once we see this, it becomes clear how, by about 1970, philosophy was ripe for the proposal that there is a multiverse of all the <u>logically possible worlds</u>.

In the third stage (Sections 4 to 8), I will sketch some of the benefits for philosophy, of adopting an explicit framework of a set of possibilities. The prototypical example of such a set---a cautious prototype, in Chapter 1's spectrum of attitudes---is the set of <u>instantaneous possible states</u> of some physical system, as postulated by some physical theory. As I shall explain, this set is called the <u>state-space</u> of the theory. But more ambitiously (agreed, much more ambitiously): one might accept maximally specific possibilities for the cosmos as a whole. These are the <u>possible worlds</u>. So one envisages a set W of all the possible worlds. The exact nature or status of these worlds thus becomes this Chapter's main concern.

One can take either a cautious or a confident attitude to them. The most confident attitude says: 'They are all equally real; the non-actual worlds are merely not "hereabouts", in much the same way that for a person in England, all the other countries e.g. France and Australia, are equally real, but merely not hereabouts.' Agreed, that is hard to believe. And indeed: almost no philosopher does believe it. But the great philosopher David Lewis, who thought hard and deeply about possible worlds, believed it. The doctrine is called <u>modal realism</u>.

Lewis argued for it at length; especially in his book, <u>On the Plurality of Worlds</u> (1986). He did not claim to have a knock-down i.e. irrefutable argument. As we discussed in Chapter 1, in philosophy such arguments cannot be expected. Rather, he argued that modal realism was on balance better than the rival, cautious, conceptions of possible worlds.

But he also agreed that most of the philosophical benefits of using a set of logically possible worlds do not require his modal realism. They can also be had while adopting much more cautious conceptions of what the worlds are.

So in the third stage, I will show how various philosophically important concepts and doctrines can be made precise in terms of the framework of possible worlds. There are many such concepts and doctrines. But I will restrict my examples to ones we will need in later Chapters.

Finally in the fourth stage (Sections 9 and 10), I turn to the outstanding question: What exactly is a possible world? This question is compulsory, for cautious conceptions of possible worlds as much as confident conceptions, in particular Lewis' modal realism. Several possible answers are defended in the philosophical literature. But to avoid anti-climax, I announce now, at the outset, that I will not settle on one answer. So the Chapter will end inconclusively, and perhaps disappointingly. For I will leave this question hanging, without endorsing any answer. But there is some consolation: the following Chapters will not depend on my having endorsed an answer. Besides, the next Chapter might help. For it will suggest a new answer, derived from quantum physics.

A final preliminary. There is another perspective on the material in the second to fourth stages of this Chapter. For the most part, I will not articulate it, since it will be clear what one would say about it. But I will be explicit at the start of the fourth stage (Section 9). In short, this other perspective focusses on the idea of a proposition rather than, as I will, on possibility. It will be clear that this difference is largely a matter of jargon, reflecting the fact that 'proposition' (and similar words one might use, like 'statement') are really terms of art, to be defined by the logician or philosopher as they see fit. So for the most part, when I talk about a possibility, or about a possible world, one could instead talk about a proposition, or (corresponding to a possible world) about a maximally specifically i.e. logically strongest proposition. But as I said: more details in the fourth stage.

Chapter 3, Section 1: The legacy of logicism: the endeavour of reduction
Since 1900 logic has been central to philosophy, in two main ways: which I take up in this Section and the next. The first way amounts to the legacy of logicism. Although logicism failed because (as we discussed at the end of Chapter 2) set theory is not really the same as logic, logicism nevertheless engendered two broader visions which have persisted. They both involve the idea of reduction; and they are the topic of this Section. I discuss them in (1) and (2) respectively.

    The first vision is about pure mathematics. And it has not merely persisted after the demise of logicism. This vision has been, in effect, proven to be true, through detailed work by various mathematicians from about 1890 to 1920. The second vision, discussed in (2), is about philosophy, especially about what the task (or at least, one task) of philosophy should be. Philosophy being controversial, this second vision remains of course unproven.

(1): The first vision is easily stated. As logicism developed, it became clearer that its task, of proving that all of pure mathematics was really logic, amounted to two sub-tasks: first, show that all of pure mathematics, e.g. arithmetic, the calculus, geometry etc., can be written in terms of a (paradox-free) theory of sets; second, show that this theory of sets is really logic in disguise. So even if---as agreed---we cannot do the second sub-task, i.e. even if set theory is not logic, we can still complete the first. And this was indeed achieved, by the collective work of various mathematicians.

    Thus by about 1910, there was a vision, endorsed by many opponents of logicism as well as by its advocates, that set theory is a universal framework in which to formulate all of pure mathematics. More precisely, the vision says: a paradox-free set theory adequate for formulating all mathematics can be written in a formal language, with precise vocabulary, rules of grammar, and of inference (as discussed at the end of Chapter 2).

    Indeed, the requisite formal language is very simple. It has exactly one basic predicate, representing the relation of set-membership. This is always written with the Greek letter epsilon, $\varepsilon$. So in set theory, 'x $\varepsilon$ y' means that x (which may itself be a set) is an element of the set y.

    Besides, the rules of grammar and the rules of inference were also very simple. They were the rules proposed for predicate logic: which had been invented by Frege in 1879. Here, 'predicate logic' comprises the logical behaviour of both (i) 'and', 'or' and 'not' (called 'propositional logic', or 'Boolean logic') and (ii) 'every' (similarly: 'any', 'all'), 'some' and 'none'. (Here, 'or' is understood inclusively, as synonymous with 'and-or'. So 'Bill is tall or blond' is true if Bill is both tall and blond.)

    Thus predicate logic is concerned with valid patterns of argument whose validity turns on the placing of these words within the argument. Here are two examples, (1) and (2); example (2) also uses some propositional logic.
        (1): Premise: 'Some As are Bs'. Premise: 'All Bs are Cs'.
        So, Conclusion: 'Some As are Cs'.

(2): Premise: 'Some As are Bs'. Premise: 'All Bs are Cs or Ds' (meaning: 'any B is a C or is a D'; not 'all Bs are Cs, or all Bs are Ds').

So, Conclusion: 'Some As are Cs or Ds' (where 'or' is again understood as inclusive).

Thus the vision had three parts. The first part is about set theory; the second about pure mathematics (apart from set theory); and the third part about how to show that the second part can be understood as included in the first part: as follows. I will label the parts (A), (B) and (C).

(A): There is a paradox-free formulation of set theory in a formal language with just one basic predicate, '… ε …', representing set-membership; and whose rules of grammar, and rules of inference, are just those of predicate logic. So in this language, the only allowed inferences are those that depend on the words listed in (i) and (ii) above, like my examples (1) and (2). Indeed this formulation of set theory is an axiomatization. All the theorems, all the truths of set theory to be appealed to, follow by these allowed inferences from a few initial axioms.

(B): Take all the accepted truths of pure mathematics, apart from set theory: the truths of arithmetic, of the calculus, and of geometry and the other traditional areas of mathematics. Here, 'accepted truths' means: claims accepted as proved by mathematicians. Of course, the usual formulations of these truths, in textbooks etc., are enormously varied, in that the different areas have their own special vocabularies. Arithmetic has the numerals '1', '2',…, the ratios (rational numbers), '2/5', '42/9', …, the signs + and × for addition and multiplication. Geometry has nouns for geometric objects, e.g. 'point', 'line', 'triangle', and predicates for relations between them, e.g. 'intersects', 'is perpendicular to'. And so on, for other areas of mathematics.

(C): Despite the sparse simplicity in (A), and the variety and complexity in (B), it is possible to give an explicit definition of each of the special vocabulary items, in each of the many areas of mathematics in (B), in terms of sets, in such a way that: once we add these definitions to the sparse and simple set theory (A), each of the claims of (B)---now understood, using the added explicit definitions, as claims about certain sets---can be derived within (A), using only (A)'s strictly limited rules of inference.

Here of course, (C) is the punch-line. It offers you re-interpretations of your traditional familiar mathematical words, e.g. the numerals '1', '2',…, '2/5',… 'intersects', 'is perpendicular to', in such a way that all the mathematical claims you accept, if thus re-interpreted, follow, by simple and compelling rules of inference, about 'and', 'or', 'all, 'some' etc., from the axioms of a simple and compelling set theory. In short: (C) shows a way to interpret (B), i.e. the truths in (B), as really "already there" in (A).

Philosophers and logicians call this a <u>reduction</u> of (B) to (A). So (C) is the claim that each of the traditional areas of mathematics (and so also: the grand conjunction of all their accepted claims) can be <u>reduced</u> to set theory. So set theory is called the <u>reduction-basis</u>.

Of course, the definitions offered of the traditional familiar words must be judiciously chosen. For if you define these words in terms of sets wholly at random, it will only be by the greatest coincidence that your beloved mathematical truths, e.g. '2+2=4', 'there are infinitely many primes', 'all equilateral triangles are equiangular', turn out to be theorems of set theory. Very probably, your haphazard definitions will render these claims as false statements of set theory; or even as not a grammatical sentence about sets at all.

On the other hand, needing to choose judiciously does not mean that there is only one choice that would work. For example, there is a great variety in which set to choose as the interpretation of the numeral '1'. But having made a choice, your choices for the other numerals, '2', '3'…, and so for other number-expressions like '2/5' etc. for the rationals, are heavily constrained. For they need to "align" or "mesh" with your choice for the numeral '1', if your accepted truths are to follow as theorems of set theory.

So let me sum up this vision. It is (C) that was achieved---proven true---by various mathematicians from about 1890 to 1920. It is a very remarkable achievement. Indeed, it is undoubtedly one of the greatest transformations in the entire history of mathematical thought.

Nowadays, this achievement is, as the saying goes, hidden in plain sight. Both research articles and pedagogic writings (textbooks) usually start by invoking the framework of set theory (almost always informally, without mentioning axiomatization), and then proceed informally, in natural language augmented with mathematical symbols. They never mention that the proofs of all the text's theorems can be formulated without loss, using the very limited rules of inference endorsed by the predicate logic.

(Of course, becoming hidden in plain sight is often the fate of major changes. They become ubiquitous, entrenched---and unnoticed. Another example in the history of mathematics is the adoption of Arabic in place of Roman numerals. The advantages for addition and multiplication are so great that we hardly ever think of adding or multiplying Roman numerals, and so we forget how cumbersome it would be.)

But in the early twentieth century, this vision, and its achievement, had a large impact on philosophy. It led to what at the start of this Section I labelled as 'the second vision' bequeathed by logicism: a vision about what the task (or at least, one task) of philosophy should be. So to this, I now turn.

(2): As discussed in Chapter 1, much of philosophy has throughout the centuries been about "conceptual house-keeping". That is: scrutinizing concepts to see if they are in order, and if so, giving an account or even an analysis of them; (and if they are misleading, rejecting or maybe revising them). One even sees this at the beginning of Western philosophy, in Plato. Socrates besets the people whom he accosts in the <u>agora</u> (market-place), with requests for definitions (analyses) of virtue, courage etc. And much of philosophy since---about many diverse topics, such as virtue, free will, knowledge, causation, number or necessity---can be read as aiming to give an account of the concept in question; and maybe even an analysis of it.

Here, 'giving an account' means describing how the concept relates to other kindred concepts (e.g. one implies the other, or one tends to cause the other); and stating what are the important accepted truths involving the concept (and of course, kindred concepts). And 'giving an analysis' means something more specific and ambitious: defining the concept in terms of previously understood concepts (and so displaying their logical connections), in such a way as to recover the accepted truths involving the concepts. And here, 'to recover' means, ideally at least: to derive, i.e. deduce, from other accepted truths invoking the previously understood concepts.

Thus we return, in the more general context of philosophy, to the above idea of <u>reduction</u>. If the scrutinized concept or concepts are considered to be in order, then we can aim, ideally, to <u>deduce</u> the accepted truths invoking them, viz. (B) in the above labelling, by adding to a previously understood and accepted body of doctrine (A), some judiciously chosen definitions, analyses, of (B)'s concepts in terms of (A)'s.

The second vision is now clear. Seeing mathematicians' achievement of reducing all of traditional pure mathematics to the sparse and simple framework of set theory and predicate logic, philosophers conceived the task of similarly reducing accepted bodies of doctrine about other matters: in particular physical theories, or even everyday propositions about the empirical world.

Of course, philosophers differed about the details of the proposed task. Russell with the programme (ca. 1910 to 1920) that he called 'logical atomism' proposed to analyse all our everyday empirical knowledge, as did Carnap with his <u>Aufbau</u> programme (1928). But their contemporary Reichenbach aimed in his 1920s work "only" to axiomatize Einstein's relativity theories. But these programmes had much in common. In particular, they agreed on the answer to the immediate question, 'What is the previously understood and accepted body of doctrine to which you propose a reduction should be made?'. Namely, a staunchly empiricist answer: propositions about sensory experience.

Thus the programmes of Russell's logical atomism, and somewhat later, the logical empiricism of Carnap, Reichenbach and others in Vienna and Berlin, should be seen as modelled on the successful set-theoretic (though not logicist) reduction of pure mathematics.

Chapter 3, Section 2: Logic as a toolbox of formal systems: modal logics
I will describe in (1) how programmes like Russell's and Carnap's led eventually to a modest conception of logic's role in philosophy; and then in (2) consider modal logic.

(1): Clearly, the reduction programmes of Russell and Carnap were very ambitious. Everyday empirical knowledge is a vast open sea. It far outstrips a single knowing mind; its content shades continuously into technical science; and we have no agreed chart for it, i.e. no agreed taxonomy breaking it down into parts appropriately (e.g. logically) related to one another. Besides, we have no agreed language in which to talk about the reduction-basis, i.e. sensory experience. In my jargon above: there is no uncontroversial 'previously understood and accepted body of doctrine'. So unsurprisingly, these programmes failed. As the Bible warns us: pride comes before a fall (Proverbs 16:18).

But programmes with a much more modest aim---for example, axiomatizing a single physical (not: pure mathematical) theory, using predicate logic and a basic vocabulary that was small, but not required to be solely about sensory experience---fared much better. A single physical theory, such as Newton's theory of gravity or Einstein's special relativity, is pretty well-defined. The textbooks largely agree in how they present it to us, and in what its special vocabulary is. And in axiomatizing it we do not need to reach for some other vocabulary, e.g. solely about sensory experience, and for some doctrine using that vocabulary, to serve as a reduction-basis. Rather, the axioms we seek will be the reduction-basis. Nor was it just philosophers like Reichenbach who undertook such efforts. Mathematicians, including great ones like Hilbert and von Neumann, did so too.

Thus arose a more modest and flexible conception of the role of logic in philosophy, which has persisted till today. Namely, as a resource, a toolbox, for formalizing various bodies of doctrine, without necessarily axiomatizing them or reducing them to another body of doctrine. Of course, the bodies of doctrine are to be chosen because of their philosophical interest. They use concepts central to everyday life and thought (like my list above: virtue, free will, knowledge etc.) and-or science (like space, time, matter, causation). And so this conception goes along with philosophers' traditional endeavour of conceptual analysis.

Nowadays, there are countless such examples of "logic in action". (Indeed there are even logics of action, as well as logics of concepts that seem more amenable than action, to a logical treatment---such as knowledge.) We already saw one example of this in Chapter 2. It was about what it is rational to believe---what principles should govern what we believe?---in addition to the indisputable requirement that we should believe the deductive consequences of what we already believe. Thus since the mid-twentieth century, philosophers have developed formal systems prescribing how to change your beliefs when you get evidence (often called 'inductive logics').

For the purposes of this Chapter, the most important example is of course: the logic of modality. (Recall that 'modality' is jargon for the topic of necessity, possibility and impossibility.)

(2): Aristotle himself initiated modal logic, by discussing such principles as that necessity implies truth. That is: if a proposition is necessary (must be true), then it is in fact true. And similarly, truth implies possibility: if a proposition is in fact true, then it is possibly true. For the actual situation counts as one of the possibilities. (Here, we set aside the conventional rule of conversation whereby calling something 'possibly true' connotes that it is in fact false.)

The natural way to think of such principles is that the phrase 'It is necessary that …' has an empty slot or argument-place … into which a sentence 'P' can be inserted, to produce a sentence 'It is necessary that P'. So there is a valid argument: 'It is necessary that P; therefore P'. Similarly, 'P; therefore it is possible that P' is a valid argument. And as I mentioned in Chapter 2: to these valid arguments, there correspond conditional propositions that are themselves necessary. Namely: 'if it is necessary that P, then P'; and 'if P, then it is possible that P'.

Medieval logicians developed the logic of modality. But as we have seen in Chapter 2, philosophy in the modern period, i.e. from the seventeenth century, neglected logic up until the late nineteenth century. And then, although logicians like Frege and Russell took logic to be a collection of necessary truths, they showed no interest in studying the logic of modality, i.e. studying the logical behaviour of phrases like 'It is necessary that …', and 'It is possible that …'. Thus the logic of modality lay dormant until spear-headed in about 1915 by the philosopher, Clarence Lewis; (usually cited as 'C.I. Lewis': no relation of David Lewis---about whom, more shortly).

C.I. Lewis was the first person to write down formal logics of modality, called 'modal logics'. They build on the logics we noted in the previous Section. Thus recall that propositional logic comprises the logical behaviour of (i) 'and', 'or' and 'not'; while predicate logic adds to this the logical behaviour of (ii) 'every', 'some' and 'none'. C.I. Lewis proposed adding to any system of propositional logic a new symbol, which I will write as 'N(…)' (for 'necessary') which accepts a sentence 'P' in its argument-place … to make another sentence, 'N(P)': which we read as 'it is necessary that P'. (Beware: though I write 'N' for 'It is necessary that…'; it is traditional to write for this, either 'L' or a box: $\Box$ .)

It follows that 'It is possible that …' does not need a separate treatment. For recall that 'not' makes a sentence from a sentence: 'not-P' is true if P is false, and vice versa. (Any piece of language that makes a sentence from a sentence, like 'N(…)' and 'not', is called a sentence-operator.) So 'It is possible that P' can be rendered as 'not-necessarily-not-P'. That is: as 'not-(N(not-P))'. (But some expositions do use a separate symbol for 'It is possible that…'. For this, it is traditional to write either 'M' or a diamond: $\Diamond$ .)

So far, so straightforward. But building such a system of modal logic soon leads to interestingly controversial issues. For sentence-operators can be iterated. So what should we say about 'NN', in particular in relation to 'N'? One may well be content that the argument 'N(N(P)); therefore N(P)' is valid, whatever our choice of proposition P. (For it is itself an instance of our previous valid form: 'N(P); so P'. Into this valid form, one inserts 'N(P)' in place of 'P'.) But what about the converse argument, i.e. the argument: 'N(P); therefore N(N(P))'? Is this second form of argument valid, for all choices of P?

On such questions, C.I. Lewis himself took a liberal view. He developed various systems of modal logic, that obeyed various sets of principles, while sharing those I began with. Namely, the principles that 'It is necessary that P; therefore P' is a valid argument, and that 'It is possible that P' is rendered as 'not-necessarily-not-P'.

Matters become even more controversial when one considers how 'N' should behave in relation to the 'every', 'some' and 'none' of predicate logic. Thus suppose 'is F' is some predicate, e.g. 'is red' or 'is a horse'. And let us suppose that 'For every object, it is necessary that the object is F' is true. Does it follow---is it valid to infer---that 'It is necessary that for every object, it is F', i.e. 'It is necessary that every object is F'?

There is good, though I think not compelling, reason to deny this. For the premise is naturally read as about all actually existing objects: and as saying of each of them that it is necessarily F, i.e. that however the world happened to be, the object in question would be F. Notice here how natural it is to say 'world' i.e 'possible world'. But the conclusion is naturally read as: however the world happened to be, every object in that world would be F. So if we envisage that the world could contain objects that it actually does not contain, then the way is open to denying that the inference is valid. For we can admit the premise, that all actually existing objects must be F, but insist that there could be yet other objects: some of which, in some worlds, are not F.

On the other hand, this reason is not compelling. For it seems tenable that the actual world is "privileged" among all possible worlds, in being "the ultimate resource" for objects. That is: any object that possibly exists, actually exists. So the idea is: "no newcomers are allowed to come into view, as my mind's eye goes from the actual world to another possible world."

So the interplay between modality and the notion of object is controversial. And the controversies show up in questions about which principles combining the modal operators with the 'every' etc. of predicate logic we should accept. These controversies were pursued by C.I. Lewis and others (including Carnap) in the mid-twentieth century. They were also much clarified and enlivened in the 1960s by the work of David Lewis, Saul Kripke, David Kaplan and others: all of whom emphasized the semantics of modal logic. This semantics explicitly invoked possible worlds, and so made vivid the central question of this Chapter: what exactly are possible worlds? And as we shall see in Sections 5 to 7, this semantics also led to detailed proposals about the semantics of natural languages.

Furthermore, some questions central to philosophy turn on examples of principles of the above sort, i.e. principles combining modal operators with the 'every' etc. of predicate logic. In Section 8's discussion of materialism and physicalism, we will see an example of this which uses the principle I have just discussed: i.e. the principle that for a property F, 'if everything is necessarily F then it is necessary that everything is F'. In Section 8, the property F in question will be 'being material' (or in the jargon of philosophy: being concrete, as against abstract). As we will see, it is a problematic property.

This completes this Chapter's first stage: a summary of logic's role in philosophy up to about 1970, especially the development of modal logics. As I announced in the Chapter's Preamble, we are now ready to discuss the proposed multiverse of possible worlds, in three further stages.

Chapter 3, Section 3: Up to our necks in modality
In this Section---which is the second stage of the Chapter---I will argue that in order to state what we believe to be true, whether in everyday life or in technical science, we need to accept non-actual possibilities. Then the main question for the rest of the Chapter will of course be: exactly what does this commitment involve?

(1): Let us begin with our beliefs in everyday life. Consider some belief of yours that is true. It can be utterly mundane, e.g. that grass is green. Then the negation of what you believe, the proposition that grass is not green, is false. It represents a non-actual possibility: but what exactly is that?

There is a temptation to dismiss this question, saying that after all, grass could not fail to be green, thanks to its genetic make-up encoding that it produce chlorophyll, i.e. the green pigment essential for photosynthesis. But this dismissal is unconvincing. For suppose we agreed that grass must be green, and also that there is no need to accept the impossibility, grass not being green, as some sort of ghostly non-fact---we can take the impossibility to be nothing at all. Nevertheless, there are surely countless everyday propositions that are in fact false but could be true. Suppose I stay at home tonight: then the false proposition 'I go to the cinema tonight' surely could be true. (For this example, it does not matter whether I have free will, i.e. whether I could freely choose to go to the cinema. The example only needs that 'I go to the cinema' could be true.) So there is a way the world could be that makes this proposition true. So accepting that it might have been true commits us to such 'ways', i.e. to non-actual possibilities---in some sense. Besides, some of these propositions that are in fact false but could be true are among our beliefs---yours and mine. So we cannot duck out of the issue by just focusing on true beliefs. For any of our false beliefs that could be true has as its content, i.e. what it represents about the world, a non-actual possibility.

This discussion may seem suspiciously abstract. Let me make it more vivid by giving two main ways in which our beliefs invoke non-actual possibilities. The first way concerns deliberation and decision. Suppose a person hesitates between two options for action, deliberating which to do, and then does one. The options could, again, be utterly mundane: for example, which of two keys to try so as to unlock a door. We cannot understand the process of deliberation, what the person thinks, purely in terms of the one actual course of events that ends in, say, trying the bigger key. To explain the process of deliberation and the eventual action, we need to attribute to the person beliefs, some of which are about non-actual possibilities. For suppose the bigger key is the wrong one. So the proposition 'the bigger key fits' is false. It represents a non-actual possibility: but the person believes this proposition and acts on it.

Examples like the choice of key are the bread-and-butter of a discipline, decision theory, that lies at the interface of philosophy with economics and psychology. We shall meet decision theory again in the next Chapter: for it has a surprising application in support of Everettian quantum theory, i.e. the quantum multiverse. But for the moment, I will just state decision theory's general description of a deliberating person, so as to bring out its ubiquitous use of non-actual possibilities.

Decision theory assumes that a deliberator has:
  (i) various degrees of belief, i.e. subjective probabilities, about various possible states of the world, i.e. degrees of belief in propositions about the world;
  (ii) desires of various strengths that various such propositions be true; and
  (iii) a set of options for action, which are again taken as propositions---propositions that the person can at will make true (like trying the bigger key).
Decision theory then formulates principles that prescribe which option for action is best for the deliberator. A common idea of these principles is that the best option has the highest 'score'. Here, a 'score' is defined as the weighted-average strength of the desired propositions (ii), where the average is computed with degree-of-belief weights given by (i). (Of course, one should also allow for first-equal scores: then the best option is any of the options with the highest score.)

This common idea can be made precise in various ways, that in some cases disagree about which option they prescribe. But we need not discuss these disagreements and the ensuing controversies in decision theory. For us it is enough that, as the common idea shows: when a person decides and acts, they are up to their necks in modality.

Besides, this involvement with modality holds good for propositions about the past, just as much as for those about the future. For think of memory rather than decision-making. Suppose that yesterday I stayed home and learn today that I missed a good film at the cinema, and I regret not going. In such a case, my mind is again focused on a non-actual possibility.

Turning from decision-making and memory to technical science: it also is up to its neck in modality. Of course, decision theory itself counts as science. But let me stress examples in physics. For again, this will help us prepare for the next Chapter.

(2): Mention of subjective probabilities prompts an obvious suggestion. Namely, <u>chance</u>: Chances are objective probabilities that are made true by the subject-matter rather than by the state of mind of a person thinking about it. As I mentioned in Chapter 1, a standard example is radioactivity: e.g. the chance of this Uranium atom decaying in the next hour. Again, the very concept of chance commits one to non-actual possibilities, going beyond the one actual course of events. For chance requires a range of future alternatives: in my atom example, just two---atom decayed after an hour, and atom undecayed.

But even without probabilities, physics endemically invokes non-actual possibilities. This occurs in every physical theory: from the most elementary, such as Newtonian mechanics, to the most advanced, like quantum theory and general relativity. (And it occurs in speculative theories, like Chapter 5's cosmological theories and string theory, as much as in well-established theories.) To explain this, I will introduce some physics jargon, which will also be useful in later Chapters; and then consider the simplest theory, Newtonian mechanics (which is familiar from Chapter 2).

Any physical theory describes a certain kind of object by ascribing to it numerically measurable properties like position or momentum (i.e. mass times velocity) or energy. In the jargon of physics, the objects are called <u>systems</u>; their properties like position etc. are called <u>quantities</u> (also: 'magnitudes', but I will not use this word); and the amounts or degrees of such properties that are ascribed are called <u>values</u> (almost always real numbers). (In the jargon of philosophy, the quantities are <u>determinables,</u> and each of their values is a <u>determinate</u>. The standard philosophical example of a determinable is colour; of which a particular shade of scarlet is a determinate.) Then a <u>state</u> of a system, according to a physical theory that describes the system, is a list, or conjunction, stating what are the system's values for the various quantities that apply to it.

The state of course changes over time, as the values of the various quantities go up or down. So the state is also called the <u>instantaneous state</u>. A physical theory gives descriptions of these changes. In most theories (including all that this book will discuss), the theory provides an equation stating exactly how the state (the values of all the system's quantities) changes over time. This is the system's <u>equation of motion</u>. Typically, it fixes the rate of change of some chosen quantity (or quantities) of interest, as a function of the values of that quantity, and usually also other quantities, at some initial time. Given those other values, and thereby the rate of change of the chosen quantity, one then solves the equation so as to find the value of the chosen quantity at later times. In short, one predicts the future values of that quantity, on the basis of some present state.

Newtonian mechanics is of course the archetypal case. Imagine that a small solid object, say a sphere, is our system of interest. In Newtonian mechanics, this is usually called a 'body'. If we know the forces that are now, and that will later, be exerted on the sphere (say by other bodies, e.g. gravitational forces or electric forces), and we also know the sphere's present position and momentum: then the equation of motion for its position can be solved. That is: the position at later times (and so also the momentum at later times) can be calculated.

Agreed, two qualifications are needed. We already glimpsed the first, in Chapter 2. When two bodies collide, what happens is very complicated. They usually distort each other, or even break up, so that describing what happens often outstrips the resources of Newtonian mechanics---for example, because the collision generates heat. So let us set aside collisions: for example, by imagining the sphere is in empty space, a vacuum, and is far away from all other bodies.

Secondly, even apart from collisions, the sphere's motion can be influenced by motions internal to the sphere, for example if it is spinning or is not completely rigid. So (as I mentioned in Chapter 2) mechanics often idealizes the situation. We imagine the sphere is so small and rigid as to be effectively extensionless: a <u>point-particle</u> (also called a 'point-mass'). The instantaneous state of such a point-particle, sufficient for solving the equation of motion, is indeed just its position in space (so three real numbers, for its x-, y- and z-coordinates) and its momentum (again, three real numbers, for its mass times its speed in each of the x-, y- and z-directions). That is: the state of a point-particle is given by an ordered set of six real numbers: a 6-tuple. So for a mass m, and writing $v_x$ for the speed in the x-direction and so on, we could write this 6-tuple of all position and momentum values as: $(x, y, z, mv_x, mv_y, mv_z)$.

Here, what matters most is not these qualifications, but the fact that Newtonian mechanics explicitly postulates the set of <u>all possible</u> instantaneous states of the sphere. And similarly for other systems that the theory describes.

For the simplest possible system, a point-particle, that means: the set of all 6-tuples of real numbers. Unlike the set of triples of real numbers, which we of course visualize as familiar three-dimensional Euclidean space, this space cannot be visualized. We should instead think of its structure as follows: at each point of physical space, i.e. at each possible position of the point-particle, we have attached a separate copy of the set of all triples of real numbers. This copy represents all possible triples of momenta in the three spatial directions, that a point-particle at that location in space could possess. It is a dizzying idea.

Besides, when we consider more and more complicated systems, the set of all possible instantaneous states rapidly becomes very intricately structured. Even for two point-particles, which we label '1' and '2', with masses $m_1$ and $m_2$, we need 12-tuples of real numbers: which we could write as $(x_1, y_1, z_1, m_1v_x, m_1v_y, m_1v_z; x_2, y_2, z_2, m_2w_x, m_2w_y, m_2w_z)$. (Here, I use 'v' for speeds for the first particle, and 'w' for speeds for the second particle.) But to set aside collisions, the two triples representing particle positions, $(x_1, y_1, z_1)$ and $(x_2, y_2, z_2)$, must be different. So the structure of this set is: for every pair of distinct positions throughout physical space, we attach to each position in the pair, a copy of the set of all triples of real numbers, representing all possible triples of momenta for a particle located there.

And so it goes. To write down Newtonian mechanics, we need to mention these sets of possible instantaneous states, endowed with their intricate structures. Although these sets are of course not physical space, nor located in physical space---one would naturally call them pure mathematical entities, albeit usefully applicable to physical systems---they are called 'spaces': more specifically, <u>state-spaces</u>. Calling a structured set a 'space' (and its elements 'points') is ubiquitous in mathematics: the rationale is that often the structure is suggested by our visual intuitions about physical space, or even by precise geometric ideas like distance.

Similarly for all other physical theories: both the classical theories developed between 1700 and 1900 of light, electricity and magnetism, and of heat; and their twentieth-century descendants, which adapted their ideas and techniques to quantum theory and relativity theory. All these theories postulate, for each system they describe, an intricately structured set of all possible instantaneous states of the system. This set is again called a <u>state-space</u>. Of course, for these theories the quantities involved will in general not be just position and momentum, as in our example of Newtonian mechanics. Quantities such as energy, or electric field might be included. (Chapter 4 will give more details about the state-space for a quantum system.)

In short: the theories simply cannot be written down without describing this space. Thus I rest my case that physics is, as the catchphrase goes, up to its neck in modality. In each case, the system concerned is like a toy-model of the universe, i.e. a very simple way the world could be, according to the theory. For example, according to Newtonian mechanics, a system of two point-particles is a toy universe: the instantaneous possibilities for such a universe are the points in the two-particle state-space. And similarly for a system of five, or seventeen, or any number of point-particles. Each is a toy Newtonian universe, whose possibilities are the points of the corresponding state-space.

(3): We can also now readily see how useful the idea of a state-space is: again, in any of these theories. A sequence of instantaneous states is a possible history of the system. (Here, 'history' means not just the system's past states, but includes future states, so that a history is an entire "life-story" of the system.) We can think of this as a curve in the state-space. Then the structure of the state-space, especially its geometric structure like distance between points, helps us to understand the behaviour of these curves, i.e. these possible histories. For example, that two curves converge represents the two histories becoming more similar, i.e. the two systems' values for quantities becoming closer.

In particular, we can now state the idea of <u>determinism</u>. There are various precise formulations, but the general idea is of course that the state at one time determines the state at other times. So one common formulation is that any state in the state-space determines the sequence of states for all future, and indeed all past, times. In terms of curves in the state-space: through any point of the state-space, there is a unique curve to the future, and indeed to the past.

Again, Newtonian mechanics is the archetypal case; (setting aside collisions, as I did above). Consider again the case of a point-particle at some position, with some momentum, at a given time (`now'). And suppose the forces that are exerted on it, not just at the given time (`now') but throughout the past and the future, are specified. Then according to Newtonian mechanics, there is a unique history or curve in the state-space that passes through the particle's present instantaneous state. (To be precise: this claim assumes not only that the forces are given throughout time, but that they satisfy some "good behaviour" properties.) So we say that Newtonian mechanics is a <u>deterministic theory</u>.

But I stress that many other theories are also deterministic: and not just non-quantum theories---in the next Chapter, the Everettian version of quantum theory will be deterministic. (I will return to determinism later in this Chapter, in Section 8, when I discuss an important philosophical notion: <u>supervenience</u>.)

Chapter 3, Section 4: A philosopher's paradise
So much by way of arguing that our everyday and our scientific beliefs commit us to non-actual possibilities. I turn to this Chapter's third stage. In the next few Sections (4 to 8), I describe how a set of possibilities gives us a framework for formulating many philosophically important ideas and doctrines. (I will restrict my examples to ones we will need in later Chapters.)

By and large, the benefits of such a framework can be had, even with only a cautious or modest conception of the possibilities. For example, they can be the elements, the states, in the state-space of a physical theory, so that the system concerned is a "toy-universe". And even if, more confidently and ambitiously, one accepts a vast set of possibilities for the cosmos as a whole, the <u>possible worlds</u>, still the benefits can be had, by and large, without addressing the question 'What exactly <u>are</u> the possible worlds?'. In particular, they can be had without endorsing Lewis' modal realism: a flexibility that Lewis himself emphasized. (As announced in the Chapter's preamble, I will address this question only in the Chapter's fourth stage, Sections 9 and 10.)

So in effect, the next few Sections are an advertisement for using the framework of possible worlds---whatever exactly they are. Thus Lewis called this framework 'a philosopher's paradise'; and I concur. (The phrase deliberately echoes the achievement I lauded in Section 1, of formulating all of pure mathematics as set theory. For the mathematician Hilbert called set theory 'Cantor's paradise', after Georg Cantor who was the main inventor of set theory.) But again: Lewis allowed---and I agree---that most of the benefits of using possible worlds do not require his modal realism. After all, he called it 'a philosopher's paradise', not 'a modal realist's paradise'.

In Section 5, I will take as my first example of how useful possible worlds are, <u>semantics</u>. More specifically: a scheme called <u>intensional semantics</u>, which is inspired by ideas of Frege and Carnap about how words gets their references in the world. It is clearest to start with Frege, who expresses the ideas without regard to modality or possible worlds. It was Carnap, and later writers like Montague and Lewis himself, who adopted possible worlds. (As I mentioned at the end of Section 1, Frege and his contemporaries like Russell did not think about modal logic.)

To make this semantics vivid, and to reflect the intentions of its proponents, I shall explain it with examples from natural language, and so invoke possible worlds representing the cosmos as a whole. So we will be envisaging a vast set W of all the logically possible worlds.

But as I suggested above, readers too cautious for such examples, and the worlds they invoke, could---and I say: should---still accept the scheme's ideas for some modest fragment of language with correspondingly modest possible worlds. On this cautious or modest approach, the obvious cases are: the languages and claims of physical theories; and their state-spaces. In such cases, the possible worlds will be the instantaneous states; or if change over time is a topic, the possible worlds will be the system's possible histories (curves through state-space).

Chapter 3, Section 5: Paradise, Part I: Intensional semantics
I proceed in three stages. In (1), I present Frege's basic ideas about meaning for simple sentences, without regard to modality. In (2), I extend this to truth-functional compound sentences. Then in (3), I bring in modality, i.e. possible worlds.

(1): Frege's first idea is that the meaning of a word or phrase has two main aspects. The idea applies to proper names like 'Plato' and 'Copenhagen'; and to definite descriptions (i.e. 'the F' phrases) like 'the most famous pupil of Socrates' and 'the capital of Denmark'; and to predicates like 'is red', 'walks', or 'is a horse'.

The first and more obvious aspect is the <u>referent</u> (in German: <u>Bedeutung</u>). This is the object or objects in the world that, as we say, the word or phrase denotes or refers to. For my examples of proper names and definite descriptions, these are, respectively: a human being, and a city. By 'referent', Frege means the "concrete" object---a human body, a conurbation etc.--- located in space and time, and with its countless properties. Such properties can include: being tall, for a human; being populous, for a city; being famous, for a human or a city. Thus the referent is not some feature of the object that is connoted or signalled by the word. Nor is it someone's, e.g. the speaker's, ideas or beliefs about the object. The referent is the object itself. So one and the same object, i.e. referent, can be referred to in diverse ways. We say: 'Plato is the most famous pupil of Socrates', 'Copenhagen is the capital of Denmark' etc.

It is these ways of referring that are the second aspect of meaning, according to Frege. He calls it 'the mode of presentation' by means of which the word presents the referent to us (i.e. draws our attention to the referent). The idea is clearest for definite descriptions, like 'the capital of Denmark'; especially those that do not include a proper name: for example, 'the tallest human alive today'. Suppose it happens to be a tall German doctor called Gustav Lauben, who lives in Hamburg and is the best-known doctor there. Then clearly, 'the tallest human alive today' presents Lauben to us in a different way than 'the best-known doctor in Hamburg'.

Frege's jargon for these modes of presentation is '<u>sense</u>' (in German: <u>Sinn</u>). So these two definite descriptions have different senses. This is of course why 'the tallest human alive today is the best-known doctor in Hamburg' conveys useful information, going far beyond saying that a certain person is identical with themself. Frege also argues that proper names have senses, though they are vaguer and more idiosyncratic than the senses of definite descriptions. Thus for me, 'Plato' might have the sense: the most famous pupil of Socrates; while for you, it has the sense: the teacher of Aristotle. (Agreed: for Frege to fully explain along these lines how each of us refers by saying 'Plato', he must hold that we each associate an appropriate sense with 'Socrates' and 'Aristotle', respectively. In fact, the name 'Gustav Lauben' is from Frege's example, which he uses to expound this very topic. It occurs in one of his most famous essays, called 'The Thought: a logical inquiry'.)

Similarly, says Frege, for predicates. A predicate has instances: the objects it is true of. The set of instances is called the predicate's <u>extension</u>. Then Frege says that the extension, i.e.

the set of instances, is the referent of the predicate. (Here, I have simplified the history. In fact, Frege said that a predicate's referent is a special notion of his own, which he called a '<u>concept</u>' (in German: <u>Begriff</u>). Then he added that a concept has an extension. But according to Frege, two predicates refer to the same concept if and only if they have the same extension. In philosophical jargon, we say that Frege's concepts are <u>individuated</u> by their extensions. As a result, most systems of semantics in Frege's spirit simplify as I have done: they say the referent of a predicate is its extension.)

But the same set of objects could be the referent of another predicate. One standard example assumes that all and only those animals that have a heart (i.e. a pump for a circulatory system for nutrients) have a kidney (to remove waste products). That assumption can of course be questioned, depending on the meanings of 'heart' and 'kidney'. But let us accept it. Then the predicates 'has a heart' and 'has a kidney' have the same set of instances i.e. extension: what Frege calls the 'referent'. But indeed, the predicates present the referent in different ways; and intuitively, they have different meanings. Thus Frege says they have different senses. And again: that is why 'an animal has a heart if and only if it has a kidney' conveys useful information, going beyond saying that a certain set of animals is self-identical.

Frege then puts these assignments to words, of senses and thereby of referents, to work in a <u>compositional semantics</u>. That is: he gives an account of how the senses and referents of individual expressions combine to determine (i.e. to uniquely specify) the senses and referents of the composite expressions in which they occur. Think for example of how the senses of 'tall', 'human' 'alive' etc. combine to fix the sense of 'the tallest human alive today'. Similarly: just as a proper name and a predicate combine in a simple sentence, such as 'Plato is a teacher', 'Dr Lauben walks', so also their senses combine to make a proposition. And this proposition is true if and only if the referent of the name is in (i.e. is an element of the set that is) the referent of the predicate; and otherwise, it is false.

(2): Frege then extends the ideas of referent and sense to propositions and to sentences. Thus he proposes that the <u>truth-value</u>, True or False, i.e. the property or status of being true/false, of any proposition is the referent of the sentence expressing it. Thus the referent of any true sentence is the truth-value True; and the referent of any false sentence is the truth-value False.

Frege's rationale for this proposal is, in part, that this promises a smooth treatment of compound sentences, like a conjunction 'P and Q' or a disjunction 'P or Q'. Thus the truth-value of 'P and Q' is true if and only if both sentences are true. Here, we think of 'and' as a <u>sentence operator</u>, '… and …'. That is: '… and …' accepts two sentences 'P' and 'Q' into its two slots or argument-places, to produce a sentence 'P and Q'. Similarly, '…or …' is two-place sentence operator.

Both operators are <u>truth-functional</u> in the sense that the truth-value of the resulting sentence ('P and Q', 'P or Q') is completely determined by the truth-values of the pair of input sentences. Thus 'and' is associated with a <u>function</u>, sending the pair (True, True) to True, and each of the other three pairs, viz. (True, False), (False, True) and (False, False), to False.

Here I use '<u>function</u>' in the mathematical sense: a rule that sends each appropriate "input" (usually called an 'argument' of the function) to an "output" (usually called the 'value' of the function for the given argument).

Thus Frege can propose that the referent of 'and' <u>is</u> this function, from pairs of referents of sentences as inputs/arguments, to referents of sentences, i.e. True or False, as outputs/values. And similarly for disjunctions: the referent of 'or' (in our inclusive and-or sense) is the function taking three of the pairs of truth-values, i.e. all except (False, False), to True. Functions like this, that map truth-values, or pairs of them, or even triples etc., to truth-values are called <u>truth-functions</u>.

So according to Frege, the referent of 'and' is a truth-function. This function can be exhibited in a <u>truth-table</u> in which each row shows, for a pair that is an argument of the function, what is the corresponding value. Writing 'T' for True and 'F' for False, we have:

| P | Q | P and Q |
|---|---|---------|
| T | T | T |
| T | F | F |
| F | T | F |
| F | F | F |

And similarly for 'or'. In our inclusive and-or sense of 'or' the truth-table displaying the truth-function, i.e. the Fregean referent of 'or', is:

| P | Q | P or Q |
|---|---|--------|
| T | T | T |
| T | F | T |
| F | T | T |
| F | F | F |

By the way: we will see shortly that we need an innocuous generalization of the idea of a function, viz. to allow that for some arguments, the rule produces no output, no value. It is simply silent: this is called a <u>partial function</u>.

(3): So far, so Frege. I have not mentioned possible worlds at all; and I have invoked the actual world only "in the background", namely as making true a sentence such as 'Plato is a teacher'.

But here enter Carnap, and his followers like Lewis and Montague. They show how these Fregean ideas, about a sense being a mode of presentation of a referent, and using functions in a compositional semantics, can be smoothly developed in a framework of possible worlds.

For example: the capital of Denmark is in fact Copenhagen. But it might not have been. It could have been Aarhus, or Odense. Following Carnap, we understand this as: in some possible worlds, but not the actual one, the capital is Aarhus; while in yet others, it is Odense. Thus the referent of the definite description, 'the capital of Denmark', varies from world to world. (On the other hand, proper names like 'Denmark' seem, at least usually, to have the same referent in the various worlds.) Similarly, of course for predicates. Plato might not have been a teacher; and so the referent of 'is a teacher', i.e. the predicate's set of instances, varies across the worlds. And so on, e.g. for 'has a heart'.

All this can be neatly formulated in terms of functions. Since at a possible world, 'the capital of Denmark' denotes a city (in Denmark), we can say that the sense of 'the capital of Denmark' is a function whose arguments (inputs) are possible worlds, and whose value (output) for a given world as argument is the city in that world which is the seat of government for Denmark. And for a proper name like 'Denmark' with, we may suppose, the same referent across the worlds, we can say that the sense is again a function from worlds as arguments to objects, viz. countries, within the "argument-world". It is just that for a proper name, this function is constant: it always outputs the same value. And again, similarly for predicates. For example, the sense of 'is a teacher' is a function from worlds as arguments to the set of teachers within the "argument-world".

Here, I should make two clarifications. The first, (1), is rather technical, and not important for us. But the second, (2), is philosophically important.

(1): Agreed: we need to allow that at some (presumably vastly many) worlds, there is no country Denmark; or there is such a country, but it has no capital (seat of government). So at many worlds, a name such as 'Denmark', or a definite description such as 'the capital of Denmark', simply has no referent. Similarly for predicates: at a possible world with no animals with circulatory systems, the predicates 'has a heart' and 'has a kidney' will have no instances. (Here again, I assume, so as to make the point as simply as possible, that we take the meanings of 'heart' and 'kidney' to require a circulatory system.)

But this pervasive scarcity, across all the worlds, of referents causes no trouble. We simply use the idea mentioned above of a <u>partial function</u>. That is, we say that the sense of a word (a proper name, a definite description etc.) is a partial function: worlds are the arguments, but for many arguments, the function produces no output, no value. Agreed: as a result, the sense of a compound expression (such as a definite description) in which the given expression occurs will also in general be a partial function. Besides, one will need some sensible rules about e.g. what should be the truth-value (referent, for Frege) of a sentence at a world that contains no referent of some proper name within the sentence.

But there are such sensible rules---and we need not consider them here. I turn to the second, philosophically important, clarification.

(2): Beware of the preposition 'at'! That is to say: it is tempting to think that in this semantics, a phrase such as 'the referent of 'the capital of Denmark' at a given world' means: the city that within the world is called by some speakers in that world 'the capital of Denmark'. That is <u>not</u> so. The semantics being provided is a semantics for <u>our</u> language (in my examples, English), as we actually speak it. What language is spoken by people within a possible world is in general not relevant to the semantics of our language; and in particular, it is not relevant to how facts about a possible world make various sentences <u>of ours</u> true at that world.

Here, I say 'in general' because I agree: some of our sentences, albeit rather long and contrived ones, are indeed about a language that could be spoken---if you like, a variant of English. So according to our possible world semantics, such sentences are about a language that is spoken by people within a possible world. For example, one such long and contrived sentence is: 'People could have spoken a variant of English that used the name 'Denmark' for Sweden (but without other changes); in which case their sentence 'Stockholm is the capital of Denmark' would be true in their language.' But these contrived sentences do not alter the general point. Namely: this semantics---though it invokes other worlds, in some of which people use our words but with different senses---is a semantics for <u>our</u> language. That is: it is a semantics for <u>our</u> language as we actually speak it, with our senses.

Though this clarification is straightforward, it is important. For as we will see later on in this Chapter (Section 9:A), the erroneous temptation goes along with a wrong answer to our main philosophical question: what exactly are the worlds?

So much by way of the two clarifications. To finish this exposition, note that just as Frege put his assignments to words, of senses and thereby of referents, to work in a compositional semantics referring only to the actual world: so also the scheme proposed by Carnap, Lewis et al., with senses as partial functions, gives a compositional semantics, in which senses get composed together according to the syntactic structure of the composite linguistic expressions. In particular, the sense of a whole sentence i.e. the proposition it expresses, is naturally taken as a function from worlds to the two truth-values, True and False. But which worlds get sent to True, and which to False, depends on the senses of the sentence's parts.

Thus consider the sentence 'Plato is a teacher'. Its sense sends a world to True provided that the sense of 'Plato' takes the world to an object (in the world) that is in (i.e. is an element of the set that is) the output of the sense of 'is a teacher', for that world as input. And similarly for compound sentences. The sense of a conjunction 'P and Q' sends a world to the output of the truth-function that is the Fregean referent of 'and' (recall Part (2) of this Section above), for inputs that are the truth-values at that world of 'P' and 'Q', i.e. are the outputs of the senses of 'P' and of 'Q'. (Another way to think of this is to identify a proposition with the set of worlds in which it is true. Then the sense of 'and' can be stated in a less cumbersome way: it is precisely the operation of intersection on sets of worlds.)

Finally, a note about jargon. You will ask: Why is this scheme called '<u>intensional semantics</u>'? The answer is that Carnap suggested using 'intension' instead of 'sense', and 'extension' instead of 'referent'. So a more informative, but long-winded, label would have been 'semantics by intensions and extensions'; but the single adjective 'intensional' was adopted.

In any case, Carnap's jargon has become widespread. In particular, it is well-nigh universal usage to call the set of instances of a predicate its 'extension'. This usage is adopted even by those who are wary about intensional semantics. And setting aside talk of possible worlds: this usage is adopted even by those who are wary about Frege's notion of sense as applied (e.g. by Frege himself) to just the one actual world.

So much by way of sketching intensional semantics, especially as it applies to names, definite descriptions and predicates. In the next two Sections, we will see how it can be readily extended to treat two further topics. First: modality, so as to give semantics for expressions like 'It is necessary that …'; and second: counterfactual conditionals, i.e. if-then sentences, whose antecedent (after the 'if') is contrary to fact, i.e. is actually false.

Chapter 3, Section 6: Paradise, Part II: Modality and laws of nature
Intensional semantics, with its set W of logically possible worlds, also treats sentences with modal locutions, like 'It is necessary that …', 'It is possible that …', or the corresponding adverbs, 'Necessarily, …' and 'Possibly, …'. As we discussed in Section 2 of this Chapter, a sentence P is to be inserted in the place marked by dots ….. . So these are sentence operators: they make a sentence as "ouput" from a sentence P as "input".

But note that unlike 'and', 'or' and 'not' (discussed in Section 5), they are not <u>truth-functional</u>. For suppose P is true but contingent, i.e. could have been false, e.g. 'I stay at home tonight'; while Q is true and necessary, e.g. '2+2=4'. Then 'Necessarily, P' is false, while 'Necessarily, Q' is true. So the truth-value of the output of the sentence operator 'Necessarily, …' does not depend solely on the truth-value of the input.

To give a semantics of these operators, the main idea will of course be the intuitive (and Leibnizian) one which we already discussed. Namely: 'Necessarily, P' is true at a world w if and only if P is true at all the worlds w in W; and 'Possibly, P' is true at a world w if and only if P is true at some world w in W.

This idea gets developed in various ways. For example, one considers what is the sense or intension of 'Necessarily, …', analogous to the sense of 'and' being the operation of intersection on sets of worlds. And (as I mentioned in Section 2) one considers how this operator relates to other logical words like 'some' and 'all'.

But for this book's purposes, the development that matters is about <u>restricting</u> the set of worlds that a sentence operator, 'L(…)' say, requires one to check (for the truth at that world of the argument/input proposition P) in order for 'L(P)' to be true. So here, 'L(…)' is not short for 'Necessarily, …' or 'Possibly, …': it is just my notation for some other operator that we will interpret in terms of a subset of W, not the whole of W. (So it would not be appropriate to call such an operator 'Necessarily, …', 'Possibly, …' etc. But philosophers still use the word 'modality'. That is, they say such an operator represents a <u>restricted</u> notion of modality.)

One philosophically important example of such a restriction is the idea of a <u>law of nature</u>. (In Chapter 1, Section 4, this was an example of a concept that is philosophically contentious; but that contentiousness will not undermine any points here.)

Thus someone might say: 'it is logically possible for you to fly to the Sun in less than 8 minutes, but it is not physically possible, i.e. it is not compatible with the known laws of physics'. Or they might say: 'it is physically possible for you to fly to the Sun in 8 hours (namely, by going at one sixtieth of the speed of light), but it is not practically possible, i.e. it is not compatible with present technology---supplies of rocket fuel, funding etc.'

Such examples prompt the idea of abstracting from the laws of physics, or another science, that we happen to know (or at least: that we believe we know). So for the example of flying to the Sun, the idea is to go beyond what I dubbed 'known laws of physics'. After all, 'known', as I used it above, is a weasel-word. Agreed, our confidence that a person cannot travel faster than light is a central claim of an extraordinarily well-confirmed theory (Einstein's relativity theory) which we can hardly imagine being overturned by future physics. Nevertheless: strictly speaking, 'known' implies 'being true'. And we must accept that all laws as at present formulated, even the laws of relativity theory, are fallible.

Thus such examples suggest that we have a notion of a <u>law of nature</u>. That is, roughly speaking: the notion of a proposition that: (i) is perhaps not formulated by us---and might <u>never</u> be formulated by humans---but that: (ii) is true about the cosmos (the actual one!), and is deeply informative about the way the cosmos "works". This last phrase is intended to set aside the countless true propositions we never have and never will formulate, that are dull, maybe arcane, matters of happenstance: as it might be, that all the children living on my street have prime-number birthdays.

Philosophers differ about how to make precise the phrase, 'deeply informative about the way the cosmos works'. One popular suggestion is by David Lewis, building on ideas of John Stuart Mill and Frank Ramsey. In short, it is that 'deeply informative' means being both logically strong and simple. But we do not need the details of this suggestion, or of other rival suggestions. We just need the idea that the laws of nature form an elite minority of the countlessly many true propositions about the cosmos, and that the conjunction, L say, of all these laws is thus an elite proposition that is deeply informative about the way the cosmos works.

The Humean tradition (cf. Chapter 1, Section 4; Chapter 2, Section 5) then suggests: although this conjunction L is true, it is <u>contingent</u>. It is not necessarily true. For example, consider the classical theory of electricity and magnetism, formulated by Maxwell in the late nineteenth century. This theory, embodied in Maxwell's famous equations, is extraordinarily successful. But it is in fact *not* true: for we live in a quantum world. But this theory *could* have been true. That is: there are logically possible worlds that are exactly and accurately described by the theory.

A note for <u>afficionados</u>: To make this more precise and more convincing, let me keep matters simple by imagining that there is no massive or charged matter---for agreed, matter is quantum. So I imagine just some configuration of electric and magnetic fields, propagating across spacetime, e.g. the spacetime of special relativity (called 'Minkowski spacetime'), obeying Maxwell's equations. That is indeed logically possible: physicists call it 'a solution of Maxwell's equations (in vacuum)'.

Thus with the set W containing all the logically possible worlds, we conclude with Hume that the conjunction L of all the actually-true laws of nature is contingent. That is: the set of worlds where L is true is a subset of W. Then 'physical possibility' corresponds to being true in some world that is in this subset of worlds.

Now we can easily make sense of our opening example. It was the sentence: 'it is logically possible for you to fly to the Sun in less than 8 minutes, but not physically possible, i.e. not compatible with the known laws of physics'. We assume that 'No object can move faster than light' is indeed a contingent law of nature, i.e. a conjunct in the long conjunction L. Then 'it is logically possible, but not physically possible, for you to fly to the Sun in less than 8 minutes' is indeed true. There is a logically possible world---but not a world making L true---in which you fly to the Sun in less than 8 minutes.

A final note about jargon. Nomos is the Greek word for 'law'. So the restriction of modality to what conforms to the laws of nature is sometimes called nomic modality (also: nomological modality.

Chapter 3, Section 7: Paradise III: Counterfactual conditionals
My next example of the philosophers' paradise is counterfactual conditionals. These are propositions of the form, 'If P were so, then Q would be so'. Here the phrase 'were so' signals that the antecedent P is actually false ('contrary to fact': hence the name). To say it in terms of possible worlds: P is false at the actual world.

The discussion will have two stages, in (1) and (2) below. The first stage is uncontroversial: I will report these propositions' curious logical behaviour, which was noticed by philosophers and logicians in the 1950s and 1960s. The second stage, in (2), is more controversial: I will report the proposal, made by Lewis and Stalnaker in about 1968 (independently of each other), that we should understand this behaviour in terms of degrees of similarity between possible worlds. This will amount to a generalization of the last Section's idea of restricted modality. For the proposal will be that to formulate what makes a counterfactual true---in the jargon: to give the truth-condition of a counterfactual---we must invoke, not a single subset of the set W of all worlds, but a collection of such subsets, where the collection gets defined in terms of similarity between worlds.

(1): So first: the curious logical behaviour. A conditional connective, 'if…, then…', is a sentence operator. Like 'and', it accepts two sentences as "inputs", and "outputs" a third sentence. The intuitive idea of a conditional, of 'if…, then…', suggests several logical principles which one naturally expects the connective to obey.

One example is transitivity. This is the principle that, writing the connective as →, the following inference is valid, for any sentences P, Q and R: 'P → Q; Q → R. So, P → R'. One naturally says: surely, any 'if, then' will obey transitivity. For it accords with the idea that the truth of a conditional goes along with an argument being valid, or anyway in some sense good or plausible; and such arguments can be concatenated to give valid or good arguments.

Another example is strengthening the antecedent. This is the principle that given a true conditional, adding a conjunct to the antecedent (usually making it logically stronger) yields another true conditional. That is: one expects the following inference is valid, for any propositions P, Q and R: ' P → Q. So, (P and R) → Q'.

But many examples show that the counterfactual conditional violates these principles. (It also violates several other principles that are, at first sight, equally plausible for any conditional connective.)

Here is one example showing that counterfactuals do not obey transitivity: an example from the Cold War in 1950s USA. So we need to recall that J. Edgar Hoover was then (in the actual world!) head of the FBI, and an ardent anti-Communist. Then the first two statements below are true. Or at least we can take them to be, in some conversational context that determines what possibilities are relevant or likely. But the third is false; or at least we can take it to be.

1) If J. Edgar Hoover were Russian, he would be a Communist. (The idea here is: Hoover's ambitious but conformist temperament is retained under the supposition that he grows up in Russia.)
2) If J. Edgar Hoover were a Communist, he would be a traitor. (The idea here is: under the supposition that Hoover is a Communist, we still imagine him as an American citizen living in the USA, indeed perhaps as head of the FBI.)

3) If J. Edgar Hoover were Russian, he would be a traitor. (The reason this is _false_, or at least we can take it to be, is exactly as in 1): Under the supposition that Hoover grows up in Russia, his ambitious but conformist temperament is retained; and so under this supposition, there is no reason to think he is a traitor to the Communist one-party state.)

And here is an example showing that counterfactuals do not obey the principle of strengthening the antecedent. The first statement below is true, the second false. (Or again, at least we can take them to be true and false respectively, in some conversational context.)
4) If I were to strike this match on the side of the matchbox, it would ignite.
5) If I were to strike this match on the side of the matchbox and the matchbox was wet, it would ignite.

(2): How should we explain such strange logical behaviour? There is a natural proposal, due to Lewis and Stalnaker, for what 'If P were so, then Q would be so' means; and this proposal explains the logical behavior. Namely: Lewis and Stalnaker propose 'If P were so, then Q would be so' means 'In the world or worlds that are most similar to the actual world while making P true, it is also true that Q'. (As I mentioned; this specification of meaning in terms of what would make a proposition true is called a 'truth-condition'.)

Lewis and Stalnaker differ about the details of this proposal. The main difference is that Stalnaker proposes that for any world $\underline{w}$ (in particular, the actual world) and any proposition P that is not true at $\underline{w}$, there is a unique world that is most similar to $\underline{w}$ while making P true; whereas Lewis proposes, more cautiously, that relative to any world $\underline{w}$ (in particular, the actual world), other worlds are ordered by their similarity to $\underline{w}$, but in this ordering, two worlds might well be equally similar to $\underline{w}$. This makes the proposal readily visualizable, if we think of worlds as dots on the page that are closer together, the more similar the worlds are. Thus Lewis envisages that around any world $\underline{w}$, we can draw a sequence of concentric circles that have (the dot representing) $\underline{w}$ as their common centre. As we go out from $\underline{w}$, we successively include worlds that are more and more dissimilar to $\underline{w}$.

But these differences of detail do not affect the main point: that if we accept the proposed truth-condition for 'If P were so, then Q would be so', then the strange logical behaviour is readily explained. For different antecedents, i.e. different counterfactual suppositions P, will ``carry us'' to different worlds making P true, at which we then ask whether Q is true. So suppositions that are more outlandish, more different from actuality, will carry us to worlds more dissimilar to the actual world, that are represented by dots in a bigger circle. And these explanations are readily visualized.

Here is how this goes, with diagrams, for our two examples. For simplicity, I will adopt Stalnaker's proposal that for any world $\underline{w}$, and any proposition P not true at $\underline{w}$, there is a unique world that is most similar to $\underline{w}$ while making P true. (There are analogous, and equally visualizable, explanations of various other curious logical behaviours.)

The failure of transitivity in the Hoover example is due to it being more outlandish, more different from actuality, to imagine Hoover growing up in Russia, than his being a Communist within the USA. So a world where he is Russian and Communist (and so not a traitor) is more dissimilar from the actual world then a world where he is American and Communist (and so a traitor). To make a diagram to display this, with worlds as dots on the page: we put the dots closer together, the more similar the worlds are. It is also usual to write @ for the actual world, $\underline{w}_1$, $\underline{w}_2$ etc. for other worlds; and to write beside each world the propositions that are true at it. The counterfactual conditional is usually symbolized with a box-arrow, $\square\rightarrow$. Hence, with 'R' for 'Hoover is Russian', 'C' for 'Hoover is Communist' and 'T' for 'Hoover is a traitor', we have the following diagram. The distances between the dots, i.e. the varying amounts of similarity between the worlds, make the two counterfactuals 'R □→ C' and 'C □→ T' true at the actual world @; and make 'R □→ T' false at the actual world @.

| @ | . w₁ | . w₂ |
|---|---|---|
| not-R, not-C, not-T | not-R, C, T | R, C, not-T |
| R □→ C , C □→ T | | |
| not-(R □→ T) | | |

In an analogous way: counterfactuals do not obey 'strengthening the antecedent' because strengthening the antecedent, from 'P' to 'P and R', can make the antecedent carry us to worlds more dissimilar to the actual world than does P (i.e. more dissimilar to the actual world than the most similar P-world(s)). Indeed, in everyday life we make sure that matchboxes stay dry, and so the antecedent of 5) above is more outlandish than the antecedent of 4). So we adopt an obvious notation: 'S' for 'I strike the match', 'I' for 'the match ignites', and 'W' for 'the matchbox is wet'. Then the obvious diagram takes all these three propositions to be actually false, and so we have:

| @ | . w₁ | . w₂ |
|---|---|---|
| not-S, not-I, not-W | not-W, S, I | S, W, not-I |
| S □→ I , not-([S & W] □→ I) | | |

Chapter 3, Section 8: Paradise IV: Supervenience: materialism, physicalism and determinism
My last example of 'the philosophers' paradise', i.e. the uses of possible worlds, is the notion of <u>supervenience</u> (also known as '<u>determination</u>'). This notion is important in many philosophical discussions. But for this book's purposes, I need to describe how it is useful for formulating (and so also assessing) just three ideas: materialism, physicalism and determinism. As we will see: for both supervenience in general, and for its application to these three ideas, one can be either confident or cautious in the sense of Chapter 1.3. In (1), I introduce supervenience. Then in (2), (3) and (4), I discuss materialism and physicalism. (5) discusses determinism.

(1): Supervenience is a relation describing how one set of properties and relations determines another. In philosophy, 'attribute' is sometimes used as an umbrella term for 'properties and relations'; but I shall just say 'properties', for short. All the properties in each set are about some single subject-matter or topic. Because of this kinship between the properties, it is common to call the sets 'families'. So philosophers talk of supervenience as a relation between families of properties (or attributes). Thus the objects in the subject-matter can be described by which properties they have; and the set of properties amounts to a taxonomy or classification-scheme for the objects.

For example: think of botanical taxonomy as a subject-matter or topic. It classifies objects as plants or non-plants, and then further classifies the plants as daffodils, roses etc. So botanical taxonomy can be presented as the family of properties: being a plant, being a daffodil, being a rose etc. And the set of all the botanical-taxonomic facts is just the classification of all appropriate objects (especially the plants) with respect to this family: the assignment of each object to its botanical pigeonhole.

Similarly for other subject-matters: including larger, more encompassing ones such as, for our example, biology (or better: biological taxonomy). We can think of the set of all the biological facts as the classification of all appropriate objects (all organisms) in terms of all the many biological properties.

So supervenience is to be a relation between sets of properties. Or in alternative jargons: a relation between subject-matters or taxonomies or classification-schemes. What relation? The answer is: the classification of any of the objects using one set of properties implies how it is classified by the other set. This is as the word 'determination' (the alternative jargon to supervenience) suggests: the classification of an object using one set of properties <u>determines</u> (also 'fixes': in the sense of 'makes unique', not 'repairs') its classification by the other set.

By taking facts as given by what properties objects have, we can also put this in terms of facts. Thus supervenience is: all the facts about one family of properties, one subject-matter, $F_1$ say, are fixed by all the facts about another family $F_2$. In other words: specifying all the facts about $F_2$ involves, <u>ipso facto</u>, specifying all the facts about $F_1$. We say: $F_1$ <u>supervenes on</u> $F_2$. We also say: $F_2$ <u>subvenes</u> $F_1$. (Thus 'subvenes' is a synonym for 'determines'.)

Here is an example that is standard, since it is uncontroversial. At least, it is uncontroversial by the standards of philosophy. The objects in question are pictures. Then it is very plausible that the aesthetic properties of pictures---the classification of them as being beautiful, being well-composed, having a dark palette etc.---supervene on their pictorial properties, i.e. the properties about how exactly the paint, and of what kind, is distributed on the canvas or paper: being an oil painting, having magenta in the top left square centimetre etc. So the idea is: any two pictures that match in all (<u>all</u>, not just some) of their pictorial properties must also match in all their aesthetic properties (again: <u>all</u>, not just some). That is: two pictures that are replicas of each other as regards pictorial properties, must also be replicas as regards aesthetic properties. So both are beautiful, or both are ugly; and both are well-composed, or both are badly composed; and so on. The pictorial properties subvene or determine the aesthetic properties. To put it the other way around: pictures cannot differ from one another as regards some aesthetic property, without also differing as regards some pictorial property. In a slogan: no aesthetic difference without a pictorial difference. This is what it means to say that aesthetic properties <u>supervene on</u> pictorial properties. (Agreed: the example is not wholly uncontroversial. For example, one might claim that being an original is an aesthetic property of a picture, not a "merely historical" property; and if so, aesthetic properties certainly do not supervene on the pictorial properties.)

(2): There are various major topics in philosophy where the question, whether a certain family of properties or subject-matter supervenes on a certain other one, is central. One is the relation between mind and matter (sometimes called the 'mind-body relation'). Do mental properties of sentient animals (like seeing yellow in the top-left of the visual field, or feeling hungry, or hoping for a sunny day) supervene on their natural scientific properties, i.e. their panoply of physical, chemical and biological properties? (Here, I use 'natural science' to include only physics, chemistry and biology, i.e. to exclude psychology and other sciences.) That is: If two animals matched as regards all their physical, chemical and biological properties, must they also match as regards seeing yellow in the top-left of the visual field, and also as regards feeling hungry?

Saying 'Yes' to this question is often called <u>materialism</u>. The idea is that all the facts about <u>matter,</u> as explored by the sciences of physics, chemistry and biology, fix all the facts about an animal, even the facts about its mental life. Nowadays, this is very widely endorsed. But agreed: in the nineteenth century, it was reasonable to deny it. It is only with the cumulative successes of physiology, molecular biology and neuroscience in describing mental states that the idea of special mental causes (of at least some such states) has died away.

Analogous comments apply to the dependence, nowadays evident, of biology on chemistry, and of chemistry on physics. That is: in the nineteenth century, it was reasonable to believe in what were called 'vital forces': causal factors occurring only in living organisms that "rode free" from their underlying chemical and physical descriptions. But the successes of physiology, e.g. its explanations in physico-chemical terms of the nerve impulse, muscle-contraction and vision, put paid to vital forces. And until 1930 or even later, it was reasonable to believe that chemical phenomena, in particular chemical bonding, would not be explicable by the physics of atoms. But since 1930, quantum theory has achieved ever more precise descriptions and explanations of chemical phenomena: in a way that was impossible---indeed, provably impossible---according to the earlier classical physics.

(3): Thus arises the doctrine of <u>physicalism</u>. This is a strengthening of materialism, that gives physics a pre-eminent role, as compared with the other sciences. So as a claim of supervenience, physicalism amounts to: all the facts described by chemistry, biology and the other sciences, in particular psychology, are determined (fixed) by the panoply of all the physical facts.

Obviously, this sketch of materialism and physicalism as claims of supervenience leaves a lot to be made precise. What exactly are the sets of objects being described by the two (or more) subject-matters or taxonomies, of which one is said to supervene on the other? And what exactly are the sets of properties (and relations) defining the subject-matters, or taxonomies. For example, what exactly is the set of physical properties? As you would expect, different philosophers give different answers to these questions: influenced, usually, by different judgments about which precise concepts of e.g. 'physics' or 'all sentient animals' make for a supervenience thesis that is---not obviously true, but---debatable enough to be worth assessing for truth. And in debating such formulations, there are choices about whether to be confident or cautious in the sense of Chapter 1, Section 4. For example: would you be confident or cautious about a firm distinction between physical properties and other ones?

But we do not need to go in to the details of these debates. Here, I only want to describe one main way that possible worlds help us to be precise in formulating such supervenience claims. This concerns what one might call the '<u>modal range</u>', or '<u>modal extent</u>' of the claim.

For consider the actual world, i.e. the actual cosmos spread throughout all space and all time, past present and future. (We adopted this use of 'actual world' at the start of Chapter 1.) And suppose we take a supervenience claim as being only about actual objects: we set aside possible worlds. That is: suppose we say that two actual objects that match exactly as regards all the properties in a set (family) $F_2$ also match as regards all the properties in another family $F_1$.

Then there is likely to be problem. For the families of properties $F_2$ we are concerned with are bound to be rich, i.e. to make fine distinctions. Recall our examples: all pictorial properties; all material properties as described by physics, chemistry and biology; all physical properties. In all these examples, the idea is that the family $F_2$ has to be rich, in order to have a chance of fixing all of $F_1$. And there is the rub. For any reasonably rich taxonomy (family of properties), the actual objects are likely to be a very varied set. Agreed: two actual objects often match for some property; both have it or both lack it. But it is very likely that no two actual objects match for <u>every</u> property in $F_2$. And if so, the supervenience claim restricted to actual objects---'if they match in this way, then they also match for all of $F_1$'---loses its force, or content. For the antecedent 'they match in this way' is never true. (Philosophers and logicians call this 'vacuous truth'.)

The answer to this problem lies in recognizing that the basic idea of supervenience is modally involved. It is not an assertion only about a case of two actual objects matching for all of $F_2$. For as we have just seen: for most of philosophy's interesting supervenience claims, there are no such cases. Rather, it is about <u>trans-world matching</u> of objects. Thus we again see the theme of Section 3: that throughout our thought and language, both everyday and scientific, we are up to our necks in modality.

To show how supervenience is about trans-world matching of objects, let us take as an example physicalism; and what it says about, say, the mental life of a cat supervening on its physical state, i.e. on all its physical properties. For illustration, I again take just one mental property (i.e. in the family $F_1$): viz. seeing yellow in the top-left of the visual field. Then physicalism says, in particular: if there were a replica of this actual cat that is now seeing yellow in the top-left of its visual field, and this replica was 'physically perfect' i.e. utterly matched all the actual cat's physical properties, then the replica would also see yellow in the top-left visual field.

      Besides, the usefulness of possible worlds for formulating supervenience claims is not limited to providing possible objects: e.g. cats that are atom-for-atom replicas of some actual cat. There is also the question of whether the supervenience claim being considered, e.g. materialism or physicalism, is propounded as contingent or as necessary. And if it is propounded as contingent, that means in a possible worlds framework: true in some possible worlds but not all. And this prompts the further question: across exactly what set of worlds is supervenience claimed? For example, for physicalism: across exactly what set of worlds must an atom-for-atom replica of some actual cat utterly match all the actual cat's mental properties?

      Again, we do not need to take a view about the answer. What matters for us are three points, all of which echo some previous themes.

(4): First: most philosophers do indeed formulate materialism and physicalism as logically contingent claims, not necessary ones. This is of course because the success, since 1800, of the natural sciences, and especially of physics, in describing and explaining phenomena lying outside their original scope---as illustrated above---was undoubtedly contingent. It did not have to be so. We might have discovered vital forces underpinning metabolic processes, or phototropism in plants, or what-not. And we might have discovered distinctive chemical forces that explained bonding, chemical valences etc., independently of the electron orbitals around atoms' nuclei. This contingency---this happenstance of a "one-way street" for 200 years, from the other sciences towards chemistry and then on to physics---makes it very natural to formulate materialism and physicalism as contingent claims.

      Second, answering the question 'Across exactly what set of worlds is supervenience claimed?' leads us back to the idea of a law of nature. For one natural answer is: 'the set of worlds that share with the actual world their laws of nature---the nomically possible worlds'. (This answer is natural, but by no means compulsory. For as we discussed in Chapter 2 and Section 6 in this Chapter: one might well be cautious, rather than confident, about the very idea of a law of nature.) So for example, it is natural for a physicalist to say: 'I accept the idea of a law of nature, and I believe they are all contingent. And I claim that physical matching of any two objects implies their <u>total</u> matching, across the set of nomically possible worlds.'

      Third: my last point returns to the theme at the end of Section 2, about the principle, about some property F, that if all objects are necessarily F, then necessarily all objects are F. This principle's connection with materialism becomes clear when we take as the property F in question, being material: that is, being made of matter. So for this property, the principle says that if all objects are necessarily material, then necessarily all objects are material. This conditional obviously relates to the formulation of materialism, and in particular to its modal range. Thus one can imagine a materialist who believes the antecedent of the conditional. They believe that all actually-existing objects are material, and that for each such object, it could not have been immaterial. As they might put it: this actual rock and plant and animal are each of them material; and though each might have had other features (this rock might have been heavier, this plant taller etc.), none of them could have been immaterial---on pain of not being that very object. But if this materialist takes materialism to be a contingent supervenience thesis (my first point above), they will probably deny the consequent of the conditional. That is, they will deny that necessarily all objects are material. For once they consider possible worlds beyond the range of their supervenience thesis---as it might be: worlds that lack the actual world's laws

of nature (my second point above)---they may well allow that some such worlds contain immaterial objects.

Again: in this book, we do not need to pursue the intricacies that these three points reveal about the formulation, and assessment, of materialism. For us, these points teach two relevant morals. First, they show the value of the possible worlds framework for articulating the various philosophical issues, and relating them to each other.

Second, these points (especially the third) bring out a feature that will be important for Section 9's question, about what exactly a possible world is. Namely, the feature that the property I called 'being material', 'being made of matter' is vague. If the materialist uses it (or some similar phrase) to formulate their materialism, should they take it to require matter of the known kinds---whatever they might choose to mean by 'known'? For example, does it include the dark matter that cosmologists nowadays believe in, though they do not know what it is composed of, nor what laws it obeys? In this book, I will not need to resolve this vagueness (not least because materialism is not our main topic). But in Sections 9 (Part C) and 10, we will see that this vagueness causes trouble for an otherwise attractive account of what exactly a possible world is.

(5): As a final illustration of the power of possible worlds, I turn to <u>determinism</u>. I briefly discussed this at the end of Section 3 above. We saw that all physical theories postulate a space of instantaneous states of the system they describe, so that a possible history of the system (i.e. life-history, comprising both past and future) is represented by a curve through the state-space. Thus I reported the idea of determinism, as follows. A physical theory is <u>deterministic</u> if the state of the system at one time determines its state at all past and future times. In terms of histories as curves through the state-space: through any point in the state-space, there is a unique curve to the future and the past. (More precisely: this uniqueness holds good, once we specify the external influences that the system is subject to during the past and future. Recall the need, at the end of Section 3, to know the forces on the point-particle.)

It is now clear that, like physicalism, determinism is a supervenience claim---as my word just now, 'determines', rightly signals. For in view of the word 'determine', the definition just given means: any two systems (of the sort that the theory describes) that match exactly in their states at one time (and in the external influences they are subject to), match exactly in their states at all past and future times. This is clearly a statement of supervenience. Namely: the past and future states of the system concerned supervene on its present state---fixing the latter implicitly fixes the former.

Besides, we here see again the contrast, being confident or being cautious, about a concept; (cf. Chapter 1, Section 4). What I just said gave only a cautious construal of determinism as a property of a given theory that applies to a given type of system; and the possible worlds involved were cautious ones, viz. instantaneous states of the given type of system.

But one might be more confident. Thus suppose we accept the idea, not just of the laws of a given theory, but of a <u>law of nature</u>. (Recall the discussion in Chapter 1, Section 4 and Section 6 above.) Then we can think of the conjunction of the laws of nature at a given possible world <u>w</u> as 'the theory of <u>w</u>'. (We might call it 'the <u>theory of everything</u> at <u>w</u>'. But nowadays, the phrase 'theory of everything' is always used for an ambitious and more specific idea: an idea that is like physicalism, as defined above. Namely: the facts described by a single theory of physics might determine all the facts of all the sciences. But 'the theory of <u>w</u>', as just defined, might well not be a physical theory.)

Given this notion of the theory of a given possible world, we can define what it is for an entire possible world to be deterministic. It is for the theory of that world to be a deterministic theory, in the previous sense. But here, the space of possibilities will be the confident (ambitious) space of all possible worlds: not a cautious (modest) state-space of a single theory, such as Newtonian mechanics. So we say that a world w is deterministic if: for any possible world that also makes true the theory of w (i.e. all the laws of nature at w), and whose state at some time matches exactly the state of w at some time---the two worlds match exactly at all times, i.e. to both past and future of the assumed matching. (Note incidentally a benefit of accepting the idea of the theory of an entire possible world. Since by definition it cannot be subject to external influences---in physics jargon: it is a closed system, not an open one---the definition of determinism as supervenience of past and future on the present does not need to include the qualification about specifying such influences.)

Let me sum up this discussion of determinism. If we confidently accept the idea of the theory of a possible world, then determinism of a world is, again, supervenience. Namely, supervenience of all the past and future states, of an entire possible world, on its present state.

Chapter 3, Section 9: Existential *angst*: what are possible worlds?
So much by way sketching the philosophical benefits of using possible worlds. So much by way of tasting the fruits in the philosophers' paradise. I turn, in this Section and the next, to the fourth and final stage of this Chapter. That is: to the question which I announced in the Chapter's preamble: What exactly is a possible world?

As I said there: this question is compulsory, for cautious conceptions of possible worlds as much as for confident conceptions; and several possible answers are defended in the philosophical literature. It is also agreed to be a very hard question. Though we can readily agree that our thought and language, everyday and scientific, continually invokes non-actual possibilities (cf. Section 3 above), what exactly they are is an open, and stubbornly difficult, question. Hence this Section's title says: *angst*. Besides, focusing on this question illustrates Chapter 1's announcement that my discussion of each of the three multiverse proposals will end by urging an open philosophical problem that the proposal prompts.

So unsurprisingly (and as I admitted in this Chapter's preamble): I cannot honestly urge one answer as correct. I will instead address the question by, first, refuting two tempting suggestions (in Parts A and B). They are tempting, but definitely false. And they are suggested, I am sorry to say, by proposals from renowned philosophers: Berkeley and Wittgenstein. Then in Part C, I will discuss a third suggestion that fares better. But it is still, I fear, wrong. The upshot (in the next Section) will be that the Chapter ends where it began: by stressing that Lewis' modal realism is a coherent intellectual possibility, even though probably, you---like I---find it incredible, in the literal sense.

Chapter 3, Section 9: A: Acts of Imagination?
One natural suggestion is that a non-actual possibility is something we imagine. But here, we have to be careful to distinguish the event or state of affairs---a person, say you, imagining that P---from the proposition P being imagined to be true. The distinction applies not just to imagination, but to many mental acts, such as hope, belief, desire, regret. Thus we say that 'John imagines/hopes/believes/desires/regrets that P'. Philosophers have a jargon for this distinction. They say that to imagine, to hope, to believe etc. are propositional attitudes; and that the proposition P "on which the mind is focussed" is the content of the attitude; (i.e. the content of the event or state of affairs of John imagining etc.).

This distinction is clear enough. But it makes trouble for the 'something-we-imagine' suggestion. There is a dilemma: the first horn makes no progress, and the second is definitely wrong. (But there will be some good news as a consolation: each horn will teach a philosophical lesson.)

Suppose, first, that the suggestion is: the non-actual possibility is the content or proposition. This suggestion may well be right. For as I said at the end of this Chapter's preamble, the question what is a possibility is tantamount to the question what is a proposition. (I touched on this again at the start of Section 3, when I remarked that the content of any false belief such as 'I go to the cinema tonight' (assuming I in fact stay home) represents a non-actual possibility.) So then our question becomes: what exactly is a proposition?

Now we see that we are no further ahead. For (as I also said) 'proposition' and similar words like 'statement' are terms of art, with no agreed precise meaning: it is up to each philosopher or logician to say what they mean. Nor does the framework of intensional semantics (reviewed in Section 5) help answer the question. There we saw how it systematically portrays how propositions (taken as the Fregean senses of sentences), the Fregean senses of words and phrases, truth-values (True and False) and possible worlds all relate to each other in a compositional semantics; and how the various senses get expressed by language. But that survey gave no opinion about what a possible world, or more generally a non-actual possibility or proposition, actually _is_. So we are no further ahead.

(To further justify a little this verdict of "no progress", let me sketch the kind of trade-off between taking as basic possible worlds or propositions. Thus in Section 5, possible worlds and truth-values were basic posits, not further analysed. Thus at the end of that Section, a proposition was taken as a mathematical function from possible worlds to the set of two truth-values. But agreed: one might instead take propositions, or Fregean senses of sub-sentential words and phrases, as basic, and build possible worlds from them, using the language of functions; or more generally, using set-theory. For example, a possible world might be taken as a maximally logically strong ('maximally opinionated') proposition. But the question, of the nature of the basic posits, would remain.)

Suppose, on the other hand, that the suggestion is: the non-actual possibility is the event or state of affairs of imagining. That certainly makes the non-actual possibility unproblematic, and "down to earth" as a short-lived episode (of a mind or brain) within the actual world. But it is definitely wrong. For obviously, there are countless non-actual possibilities that nobody ever actually imagines.

Besides, this suggestion implies that any non-actual possibility has as a necessary concomitant, as an implication, the existence and imaginative activity of a mind. Which is not so. Here we return to the clarifying comment (2) at the end of Section 5 above. There, I stressed that possible worlds provide a semantics for our language as we actually speak it; and that (setting aside some contrived sentences about how we might have spoken), how people in other worlds---if there are any---speak is in general irrelevant to the semantics of our sentences. This also means that a possible world with no people, indeed no animals, or other sentient beings, is entirely coherent. Such a world can make true a proposition of our language, such as 'the world consists entirely of five boulders of granite floating in a Newtonian space, without any living or conscious being'. No sentience—in particular, no visualisation of the boulders---is needed within the world. In short: the idea of possibility as such does not imply the existence and imaginative activity of a mind.

Incidentally, here we also see the flaw in claims made by the eighteenth-century idealist philosopher Berkeley, in his <u>Treatise concerning the Principles of Human Knowledge</u> (1710). Berkeley claims that: (i) we cannot imagine an unperceived object; and therefore (allegedly!) that (ii) any object must be perceived. (He sums this up in a famous slogan, that for an object, to be <u>is</u> to be perceived; in Latin, <u>esse est percipi</u>.)

The flaw lies in an equivocation. If (i), i.e. 'we cannot imagine an unperceived object', means 'we cannot imagine a possible world containing an unperceived object', then (i) is false. (Just think of the boulder world, above.) But I am willing to concede: so understood, (i) implies (ii). That is: if (i) were true, (ii) would also be true. If on the other hand, (i) means 'we cannot imagine ourselves within a possible world without perception', then I can concede that (i) is true.

Indeed, it is necessary if 'ourself' implies being able to perceive. But it by no means implies (ii). (Again: just think of the boulder world.)

This ends my rebuttal of appealing to imagination as the way to understand possibility. I turn to my second tempting, but wrong, suggestion.

Chapter 3, Section 9: B: Combinations?

Here the "culprit" will be Wittgenstein, in his early work, the <u>Tractatus Logico-Philosophicus</u> (1921); a work which he later disavowed, partly for the reasons I will present. Indeed, we will see that for our purposes, he is "more guilty" than Berkeley. For he does not just make claims that prompt the false suggestion: he explicitly makes the suggestion.

The idea of the suggestion is, at the start, modest. It proposes that we should lower our sights about understanding what a possibility, or possible world, or proposition, really is; and assuming we accept these notions, we should focus instead on the following question---which, admittedly, is vague: How can a proposition be <u>necessary</u>? What explains that?

This echoes the questions we pursued at the end of Chapter 2 (Section 8) and the start of this Chapter (Section 1). Namely: What is pure mathematics really about? What is its subject-matter? And can it be reduced, as logicism claimed, to logic?

It is in the context of those questions that Wittgenstein, in the <u>Tractatus</u>, suggested that for any necessary proposition, whatever its subject-matter, its necessity is exactly like that of what propositional logic calls <u>tautologies</u>. These are special sentences whose necessity can be agreed by all parties to be utterly unproblematic. For they are defined as those sentences built from others using 'and', 'not' and 'or' (which, as discussed in Section 5 above, are <u>truth-functions</u>), with the feature that whatever the truth-values of the component sentences, the compound sentence <u>must</u> be true---just because of the order in which the truth-functions, such as 'and', 'not' and 'or', have been applied to the components. Examples starting with one component sentence include: 'P or not-P', and 'not-(P and not-P)'. An example starting with two components, P and Q, is: '(P or Q) or (not-P)'. (Here 'or' is understood, as usual, in our inclusive and-or sense.)

The simple diagrams called <u>truth-tables</u>, introduced in Section 5, make the idea clear. We give each component sentence a column, and underneath we assign a row to each combination of truth-values that could occur; and then we apply the truth-functions to calculate, in each row, what is the truth-value of the compound sentence. We say that the compound sentence is a <u>tautology</u> if it comes out True in every row.

Thus to calculate the truth-table for '(P or Q) or (not-P)', we need four rows: one for 'P true and Q true', one for 'P true and Q false', one for 'Q true and P false' and one for 'P false and Q false'. Then, by applying the truth-functions 'not' and 'or' appropriately, we calculate that '(P or Q) or (not-P)' comes out True in every row. As follows:

| P | not-P | Q | P or Q | (P or Q) or not-P |
|---|-------|---|--------|-------------------|
| T | F     | T | T      | T                 |
| T | F     | F | T      | T                 |
| F | T     | T | T      | T                 |
| F | T     | F | F      | T                 |

So we think of each row, each combination of truth-values for the component sentences, as a 'way the world could be', as described by those sentences. In short: it is a toy-model of a possible world. Then calculating that the truth-value must be True in every row explains why the whole sentence is necessary, in a completely unproblematic way.

The crucial word here is 'combination', as in 'combination of truth-values'. No necessity, nor any other unexplained modal status or mutual logical relation, is attributed to the component sentences. They can be true or false, quite independently of each other: all combination of truth-values are genuinely possible. This is called their being <u>logically independent</u>. So the idea is: whatever their combination of truth-values, the placing of 'and', 'not' and 'or' in the whole sentence forces it to be true: i.e. true in that row, that combination.

      Thus Wittgenstein proposed that all necessity had this lucid <u>combinatorial</u> origin and explanation: that all necessary propositions are really---could be analysed into---tautologies.

      So far, this is entirely programmatic: a mere declaration. Indeed, there are three large kinds of necessary propositions, whose necessity seems to have little if anything to do with the placing of 'and', 'not' and 'or' in any sentences.

      First: the truths of pure mathematics, like '2+2=4', 'there are infinitely many prime numbers', 'equilateral triangles are equiangular', seem to be necessary. But this necessity seems to be very different from---and much more problematic than---the placing of 'and', 'not' and 'or' in any sentences. Recall from Chapter 2, Sections 7 and 8, the struggles since Kant to explain the necessity of mathematics; and in particular, the growing separation of pure and applied mathematics, with the former accorded a special <u>non</u>-empirical subject-matter, such as numbers and geometric figures. And though the logicism of Frege and Russell tried to reduce pure mathematics to logic, they really succeeded in reducing it to set theory. Agreed: that was a great achievement, as extolled in Section 1 above. But set theory is not logic: it has a special non-empirical subject-matter, viz. sets.

      Second: there are propositions (about any subject-matter) whose necessity turns upon the placing of logical words <u>other than</u> 'and', 'not' and 'or': especially those other logical words, 'all', 'any', 'some' and 'none', that---as also explained in Section 1 above---are studied in predicate logic. For example, consider the necessary proposition (for any predicates A and B): 'Either it is not true that everything is both A and B, or something is A'. Its necessity is a matter of the placing of, not just 'and', 'not' and 'or', but also 'everything' and 'something'. So even if we set aside pure mathematics i.e. the first kind of necessary proposition above (whether or not it is really set theory), here is another kind a necessary proposition that is not a matter of tautologies. Besides, we saw in Section 2 that the logical words on which propositional and predicate logic focus---'and' and its brethren, 'all' and its brethren---are not all the logical words, i.e. words on the placing of which the necessity of a proposition (or the validity of an argument) can depend. We saw in particular that C.I. Lewis developed modal logic, about the logical words 'It is necessary that …' and 'It is possible that …'.

      (An incidental comment. Here, the propositional form 'Either not-P or Q' is logic's much-used "weaker cousin" of the form, 'If P then Q'. Thus one readily agrees that the proposition 'If everything is both A and B, then something is A' is necessary; whereas my example above is a tongue-twister, whose necessity is hard to see. But I use the tongue-twister form, 'Either not-P or Q', to avoid subtleties about the meaning of the English 'If…, then …': subtleties that are not needed in this book, but which are rather like those we saw for the counterfactual conditional in Section 7.)

      Third: there are propositions whose necessity depends, not upon the placing of any logical words in the sentence expressing them, but upon relations between the meanings of other words. For example, consider: 'All bachelors are unmarried', and 'A vixen is a female fox'. (Though philosophers use 'analytic' in different senses, the common thread in their usages is undoubtedly the idea of being true in virtue of the meanings of words. So these sentences would certainly count as analytic.)

      To sum up: in order to show that all the necessary propositions of these three kinds are really tautologies, one would have to show, somehow or other, that they are built up from---can be analysed into---component propositions that are logically independent, i.e. component propositions for which every combination of their truth-values is genuinely possible. That would be a programme of <u>reduction</u>, in roughly the sense of Section 1 above.

Wittgenstein in the Tractatus was committed to such a programme. He was of course influenced by the logicism of Frege and Russell, and so made some suggestions about how to cope with the first and second kinds above. For example, maybe 'all' could be reduced to the idea of conjunction, though a possibly infinite one; (and similarly 'some' to a possibly infinite disjunction). But there were few details. The main lacuna is his silence about the third kind; the difficulties about analysing them as tautologies is just glimpsed at assertion 6.3751 of the Tractatus. Indeed, this shortcoming was one of the main reasons why Wittgenstein, a few years later, abandoned its claims.

Furthermore, no one else has succeeded in this combinatorial approach to explaining necessity. In particular, I note that its prospects do not improve if we adopt a cautious conception of possible worlds, suggested by the state-spaces of physical theories. The problem is simply stated. The different values of a quantity---whether a physical quantity like position of a point-particle, or a psychological quantity like 'magenta in the top-left of the visual field'---obviously exclude one another. And this means that the propositions ascribing such values cannot be logically independent. Far from it: that values exclude each other means precisely that each proposition ascribing a specific value implies the negation of each of the other propositions ascribing specific values.

Chapter 3, Section 9: C: Sentences and sets?
I turn to my third suggestion about what a possible world is. As I announced, it fares better than the first two. But I fear (following Lewis' critique of it) that it too is wrong.

The idea is that a possible world is like a novel: that is, the sentences, rather than the propositions they express. (Saying 'the propositions expressed' would get us no further ahead, as we discussed under the first suggestion.) At first sight, the advantage of this suggestion is that a sentence is an unproblematic object to believe in. For it can surely be taken as the set of all its physical inscriptions in pencil, ink etc., and all the events of its being spoken.

But being more precise brings difficulties. Surely most possible worlds that our thought and talk invokes (cf. Section 3) never get represented in even one inscription or utterance of a sentence. And setting aside whether there is an actual inscription or utterance: surely most cannot be represented in all their myriad details by a finite sentence, even of a richly expressive language like English. And surely, infinitely long sentences do not exist, i.e. exist in the actual world.

These difficulties prompt one to generalize the idea from sentences of an existing language such as English, to sets of actual objects, and also of actual properties and relations. For set theory provides countlessly many sets, many with very intricate structures; (since the operation of making a set out of some given sets---'putting a curly bracket around them'---can be iterated endlessly). Here we return to the discussions in Chapter 2, Section 8 and Section 1 above, about set theory as a lingua franca for expressing all of pure mathematics. The idea now is, in effect, that it is a lingua franca for expressing anything.

Thus the suggestion is that: (i) possible worlds are sets of a certain kind built from actual constituents, i.e. actual objects, properties and relations; and (ii) the structure of such a set, i.e. the pattern of curly brackets by which it is built up, encodes how it represents a possibility, in a manner similar to that in which the grammatical structure of a sentence encodes how it represents. (Recall Section 5's idea of compositional semantics.) Or in other words: the idea is that the set's structure exactly mirrors the structure of the possible world. And this makes the set the preferred official proposal for being the possible world.

To many philosophers, this suggestion has seemed promising. One main reason they find it attractive is that it seems to secure for us a deeply felt contrast between the actual world and all other worlds, viz. that the actual world is "concrete" while non-actual worlds are "abstract". As the scare-quotes show, these two words are philosophical jargon, and vague: 'concrete' does not mean 'cement'! But many philosophers, though they accept sets built from actual objects etc. as being themselves legitimate objects, think of sets as abstract objects; and they also think of ordinary actual objects as not abstract---which they dub 'concrete'. So to these philosophers, the suggestion that any non-actual possible world <u>is</u> a set, while the actual world is not a set, seems to classify worlds rightly, as regards concrete vs. abstract.

But I fear that this suggestion does not work. There are several objections; but I shall present just two. (They are urged by Lewis himself, along with others. My presentation summarizes some passages in his <u>On the Plurality of Worlds</u>; viz. in Section 1.7 and Section 3.2, p. 150 f.)

The first objection is that the advantage just mentioned, of classifying worlds rightly as regards concrete vs. abstract, is spurious: for three reasons, of which I think the third is more important.

(i): Since there will be set-theoretic representation, a mock-up or replica, of the concrete actual world, just as there are of non-actual worlds, intensional semantics will presumably specify the actual referents of our words as ingredients (barely visible, deep in a forest of curly brackets) of the actual world's mock-up. But that apparently conflicts with the idea that we refer to concrete objects.

(ii): Having the actual world be concrete, and all the other worlds abstract, is an "absolute fact", holding across all the worlds, rather than a proposition that is true at individual worlds. This makes for a conflict with the idea that much of what is actually true is contingently so, i.e. that it is contingent which world is actual. We should no doubt hold fast to that idea: it is at the root of our commitment to other possibilities, at the root of our being up to our necks in modality (Section 3). But if the actual world is concrete, and all the others abstract, the idea that "things could have gone differently" requires that an abstract item could have been concrete---which most philosophers who endorse an concrete/abstract distinction would resist.

(iii): The words 'concrete' and 'abstract' are not just vague, in the sort of way that being red, or being bald, are vague: namely, made precise by specifying a position somewhere in one, or perhaps a few, reasonably well-defined spectra, like hue, brightness and number of hairs on the head. These words are not so much vague as ambiguous, in the sense that making them precise is not a matter of position in one or a few spectra. Thus different philosophers make 'concrete' precise in very different ways. Some only give examples. The paradigm much-used example is tables and chairs (wittily dubbed 'medium-sized dry goods' by the philosopher J.L. Austin), but one surely should add much smaller and much bigger "ordinary objects", such as bacteria and stars. On the other side, the paradigm much-used example of 'abstract' is sets, and other objects of pure mathematics such as numbers and geometrical figures. But listing examples gives no indication of where the boundary lies. And the general notions often invoked cut across one another. Thus should 'concrete' be taken to mean being material? As I mentioned in Section 8, this is itself ambiguous; (matter of a familiar sort? having mass? having energy?). Or should 'concrete' be taken to mean being located in space and time? Or as being either a cause or an effect (or both)? Besides, a good case can be made that those paradigm abstract objects, sets, can be concrete according to some of these proposals. For example, why not say that a set of ordinary objects each of which is located in space and time is multiply located at the places where the objects are? And some philosophical accounts of causation take causes and effects (i.e. events, the relata of causation) to be sets. To sum up this discussion: the concrete/abstract distinction is very unclear. Of course, this is not the place to try and "clean it up". But all I need here is the point that because it is unclear, the suggestion above, that non-actual worlds are sets, cannot claim any substantive merit in classifying such worlds as abstract.

The second objection is that the suggestion <u>assumes</u> the notion of possibility; it does not analyze or explain it. For if a possible world is a set of sentences, then they must be consistent with each other, i.e. possibly all true. (Equivalently, a possible world taken as a long conjunction must be possibly true.) But there is no unproblematic, in particular no syntactic, test for consistency. For inconsistency is not just a matter of the set containing 'P' and 'not-P', for some P. (Equivalently: not just a matter of the long conjunction containing 'P' and 'not-P' as conjuncts.) That is only one, very simple, way to be inconsistent. Relations between the meanings of non-logical words provide many other examples. Here we return to the third kind of proposition that beset the early Wittgenstein's combinatorial account of necessity (cf. Part B above). Think of 'Fred is a married bachelor', or 'A male vixen got into the chicken-hutch'. And there is no reason to think that we can somehow analyze all our language, so as to devise a syntactic test for consistency. (In particular, there is no reason to think as the early Wittgenstein did that any proposition is a truth-function of a set of logically independent propositions, so that consistency can be tested by truth-tables: i.e. ascertained by finding at least one row with a 'T'.)

  Nor does it help to move from sentences to sets. Just as there are sets of sentences, or conjunctions, that are inconsistent without "wearing it on their sleeve", i.e. without a syntactic sign of it: so also there are countless sets that, once we endeavour to interpret each of them as representing a possibility, in fact represent an impossibility---without the structure of the set encoding any sign of it. So again, the suggestion assumes, but does not explain, the notion of possibility.

  Here is a simple example. Consider: 'Butterfield is in Rome in August 2024.' That is false, but possibly true. Following the suggestion, we are to apply set theory to actual objects and properties so as to build, with appropriate representational conventions, a set-theoretic "mock-up" or "replica" of this possibility.

  Let us adopt very simple representational conventions, as follows. (They probably work smoothly only for very simple examples, but that will not matter.) We take a period of time, such as August 2024, to be a spacetime region: for example, the Earth during that month. (I set aside the need for further conventions about where and when on Earth, the month begins and ends.) In terms of semantics (Section 5), the referent of 'August 2024' is the spacetime region. And let us for simplicity take a city, such as Rome, during a period of time to <u>be</u> the set of all its contents during that period, or any part of the period. So there is a set we can label by the description, 'Rome-in-August-2024'. This set in the actual world contains e.g. Pope Francis, and the actual Italian Prime Minister; but not Butterfield.

  But with our representational conventions, we can still represent very simply the possibility that Butterfield is in Rome in August 2024. For recalling how intensional semantics gives descriptions like 'the capital of Denmark' different referents at different worlds (cf. Section 5), we see that, according to the suggestion: this possibility just <u>is</u> a set-theoretic fact. Namely: it is the fact that Butterfield is a member of (the set that is) the referent of 'Rome-in- August-2024', at various worlds. (Amongst these worlds, those most similar to the actual world, according to our prevailing criteria of similarity, will no doubt "retain" most of Rome's actual contents during August 2024, e.g. Pope Francis. "There is room in town for both of us".)

  So far, so good. So far, the suggestion that a possibility <u>is</u> a set has held up well. For I have exhibited a set that is appropriately structured to <u>be</u> the possibility that Butterfield is in Rome in August 2024.

  But the problem for possibilities as sets is parallel to that for possibilities as sentences (or sets of them). Namely: given our representational conventions (about periods of times, about cities as sets of their contents etc.), we are equally committed to countless sets that represent an impossibility, with no sign of why they do---and with no hope of evading the problem, by some change of representational conventions.

For example, I take it to be impossible that I am a fried egg. So 'Butterfield is a fried egg' is necessarily false. Yet there are countless sets that, in an exactly parallel manner to the previous example, put me in the extension (set of instances) of the predicate 'is a fried egg'.

Here I admit: <u>if</u> we assume we have in place a framework of intensional semantics that respects the meanings of our words, so that all assignments of extensions to predicates at the various worlds are genuinely possible, and are not ruled out like a married bachelor, male vixen or human fried egg, <u>then</u> indeed, all will be well. That is: <u>ex hypothesi</u>, the sets mentioned by our semantics as representing possibilities, e.g. properties that a man or a fox could have, will succeed in doing so. The sets will not "lead us astray" by making impossibilities appear possible, masquerading in an appropriately structured nest of curly brackets.

But of course even if we make this assumption, the basic problem remains: as it did for sentences, rather than sets. Namely: assuming such a framework of meaning-respecting intensional semantics means <u>assuming</u>, not explaining, the notion of possibility.

We can sum up this objection to possible worlds (or possibilities) being sentences or sets, as follows. Saying that the sentence 'Butterfield could be in Rome in August 2024' <u>is made true</u> by the existence of a certain set looks plain wrong. For by parity of reasoning, one would have to also say that 'Butterfield could be a fried egg' <u>is made true</u> by the existence of an equally legitimate set.

Chapter 3, Section 10:  Lewis' modal realism
So I end with what began this Chapter: Lewis' modal realism. Lewis believes that:
- (i) all the possible worlds are equally real;
- (ii) the actual world is in no way special, except from our standpoint within it; and
- (iii) although we use 'actual' (and 'real' and similar words) restrictedly, for the actual world, 'actual' is like the word 'here': it is what philosophers call an <u>indexical,</u> i.e. it is a word whose referent depends on the context of utterance---but for 'actual' the relevant aspect of context (with respect to which one asks for the referent) is the world, not the spatial place.

So this is the philosophical multiverse, <u>par excellence</u>.

As I said in the preamble to this Chapter: Lewis does not claim to have an irrefutable argument in favour of his view. His extended defence of it (especially in <u>On the Plurality of Worlds</u>) claims only to show that on balance, it is more credible than rival views. He gives several of these rival views a good run for their money. (This includes the last Section's 'sentences and sets' suggestion, which is roughly equivalent to what he calls '<u>linguistic ersatzism</u>'.) His defence also includes much else. Here I just briefly report three main aspects, out of many.

(a): He replies in great detail to various objections to his view (several of which he himself thought of). In particular, he "takes the sting" out of the objection that according to him, the possibles worlds are each concrete, just like the actual world is concrete---and that surely non-actual worlds should be classified as abstract. Namely, as I reviewed in Section 9.C: the concrete/abstract distinction is so unclear, that this 'surely' claim stumbles.

(b): He defends his view about what it is for an object to be in two worlds, as in the previous Section's closing example of Butterfield and Rome. In short: he denies that the selfsame object can be in two worlds. Instead, given an object in one world, another world may contain a suitably similar object: which Lewis calls a '<u>counterpart</u>' of the given one.

(c): He also explains, using his persuasive account of causation, why there is no causation between worlds, i.e. why no event in any world is a cause of an event in another. So his views satisfy our requirement, at the end of Chapter 1 (Section 6), that advocacy a multiverse should not be undermined by the bewildering idea that most of one's readers or hearers are in another universe.

I shall say more about (a) to (c) in Section 2 of Chapter 6.

But to conclude: let me try to live up to Chapter 1's announced standards of being honest about what one can believe, and self-aware about one's intellectual temperament. I must admit that (like most philosophers) I simply cannot believe Lewis' view.

So for me, concerning the question what a possible world exactly is: the jury is still out. So this Chapter is inconclusive, and perhaps disappointing. But there is some consolation: the next two Chapters will not depend on my having endorsed an answer to this question. And the topic of the next Chapter, the Everettian interpretation of quantum theory, will suggest another answer, another conception of what a possible world exactly is. I myself do not find it persuasive; but it is certainly worth considering, and I will do so in Chapter 6.

In any case, this is not the place to further expound or assess Lewis' views. For this book's purposes, it suffices to have established in this Chapter the following three main points. Namely: (i) we are, in our thought and language, up to our necks in modality; (ii) logicians and philosophers have developed detailed frameworks for describing and analysing modal concepts; and (iii) nevertheless, the basic question, 'what exactly is a possibility, or a possible world?', still remains stubbornly difficult.

So I shall end with a glimpse of "Lewis in action". He was a very active philosophical correspondent; and in a letter of 15 June 1984 to the cartoonist Roz Chast, he asked to use her witty cartoon 'Parallel Universes' from <u>The New Yorker</u> magazine as a frontispiece of his book, <u>On the Plurality of Worlds</u>. (The cartoon represents what it calls 'our universe' by a realistic scene of a woman baking cookies; and three successively more dissimilar universes by weirder and weirder analogues of that first scene. Unfortunately, the cartoon was not used in the book.) In this letter, Lewis gave a vivid and witty summary of his modal realism. He wrote:

Dear Roz Chast,

I'm writing to explore the possibility of using your 'Parallel Universes' as a frontispiece in a forthcoming book of mine about possible worlds.

I have gained some notoriety among philosophers by claiming that this world we are part of is just one of many possible worlds; in no way is it special, except from the standpoint of us who inhabit it. It turns out that systematic philosophy goes more smoothly if we suppose that there are many worlds, and I take that to be a good reason why we should believe that there are. My views are highly controversial, to put it mildly; I think 'crazy' is how many would prefer to put it. For years, I've been helping myself to the other worlds when I wrote about one or another philosophical problem. But I never wrote at length about what it mean to believe in them, and why we ought to. Now I have. I've written (well, almost finished writing) a book titled <u>On the Plurality of Worlds</u>. … It is written in prose, not math; but I fear that it still will be a book mostly for specialists, because it presupposes familiarity with a good deal of recent philosophical writing. I'd be glad to send you a copy of the manuscript if you like, but I didn't want to inflict it on you uninvited.

When 'Parallel Universes' appeared, it put many philosophers who saw it in mind of my notorious views. And rightly so: I do claim that there are four such universes. So I thought it would be quite appropriate and fun if your cartoon could appear as a frontispiece in my book. It would please me very much if that could be arranged. …

… It wouldn't do for me to use it if some other author on possible worlds already has. Of course, there <u>are</u> infinitely many other authors who are using it; but I hope all of them are safely off in other worlds, and no thisworldly author has beaten me to it!

Thank you very much for considering my request. And thank you also for the enjoyment that 'Parallel Universes' has given me. …

Sincerely,
David Lewis.

Chapter 3: Notes and Further Reading

Since this Chapter's topic, logic, has been centre-stage in philosophy for a century, there is an enormous literature. But like for Chapter 2, my main suggestion for reading is some of the original masterpieces. As I said for Chapter 2: though daunting, one should at least dip into them. For such masterpieces, I will emphasize Frege and Lewis: who---as is clear from the Chapter---are my heroes.

      For Frege, I suggested in the Notes for Chapter 2, his Foundations of Arithmetic (1884) and the selection of his writings, The Frege Reader, ed. M. Beaney, Wiley-Blackwell 1997. More specifically, for the material in Section 5 of this Chapter, I recommend two of his great essays (which are in The Frege Reader). (i): 'On Sense and Reference', which was also reprinted (translated) in The Philosophical Review in 1948, and is available at the JStor archive of learned journals, i.e. at: https://www.jstor.org/stable/2181485. (ii): 'The Thought: a logical inquiry': which was also reprinted (translated) in Mind in 1956, and is also available at the JStor archive, i.e. at https://www.jstor.org/stable/2251513. I should add, about (ii), that in the title, the word 'Thought' (German 'Gedanke') is Frege's term of art for what I, and most analytic philosophers, call a proposition: in short, the content or meaning of a sentence. (Frege's term 'Gedanke' is unfortunate since he intended his notion to abstract away from the psychological aspects of meaning.)

      For Lewis, the pre-eminent reference for this Chapter is his On the Plurality of Worlds (Blackwell, 1986). As I said in Section 4, the phrase 'a philosophers' paradise' is his. It is the title of that book's Chapter 1; which develops the themes of my Sections 5 to 8. The book's other Chapters (Chapters 2 to 4 respectively): (i) answer objections, (ii) rebut accounts of possible worlds that are rivals to his modal realism, and (iii) develop his counterpart theory account of what it is for an object to be in two worlds (as mentioned in (b) of my Section 10).

      Further details are in some other masterpiece papers by Lewis. His main exposition of intensional semantics (my Section 5) is in 'General semantics' in the journal Synthese (1970); and available in the JStor archive at: http://www.jstor.org/stable/20114749. A companion paper synthesizing this semantics with his account of the pragmatic and social aspects of language (about which I have said nothing) is 'Languages and Language', in Minnesota Studies in the Philosophy of Science, volume 7, ed. K. Gunderson (1975). Both these papers are reprinted in Lewis' first collection of selected papers, called Philosophical Papers, volume I (Oxford University Press, 1983); which is available at: https://academic.oup.com/book/36015.

      Furthermore, these papers, together with all (so far as I know!) of Lewis' papers, can be downloaded from a website built by Andrew Bailey; (which also contains details of his books and book reviews). It is at: https://andrewmbailey.com/dkl/

      Of his books other than On the Plurality of Worlds, the one most relevant to this Chapter is his Counterfactuals (Blackwell, 1973). It expounds his version of the analysis I summarized in Section 7, invoking similarity between possible worlds. It also gives (in its Section 3.3) Lewis' first statement of his account of laws of nature, which is Humean and kindred to ideas in Mill and Ramsey (my Section 6 above): which has been very influential in philosophy of science.

      Finally, Lewis' philosophical correspondence has been published by Oxford University Press: in two volumes, edited by H. Beebee and A. Fisher. Volume 1 (about causation, modality and ontology: and containing the letter to Roz Chast, which I quoted in Section 10) is at: https://global.oup.com/academic/product/philosophical-letters-of-david-k-lewis-9780198855453?lang=en&cc=gb.
And Volume 2 (about mind, language and epistemology) is at:

https://global.oup.com/academic/product/philosophical-letters-of-david-k-lewis-9780198855842?cc=gb&lang=en&q=Industrial%20Policy%20and%20Development:%20The%20Political%20Economy%20of%20Capabilities%20Accumulation&tab=overview

   Apart from Frege and Lewis, I should list under 'masterpieces', the books by Berkeley and Wittgenstein mentioned in the first two proposals about what a possibility really is (i.e. Section 9, Parts A and B). They are: G. Berkeley, <u>Treatise concerning the Principles of Human Knowledge</u> (1710), which is available in many editions, including online, for example at:
https://www.cambridge.org/core/books/berkeleys-a-treatise-concerning-the-principles-of-human-knowledge/DAB1D1CB81E7D0659900B4CDF270E3C2
and L. Wittgenstein, Tractatus Logico-Philosophicus (1921) , which is also available in many editions, including online at the Internet Archive, at:
https://archive.org/details/tractatuslogicop1971witt/page/n5/mode/2up

Turning to secondary reading: again, there are excellent entries about the topics of this Chapter in the internet resources suggested in the Notes for Chapter 1. For example, the entry in <u>The Stanford Encyclopedia of Philosophy</u> on 'Possible Worlds', discusses Lewis' modal realism in its Section 2.1, and combinatorialism (cf. my Section 9, Part B) in its Section 2.3. It is at:
https://plato.stanford.edu/entries/possible-worlds/
   As to books, an excellent monograph on Leibniz's views is: Benson Mates, <u>The Philosophy of Leibniz: Metaphysics and Language</u>, Oxford University Press 1989; available at Oxford Scholarship Online, and on the Internet Archive at
https://archive.org/details/benson-mates-the-philosophy-of-leibniz-metaphysics-and-language
   Looking beyond logic to the philosophy of science, specifically to the topic of determinism (cf. my Section 8): an excellent survey of the issues is J. Earman, <u>A Primer on Determinism</u> (Kluwer, 1986). Earman's work throughout the philosophy of science, especially of physics, has been magisterial. This book is available at:
https://sites.pitt.edu/~jearman/Earman_1986PrimerOnDeterminism.pdf
and other books and papers can be downloaded from the parent site,
https://sites.pitt.edu/~jearman/

So much by way of masterpieces, and secondary reading, in the philosophy of logic and language. Going beyond this, I will here just give a bit more detail about just two of this Chapter's themes, as follows. (1): The subject-matter of pure mathematics; this will develop themes in my Sections 1 and 9 Part C. (2): The endeavour of reduction (cf. my Section 1).
   My rationale for these two choices is that (1) will lead to a brief discussion of another "Pythagorean" multiverse proposal; while (2) will connect reduction with other philosophical themes such as supervenience (cf. Section 8) and emergence, which will figure in the next Chapter.

(1): This Chapter touched on the philosophical question, what is the subject-matter of mathematics, at two places. Section 1 reviewed the achievement of the early twentieth century in casting all of pure mathematics as part of set theory; and Section 9 Part C criticized the concrete/abstract distinction as being very unclear---so that in particular, saying that sets are abstract might not prevent them from being located in space and time, or from being causes or effects.
   The point now is that those two discussions pull in opposite directions---and that seeing the tension between them can prompt a "Pythagorean" view of the nature of mathematics. For the first discussion consolidates the late nineteenth century distinction between pure and applied mathematics (cf. Chapter 2, Sections 7 and 8): a distinction that seems, when one hears philosophers say 'concrete' and 'abstract', to be precisely a distinction between studying abstract objects (as in pure mathematics) and studying concrete objects (as in applied mathematics). So since the second discussion criticized the concrete/abstract distinction as unclear, what should we conclude about the validity of the distinction between pure and applied mathematics?

This is a live question in the philosophy of mathematics. We philosophers do <u>not</u> have an agreed uncontroversial view of what the objects of mathematics---numbers, geometrical figures etc.---<u>really are</u>, notwithstanding the twentieth century's achievement in showing that by adopting appropriate definitions of them as sets, one can recover, i.e. derive, the sentences taken as true in mathematics: sentences like '2+2=4', and 'all equilateral triangles are equiangular'---albeit now understood as about certain sets, not about numbers and triangles as <u>sui generis</u> entities.

Of course, this is not a book about the philosophy of mathematics. Fortunately for me. So I do not need to justify an answer to the above question; or to related ones, like (i) how best to repair (i.e. make precise) the concrete/abstract distinction, or indeed the basic question (ii) what exactly is a number, or a triangle?

But in philosophy, perhaps more than any other discipline, one question leads to another. And indeed: if one rejects the concrete/abstract distinction, one may be tempted by what I called a "Pythagorean" view of the nature of mathematics. Namely, that the world, i.e. the actual world of tables and chairs ('medium-sized dry goods') and bacteria and stars (cf. Section 9 Part C), <u>is</u> mathematical. That is: the world is, not just accurately described by mathematics, but is <u>made of</u> mathematical objects. Numbers and triangles are literally <u>in</u> the actual (allegedly concrete!) world.

And for us, with our focus on multiverse proposals, this Pythagorean view is relevant in two ways. First: if true, it would alter (but not necessarily solve) the problem that in Section 9 Part C we saw confront the suggestion that non-actual possible worlds are sets ("abstract") while the actual world is not a set ("concrete"). The problem was that set theory is too "profligate" in its ability to construct sets. That is: under whatever representational conventions we adopt, there are bound to be sets that represent impossibilities (like Butterfield being a fried egg) in just the same way that there are sets representing possibilities. So on the Pythagorean view, the status of this problem---solved? still recalcitrant?---will depend on what the Pythagorean says about the constructive, or generative, power of its in-the-actual-world mathematical objects. Maybe it can somehow avoid being profligate and representing impossibilities. But the jury is out.

Second: I should report that the popular book advocating a multiverse (in several senses), which I recommended in the Notes at the end of Chapter 1, also advocates this Pythagorean view of mathematics. The book is Max Tegmark's <u>The Mathematical Universe</u> (2014). In a review of it, I criticized Tegmark's proposals, especially his Pythagorean view, at some length. I will not here repeat the details of my criticisms; (for the review is also cited in those Notes, and is on the internet). But to help orient the reader, I will just summarize his claims, and the "core" of my critique.

First, Tegmark advocates the cosmological multiverse and the Everettian multiverse; which I will treat in the next two Chapters. He labels these multiverses as 'Levels'; and he distinguishes within the cosmological multiverse whether the laws of physics vary across the different universes that are contained in the multiverse. (More details about this idea in Chapter 5.) So Tegmark labels the cosmological multiverse as comprising 'Level 1' and 'Level II'. And he labels the Everettian multiverse as 'Level III': more details in Chapter 3. But the relevant point here is that he then goes on to advocate, not just that his cosmological-cum-Everettian multiverse <u>is described by</u> mathematics, i.e. instantiates a mathematical structure; but that it <u>is</u> mathematics. This is of course the Pythagorean view discussed above.

He also says that all mathematical structures exist: including all the structures large and intricate enough to encode or represent, as we would naturally say---though Tegmark would say: be---various possible cosmological-cum-Everettian multiverses. (Tegmark does not say how

'various possible' should be understood; but nevermind that here.) So the upshot is that the cosmological-cum-Everettian multiverse that Tegmark first advocated, and labelled Levels I to III, is just one of countlessly many mathematical structures. They are all equally real: just as real as the multiverse at Levels I to III which he first advocated. Thus Tegmark is claiming that all of reality, comprising the physical and the mathematical, is a mathematical multiverse: which he labels 'Level IV'.

So much by way of summary. The "core" of my critique lies in the fact that even if one is a Pythagorean like Tegmark, the distinction between pure and applied mathematics remains. One sees the distinction in play, in the idea of a <u>physical quantity</u>.

All agree that when physics describes the world using mathematics, it does not just attribute a pure ("raw") number (or similar quantitative measure or magnitude) to "bits of reality". The attribution is always of some number of units of a physical quantity: 5 units of energy, 7 units of angular momentum, 9 units of electric charge etc. Without mention of the quantity concerned, the description is so incomplete as to be meaningless, e.g. 'This object has number 5'. Thus even if numbers and the other objects considered part of the subject-matter of pure mathematics are in the world, as the Pythagorean claims, nevertheless there is undoubtedly more to the physical world than these numbers etc. Namely, the pattern of occurrence of the quantities; (where 'pattern' includes their relations to one another, and the relations of their values, as stated in the laws of a physical theory).

On the other hand, no such quantity gets mentioned in a work of pure mathematics. That is, the enterprise of pure mathematics as it is conceived today wholly disregards which quantities physics needs (energy, charge) and which it does without (such as erstwhile contenders like caloric). (If like me you endorse the Humean doctrine from Chapter 2 that physics is contingent, you may well see (as I do) this disregard as part and parcel of pure mathematics' enterprise being to formulate, and justify by proof, necessary propositions.)

In this way, the distinction between pure and applied mathematics is, I think, mandatory in the light of the development of logic and mathematics in the last 150 years---as we reviewed at the end of Chapter 2 and in this Chapter. Indeed, it is a defect of Tegmark's book that he does not connect his multiverse proposals to this development, nor to its philosophical ramifications such as this Chapter's philosophical i.e. modal multiverse. In any case: the upshot for Tegmark is that we can accept his "Level IV" claim that all mathematical structures exist . . . but if this means <u>pure-</u>mathematical structures in the modern sense, i.e. regardless of physical quantities, then his claim has little bearing on the debates about the physical multiverses of his Levels I to III. And accordingly, the rest of this book will set his Level IV aside.

Finally, a note about using the label 'Pythagorean' for the view that the empirical world is made of numbers. The historical Pythagoras (ca. 572 – 497 BC) is lost in the mists of time; see for example the entries 'Pythagoras' and 'Pythagoreanism' in <u>The Stanford Encyclopedia of Philosophy</u>. So for our purposes, the label is admittedly anachronistic: it is just rooted in the fact that Pythagoras seems to have led a sect of "number-mystics".

The broader theme here is of course the fact that philosophers have for centuries---from Plato to Russell---been preoccupied by the nature of mathematics, especially as a realm of certain knowledge apparently not derived from experience. Cf. again the end of Chapter 2; and from its Notes, the three essays by J. Bennett, M. Burnyeat and I. Hacking in <u>Mathematics and Necessity: Essays in the History of Philosophy</u> (ed. T. Smiley).

(2): In Section 1, I reviewed the idea of <u>reduction</u> of one theory to another: understood as adding to the latter ("reducing") theory, some judiciously chosen definitions of the terminology of the former ("to-be-reduced") theory that enable a deduction of the claims of the latter (those claims now reinterpreted through the definitions). In the first half of the twentieth century, the paradigm example was the reduction of pure mathematics (arithmetic, algebra, geometry etc.) to set theory: which, as I said, was very influential in the philosophy of science, as a template for how a pair of scientific, in particular physical, theories might be related.

Of course, not all pairs. Reduction might fail: one theory could be irreducible to another. And there seem to be other important inter-theoretic relations apart from reduction and its denial. In particular: a theory might <u>supervene</u> on, or be <u>determined</u> by, another theory (in senses of 'supervenience' and 'determination' like those in Section 8); and a theory might be <u>emergent</u> from another (a topic which will figure in the next Chapter).

Hence there is nowadays a large literature in philosophy of science about these various inter-theoretic relations. Since I have written three articles that each try to both survey the situation, and to argue for some topical, perhaps contentious, claims: I recommend them here.

The first two (from 2011) form a pair, focused on the relations between reduction, supervenience and emergence. The third relates reduction to a doctrine, <u>functionalism</u>: which was first introduced as a label for a position in the philosophy of mind, but was recently brought in to the philosophy of physics. (The position is vague, but normally associated with a failure of reduction, though compatible with supervenience. I aim to show that work of Lewis (the same one!) makes functionalism more precise, and shows it to be compatible with reduction.)

The papers are:

(i): 'Emergence, Reduction and Supervenience: a Varied Landscape', <u>Foundations of Physics</u>, <u>41</u>, 2011, 920-960. Available at: http://arxiv.org/abs/1106.0704: and at: http://philsci-archive.pitt.edu/5549/

(ii): 'Less is Different: Emergence and Reduction Reconciled', in <u>Foundations of Physics</u>, <u>41</u>, 2011, 1065-1135; http://arxiv.org/abs/1106.0702;  and at: http://philsci-archive.pitt.edu/8355/

(iii): 'Functionalism as a species of reduction' with H. Gomes. In <u>Current Debates in Philosophy of Science</u>, ed. C. Soto, Springer: Synthese Library 477 (2023), pages 123-200. Available at:
https://arxiv.org/abs/2008.13366; http://philsci-archive.pitt.edu/18043/

# Chapter 4: All the worlds encoded in the quantum state of the cosmos

This Chapter expounds the multiverse proposed by the Everettian interpretation of quantum theory. The first half is largely independent of previous Chapters' philosophical discussions. But philosophical themes will emerge as the Chapter unfolds.

The Chapter proceeds in four stages. First, I introduce quantum theory (Chapters 4.1 to 4.3). I build on the last Chapter's discussion of state-space (Chapter 3.3), so as to emphasize how strange the conception of quantum state is. This leads, in the second stage, to the measurement problem, symbolized by Schroedinger's cat (Chapters 4.4 and 4.5). This problem has no agreed solution. But I will, in the third stage, (Chapters 4.6 to 4.10) develop just one approach: the Everettian interpretation, with its multiverse. In this approach, a physical process called <u>decoherence</u> will be crucial.

These first three stages will all emphasize what one might call 'synchronic issues': issues about the quantum state at a single time. The topic of time, or diachronic issues, will enter only at the last stage (Chapters 4.11, 4.12), which focus on how the Everettian treats probability. There, I will press one philosophical question that this multiverse raises---what exactly is objective probability?

<u>Chapter 4.1: What is matter? From lumps in the void to fields</u>
So far, our rapid review of physics has consisted of: (i) the rise of mechanics, especially Newton's theory of gravity (Chapter 2.3 to 2.6), and the idea of a <u>state-space</u> (Chapter 3.3).
To understand the Everettian multiverse, we need to understand how in quantum physics, the notion of state is very different from the notion in classical physics.

To prepare for that, the clearest and most vivid route is to review how our conception of matter developed historically, from ancient times to the end of the nineteenth century.

In this development, the main theme is that the idea of matter as a lump of stuff surrounded by void (vacuum) gave way to the idea of a <u>field</u> that pervaded all of space. (Here, I hasten to explain that 'field' has nothing to do with fields of wheat etc. in the countryside.) A field is, rather, there being a physical quantity associated with each place in space; and a state of the field is therefore an assignment of a value to the quantity at each place. An elementary example is the temperature of the air throughout a room: hotter here, cooler there. Strictly speaking, temperature does not make sense at an extensionless point of space: it is an average property of the air in a small volume, say a cubic millimetre, around the point. But let us idealize, and speak of a temperature at every point of space in the room. Then an assignment of temperature values to all points is a state of the temperature field.

So let us begin with lumps in the void. We discussed this conception of matter in Chapter 2.2 and 2.3. We saw that it was advocated not just by ancient atomists like Democritus and Lucretius, but by many seventeenth-century mechanical philosophers, including Newton himself. In particular, we discussed how non-obvious, indeed unclear, it is: as regards both how it might explain the very varied phenomena we see around us, and how such lumps might interact (whether by contact-action, or by action-at-a-distance).

Then in Chapter 3.3, we noted how complicated a collision between two such lumps really is. This led to the idea of the <u>point-particle</u>: mass concentrated at an extensionless spatial point. So on this conception, ordinary objects are clouds, more or less dense, of such point-particles. This idea was introduced as an idealization by Euler (1707-1783), and then advocated as physically real, i.e. the true nature of matter, by Boscovich (1711-1787).

Again, we should pause over how non-obvious, even problematic, the idea is. For any point-particle, its density, i.e. the ratio of mass to volume, is infinite. So if one advocates point-particles with different masses, one must accept different sizes of infinity, in order to describe their mass-densities. Besides: how do point-particles exert force on each other? And what happens if they ever collide? Boscovich himself---writing in an era when Newton's theory of gravity with its action-at-a-distance was accepted---suggested that at very short distances, a repulsive force, that is ever stronger at shorter distances, comes in to play and overcomes the particles' gravitational attraction: so that collisions never occur. But whatever you say about collisions, the question arises: can you give a good account of the contact and interaction of ordinary objects?

Difficult questions like these suggest a rival conception of matter, as <u>continuous</u>. On this view, there is no void anywhere, not even on the tiniest length-scales: matter fills space completely. As we mentioned in Chapter 2, Descartes endorsed this conception. He explicitly identified matter and extendedness; (while in his metaphysics, mind was essentially unextended). During the eighteenth century, this conception went on being developed. Indeed, the exact mathematical description of how continuous matter moves, and how one part of it exerts forces on, and responds to forces from, various other parts, is a very subtle affair. It requires a lot of advanced calculus, as well as physical insight. Unsurprisingly, there was, from the time of Euler onwards, a century-long struggle to achieve this description.

The result, in brief, is to describe matter as a field, in the above sense. For think of a continuous piece of matter. To begin, let us suppose for simplicity that the matter is utterly rigid. That is: the distance between any two of its material parts, no matter how tiny, remains constant over time. So think of a metal bar, and set aside your knowledge that it has layers of microscopic structure, i.e. crystals, atoms etc. Although it is rigid, properties such as mass-density and temperature may well vary across its expanse; and so these properties call for a field description. Besides, for continuous matter that is not rigid---that can be deformed (like an elastic solid: think of a pencil eraser) and-or compressed (like a liquid or gas)---the positions and velocities of its material parts are not "locked-in-step" together. So the parts' positions and velocities, as well as of their density and temperature, also call for a field description. No wonder that some advanced calculus is required.

In the nineteenth century, electricity and magnetism "went the way" of continuous matter such as fluids. That is: it turned out that, whatever the ultimate micro-structure of matter was (point-particles or continuous), electric and magnetic forces take time to propagate across space, between, say, positively and negatively charged matter. As discussed in Chapter 2.3, this is unlike the gravitational force, as it had been described by Newton. According to him, gravitational force propagates instantaneously and unmediated, i.e. without need of an intervening medium.

Besides, these propagating electric and magnetic forces call for a field description. That is: one needs to attribute to each point of physical space, a vector, i.e. a line-segment in physical space (given by three real numbers relative to a coordinate system at the point) which is the (value of the) <u>electric field</u> at that point. This vector represents the electric force that would accelerate a stationary point-like electric charge, if the charge were at that point. And similarly for the magnetic field: though with the difference that it represents the force felt by a moving electric charge. (Here, the phrases 'if it were at', and 'would accelerate' signal a counterfactual conditional. This is another example of science being up to its neck in modality: cf. Chapter 3.3, 3.7.)

Besides, this field description of the electric (or the magnetic) field is not "just" a very convenient way of stating an infinite conjunction of counterfactual conditionals: namely, as a mathematical function from spatial points to vectors located there. For Maxwell (1831-1879), in his stupendously successful theory of the electric and magnetic fields, showed that there is much more to these fields that their describing how a charge would accelerate.

His theory unified electricity and magnetism as two aspects of a single field: the electromagnetic field. It also showed light (and later: radio waves etc.) to be waves in this field. That is: light is an oscillating pattern of electric and magnetic vectors at points of apparently empty space. It is a pattern that propagates, like a wave-form on the surface of the ocean. But it propagates at the speed of light.

Furthermore, this field has energy and momentum: quantities previously attributed only to matter, i.e. to stuff that had mass. That is: the field can convey energy and momentum from one place to another. Thus when you listen to the radio, your aerial is energetically excited, i.e. given energy, by the arriving pattern in the electromagnetic field; and the pattern of excitation is then decoded and amplified into sound.

To sum up: by the end of the nineteenth century, classical physics had a broadly dualist ontology of matter and field. The picture was that matter with mass (and with energy and momentum) is localized in space. It was unknown, and controversial, whether it consisted ultimately of point-particles or of continuous, space-filling, matter. But in the space between localized pieces of matter, there was: not just Newtonian gravity, with its action-at-a-distance; but also an all-pervading electromagnetic field that is the medium by which electromagnetic interactions between bits of charged matter occur, and that also itself possesses energy and momentum.

In the twentieth century, this dualism was overcome---with all-pervading fields getting the upper hand. This happened in various ways. But we need only state two.

First: Einstein's relativity theory (from 1905) identified mass and energy; so that one speaks of 'mass-energy'. So the quantity, mass, that had from Newton onwards been attributed only to matter, was now seen as also an attribute of the electromagnetic field.

Second and more important for us: from the mid-1920s, quantum theory replaced classical physics' matter---even a single point-particle, not only extended matter---by a field. But it is a very strange field even for a single point-particle. And it replaces the classical electromagnetic field by another field that is also strange, in a way exactly parallel to the strange field that it postulates for a piece of matter such as a point-particle.

This strangeness is the source of all the problems about interpreting quantum theory---and it will dominate this Chapter.

Chapter 4.2: The quantum state: probabilities for classical alternatives
The clearest way to grasp this strangeness is to go back to the idea of that a theory attributes to the physical systems it describes, instantaneous states.

We saw in Chapter 3.3, that in classical physics, specifically Newtonian mechanics, the state of a point-particle is given by an ordered set of six real numbers, a 6-tuple: three numbers for its position in space, and three for its momentum. This meant that the state-space of a point-particle, that can be anywhere and have any momentum, is the set of all 6-tuples of real numbers. This is a six-dimensional space: where we use the word 'space' because, although this is not physical space, we can use geometrical ideas in describing it. And we saw that for more complicated systems, the state-space rapidly becomes more complicated and intricately structured. Even if we set aside all the momenta, and consider only the positions of the component parts---which is called the configuration of the system---the space of configurations (called 'configuration-space') rapidly becomes complicated.

Now that we have the idea of a classical field, we can also talk about an instantaneous state of such a field. Think for example of the electric field throughout 3-dimensional physical space. Following the discussion above (Chapter 4.1), its state is of course the assignment of an electric field vector at each spatial point. Such a state is also called a <u>field-configuration</u>. So this requires infinitely many real numbers to specify it: because, for each of infinitely many points in physical space, we must specify three real numbers. We say the field's state-space, i.e. its set of instantaneous states or configurations, is an <u>infinite-dimensional</u> space. (Again, we say 'space' and 'dimensional' because we can again use geometrical ideas: much of the intuition, and precise results, about finite-dimensional spaces carries over to infinite-dimensional spaces.)

Now we can state how quantum theory is strange. It lies in a striking contrast between states in classical physics and states in quantum physics. This contrast applies equally to (i) a point-particle, and to (ii) a finite set of them---any such set would have a finite-dimensional classical state-space---and even to (iii) a field (which has an infinite-dimensional classical state-space).

In short, the contrast is this. A classical state is an assignment of specific values to appropriate quantities. For our purposes here, we can neglect ideas about momentum, and focus only on position and similar quantities, i.e. on configurations. So a classical state is an assignment of specific values: either to the positions of a material object's component parts, or to field-quantities such as the electric field at all the points of physical space. But …

A <u>quantum state</u> is an assignment of a "square root of a probability" to <u>every</u> <u>possible configuration</u> of the corresponding classical system!

So a quantum state is a function in the mathematical sense. Its inputs (as we discussed: also called 'arguments') are the classical configurations, and its outputs are "square roots of probabilities". Here, what matters most---and what is most revolutionary about quantum theory---is, not the curious "square root of probability" outputs (which I will discuss shortly), but: the fact that a <u>single</u> quantum state mentions, i.e. takes as its domain of inputs, <u>all</u> possible classical configurations.

This fact will be the origin of both the measurement problem, and of the Everettian proposal about how to solve it.

Even for a point-particle, the proposal is hard to get one's mind around. For classical physics posited a point-particle. Its possible configurations were its possible spatial positions. Agreed, the theory is involved in modality (as discussed in Chapter 3.3). But only one configuration is actual: where the point-particle happens to be. Now quantum physics tells us: there are no such point-particles, each with a single actual position. Each such is replaced by what gets <u>called</u> a 'quantum particle'. But this entity hardly deserves the name 'particle'. For it has no single position. Indeed, it seems thoroughly smeared out in space. For the actual state of this entity, at some time, is an assignment to each point of space---i.e. to each possible configuration of yesterday's classical point-particle---of a number, which (once squared) gives a probability. In short, the state of this so-called quantum particle is a <u>field</u>.

But it is not a field like Newton's gravity, or Maxwell's electric field. For the so-called particle is not, as I put it, 'thoroughly smeared out in space', in the sense of being a cloud of mass or of electric charge. It is a field of probabilities (or rather, of their square roots). This field, this function on classical configurations that assigns to each configuration a square root of a probability, is called a <u>wave-function</u>. It is almost always written as a Greek letter, especially $\psi$ (pronounced 'psi') or $\phi$ (pronounced 'phi').

Besides, where classical physics posited two point-particles, and so configurations that are 6-tuples, and so a six-dimensional configuration space: quantum theory says the state is a wave-function on this six-dimensional space. So the 'smearing' of what is (undeservedly) called the 'quantum two-particle system' is a smearing, not in physical space, but in the abstract space of 6-tuples. And so on, for the quantum replacements of more complicated classical systems. That is: the quantum state, the wave-function, has as its domain of inputs (its arguments) the more complicated classical configuration space.

So far, I have summarized the mathematical idea of the quantum state as a function on classical configurations. But the picture gets yet stranger, when we ask what is the physical meaning of this function. Again: stranger, even for a point-particle---or rather for what replaces the classical point-particle and is honorifically labelled 'quantum particle'. For one asks: probability of <u>what</u>? And the answer is a mouthful, that refers to the outcome of a possible measurement, if you were to undertake one, on the system.

For the answer is, for a quantum particle: for each place (i.e point) <u>x</u> in space, the value (output) of $\psi$ at the argument <u>x</u> gives the probability, were you to measure the quantity position on the system, that you would get the outcome 'It is at <u>x</u>'. (Strictly speaking, the value is what I called the curious "square root of probability". But this idea of square root is a minor aspect, which we can postpone to the next Section.)

Equivalently, we can think of measuring a quantity with just two values 'Yes' and 'No' (or if you prefer: '1' and '0') that is defined in terms of the place <u>x</u>. In effect, to measure this quantity is to ask the system the question 'Are you at <u>x</u>?'. Thus the value of $\psi$ at the argument <u>x</u> gives the probability of getting the answer 'Yes' to this question.

The reason why I call this answer 'strange' is that it means that the basic interpretation of the theory's most central mathematical notion, its very concept of state, is in terms of measurement. For think what this implies. Suppose I ask the quantum theorist what their theory of, say, an atom, written in their mathematical language, means in physical terms. I ask: what information about the atom is contained in this mathematical notion $\psi$ that they ascribe to the atom? And their official reply is that $\psi$ gives probabilities of measurement outcomes: measurements using an apparatus that (for an experiment on an atomic system) is typically more than a million million million times bigger than the system being measured.

One naturally asks: how can this interpretation of $\psi$ possibly hold up? For it invokes systems, viz. measurement apparatuses, that are not only utterly different from the system we are concerned with, but also vastly larger---and vastly varied. Can such a grossly extrinsic conception of state, for e.g. an atom, really be true?

To this, the short answer is that until now, more than eighty years after this conception of state was formulated, it is indeed still unrefuted. It is unrefuted for the simple but all-important reason that calculating with it, with due care, delivers the right answers to countless experiments---right answers that underpin countless modern technologies. But on the other hand: not only does every newcomer, every student of quantum theory, find this conception of state very hard to believe---indeed, bewildering. Also, most physicists and philosophers who consider in detail this conception, and the questions it raises, conclude that it is <u>not</u> satisfactory.

More precisely: either they conclude that though unsatisfactory, this conception is the best we can now do, and we must hope that the future will bring insight, maybe even a whole theory replacing quantum theory; or they conclude that we already have some special account of the mathematics of quantum theory, and-or how we apply this mathematics to the empirical world, that vindicates this conception. But there are many such special accounts, which get called 'interpretations of quantum theory'. There are about half a dozen main ones, each with many distinctive varieties. The debate between them still rages, decades after quantum theory was formulated---and one such interpretation is the <u>Everettian interpretation</u>, with its multiverse.

But before discussing that, there is more to say about this strange conception of state.

Chapter 4.3: Amplitudes and quantum fields
We can sum up the exposition so far, in terms of how quantum theory replaces the classical physical description of two or more particles. For two particles, the quantum state is an assignment, to <u>each pair of points</u> of physical space, of a number which (once squared) gives the probability, were you to measure the two position quantities, of getting the answer "Yes" to the two specific questions, "Is one of the particles here?", for the two points. And similarly for how quantum physics replaces classical physics' description of <u>N</u> particles. The state is an assignment to each N-tuple of points of physical space, <$x_1,y_1,z_1,x_2,y_2,z_2,\ldots,x_N,y_N,z_N$> (i.e. each sequence of 3N real numbers) of a square root of a probability.

There are two further comments to make. The first comment is about quantities other than the system's position; the second comment is about how quantum theory treats fields rather than particles.

(1): Recall that I called the wave-function's outputs, i.e. the values of the function, "curious square roots of probabilities". The explanation is that there is a kind of number which this book has so far not mentioned, called a <u>complex number</u>. In effect, a complex number encodes a pair of real numbers in ways that are fruitful. In particular: taking the square of a complex number delivers a third real number; (in almost all cases, different from both the given real numbers). So the values of the wave-function are complex numbers. They are called <u>amplitudes</u> (also: <u>probability amplitudes</u>).

Using complex numbers is fruitful for quantum theory because it underpins the treatment of quantities apart from position. Recall how our interpretation of the wave-function, above, was in terms of probabilities for outcomes of measurements of position. I said nothing about other quantities such as momentum. But it is natural to expect quantum theory's conception of state to say something about them. Indeed it does, by encoding the extra information in its use of complex numbers, rather than real numbers.

Amazingly, the system's wave-function gives, for <u>any</u> quantity (momentum, energy, what-not), the probabilities of the various possible outcomes of measuring that quantity on the system. So the conception of state is again bewildering, as regards any quantity. For measuring any quantity on e.g. an atom will involve an apparatus that is vastly larger than the atom. But I should also note that the mathematics of how the wave-function encodes all the probabilities for all the quantities is unified and very elegant: indeed very geometrical. For calculating probabilities for various different quantities turns out to be a matter of expressing a vector (not in physical space, but in an abstract space) as a sum of vectors, in various different ways. I will say a bit more about this in the next Section.

(2): Finally, let me return to the idea of fields. I introduced this idea for classical fields. Recall the classical description of a fluid, taken as being made, not of atoms jostling each other in a void, but as made---on all length-scales, no matter how minuscule---of extended stuff. Or recall the electric field, whose state is an electric vector at each point of physical space. But then we learnt that the quantum replacement of a classical point-particle is a field of (square roots of) probabilities, defined on configuration space. So one naturally asks: what about the quantum replacement of a classical field, such as the electric field?

Amazingly, the same strange idea of state works again; as follows. We saw that the configuration of a classical field is given by an infinite number of real numbers (not by 3N real numbers for some whole number N). Thus the quantum replacement of such a field has as its state an assignment, to each possible configuration of the classical field (one of which, we used to imagine, is physically real), of a complex number: a probability amplitude. This quantum replacement is called, of course, a <u>quantum field</u>; and the theory of them is called <u>quantum field theory</u>.

Note the dizzying mathematical abstraction. The configuration space of the classical field was itself infinite-dimensional; and now quantum theory posits a state space of functions, each of which has as its arguments that infinite-dimensional space. This makes quantum field theory much more complicated, mathematically, than the theory of quantum particles, i.e. the theory of wave-functions on finite-dimensional classical configuration spaces.

Besides, quantum field theory gives a supremely successful replacement not just of what classical physics called fields, like the electric field, but also of what classical physics called 'particles'. So there is here another step of conceptual novelty---and again, a dizzying one. For quantum field theory revises what I have said so far about quantum particles---though 'revises' means here 'extends' rather than 'overturns'.

Thus consider the electron. As we have seen: classical physics treats it as a point-particle, located in space at some actual position, moving with some momentum. And similarly for a pair, or any finite number $N$, of electrons. In short: once you fix the number $N$, a classical state-space is defined. And so far in this Chapter, we have learnt that quantum physics replaces this with something probabilistic. Namely: the state provides, for any quantity, a probability distribution over its possible values, where 'values' are understood to be outcomes of a possible measurement. But we can deduce all these probability distributions from a single representation of the state, the wave-function. And as in the classical case: once you fix the number $N$, the quantum state-space---which, to repeat, consists of wave-functions assigning amplitudes to each possible configuration of $N$ classical point-particles---is defined.

So much by way of summarizing the story so far. Quantum field theory goes beyond this. Its state-space is "even bigger" than those we just mentioned. Thus we have so far mentioned: the space of wave-functions for one particle (i.e. complex-valued functions of position in physical space), the space of wave-functions for two particles (i.e. complex-valued functions of pairs of positions in physical space), … , the space of wave-functions for $N$ particles (i.e. complex-valued functions of $N$-tuples of points of physical space). But for quantum field theory, the state-space contains all these infinitely many spaces of wave-functions. More precisely: there is a way of adding together two or three or … even infinitely many such spaces. (The definition of how to add together spaces is in terms of all the ways of adding together elements of the spaces.) So for quantum field theory, the state-space is the sum of all the spaces just mentioned, for all positive whole numbers $N$. This vast sum state-space is called Fock space.

Thus quantum field theory envisages the number of particles as a property of the system, that can vary from one state to another. That is: it envisages the number of particles as a quantity of the system. This amounts to treating the system as a field; and to treating the quantum particle, discussed above, as itself a state: a state of this field.

So quantum field theory's conception of an electron is that it is an excitation (an "agitation" with an associated energy) of the field. Similarly, two particles are a pair of such excitations; and so on. Thus the treatment of, say, five electrons that a quantum theorist first learns (using wave-functions on 5-tuples of points of physical space, as in Section 2 above) turns out to be just the five-particle part of a treatment that encompasses any number of electrons. In the jargon: the elementary quantum state-space for five electrons is a subspace of Fock space.

Again, the degree of abstraction is dizzying. But so is the empirical success. The quantum field theoretic treatment, both of what classical physics called 'fields', e.g. the electric and magnetic fields, and of what classical physics called 'particles', e.g. electrons, yields vastly many precise predictions that have been confirmed. So much so, that quantum field theory is now regarded as the lingua franca of physics.

And yet . . . questions remain. For the official interpretation of the state of a quantum field is exactly parallel to what I reported, and questioned, in the last Section, for the quantum theory of a fixed number of quantum particles. Thus for the electric field, the state (i.e. the function on configurations of the classical electric field) encodes probabilities for the various possible values of any quantity, such as energy or momentum of the field, that you might decide to measure on it. So our interpretative worries at the end of the last Section---can such a grossly extrinsic conception of state really be true?---persist.

Chapter 4.4: The measurement problem: Schroedinger's cat
The worries raised in the last Section about how to interpret the quantum state can be sharpened into an argument, whose conclusion is that quantum theory makes a wealth of flagrantly false predictions about the macroscopic world around us. This argument is called the measurement problem (also: 'the reality problem'). It is vividly illustrated---indeed, symbolized---by Schroedinger's cat: which is a thought-experiment presented by Schroedinger in 1935.

So in this Section, I will expound the measurement problem, and then the cat. To do this, I first need to explain: (1) the idea of superposition, and (2) how the famous equation of motion of quantum theory, the Schroedinger equation, "preserves superpositions".

(1): So far, we have seen that a quantum state prescribes, for any quantity, a probability distribution over the possible outcomes of measuring that quantity (on a system in the given state). For example, a quantity might have four possible values, whose probabilities are 1/4, 1/3,1/3, 1/12: (these add up to 1). Of course: a probability equal to 1 for one value, with probability 0 for any other value, counts as a legitimate probability distribution. It is called a trivial distribution (though 'dogmatic' would be a better name). And for a given quantum state, there are quantities that get such a distribution: one outcome is ascribed probability 1, all the others probability 0. Quantum theory has jargon for this. We say that: the state is an eigenstate of the quantity; (and the quantity is an eigen-quantity of the state---though this second word is less common); and the outcome that is "favoured", i.e. gets probability 1, is the eigenvalue.

As to the more general situation, viz. each outcome getting a probability less than 1 (maybe some get 0, but all together, they add up to 1): the treatment of this situation turns on the idea that quantum states can be added. This is because states are functions, and any two functions that have numbers as their values can be added by, at each individual argument, adding their values. That is: given any two functions f and g defined on the same domain, i.e. the same set of arguments, their sum is defined to be the function, written f+g, whose value for any argument, a say, is the number f(a)+g(a). (This is called 'point-wise addition' of the functions. But 'argument-wise' would be a better name, since there is no need for the arguments to be points in a space, in particular in physical space.) In particular, this is true for quantum states i.e. functions $\psi$ and $\phi$. We can add them argument-wise, writing $\psi + \phi$.

This addition of functions is exactly parallel to how we add vectors; which therefore help to give one geometrical intuition. Think of vectors as directed line segments, with their "tails" at the origin, in the plane or in three-dimensional space. We define addition for them by adding them nose-to-tail, and this corresponds to adding their respective coefficients. Thus in three-dimensional space (recall the triples in Chapter 3.3): one adds the two x coefficients to get the sum-vector's x coefficient; and similarly one adds the two y coefficients to get the sum-vector's y coefficient: and similarly for the z coefficients. So given two vectors $(x_1, y_1, z_1)$ and $(x_2, y_2, z_2)$, their sum is defined to be the vector $(x_1+x_2, y_1+y_2, z_1+z_2)$. In short: we add corresponding coefficients of a vector, just like we add corresponding values of a function.

Returning to quantum theory: if you add two eigenstates for a quantity that ascribe probability 1 to two different eigenvalues as possible outcomes of a measurement (i.e. the eigenstates disagree!), the result is a state that ascribes probability 1/2 to each eigenvalue. This state is called a superposition of the two given eigenstates. Again: we write a '+' sign for addition of states.

This example was one of equal weighting. (One could also call it '50-50 weighting'.) But we can also increase one weight and decrease the other. This is like lengthening one directed line-segment and shrinking the other, before we add them, nose-to-tail. The resulting state is again called a 'superposition', and is written with a '+' sign. It gives the two eigenvalues different probabilities. And again: the probabilities that the state prescribes are the squares of its values for the two possible outcomes i.e. the two eigenvalues. So writing $\sqrt{\ }$ for 'square root' (and setting aside the topic of complex numbers, cf. (1) in Section 3): the values might for example be $\sqrt{(6/10)}$ and $\sqrt{(4/10)}$. Besides, thinking of the state as a vector rather than a function: it has coefficients $(\sqrt{(6/10)}, \sqrt{(4/10)})$.

Note that for any given quantity, almost all states are superpositions for that quantity: its eigenstates are a small, special set of states. In other words: For any given quantity on our system, almost all the probability distributions that quantum theory envisages for it are non-trivial. That is, more than one possible value (eigenvalue) of the quantity gets a probability greater than zero.

(2): Just as classical theories have equations of motion that describe how the system's state changes over time (cf. Chapters 2.6, 3.3, 3.8, 4.1), so does quantum theory. Its equation is called the Schroedinger equation; (he published it in 1926). It has two crucial features.

First: it is deterministic, in the sense of our previous discussions (Chapter 3.3, 3.8). That is: given the system's quantum state at a time, and the forces exerted on it in past and future, the equation prescribes what the state is at all other times. As we put it there: the equation prescribes a unique curve through the quantum state-space, i.e. the space of wave-functions. Later (in Sections 5 and especially 11), we will confront the obvious question: how can this determinism be reconciled with quantum theory's use of probabilities?

But here in this Section, what matters is the second crucial feature of the Schroedinger equation. Namely: it preserves the addition structure of quantum states. That is:

if (i) a system's quantum state, $\psi$, would evolve (i.e. change), in say five seconds, to a state $\psi'$, and

(ii) another of its states, $\phi$, would evolve in that same five seconds to a state $\phi'$, then:

any superposition state of $\psi$ and $\phi$, got by for example "lengthening" $\psi$ and "shrinking" $\phi$, e.g. the state $(\sqrt{(6/10)}\mathbf{psi} + \sqrt{(4/10)}\phi)$, would evolve in that same five seconds to the corresponding superposition, i.e. to $(\sqrt{(6/10)}\psi' + \sqrt{(4/10)}\phi')$.

In short: quantum states' addition structure---including the numbers that are the weights, $\sqrt{(6/10)}$ etc.---is preserved under time-evolution. This property of the Schroedinger equation is called linearity: the Schroedinger equation is linear. (Here, for simplicity, I have written familiar real numbers, $\sqrt{(6/10)}$ and $\sqrt{(4/10)}$, for the weights, rather than the complex amplitudes i.e. 'curious square roots of probabilities', as mentioned in (1) of Section 3. The point about linearity is unaffected.)

So far, this Section has just done some stage-setting: the addition of functions and vectors, and the jargon of eigenstates, eigenvalues, superpositions and linearity. But now quantum theory, in its orthodox formulation, makes an interpretative claim. It is a very important claim, since it is restrictive---and it leads directly to the measurement problem.

Namely: quantum theory says that if for a given quantity, the state is a superposition for that quantity (so: not an eigenstate---it ascribes a non-trivial, "non-dogmatic", distribution over possible measurement outcomes), then the system has no value whatsoever of the quantity.

The paradigm case is the quantity position, for a quantum point-particle: as in Section 2 above. The state is a wave-function $\psi$ that assigns a complex number to positions in space, whose squares give probabilities of outcomes of position measurements. In picturesque language: the state gives probabilities of getting the answer 'Yes' ('1') to asking the system 'Are you at position $\underline{x}$?'. (From now on, I will say 'system' rather than 'point-particle' or 'particle' since the word 'particle' strongly connotes a definite position (and definite momentum etc.): which, as I will stress, quantum theory denies.)

Now imagine that the value of $\psi$ is non-zero only in two separated spatial regions, which we call 'L' and 'R'. Here, 'L' and 'R' are mnemonics for 'Left' and 'Right'. For nothing here will depend on space being three-dimensional; so we may as well imagine it as one-dimensional--- "life on a railway line". So if we draw a graph of the values of $\psi$ (more precisely: their squares), it looks like two humps with a flat line between them. (Think of the road-sign for 'Bumps in the road ahead.')

So quantum theory says that $\psi$ is a superposition of two states. One of these two states is an eigenstate for being in L. So this state says: the system would with probability 1 be found in L, if measured. The other state is an eigenstate for being in R; so this state says, correspondingly:

the system would with probability 1 be found in R, if measured. This superposition can be written with a '+' sign for addition of states. We could write mnemonically: $\psi$ = in L + in R.

It is also common to write states between a vertical line and an angle bracket; (a notation invented by Dirac (1902-1984), one of the great quantum physicists). So for the equal or '50-50' weighting of L and R, we write: $|\psi> = |$in L$> + |$in R$>$.

But, says quantum theory: a system in state $\psi$ has literally no position at all. So the system exists---but it has no location. It only has dispositions to be found in L, or in R, if we were to measure position.

Similarly for other quantities, such as momentum. A system might be in a superposition of momentum eigenstates, for two different values (measurement-outcomes) of momentum, say '1' and '2'. So for equal weighting of '1' and '2', we can write the state as: $|$momentum 1$> + |$momentum 2$>$. Again, quantum theory says that a particle in this state has literally no momentum at all.

To sum up: since for any given quantity, almost all states are superpositions, quantum theory's denial that systems in superpositions have a value (for that quantity) makes the lack of values <u>endemic</u>. Obviously, this situation prompts the question: how can this lack of values be reconciled with the apparent fact that objects <u>do</u> have values for position, momentum and other quantities such as energy?

Besides: recall how classical physics gives supremely successful descriptions and explanations of the physical behaviour of macroscopic objects, by ascribing them definite values of position, momentum etc., subject to equations of motion (cf. Chapters 2.6, 3.3, 3.8, 4.1). Once we recall this, the question becomes more pointed: how can quantum theory's denial of values be reconciled with the supreme success of classical physics?

This question is now easily sharpened in to an argument: an argument that the lack of values contradicts countless facts that macroscopic objects have definite values. We only need to transmit the lack of values from the microscopic realm of electrons, atoms etc., for which quantum theory is indeed successful, to the macroscopic realm of tables, chairs etc., where classical physics with its definite values is successful. This is done by describing, within quantum theory, a measurement of a quantity on a microscopic system (say, an electron) that is in a superposition (for that quantity). In such a description, we can see how the lack of values is transmitted to the macroscopic realm---surely contradicting countless facts of definiteness.

So let us assume we have a measurement apparatus for measuring an electron's momentum that is <u>reliable</u> on each of the two eigenstates, $|$momentum 1$>$ and $|$momentum 2$>$, in the following sense. Starting the apparatus in an appropriate 'ready' state, the state of the pair of systems changes over time, so that at the end of the measurement interaction, the pointer on the apparatus reads the corresponding eigenvalue of momentum. Thus we think of the pointer as being, at the end of the measurement, in front of the digit, '1' or '2', painted on a dial.

Let us use an arrow, $\rightarrow$, to symbolize the change of state over the period of the measurement. And let us write the state of the pair of systems by simply juxtaposing their states on the paper. This state of the pair is, in the jargon of logic or philosophy, the conjunction of the two components' individual states. In physics jargon, it is called a <u>product state</u> (or 'the product of the individual states').

Then our assumption of reliability, that measuring either of the electron eigenstates yields a veridical reading at the end, can be written as:

$\quad\quad\quad |$momentum 1$> |$ready$> \;\;\rightarrow\;\; |$momentum 1$> |$reads '1'$>$

and

$\quad\quad\quad |$momentum 2$> |$ready$> \;\;\rightarrow\;\; |$momentum 2$> |$reads '2'$>$ .

So far, so good. But now suppose the electron's initial state is a superposition: for example, of our two eigenstates, $|$momentum 1$>$ and $|$momentum 2$>$. Then the composite system of electron and apparatus has the conjunction or product state, as its initial state:

(|momentum 1> + |momentum 2> )|ready>. This can also be written as: |momentum 1>|ready> + |momentum 2> |ready>.

Then because the Schroedinger equation is linear, we must accept:

(|momentum 1> + |momentum 2> )|ready>  →
|momentum 1> |reads '1'> + |momentum 2> |reads '2'> .

This is the punchline. For consider the state of the composite system after the measurement: i.e. the second line, or right-hand-side, of this formula. Consider what it says about the apparatus, in particular the position of its pointer. The main point to notice is that it is <u>not</u> an eigenstate of the quantity pointer-position, i.e. of the quantity position, for the pointer.

(There is also another point to notice about this state. It is not a product state: it is not a conjunction of states for the components. Such a state is called <u>entangled</u>; and the theory allowing for such states is called <u>entanglement</u>. We will return to this in Section 4.7.)

So quantum theory denies that the pointer has any position. The situation is like in the example above, $|\psi> = |in\ L> + |in\ R>$. Namely: the pointer only has dispositions to be found in front of the numeral '1' on the dial, or to be found in front of the numeral '2' on the dial---if we were to measure it.

But the pointer <u>not</u> having any position surely contradicts the fact that macroscopic apparatuses give definite readings. Besides, the argument is so simple---depending only on microscopic superpositions and the Schroedinger equation being linear---that it suggests more generally that orthodox quantum theory's denial of values in the microscopic realm will contradict countless facts of definiteness about the macroscopic realm.

So this is the measurement problem. As I said at the start of this Chapter, it has no agreed solution. So our first job, in the next Section, will be to consider some possible solutions: including the Everettian proposal, on which we will then focus.

But before that, it is worth summing up the measurement problem, with Schroedinger's own description of his eponymous cat. In his paper of 1935---which is still worth reading for many reasons, some of which we will touch on later---he writes (at the end of Section 5 of his paper: translation from 1980):

One can even set up quite ridiculous cases. A cat is penned up in a steel chamber, along with the following diabolical device (which must be secured against direct interference by the cat): in a Geiger counter there is a tiny bit of radioactive substance, so small, that perhaps in the course of one hour one of the atoms decays, but also, with equal probability, perhaps none; if it happens, the counter tube discharges and through a relay releases a hammer which shatters a small flask of hydrocyanic acid. If one has left this entire system to itself for an hour, one would say that the cat still lives if meanwhile no atom has decayed. The first atomic decay would have poisoned it. The wave-function of the entire system would express this by having in it the living and the dead cat (pardon the expression) mixed or smeared out in equal parts.

It is typical of these cases that an indeterminacy originally restricted to the atomic domain becomes transformed in to macroscopic indeterminacy.

Chapter 4.5: Solving the problem: the usual suspects
There are three main strategies for addressing the measurement problem. It will be clear that each is a broad church that includes many versions; and that jointly, they are exhaustive. So one has to endorse one of the three. But I will not try to formulate the strategies precisely, and will only give two or three examples of each strategy. Nor is this trio of strategies, and their versions, original to me. Many surveys of the measurement problem give a similar trio. Hence this Section's title: the usual suspects.

I especially admire the survey by Bell, which I will follow (1986). It is a brilliant, and equation-free, introduction to quantum theory and its interpretation. Bell describes, for each of his three strategies, a pair of versions; and for each pair, he gives what he wittily calls a <u>romantic</u> version, and an <u>unromantic</u> version. Unsurprisingly, the Everett interpretation will be the romantic version within its pair.

Bell also makes it clear that for each pair, he prefers its unromantic version. That is perhaps disappointing. But as I discussed in Chapter 1.4, we each have an intellectual temperament which is hard to change, and about which we are obliged only to be self-aware. (John Bell (1928-1990) was a profound quantum physicist, as well as a gifted writer. He also originated what is now called 'Bell's theorem': it is about how correlations between quantum systems defy a very natural form of probabilistic explanation.)

The first strategy is to reject, somehow or other, the formulation of the problem. That is, one rejects the premises of the argument leading to the contradiction. The main idea must be to deny that the quantum state is "physically real", in any sense of that phrase that makes the final post-measurement state (at the end of the last Section) contradict the macroscopic pointer having a definite position.

Bell calls his two versions of this strategy: 'pragmatism' (the unromantic version) and 'complementarity' (the romantic version). Here, 'pragmatism' means---not the philosophical tradition launched more than a hundred years ago by American philosophers such as Peirce, James and Dewey---but 'being practical'. That is: using the theory to calculate probabilities of outcomes, without taking its words and concepts to describe any reality other than those outcomes. (Philosophers often call this 'instrumentalism': the theory is an instrument, a tool for predicting observable facts, but not a description of the world, especially not the world beyond observations.) Obviously, pragmatism, in this instrumentalist sense, shades in to simply not wishing to ponder whether the theory goes beyond predicting observable facts, rather than the firm view that it does not go beyond such predictions.

On the other hand, 'complementarity' is Bohr's word for his views that attempted to go beyond pragmatism and, by explicitly philosophical argumentation, to solve the measurement problem. His core idea was that since measurement outcomes must be stated using the concepts of classical physics, there is no contradiction with, in Schroedinger's phrase, 'macroscopic determinacy'. And this is so, even though: (1) the Schroedinger equation, with its linearity, is always correct; and (2) quantum systems have no values for quantities except the eigenvalue for those quantities for which they are in an eigenstate.

The other two strategies deny, respectively, these claims, (1) and (2). These were, of course, the premises of our formulation of the measurement problem. So one might say that the first strategy aims to dissolve, rather than solve, the problem; while these other two strategies accept that the problem is genuine, and then propose to solve it. Bell himself clearly prefers these two strategies (in their unromantic versions) over the first. As he puts it in another paper: 'either the wave-function, as given by the Schroedinger equation, is not everything or it is not right' (1987, p. 201). Here, 'the wave-function is not everything' means that a system has values for quantities additional to the eigenvalues ascribed by orthodoxy, i.e. (2) is wrong. And 'the wave-function is not right' means that the system is being attributed the wrong state, because (1) is wrong. In our paradigm case of the pointer: we should attribute to it a state of definite position, not a superposition of position eigenstates, and to do so, we should revise the Schroedinger equation.

So suppose we deny (1), that the Schroedinger equation is always correct. Again, this strategy comes in various versions.

The simplest version says that at the end of a measurement process (like that at the end of the last Section), the troubling superposed final state is replaced by an eigenstate of pointer position, so that indeed, the pointer has a definite position. (Usually, advocates of this version also say that the measured system goes in to an eigenstate of the measured quantity: in our example, the electron goes in to a momentum eigenstate. But we can concentrate on the measurement apparatus and its pointer.) Besides, which eigenstate replaces the troubling superposed state is said to be a matter of sheer chance, with each eigenstate occurring with a probability equal to its weight (more precisely: the square of its complex amplitude weight) within the superposed state.

Again, there is jargon: this replacement of the superposition by the eigenstate is called 'the projection postulate', or 'the collapse of the wave-function'. (But this second phrase is also used as a vivid label for the measurement problem, not just for this approach to solving it.) And the prescribed probabilities (the squares of the complex amplitude weights) are called 'Born-rule probabilities'. (This is in honour of Max Born (1882-1970), one of the half-dozen co-discoverers of quantum theory who realized this role of probabilities in a 1926 paper.)

This simplest version of denying (1) occurs in many textbooks of quantum theory. But evidently, it is vague and contentious. For when exactly is the state calculated from the Schroedinger equation to be 'replaced'? Or in other words: what is the exact definition of 'the end of the measurement process'? And since a measurement is, after all, a physical process, how is this suspension of the theory's equation of motion to be justified? Clearly, this version is close to what Bell called 'pragmatism': and we are back at our initial bewilderment that the notion of a system's state should invoke the idea of measurement, which is so extrinsic to the system.

Two other versions of this strategy, developed in response to these difficulties, are worth mentioning. The first is very speculative; the second is down-to-earth. So Bell calls them, respectively, romantic and unromantic; (and as I said, he prefers the latter).

The first says, in a slogan, that consciousness collapses the wave-function. The idea is that the Schroedinger equation only falters at the interface between mind and matter. Although the inanimate physical world may get in to states superposing macroscopically distinguishable alternatives (e.g. of positions of a pointer), once a conscious being "looks", the state changes to a macroscopically definite state.

Obviously, people will differ about how plausible they find this proposal. Someone who is a physicalist (cf. Chapter 3.8) will almost certainly reject it. Yet it might at first sight appeal to the practically-minded physicist, on the grounds that it makes the measurement problem "someone else's problem". They might say: 'surely, physics has no responsibility to describe the relation between mind and matter'. But I would say that this response is just a verdict about disciplinary boundaries, or the division of cognitive labour: about who needs to worry. Whichever discipline one takes the problem to fall within, the proposal is obviously hard to make precise, hard to gather evidence for, and indeed: hard to believe. In particular, why should these collapses of the wave-function due to consciousness respect the Born-rule probabilities?

The other version of this strategy is much more down-to-earth: as Bell says, unromantic. It seeks, as quantum theory's fundamental equation of motion, a "cousin" of the Schroedinger equation. This equation is to be chosen so as: (i) to agree with the Schroedinger equation for microscopic systems like atoms, so that it also gets the vast amount of confirmation that the Schroedinger equation has gathered over the last hundred years; and yet (ii) to disagree with the Schroedinger equation for macroscopic systems like pointers, and cats. So according to this version, the wave-function of a quantum system does indeed collapse i.e. transit to an appropriate eigenstate, in suitable---in particular: suitably large---systems. And this occurs in the inanimate world, wholly irrespective of consciousness: the collapse of the wave-function is an indeterministic physical process. So our task is to find the equation that describes these collapses precisely, in a way that meshes with the established successes of the Schroedinger equation. This will include recovering the Born-rule probabilities for experimental outcomes. In short: our task is "theoretical physics as usual: find the right equations".

In the last forty years, a great deal has been learnt about such cousin equations, both mathematically and physically. But no single proposed equation has yet won the allegiance of physicists, on either theoretical grounds or by being confirmed by experiment. So the question whether the Schroedinger equation will indeed be overturned remains open. For this, we must wait upon the future of physics.

Finally, suppose we deny (2). That is: we say that quantum systems <u>do</u> have values for quantities for which their state is not an eigenstate. The motivation for saying this is of course to keep the post-measurement state given at the end of the last Section, but nevertheless to ascribe a definite position to the pointer.

One version of this strategy is, again, "theoretical physics as usual". It was invented by de Broglie (1892-1987). In the mid-1920s, he was another of the half-dozen creators of quantum theory; and his formulation of the theory explicitly attributes values to quantities additional to eigenvalues (of quantities for which the state is an eigenstate). To be precise, let us consider a system that orthodox quantum theory (the textbook) calls '<u>N</u> quantum particles'. De Broglie proposed that this system's real physical state includes, in addition to the wave-function $\psi$ on configuration space, that always obeys the Schroedinger equation, also---our "old friends": <u>N</u> point-particles, each with a definite actual position.

In themselves, these point-particles are as described by classical physics: at any instant, each has a definite position in space, and over time each moves in a continuous trajectory. The difference from classical physics lies in how they move. Namely: at each instant, their velocity is determined by a combination of: (i) the wave-function, and (ii) the positions of the other point-particles in the system. (The position of another particle contributes to determining the given particle's velocity in an action-at-a-distance manner, similar to Newtonian gravity---except that the influence does not diminish if there is a larger distance between the particles.) The idea in (i), that the wave-function, though it lacks mass and energy, "guides" all the particles in the system, has given rise to the theory's name: <u>pilot-wave theory</u>. (Think of how a pilot guides a ship: but he does not do so by the effects of his mass or energy.)

We do not need the details of this theory. What matters for us is that the pilot-wave description of how the point-particles move gives the key idea for solving the measurement problem. (This merit became clearer in the work of David Bohm (1917-1992), who in 1952 re-discovered the pilot-wave theory.) In the last Section's measurement scenario, the pilot-wave theory takes the pointer to really be a cloud of point-particles. Then it gives a detailed description of how, in each individual measurement, this cloud of particles is guided by the wave function (which always obeys the Schroedinger equation), so as to be either in front of the digit '1' painted on the dial, or in front of the digit '2'. In short: the pilot-wave theory, by ascribing values of position additional to those ascribed by orthodox quantum theory, secures a definite outcome for each individual measurement---solving the measurement problem.

The second version of this strategy, i.e. the second way to claim that quantum systems <u>do</u> have values for quantities for which their state is not an eigenstate while the Schroedinger equation is nevertheless always correct, is the Everettian proposal---with its multiverse. It will be our topic for the rest of this Chapter.

<u>Chapter 4.6: Everett's proposal: a bluff?</u>
Let me start by stating baldly the Everettian proposal. This will show how it counts as a version of this last strategy. (From now on, I will talk about 'the Everettian', rather than Everett himself, since the proposal has developed a good deal since Everett's paper in 1957. Besides, there is controversy about whether the version of the proposal that is nowadays dominant---on which I will concentrate---matches Everett's own ideas.)

The key ideas are as follows. The cosmos as a whole has a quantum state, which always evolves according to the Schroedinger equation: indeed, a <u>very</u> grand version of the equation that describes quantitatively how all the cosmos' component parts interact, exerting forces on one another. Needless to say, no one has come close to writing down this version of the Schroedinger equation. But the Everettian proposes that it is, as usual, deterministic; so that the wave-function of the cosmos never collapses. (In Section 11, we will confront the obvious question: how can this determinism be reconciled with quantum theory's use of probabilities?)

This state is usually written as $\Psi$: where the use of the capital Greek letter (again pronounced 'Psi') is, so to speak, honorific---for no one has the faintest idea how to write it down in detail.

$\Psi$ is usually called 'the universal state', or 'the universal wave-function'. But in this book, I have adopted, since the Introduction: 'multiverse' as the name of all of reality in the most inclusive sense; and 'world' or 'universe' as the name for its "more familiar" parts, where as we discussed, each part is understood to mean 'throughout all of time and space'. So using my jargon: the cosmos' quantum state $\Psi$ is the multiverse's quantum state. But since no one says 'the multiversal state' or 'the multiversal wave-function', I will in this Chapter talk of 'the cosmos' quantum state'. Hence this Chapter's title.

Our knowledge of quantum theory, and our empirical success in applying it to small systems, suggests that superpositions will promulgate, in the way we saw in the measurement problem at the end of the Section 4. So we have good reason to think that the cosmos' quantum state is a vast superposition, i.e. a vast sum, of product states. Each of these is a long "conjunctive state" for countless component systems---not just an electron and a measurement apparatus or its pointer, but countless electrons, quarks, molecules, specks of dust etc.

Again, there is jargon: when items, like numbers or vectors or these product states, get added together, they are called <u>terms</u>, or <u>summands</u> (for 'item that gets summed'). So these product states are terms of the vast superposition.

The Everettian proposes a literal interpretation of this vast superposition. Just as this state contains vastly many product states, so also the cosmos (i.e. the multiverse) contains a plethora of Everettian 'worlds'. (These are also called 'branches'.) Each is represented by such a product state. Some (perhaps many) of these worlds are something like the macroscopic realm familiar to us: with all macroscopic objects (pointers, tables etc.) in definite positions.

But the worlds differ among themselves about these positions; and it is relative to each such world that there are extra values, i.e. values additional to orthodox quantum theory's attributing only eigenvalues. Think, for example, of Section 4's toy-model of a momentum measurement. The two possible outcomes were distinguished by two different positions of the pointer: in front of '1', or in front of '2'. The Everettian proposes that the two outcomes, the two positions, are in different worlds.

This bald statement of the Everettian proposal prompts the obvious question: 'If this is so, how come I have no evidence of the other worlds? In particular, how come I experience a single definite macrorealm: a macrorealm whose objects are definite in position, and other quantities like momentum?'.

As I see matters, there are two main Everettian answers to this question: a traditional one, which was prevalent in the literature between Everett's original paper in 1957 and about 1990; and a modern one which has been prevalent since the mid 1990s. The first answer was cast in terms of measurement processes and their outcomes, as in Section 4's toy-model. I think this first answer is unsatisfactory. It seems like a bluff, or a mere debating tactic. I will discuss it in this Section. But the second answer <u>is</u> satisfactory; although not wholly convincing. It is <u>not</u> cast solely cast in terms of measurement processes and their outcomes. It considers equally all macroscopic objects that classical physics successfully treats as having definite values for position, and other quantities like momentum. And some of these are indeed tiny, for example specks of dust that are too small to be visible; (though I will still say 'macroscopic'). This second answer secures definiteness of position etc. (relative to a world) even for such tiny specks. It does so by appealing to a very important phenomenon that I have so far not mentioned: <u>decoherence</u>. I will take it up in the next Section.

I say 'as I see matters', because since Everett's original paper in 1957, many different Everettian answers have of course been proposed---and then disagreed with on various grounds. But my claim that there have been two main answers is widely agreed.

So first, here is what I called the traditional answer. It says: the reason why appearances are indeed definite---just as much as they are on a proposal that the wave-function collapses---is that the objects involved in the problematic superposition, say at the end of a quantum measurement, each <u>split</u> into many <u>copies</u>, corresponding to the various worlds. So in our toy model of measurement with two outcomes, the apparatus' pointer splits into copies, some with the '1' outcome, and some with the '2' outcome. Authors advocating this answer differed about how to understand the splitting; in particular, about how many copies there are. Some said that for two outcomes, there are just two copies; some said that for each outcome, there are many copies, perhaps infinitely many. (I will return to this in later Sections.) But the common idea is that this splitting secures definite appearances. For appearances only "appear" within a world: and within a world, the wave-function is, by definition, the corresponding term—which <u>is</u> an eigenstate of the quantity, such as pointer position, that in order to solve the measurement problem, we want to have a definite value.

The trouble with this answer is not that the idea of splitting is plain wrong. It is that, stated so briefly, the answer is too programmatic: it raises more questions than it answers. If the splitting is a <u>bona fide</u> physical process, we need to hear details: for example, about how it can be consistent with laws like the conservation of mass or of energy. If it is somehow a conceptual splitting, without a physical description, then there is philosophical work to do, to explain what it involves. In particular, if the splitting is 'conceptual' in the sense of being a distinction made by a conscious mind, then presumably, there is no splitting in those regions of the cosmos without conscious minds. In that case, the proposal is similar to that considered in the last Section, that consciousness collapses the wave-function; and it is thereby similarly hard to make precise and to gather evidence for.

In my opinion, until about 1990 most of the Everettian literature did not adequately answer such questions. Hence my accusation that it seems like a bluff. Of course, I am not alone in my misgivings: many authors pressed such questions. In particular, Bell was very doubtful. In his 1986 paper which I recommended and summarized in Section 5, he says that the Everettian proposal 'is surely the most bizarre' and 'extravagant, and above all extravagantly vague, hypothesis. I could almost dismiss it as silly' (1986, pp. 192, 194). But as I announced above: since about 1990 (when Bell died), the Everettian literature has appealed to decoherence, which does address the questions.

That will be the topic of the next three Sections. But first, it is worth discussing---and criticising---an analogy that is sometimes suggested in defence of the suspiciously programmatic 'splitting' answer above.

The alleged analogy is that the 'splitting' answer is like what is surely the right reply to the objection (perhaps apocryphal) against the proposal by Galileo and other advocates of a heliocentric astronomy, i.e. the proposal that the Earth goes around the Sun. The idea of the objection (i.e. in defence of the traditional, Aristotelian geocentric astronomy, that the Sun goes around the Earth) is to appeal to appearances. Namely: it looks (especially at sunrise and sunset) as if the Sun goes around the Earth. So the objection is that the heliocentrists' radical proposal seems to conflict with the appearances.

The reply that Galileo and the heliocentrists are said to have made is that in fact, assuming that the Earth goes around the Sun leads to the <u>very same appearances</u>. That is: the <u>appearance</u> at sunrise is exactly the same whether you describe this as the sun rising above the horizon, or as the horizon sinking below the sun. (And similarly, of course, at sunset: the appearance is exactly the same whether you say the sun is sinking, or the horizon is rising.)

Similarly, it is alleged, for the Everettian's "splitting" answer. If we assume such a splitting, the appearances will be just as the objector says they are: that is, perfectly definite.

But I submit that this analogy has a merely rhetorical force. Agreed: the broad logic of the two disputes is the same. In both cases, the new radical proposal (Galileo's, or Everett's) replies to the objection that it conflicts with appearances, by saying: 'No, I do not: I accord perfectly well with appearances'.  But there is a big difference between the two cases.

For Galileo and the other advocates of a heliocentric astronomy can readily spell out how they accord with appearances. It is a matter of optics, i.e. the paths of light rays. In particular, it is straightforward to argue that: (i) the appearance at sunrise is a matter of the angle between a ray of sunlight and one's line of sight to the horizon increasing (and similarly, at sunset: decreasing); and (ii) this increase (respectively: decrease) depends only on a relative motion of sun and horizon. But the Everettian's idea of splitting gives no such straightforward argument for recovering definite appearances: it leads only to the questions I pressed above.

Chapter 4.7: Doing better with decoherence
However, as I said: by appealing to a process called decoherence, the Everettians can make much better sense of their proposed splitting; and since about 1990, they have done so. So in this Section and the next two, I will spell out what decoherence is, and how it clarifies what splitting involves. It will also be clear that decoherence is important for all approaches to quantum theory, not just for Everett's: any of Section 5's interpretations need to accommodate it.

There will be three stages, one in each Section. This Section gives the basic idea of decoherence.  In the next, decoherence helps make more precise the definition of an Everettian world (or 'branch'). After that, decoherence will suggest that macroscopic objects such as cats or pointers are---not aggregates of stuff, but---enduring and stable patterns. (Then in the Chapter's final Sections, I will turn to other topics, especially probability.) The plan for this Section is that I will (1) state the idea of decoherence, and then (2) state a merit, and a demerit, of it.

(1): 'Decoherence' means, in this context, the diffusion of coherence. Here, 'diffusion' means spreading: namely, spreading from the system of interest to its environment with which it is interacting.

'Coherence' is physics' jargon for some characteristic differences between (i) the probability distributions prescribed by quantum states, especially superpositions, and (ii) those prescribed by classical states. We saw already in Section 4 that a quantum superposition prescribes a non-trivial probability distribution over the possible outcomes of a measurement of the quantity concerned; while eigenstates are "dogmatic" or "opinionated"---they give probability 1 to just one outcome.

Classical states---meaning states prescribed by classical physics---also prescribe distributions, once we include probabilistic mixtures of states. Thus imagine being given, say, three states, and taking a quarter-quarter-half mixture of them. This means, for example, taking a thousand systems: of which 250 are in the first state, 250 are in the second state, and 500 in the third. Then the predicted statistics for measuring any quantity on a randomly selected member of the set of a thousand systems would be the average, with weights ¼, ¼, ½, of what the three given states prescribe.

The important point here is that superpositions cannot be understood as such classical mixtures. For although for a single quantity, a superposition and a classical mixture might prescribe exactly the same distribution: there will be other quantities (which are said to not commute with the first one), about which their distributions differ. The numerical differences, outcome by outcome, between these distributions (for this other quantity) are called interference terms.

So in short: interference terms are the signature of a state being a superposition. (The word 'interference' comes from the physics of waves. When the peaks of two water waves meet and make yet higher peak, we say that the waves interfere constructively; when a peak of one wave meets a trough of the other, and they cancel out to give a level surface, we say that the waves interfere destructively.)

So imagine a quantum system interacting with its environment. Given our interest in the measurement problem, a paradigm case is the pointer of an apparatus (or a cat) interacting with the air around it, by air molecules bouncing off it. So quantum theory tells us, of course, that the quantum state of the composite system, pointer plus air, prescribes probabilities for the various possible outcomes of measuring any quantity on either component, the pointer or the air. And in general, this composite quantum state will be a superposition of product states, so that the probabilities it prescribes will include interference terms---even for a quantity on a component system. That is: one expects, <u>a priori</u>, that the component system's state will be superposed, i.e. its probabilities will have interference terms.

Now we can state the punch-line about decoherence. Namely: according to many realistic models of how a macroscopic object like a pointer interacts with its environment like air molecules, the interaction establishes very rapidly a composite quantum state that <u>falsifies</u> this <u>a priori</u> expectation, as regards the macroscopic object. That is: very soon after the interaction starts, the composite quantum state prescribes probabilities for the macroscopic object whose interference terms are negligible. In other words: the composite quantum state determines a state for the pointer that is almost a mixture, in the classical sense above. For it differs from a mixture only by tiny interference terms.

More is true. The states within this mixture, that the pointer is "almost in", are the states that we intuitively want, in order to solve the measurement problem. For they are, roughly speaking, states of definite position: for example, position of the centre-of-mass of the pointer. Thus for Section 4's toy-model of measurement: once we augment the model by including in our analysis the air molecules, then the post-measurement state of the pointer, as determined by the state of the entire composite system (measured system and pointer and air molecules) is---neglecting tiny interference terms---a mixture of the centre-of-mass being in front of the numeral '1', and it being in front of the numeral '2'.

And yet more is true. One does not just get such promising-looking mixtures in situations of explicit measurement, involving everyday-sized objects like pointers. Nowadays, there are detailed models of much smaller objects, immersed in their environment, that give mixtures of states that give definite values to the quantities that intuitively we want to be definite. For example: a tiny dust-particle, a tenth of a millimetre in diameter, floating in outer space, will be in a mixture of states of definite position, thanks just to its interacting merely with the dim light of the stars.

(Two paragraphs above, I said 'roughly speaking, states of definite position', because in many models, the states in the mixture obtained are so-called <u>coherent states</u>. These are states whose probability distributions are sharply peaked for both position and momentum, so that a system in such a state seems almost definite in both position and momentum. I say 'sharply peaked' because the distributions have enough spread so as to obey quantum theory's Uncertainty Principle, which vetoes having absolutely precise values for both position and momentum.)

So to sum up so far: decoherence is the fast and ubiquitous process whereby, for appropriate physical quantities on a system immersed in its environment, the interference terms that are characteristic of a quantum state (a 'superposition') as against classical states (a 'mixture'), become tiny. In effect, the coherence has diffused from the system to its environment.

We will see in the next two Sections how with decoherence the Everettian can give a much better account of the splitting of worlds. I end this Section by stressing two important features of decoherence, that apply not just to Everettians, but to all approaches to the measurement problem. The first is positive; the second is a limitation.

(2): The positive feature is flexibility. The need to be flexible becomes clear when we realize that in order to solve the measurement problem, we need the classical physical description of the world to be vindicated <u>only approximately</u>, not in every detail, by quantum theory. Thus we should admit that we need only some subset of quantities, not all quantities, to have definite values. And we should allow that maybe this subset is specified contextually, even vaguely. And besides, maybe the values should only be definite within some margins of error, even vague ones.

The point now is that decoherence secures this sort of flexibility. For which quantity on the system is "preferred" (i.e. rapidly becomes definite, in that the state is a mixture of its eigenstates) is determined by the physical process of interaction---whose definition in the model can be legitimately varied in several ways. Here are three examples of such ways. One can vary the definitions of: the system-environment boundary; the time at which the interaction is taken to end; and what counts as the state being 'sharply peaked' for a quantity.

I now turn to the limitation. It is very important since it is often ignored---in the research literature as well as in popular accounts of decoherence.

The limitation is that decoherence does <u>not</u> just by itself solve the measurement problem. More precisely: it does <u>not</u> imply that in any individual case, the macroscopic system actually is in one of the states in the mixture. It implies only that the quantum probabilities for any quantity are <u>as if</u> the system were in one such state.

Furthermore, quantum theory implies that the macroscopic system is in fact <u>not</u> in one of those states. This last is a subtle point, which many textbook discussions miss. Some authors signal this point by using a special jargon. They call the mixture that the macroscopic system is in an <u>improper mixture</u>; and for a mixture for which the system in any individual case is indeed in one of the states getting mixed, they use the phrase: a <u>proper mixture</u>, or an <u>ignorance-interpretable mixture</u>.

So decoherence secures that the macroscopic is in an improper mixture. In other words, setting aside this jargon: to say instead that the macroscopic system is in any individual case in one of the states getting mixed would contradict the original hypothesis that the entire composite system including the environment is in a superposition, not a mixture.

Though this point is subtle, it is uncontroversial. In fact, it is already clear in Schroedinger's 1935 paper, in which he introduced his eponymous cat. For in Sections 10 to 13 of that paper, he discusses entanglement (indeed, he introduces this word), and expounds the point.

This limitation of decoherence can be made vivid in terms of Schroedinger's cat. Namely: at the end of the measurement and after the decoherence process, the quantum state still describes two cats, one alive and one dead. It is just that the two cats are correlated with very different microscopic states of the surrounding air molecules. For example: an air molecule will bounce off a wagging upright tail (as in a living cat), and a stationary horizontal one (as in a dead cat), in different directions.

Indeed, we can put the point in <u>cartoon</u> form. Think of the cat being alive as its having vertical legs, and a vertical tail, and a smile on the mouth. Similarly, think of its being dead as its having horizontal legs, and a horizontal tail, and a frown on the mouth. Then the usual image for the measurement problem, i.e. the Schroedinger cat paradox, is the combination of these two configurations of the cat. Using plastic transparencies, as in a traditional class-room: one would overlay one transparency, portraying the dead "horizontals" cat, on another transparency, portraying the living "verticals" cat---getting the usual paradoxical double image.

Now the advocate of decoherence points out to us that the cat is really in an environment of air molecules etc. So they point out that a realistic image for the living cat, standing and smiling, should show many air molecules bouncing off the cat with such-and-such trajectories; while a realistic image for the dead cat, lying down and frowning, should show some similar large number of air molecules bouncing off the cat, but with <u>different</u>, so-and-so, trajectories. And so the realistic final state of the entire composite system, cat plus air molecules, is the superposition of states corresponding to these two images.

All this we must accept. But it seems blatantly to <u>not</u> solve the basic problem: that quantum theory describes the cat as blurred between being alive ("verticals") and being dead ("horizontals"). This problem seems wholly unaffected by invoking the air molecules. After all, think again of plastic transparencies. Adding scattered dashes portraying such-and-such trajectories to the transparency for the living "verticals" cat, and adding different scattered dashes portraying so-and-so trajectories to the transparency for the dead "horizontals" cat, does nothing towards unblurring the image. If anything, it makes the situation worse. For it shows the air molecules being blurred, "suffering" indefiniteness of position, just as much as the limbs of the cat.

In Section 9, we will see how the Everettian can accommodate this limitation by taking the cat, and other macroscopic objects, to be enduring and stable patterns. But to prepare for that, we need more detail about the Everettian's concept of a world.

Chapter 4.8: A sketch definition of 'world'
In this Section, we will see how decoherence clarifies the Everettian's idea of worlds and thus of how they "split". I will first, in (1), use decoherence to state a sketch definition of 'world'. Then in (2), I will explain two ways in which this sketch definition needs to be improved.

(1): At the beginning of Section 6, I introduced the Everettian proposal as saying that the cosmos as a whole has a quantum state $\Psi$ which is a vast superposition, i.e. a vast sum, of product states describing countless component systems---not just an electron and a measurement apparatus or its pointer, but countless electrons, quarks, molecules, specks of dust etc. So the Everettian now needs to be more precise about how to extract from this state a set of worlds: each of them (or maybe just: some of them) like our familiar macroscopic realm, with all its tables etc. in definite positions.

In this endeavour, decoherence is a promising resource, not least because as we discussed near the end of the last Section: it is flexible. The idea will be that to get something like our familiar macroscopic realm, the Everettian will adopt a definition of 'macroscopic object', or for short 'macrosystem', and will then take the cosmos as the vast composite system consisting of:
    (i) all the macrosystems; together with
    (ii) all the countless microscopic systems that are not in a macrosystem, which
    are treated collectively as a single system, the rest of cosmos.
Accordingly, the quantum state-space of the cosmos is broken down, in the mathematics of the theory, into the state-spaces of its components: all the state-spaces of the macrosystems, and the state-space of the rest of cosmos. In the mathematics, the state-spaces combine rather like numbers being multiplied, rather than added. So the component state-spaces are called <u>factor-spaces</u>, and the state-space of the cosmos is a <u>product space</u>. I shall assume there is some vast finite number <u>N</u> of macrosystems, and so there are exactly <u>N+1</u> factor-spaces in all.

So far, so mathematical. But now decoherence prompts two distinctively physical suggestions.

First: the Everettian can legitimately define 'macrosystem' along the lines: 'any system whose interaction with its environment rapidly makes its state almost a mixture in quantities such as position (and momentum), i.e. the quantities whose values, as attributed by classical physics, gave such a successful description'. For we learnt in the last Section that nowadays physicists' models of decoherence are so many and so varied that a definition along these lines will encompass, not just everyday objects like a table or a pointer of an apparatus, but also a tiny dust-particle floating in outer space---and countless objects in between, such as a lock of hair or a dew-drop. So in effect, every object that anyone has described, or could successfully describe, with classical physics will be among the Everettian's vast set of N macrosystems.

Second: I have so far considered only a single macrosystem interacting with its environment: as the saying goes, being <u>decohered by</u> (i.e. its state becoming a mixture, due to) its environment. But such decoherence interactions are happening continuously, to each of the N macrosystems. For the rest of the cosmos is a vast, common environment for them all: for the table, the pointer, the lock of hair, the dew-drop, the tiny dust-particle in outer space. Above the atmosphere, where the air molecules give out---for our purposes: not for human breath, but as a sink into which the quantum interference terms can diffuse---the dim light of the stars can take over.

Now, with a bit more stage-setting, I can state an Everettian definition of 'world'. There are two preliminary steps, and then the definition itself. (After stating it, I will comment on how it remains only a <u>sketch</u> definition.)

(1): We adopt a definition of 'macrosystem' along the lines above. Suppose that throughout the cosmos, there are N macrosystems, as thus defined. Then we factorize (break into its factors) the cosmos' state-space into N+1 factor spaces: one factor for each of the N macrosystems (dust-particles etc.) that gets decohered by its environment, and one factor for their common environment---the vastly complicated and dispersed rest of the cosmos.

(2): Then we express the cosmos' quantum state, $\Psi$, as a superposition whose components (i.e. summands) are (N+1)-fold product states, i.e. products of N+1 states, one in each factor-space. We choose these product states by: (i) for each of the first N factors, taking an eigenstate, on the corresponding macrosystem, for the quantity that gets "preferred" (i.e. selected) by the decoherence process; and (ii) for the last factor, i.e. the (N+1)-th factor associated to the rest of the cosmos, taking what is called the <u>relative-state</u> of the rest of cosmos.

(We do not need to pause on the notion of relative-state, although it was a centre-piece of Everett's original paper (in 1957). Suffice it to say that, stated for a two-component system (rather than a N+1-component system), the relative-state is roughly: what the component system "looks like", assuming the other component system is in a given state. More precisely: it is the state that would be assigned to the component, after a projection-postulate measurement that results in the given state on the other component.)

Finally, a <u>world</u> is defined as: the physical reality described by a summand, in this way of writing $\Psi$.

In other words, again using some of the jargon we have introduced: (i) for each of the N macrosystems, we consider a component of its "post-decoherence" improper mixture (i.e. an eigenstate of the quantity selected by decoherence); and then (ii) for the rest of universe, we consider the relative state. We then consider the product of (i) and (ii): which is an (N+1)-fold product state. This product state defines a world. The world is to have as values for quantities just what that product state ascribes, according to orthodox quantum theory, i.e. just the corresponding eigenvalues.

It is clear that in a world, as I have just defined it, each macrosystem has, by construction, a definite value for the quantity on that macrosystem that was selected by the decoherence process. Thus the promise of decoherence for the Everettian lies in the fact that in many models of many sorts of macrosystem, the definite-valued quantity is calculated to be position or momentum, or something "close" to these---in short, the sort of quantity that, to solve the measurement problem, we intuitively want to have definite values.

So this definition makes 'world' precise in a way that meshes with the basic ideas of Everett's proposal. It helps the Everettian answer Section 6's accusation of bluffing.

(2): But I stress that this definition of 'world' is very much a <u>sketch</u> definition. There are at least two broad ways in which one would like to see it improved. In listing these ways, I am not being controversial. For the physics of decoherence remains an active area of research, both theoretically and experimentally. Besides, it will be clear that not only the Everettian, but anyone interested in the interpretation of quantum theory, will want to see progress about these two ways.

    First: this definition assumed a notion of macrosystem, which I initially said would be defined in terms of: 'getting a mixture in quantities … whose values, as attributed by classical physics, gave such a successful description'. Such an assumption seems suspiciously close to postulating what one wants, rather than arguing for it. I think this suspicion lessens once one sees the detail in the definition of 'world': the factorization of the cosmos' state-space, and the decoherence of each component system. But I agree that much more detail is needed. Not just more than I have given here; but more than the research literature has so far achieved. As I said: this is an active area of research.

    In broad terms: one would like to see models of decoherence that are more rigorous and of wider scope (i.e. cover more systems) and that make definite the "right" quantities. One aspect of this is a limitation of my sketch-definition, as I stated it. Namely, it takes the environment (the rest of the cosmos) to be spatially external to the macrosystems; e.g. air molecules external to an apparatus' pointer, starlight external to a dust-particle floating in outer space. But in many cases, the environment is internal. For example, an iron bar can have long-wavelength and short-wavelength vibrations (called 'modes'); and when these modes are treated quantum mechanically, they are themselves systems, and the long-wavelength modes get decohered by the short-wavelength modes. In this way, the long-wavelength modes, taken together, form a macrosystem. Thus defining the factorization of the cosmos' state-space in terms of <u>N</u> macrosystems will be a lot more subtle and complicated than just "listing" all the <u>N</u> "lumps of matter" (cf. Section 1) such as pointers and dust-particles.

    In this endeavour, one can of course appeal to the flexibility of decoherence which I praised at the end of the last Section. Thus I said that one can vary the definitions of: the system-environment boundary; the time at which the interaction is taken to end; and what counts as the state being 'sharply peaked' for a quantity. So in seeking a better definition of 'macrosystem', in terms of which to make the factorization, one can hope to exploit this variety or "wiggle-room".

    So to sum up this first issue: the Everettian hopes that the physics of decoherence will enable us to avoid taking 'macrosystem' as a primitive concept; but there is much work yet to do. I will return to this challenge at the end of the next Section.

    Second: there is an issue about the fact that our sketch definition of 'world' appeals to a factorization of the cosmos' state-space, prior to and independent of what the state of the cosmos $\Psi$ actually is---which seems wrong. This issue is independent of the first. For even if the first issue was completely dealt with by a satisfactory rigorous definition of 'macrosystem' etc., it will surely still be true that what macrosystems there are, and how many there are, will be very much a matter of happenstance, a matter of contingency. This contingency involves two points.

    The first is a matter of everyday belief, and regardless of quantum theory, Everettian or otherwise. (It is also regardless of the first issue, i.e. how to define 'macrosystem' precisely.) It goes back to Chapter 3.3, about our being up to our necks in modality. Namely: there surely could have been different macrosystems than there in fact are; not just different tables, or locks of hair, or stars, but also more mundanely, different dust-particles. Secondly, even if one agrees with the Everettian in boldly postulating a quantum state of the cosmos $\Psi$: still it is presumably a contingent fact what that state is. After all, $\Psi$ is an element of a vast state-space, with countless other elements; and nothing in the Everettian proposal, as so far stated, forbids the cosmos being instead in one of those other states.

Putting these points together, the obvious suggestion is that we should allow the factorization of the cosmos' state-space to depend on the state. By doing that, the second point---the happenstance about what the cosmos' state in fact is---might perhaps cover, <u>en passant</u>, the first point, viz. that we need to accept happenstance about what, and how many, macrosystems there are.

(I say 'might perhaps cover', not 'does cover', because we cannot now be certain that our convictions about the possible varieties and numbers of macrosystems will follow from what quantum theory turns out to say about which states of the cosmos $\Psi$ are possible. And our uncertainty is all the greater because of the first issue, i.e. our not knowing how to define 'macrosystem' precisely. )

I will return to this second issue in Chapter 6. But nothing I say there will undermine the present conclusion: that we should let the Everettian's factorization of the cosmos' state-space depend on the state---which our sketch definition above does not do.

<u>Chapter 4.9: On what there is: objects as patterns</u>
'On what there is' is the title of a famous article by the philosopher Quine about deciding what truly exists. In this Section and the next, I will pursue this theme, as regards the Everettian multiverse. This Section will be largely positive, i.e. <u>pro</u> the Everettians. I will start in (1) with an explanation of how objects being patterns implies that at the end of the Schroedinger-cat experiment, there really are two cats. Then in (2) I will state my misgivings. (The next Section will raise another objection to Everettians; and it will lead in to the last two Sections about probability.)

This Section's main idea is in the title: 'objects as patterns'. Nowadays, Everettians build on the previous Sections' account of decoherence, so as to justify their metaphor of the world splitting into many alternatives corresponding to, for example, the various outcomes of a measurement process. In this justification, their main new idea is that a macroscopic object is <u>not</u> a lump of stuff, or an aggregate of tiny lumps, or even a cloud of point-particles. (Recall Section 1 above, and Chapter 2's discussion of the seventeenth-century mechanical philosophers.) It is really a <u>pattern in the quantum state</u>: in the quantum state of the cosmos.

As we will see, this idea promises to overcome the limitation I explained at the end of Section 7 above: that decoherence does not by itself solve the measurement problem, since it does not imply that in any individual case, the system actually is in one of the states in the improper mixture obtained after the decoherence process. I made this limitation vivid in terms of Schroedinger's cat. Thus I complained that this improper mixture still describes two (contradictory) states of a cat, being alive and being dead: it is just that the two states of the cat are correlated with very different microscopic states of the surrounding air molecules.

Now, the idea that macroscopic objects are patterns will vindicate the proposal that: (i) at the end of a Schroedinger-cat experiment, there are indeed---not just two (contradictory) states of a cat---but two cats; and (ii) the two cats are in different worlds (branches), what Section 7 called 'cases'. For there are indeed two <u>patterns</u> in the quantum state of the whole system (say: the cat, the apparatus, and air). <u>Therefore</u>, there really are two cats. In other words, the Everettian claims: the final quantum state being an improper mixture describing two states of a cat, being alive and being dead, is a matter of the state encoding two patterns---and, since cats just <u>are</u> patterns, that description <u>is</u> a description of two cats.

This claim is certainly dizzying. But it is, I think, completely coherent. (My misgivings at the end of this Section will not refute it, but just make it less plausible.) It becomes clearer if we assume that the quantity selected by decoherence (i.e. having negligible interference terms in the final mixture) is a familiar one; and one which, to solve the measurement problem, we intuitively want have a definite value. Let it be our old friend: position. That assumption is reasonable. For as hinted in Section 7: in models of decoherence, the final mixture typically contains coherent states, in which both position and momentum are very close to definite in value.

Of course, there is no single position quantity for the cat. For the cat has very many parts, both macroscopic like legs and tail, and tiny like individual cells and molecules. Similarly of course for the pointer of a measurement apparatus. When we discussed its being in front of the numeral '1' on the dial, or in front of '2', we were exploiting the assumed rigidity of the pointer, so as to get by with a description using a single position quantity, viz. the position of the centre of mass.

So we need to recall from Section 2 above (and Chapter 3.3's discussion of state-spaces in classical physics) that even the <u>classical</u> configurations of a composite system, with say $\underline{N}$ point-particles as its components, form a vast state-space. With each point-particle placed somewhere in three-dimensional physical space, the classical configuration space is the set of (3N)-tuples of real numbers. For one needs three real numbers for the spatial position of each point-particle. So in all, one needs $3\underline{N}$ real numbers. The state of the corresponding quantum system is then a wave-function (with complex numbers as values) on this classical configuration space. (As we saw in Section 2, this system is called '$\underline{N}$ quantum particles'. But as I lamented there: the word 'particle' is misleading. For it strongly suggests the system is localized: whereas, according to orthodox quantum theory, it in fact only has tendencies i.e. dispositions (quantified as probabilities) to be found in various locations, if measured by a position apparatus.)

So we need to ask: how many parts should we take a cat to have? In other words: what is a good guess for the number $\underline{N}$, such that a successful quantum description of a cat can use a wave function on the set (3N)-tuples of real numbers? (Again, the factor 3 just encodes the fact that physical space is three-dimensional.) Let us for simplicity think of an atom as a single particle. (Again, we mean 'particle' in the Pickwickian quantum sense that we take merely to define a location in physical space, in our effort to estimate the number $\underline{N}$). But chemistry teaches us that atoms are minuscule. For example, in twelve grams of carbon, the number of carbon atoms is 6 followed by twenty-three noughts. This is written: $6 \times 10^{23}$. (This is called 'Avogadro's number'.)

So for a cat weighing say a thousand grams (1 kilogram: so roughly $12 \times 100$), the number of atoms---most of which will weigh less than carbon---will be enormous. It will be about $6 \times 10^{23} \times 100$. Which is roughly: $1000 \times 10^{23}$. So we can take as a guess for $\underline{N}$: $10^{26}$. So even with our simplifying assumption to think of an atom as a single particle, the classical configuration space is stupendously large. It consists $(3 \times 10^{26})$-tuples of real numbers. So it has dimension $3 \times 10^{26}$: a vast number—in which the '3' is hardly worth keeping track of. So the quantum state is a wave-function whose arguments (inputs) are elements of this space. Each argument is a $(3 \times 10^{26})$-tuple, i.e. an exact <u>classical</u> configuration for all the $3 \times 10^{26}$ atoms.

I can now say how the Everettian claims that in the myriad complexity of such a wave function, there is a pattern that deserves to be called 'a living cat', and another pattern that deserves to be called 'a dead cat'. (I say 'claims', since it will be clear that there remains a lot of intellectual work to do.)

The idea is to focus on the fact that in a <u>classical</u> description, there is a set of configurations that would all count as a living cat. Indeed, there are vastly many configurations of about $10^{26}$ classical point-particles that would count as constituting a living cat. We can put the point in the cartoon form which I adopted at the end of Section 7. Think again of the cat's being alive as its having vertical legs, and a vertical tail, and a smile on the mouth. Similarly, think of its being dead as its having horizontal legs, and a horizontal tail, and a frown on the mouth. So there is a similarly vast set of configurations that would all count as a dead cat. Notice that these two sets, though both vast, do not overlap at all. No configuration of a classical cat with the point-particles composing its legs and tail aligned vertically is also a configuration in which the legs and tail are aligned horizontally.

And so---now returning to quantum theory--- the Everettian claims that the quantum state at the end of Schroedinger's experiment is, as regards the cat, a wave function with two peaks. That is: they claim that there are two regions of the classical configuration space, i.e. the set of arguments of the wave-function, where the function's value, i.e. the output or amplitude, is non-negligible.

(And the Everettian claims that for countless other configurations, the amplitude is negligible. Bear in mind that, a priori, these $10^{26}$ point-particles could be configured to be any classical object of the same total mass. For example: a puddle, or a saucepan with risotto, or a small dachshund, or any of myriadly many nameless and often monstrous combinations that only a horror movie might devise, such as a half-cat-quarter-dachshund-quarter-risotto.)

Summing the amplitudes for all the configurations in each of these two regions, we get a (square root of) a probability that is substantial. In Schroedinger's original version of the experiment (end of Section 4), where there is about a 50% probability of an atomic decay causing the release of the poison, the probability for each of the two regions is about 50%. That is: 50% for being alive, and 50% for being dead.

But whether the probabilities are 50-50, or nearly so, doesn't matter here. What matters is the claim that there are two such patterns in the quantum state: two regions of the classical configuration space where the amplitude is vastly greater (as a ratio) than it is for points outside both regions. So if we accept that a cat is such a pattern, then there really are two cats.

Note that the essential idea here is independent of quantum theory's details; (as Everettians note). For the idea is closely analogous to one which we all unhesitatingly endorse for several other physical theories. Namely, theories in which states can be added together to give a sum-state, in which the component states do not influence each other, or only do so negligibly. This is called the state being dynamically isolated from each other. Examples include the theory of water-waves, and the theory of electromagnetism. (Recall from (1) at the start of Section 4, that just as we add two vectors by adding corresponding coefficients, we add two functions such as wave-functions by adding their values for each given argument.)

For example: the water in Portsmouth harbour can get into many different states. It can can get into (i) a state which we describe as, e.g. a wave passing through the harbour's centre heading due West; or into (ii) a state which we describe as a wave passing through the centre heading due North; or (iii) into a state which is the sum of these. But does this last case (iii) imply that we face a 'Portsmouth water paradox'? Should we agonize about how the Portsmouth harbour water-system can in one place (viz. the harbour's centre) simultaneously have the contrary properties of being Westward and being Northward?

Of course not! Rather, we say that waves are patterns in the water-system. (Agreed, we call such patterns 'objects'; in the jargon of philosophy, they are often called higher-level objects.) And so we say that there are two waves, with the contrary properties, one Westward and one Northward. Since the contrary properties are possessed by two distinct objects, there is no contradiction.

Similarly for other theories whose states can be added together, with the component states not influencing each other. For example: think of the electromagnetic field in a certain region, and e.g. pulses of laser light travelling in different directions across it---as happens in a light-show at a rock concert. There is no 'laser light-show paradox'.

Analogously, says the Everettian, we should endorse this idea when it is applied to the end of Schroedinger's experiment. Thus the Everettian claims to overcome the limitation at the end of Section 7 above. For with macroscopic objects as patterns in the quantum state, not lumps of stuff, we see that solving the measurement problem does not require that in any individual case, the quantum system is actually in just one of the states in the improper mixture obtained after the decoherence process. The system is in none of these states. But each state is a pattern; and that---a pattern---is what a macroscopic object such as a cat really is. So there are two cats, with contrary properties: one in each 'case', i.e. world.

So much by way of expounding the Everettian's claim that there are two cats. Now for my misgivings.

(2): Recall that to expound the Everettian's claim, I put it in <u>cartoon</u> form (above and at the end of Section 7): with being alive corresponding to vertical legs, a smile etc. So clearly there is a vast amount of intellectual work still to do. The Everettian owes us details about which classical configurations are to count as 'being alive'. More generally, setting aside cats and Schroedinger's experiment and biology: we need to be told which classical configurations are to count as the system we are concerned with, having each of a host of properties that we routinely ascribe to macroscopic objects. 'Being alive' is of course a property ascribed in both everyday life and scientific work. But we can concede that it would be enough for the Everettian to give a "translation-manual" from the regions of their vast classical configuration space to just scientific properties; or even just properties used in physics. For example, which regions (and so which quantum states, with a peak, a non-negligible amplitude, over those regions) count as: having a density of 7 grams per cubic centimetre, or being made of lead, or being a fluid, or being a good electrical conductor? Only if the Everettian can give us such a translation-manual will it be plausible that the familiar macroscopic realm---and more precisely, the vast empirical success of classical physics---can be understood as emergent from the quantum state of the cosmos.

Obviously, this is a vast challenge. Besides, the challenge is all the greater, when we remember two points made above. First: there are countless configurations of $10^{26}$ particles that correspond to no recognizable macroscopic object, not even a horror-movie monstrosity like a half-cat-quarter-dachshund-quarter-risotto. Think of a vastly variegated and spatially very dispersed "sludge": about a cubic centimetre of cool lead here, a cubic centimetre of hot methane gas there, and so on, and, say, 100 miles away, a (more appealing) hunk of ice-cream. Nothing in our Everettian description of a quantum measurement avoids such a sludge getting a non-negligible amplitude. So why don't we ever see it?

The second point aggravating the challenge was at the end of Section 8. Namely: the example of a how a system's decohering environment is in some cases spatially internal showed that defining the factorization of the cosmos' state-space in terms of <u>N</u> macrosystems will be a lot more subtle and complicated than just "listing" all the <u>N</u> systems which we tend to think of as "lumps of matter", such as pointers and dust-particles. This point is all the sharper once we realize that the macroscopic realm which the Everettian must recover consists of patterned configurations that are a minuscule subset of a configuration space, most of whose elements are nameless and unstructured "sludge".

But I do not say that this challenge is impossible. And I stress that in fulfilling it, decoherence will undoubtedly play a crucial role. We can illustrate this by going back to the cat, i.e. to biology. Recall from, for example, your biology lessons that biochemistry successfully describes the metabolism of cats (and of course all other organisms) in a completely classical way. Its models of chemical reactions in the cell assume that the proteins, DNA-sequences etc. are localized: these molecules are modelled as minuscule cousins of the ball-and-stick models on the biochemist's table-top; (balls for the atoms, sticks for the bonds between them).

That this classical description of what is after all a quantum reality can succeed so well reflects the efficiency and ubiquity of decoherence. For the protein and DNA molecules are, on the atomic scale, very large: they often contain well over 10,000 atoms (and so are often called 'macromolecules'). And they are constantly bombarded by tiny molecules such as water molecules that decohere them, and so---in the "improper mixture sense" (Section 7)---localize them. The upshot is that at the length-scale of macromolecules and at longer lengths, a classical description of protein molecules, DNA-sequences etc. as having well-nigh definite positions can succeed. And this success is well illustrated by the models filling biochemistry textbooks.

(With these remarks about biology managing well while treating quantities like the position of a macromolecule classically, without regard to quantum theory, I do not mean to deny that some important biological processes "cue in" to quantum aspects. Examples of this, including crucial processes like photosynthesis, are nowadays a focus of research in the new field called 'quantum biology'.)

Chapter 4.10: A reversal of ideas
So much by way of expounding the Everettians' claim that indeed there are two cats, just because there are two patterns: or more generally, that indeed the world splits---there is a multiplicity of objects---at the end of a decoherence process, since the state is then an improper mixture with two or more components corresponding to macroscopic realms with different values for the quantities selected by decoherence. As I said: I think this claim is, though dizzying, coherent. And suppose we now set aside all my misgivings at the end of the last Section.

  Should we then conclude that the Everettian is home free? That is: does their solution to the measurement problem (specifically: their appeal to decoherence to justify talk of splitting) have no internal conceptual difficulties? I say 'internal' because we may prefer a rival solution---perhaps one of those reviewed in Section 4 above---for other reasons; (including perhaps reasons that boil down to our intellectual temperament, as discussed in Chapter 1.4).

  I say: No. I submit that there are two main difficulties remaining. Both are distinctively philosophical, or interpretative, rather than physical. One difficulty is about the topic of probability: I will address it in the next (and final) two Sections. In this Section, I raise a difficulty about the quantity selected by decoherence.

  I do not claim that it is a knock-down objection: it is a conceptual tension or embarrassment facing the Everettian, rather than an outright problem. But it is worth articulating for two reasons. First: so far as I know, it is not addressed in the Everettian literature. Second: there is an interesting analogy between it and a criticism of Bohr's complementarity interpretation (cf. Section 5 above) that Schroedinger made in the great 1935 paper that formulated his cat paradox (and that also, as I mentioned, analysed entanglement). I will first, in (1), state the objection; and then in (2), explain the analogy with Schroedinger's criticism of Bohr.

(1): To explain this difficulty, I need first to stress the striking conceptual unity of classical physics' successes from the time of Newton till about 1900 (reviewed in Chapters 2.3, 2.6, 3.3 and Section 1 above). In classical physics, each of a very small set of quantities fulfils two roles that are, <u>a priori</u>, disparate. Namely: (i) being postulated as basic for the description of matter's tiniest components (whether point-particles or small extended pieces of matter); and (ii) being used to describe composite systems with vastly many such components. The paradigm examples are position and momentum within mechanics. In classical physics, a single quantity's fulfilling these disparate roles (i) and (ii) was unified by various procedures: especially summing or averaging of the values for the tiny components in (i), to get the values for composite systems in (ii). The simplest case is elementary and familiar: the centre-of-mass of a composite object is a weighted average of the positions of its components, with weights equal the components' masses.

  Agreed: with the development of other branches of physics, especially the rise of field theory (Section 1 above), quantities other than position and momentum, such as electric charge and electric field, had to also be accepted. Nevertheless, classical physics manages with a very small set of quantities. (Depending on how finely you distinguish quantities, there are between about a dozen and about fifty of them.) And most of the quantities fulfilling role (i) also fulfil role (ii): again with various procedures, especially summing and averaging, unifying the roles.

Against this background, we can now see the conceptual tension or embarrassment facing the Everettian. The Everettian makes two claims that are in tension with each other. Namely, they say:

(a): Although classical physics took quantities such as position and momentum to be exact, and to always have exact values: such quantities are in fact only definable approximately, through the process of decoherence. (Jargon: philosophers might call them 'emergent'; physicists also say 'effective').

But they also say:---

(b): These classical notions are needed to define the quantum state-space. For as we saw: the quantum state is a wave-function defined on <u>classical</u> configuration space. Agreed, I have hitherto simplified. For quantum states can be represented as complex-valued functions (also called 'wave-functions') with <u>other</u> sets of arguments than components' positions. The main such alternative representation uses momentum. That is: the set of arguments for a <u>N</u>-particle system are the 3<u>N</u>-tuples of all the possible values of the components' classical momentum in each spatial direction. But here again, it is the <u>classical</u> notion, viz. of momentum, that needs to be invoked; and so the same conceptual tension arises.

So the difficulty, or tension, is that notions which according to the Everettian is really approximate (emergent, effective) must be appealed to, in order to interpret the theory at the smallest and most basic level.

As I said, I do not think this is a knock-down objection. For example, the Everettian might reply that we often learn new theories by understanding its new concepts through the prism of the old theory, but that in due course there is a gestalt-shift, so that we re-interpret the old via the new. Fair comment. But I am uneasy that today in 2025, a hundred years after the quantum revolution, we (including the Everettians!) have not yet achieved the envisaged gestalt-shift.

(2): Finally, an analogy. In Section 5's quick review of the main strategies for solving the measurement problem, I mentioned Bohr's complementarity interpretation. I can now explain how this difficulty for the Everettian is analogous to a striking criticism of Bohr that Schroedinger made in his "cat" paper of 1935.

Schroedinger's criticism is based on a historical fact about classical physics being in tension with Bohr's (as one might say: "the Complementarian's") claim that classical concepts are indispensable within quantum physics. Adopting a labelling to show the analogy with my (a) and (b) above: the fact and the claim are, respectively, as follows.

(a'): Despite the conceptual unity of classical physics' successes (described above), classical physicists did not claim that the classical quantities were indispensable for physics; nor did they claim that such indispensability would be shown by future physics. Indeed, many expected these quantities to be superseded by future physics.

But Bohr claims that:--

(b'): Classical quantities are indispensable for physics, although of course quantum physics has shown they do not always have exact values. For Bohr, the principal reason for this "lesson" from quantum theory is that position and momentum cannot be measured simultaneously with arbitrary accuracy. (This is the famous Uncertainty Principle (in it best known form). In the mathematics of quantum theory, this is represented by position and momentum <u>not commuting</u> with each other.) And by 'indispensable', Bohr means, roughly speaking: indispensable for reporting experimental results in an objective language. (But here, we do not need the details of Bohr's doctrine, or of why he held it: which are controversial.)

So the difficulty, or tension, which Schroedinger is articulating as a criticism of Bohr is that: notions which classical physicists expected to be superseded by future physics, and which according to the Complementarian are indeed really limited (a lesson taught us by quantum theory), must be appealed to, in order to interpret quantum theory. Indeed, the Complementarian says they are needed in order to report experimental results objectively.

Let me end by quoting Schroedinger's own words against Bohr; (at the end of Section 2 of his 1935 paper). He begins with (a'). He praises classical physicists' intellectual modesty about their theory (which he calls 'model' and 'picture'): in particular, about whether its quantities (which he calls 'determining parts') can be measured on a microscopic object in nature (which he calls 'natural object'). He writes:

'Scarcely a single physicist of the classical era would have dared to believe, in thinking about a model, that its determining parts are measurable on the natural object. Only much remoter consequences of the picture were actually open to experimental test. And all experience pointed toward one conclusion: long before the advancing experimental arts had bridged the broad chasm, the model would have substantially changed through gradual adaptation to new facts.'

Then he goes on to criticize what he calls the 'reigning doctrine' (i.e. Bohr's complementarity) for declaring that only familiar classical quantities (position, momentum) are measurable. He writes:

'Now while the new theory [i.e. quantum theory: JB] calls the classical model incapable of specifying all details of the <u>mutual interrelationship of the determining parts</u> (for which its creators intended it), it nevertheless considers the model suitable for guiding us as to just which measurements can in principle be made on the relevant natural object. ... This would have seemed to those who thought up the picture a scandalous extension of their thought-pattern and an unscrupulous proscription against future development. Would it not be pre-established harmony of a peculiar sort if the classical-epoch researchers, those who, as we hear today, had no idea of what <u>measuring</u> truly is, had unwittingly gone on to give us as legacy a guidance scheme revealing just what is fundamentally measurable for instance about a hydrogen atom!?'

<u>Chapter 4.11</u>: *Angst* about probability: what is objective probability?
As this Chapter's Preamble announced: I have so far emphasized synchronic issues, i.e. issues about the quantum state at a single time. I have neglected issues about time and change, except to say that I postpone till this Section the question of how the deterministic Schroedinger equation can be reconciled with quantum theory's use of probabilities. So I now focus on how the Everettian answers this question. This will raise the philosophical question: what exactly is <u>objective probability</u> (also known as: <u>chance</u>)?

The first point to make is that the question really breaks down into two problems that the Everettian faces. The Everettian literature calls them 'the <u>qualitative</u> problem of probability', and 'the <u>quantitative</u> problem of probability'. Discussing the first will lead in to the second, which I address in the next Section.

The qualitative problem is that probability seems to make no sense, if all possible outcomes of a putatively probabilistic process in fact occur. But this is what the Everettian claims, at least for quantum measurements and the other processes, such as radioactive decay (remember the poison for Schroedinger's cat), in which the quantum state evolves to include a term, i.e. a summand in the sum, for each outcome. (Here, 'outcome' was made more precise by Section 8's sketch definition of 'world'.)

In short, the Everettian's answer to this question is that even though all possible outcomes occur, there is <u>subjective uncertainty</u>. Roughly speaking, it is the uncertainty of an experimenter just before doing a quantum measurement, about the question 'Which outcome will <u>I</u> see?' Again, this answer comes in several versions, but I will develop just one main version. I begin in (1) by explaining how subjective uncertainty is invoked in some other physical theories; and then in (2) returning to the Everettian.

(1): It will be clearest to begin by explaining how, in a broadly similar way, probability is taken as subjective uncertainty, for deterministic processes of the kind familiar within classical physics, e.g. Newtonian mechanics. (Recall the discussion of determinism in Chapter 3.8.) For such a process, a unique future sequence of states is determined by the present state. (More precisely: by the present state, together with the process' deterministic law: which in mechanics would be a specification of the future forces exerted on the system.) But a person, for example an experimenter, may not know this future sequence of states in advance: either because she does not know the present state in full detail or because it is too hard to calculate from that state what the future sequence will be. Given our present interest in how to understand probabilities, let us set aside the latter cause of uncertainty, since it is a matter of calculational intractability rather than ignorance of which among several alternatives occurs. It is ignorance of this latter kind which gives scope for the idea of probability. So in the context of classical physics, probability is reconciled with determinism by subjective uncertainty: by the idea of a person not knowing which alternative really occurs, but having various <u>degrees of belief</u>, i.e. subjective probabilities (cf. Chapter 3.3), about the matter. In the context of deterministic physics: these will be subjective probabilities about what exactly is the present state.

But here, the phrases 'degrees of belief' and 'subjective probabilities' should be understood in a logically weak sense. They do <u>not</u> imply that the probabilities, i.e. the numbers ½, 1/3 etc. assigned by the person, are a matter of idiosyncratic taste or temperament: that is, are undetermined by all the objective evidence. For there is a branch of classical physics, called 'statistical mechanics', that studies composite systems with vastly many components: a large and important branch, though I have not yet had occasion to mention it. (It was developed from the late nineteenth century onwards: among its main figures were Maxwell (1831-1879) who we met in Section 1, Gibbs (1839-1903) and Boltzmann (1844-1906).) Thus statistical mechanics studies systems like a sample of gas taken as composed of classical molecules, tiny "lumps in the void". It surmounts the utter unknowability of the exact microscopic state---the exact positions and momenta of all the classical molecules---by postulating a probability distribution over the possible states, and then calculating average (also called: expected) values of quantities like energy etc. Here, the probability distribution gives the weights to be used in calculating the average. The many resulting predictions meet with great empirical success.

Now the point is: a very good case can be made for calling this distribution 'objective', even though it is <u>not</u> determined by the exact microscopic state. Making this case invokes technical notions which go by names like '<u>mixing</u>' and '<u>ergodicity</u>'. But we need not go into details. For us it is enough that some rather natural assumptions about these notions select the empirically successful probability distribution from the countless horde of mathematically possible distributions: a selection that has nothing to do with idiosyncratic taste or temperament.

Besides, a similar strategy for reconciling probabilities to determinism, and justifying them as objective, occurs in the pilot-wave theory that we mentioned at the end of Section 5. Recall that like the Everettian, the pilot-wave theorist says that the Schroedinger equation is always 'right' (as Bell vividly put it: 1987, p. 201); but unlike the Everettian, the pilot-wave theorist says a quantum system has values for quantities other than its state's eigenvalues. For there are also point-particles with exact positions. With this as background, the pilot-wave theorist goes on to say: the apparent indeterminism of quantum theory arises from the utter unknowability of these positions; and quantum theory's orthodox (and again: empirically successful) Born-rule probability distribution over those positions can be derived from rather natural assumptions about notions like mixing and ergodicity, applied to the quantum state.

To sum up: both classical statistical mechanics and the pilot-wave theory reconcile the use of probabilities with the future sequence of states being determined by all the details of the present microscopic state, by: (i) invoking the utter unknowability of all the details of that state (in short: particles' positions); and (ii) arguing that the empirically successful probability distribution over these details is not a matter of taste or temperament, but of natural physical assumptions.

(2): Now we can say how the Everettian's answer to the qualitative problem of probability is analogous to all this. They claim that also in the Everettian framework, probability can be taken as subjective uncertainty about a deterministic process. But now, it will be uncertainty about a deterministic process of the unfamiliar Everettian kind. For such a process, a unique future sequence of states for the composite system---in principle, the entire cosmos---is determined by the present quantum state (together with the Schroedinger equation encoding all the forces that are acting). Yet, says the Everettian, there can still be subjective uncertainty. But the situation differs from the classical one in that this uncertainty arises, even if we assume the person, e.g. the experimenter, <u>does</u> know the present state in all its detail, <u>and</u> also all the forces that are acting, and so also knows how to calculate the entire future sequence of states.

Here, I should clarify that Everettians have an account connecting their state of the cosmos $\Psi$ (which of course no one knows) with the various quantum states we ascribe, with great empirical success, to objects in the laboratory in real-life experiments. So although of course no one can write down $\Psi$, the Everettian framework can recover the successful real-life ascriptions of quantum states to electrons, atoms and even dust-particles. (We can skip the details of this account: suffice it to say that it uses the ideas in Sections 7 and 8.) Thanks to this account, the Everettian can recover the idea, which is realized every day in real-life experiments, of the experimenter ascribing a quantum state to a microscopic system such as an atom that is about to be measured, and thereby deducing from that state the orthodox Born-rule probabilities, i.e. those numbers.

But do those numbers deserve to be called '<u>probabilities</u>'? After all, according to the Everettian, each of the various measurement outcomes truly occurs. To this the Everettian answers: 'Yes: the experimenter is uncertain since, thanks to the impending 'splitting' during the process of measurement, she will experience, not all the outcomes, but just one---and so she can ask 'Which outcome will I see?'. And that is enough for the numbers to be called 'probabilities'.'

I think this answer is tenable. For I think the core meaning of 'probability' requires that there are various cases, with numerical weights (or if you prefer: measures or intensities) assigned to them that are to guide in some suitable way agents' beliefs and actions. But this core meaning does not require that only one case really occurs. However, I agree that this answer is incomplete, in that it raises philosophical issues: indeed, at least three.

First: the answer leads to the issue of the identity over time of persons and-or consciousnesses. For it clearly depends on taking the question 'Which outcome will I see?' to be analogous in relevant respects to the question in the context of classical physics or pilot-wave theory, 'Which alternative (among the many microscopic states compatible with my incomplete knowledge) actually occurs?'. So the analogy involves accepting that the 'I' which sees just one outcome, could---in some good sense of 'could'---see another outcome. Here, we again see Chapter 3.3's theme: that in science, no less than in everyday thought, we are up to our necks in modality. But here, the modality at issue is different from that in Chapter 3. For as we have seen: for the Everettian, what could occur i.e. the unexperienced alternative is real (if you prefer:

actual), albeit in an unseen branch. (In Chapter 6.3, I will return to this comparison of possible worlds with Everettian branches.)

The second issue is related to the first. In recent decades, quite independently of these quantum conundrums, philosophers of mind and metaphysicians have identified the need for a notion of possibility that generalizes Chapter 3's idea of a possible world. It is sometimes called a 'centred world'. The general notion of possibility is needed to understand the content of sentences that contain (and mental states that are naturally expressed using) words like 'I', 'you', 'now', 'then', 'here', 'there', 'this' and 'that'---i.e. words whose referent depends on the context of utterance (cf. Chapter 3, Section 5). Such words are called 'indexicals' or 'token-reflexives'; and the need for the generalized notion of possibility, that is shown by such sentences, is called 'the essential indexical'. We need not go into this issue in detail: both here and in Chapter 3, we have had enough to do. I just note that the uncertainty that the Everettian's answer invokes---viz. uncertainty despite full calculational ability, and full knowledge of the composite system state, and of all the forces encoded in the Schroedinger equation---invites comparison with the kind of indexical uncertainty that philosophers nowadays address using centred worlds.

The third issue is: why should this uncertainty be quantified by the Born-rule probabilities derived from the quantum state? Why are they the right, or somehow appropriate, degrees of belief for the experimenter to have about 'which outcome I will see'? So this is what the Everettian literature calls 'the quantitative problem of probability'.

Chapter 4.12: Subjective probability to the rescue?
Indeed, this problem can be made sharper by imagining that a quantum system is subjected to a sequence of measurements. This prompts a tempting line of thought that the Everettian should regard the Born-rule probabilities as wrong. This line of thought was a focus of discussion for what in Section 6 I called 'the first Everettian answer' (spanning about 1960 to 1990): which formulated the splitting of worlds in terms of measurement outcomes, and which made no appeal to decoherence. The line of thought goes as follows.

According to the Everettian, the quantum state evolves over the course of a sequence of measurements, so as to encode all possible sequences of outcomes. Formally, the final state has a term (i.e. a summand in the sum) representing each sequence of outcomes. For example, consider a toy-model in which there are ten measurements, each with two outcomes (say, H and T, for 'heads' and 'tails'). Then there are $2^{10} = 1024$ sequences of outcomes; and so the Everettian must say there are 1024 terms in the final quantum state.

Since according to the Everettian, each such sequence actually occurs, it seems at first sight that the Everettian probability of a sequence should be given by the naïve counting measure. That is, the Everettian should say: each sequence has probability 1/1024. And so more generally, it seems that the probability of an event corresponding to a set of sequences, such as three of the ten measurements having outcome H, is the sum of the basic probabilities (each equal to 1/1024) of the sequences in the set. But this amounts to assuming that the two outcomes H and T are equiprobable; (and that the measurements form independent trials in the sense of probability theory). And this spells disaster for the Everettian. For the counting measure probabilities bear no relation to the quantum Born-rule probabilities, and so the procedure of counting Everettian worlds just by their sequences of outcomes seems to conflict with quantum theory's treatment of probability.

So much by way of sharpening the quantitative problem. I will now report what is nowadays the best-known and most developed Everettian answer to it. (But the Notes and Further Reading will also mention a recent alternative.) This answer has two parts. The first part builds on the preceding Sections' material; but the second part is wholly novel, and will occupy the rest of the Section.

The first part is to point out that decoherence, thanks to its flexibility, refutes the toy-model with its naïve counting measure. (Recall the end of Section 7, and Section 8.) That is: on any precise definition of 'world' for the systems concerned, there will be many trillions of worlds, wholly independently of the number of kinds of outcome registered by the measurement apparatus (in my example: just two, H and T). And more important: because one can vary the exact definitions of decoherence's crucial notions (like 'system-environment boundary'), there is <u>no</u> definite number, not even in the trillions, of worlds which we need to—or could!---appeal to, in order to give an account of probability in terms of counting. In short: the naïve counting measure is a mirage. It is woefully ill-defined, and the Everettian can and should just reject it.

The second part of the Everettian's answer is a remarkable recent development, that is wholly unlike anything in the previous discussion (either by me in this Chapter or in the Everettian literature I have so far drawn on). In terms of the previous Section's discussion, it is an analogue of the arguments within classical statistical mechanics and pilot-wave theory that I mentioned. Recall that they justify those theories' probability distributions, not by their empirical success, but by their following from natural assumptions. Analogously, Everettians have recently developed theorems that justify the Born rule, not by its empirical success, but by its uniquely following from certain general assumptions. But there are also two striking differences between the two cases.

(1): The arguments within statistical mechanics and pilot-wave theory make assumptions about how the system changes over time, albeit general ones. In the jargon: the assumptions about mixing and ergodicity are assumptions about dynamics. But the assumptions of the recent Everettian arguments do not refer to how the system changes: they are synchronic, or kinematic. We do not need details: but in short, they turn on the linear structure of the quantum state-space.

(2): The second difference is even more striking. The arguments within statistical mechanics and pilot-wave theory, and their assumptions, make no appeal to general principles about subjective probability; in particular, to principles about how rationality should govern a person's subjective probabilities. That is as we saw in the previous Section. Although I introduced the reconciliation of probability with determinism in classical physics by invoking subjective probabilities about what is the exact state, it is details of physics, such as assumptions about how a system changes over time, that are the dominant considerations determining which probability is correct; (and in particular, which probability is empirically successful). But the Everettians' assumptions <u>are</u> indeed about what principles of rationality should govern a person's subjective probabilities.

This is very remarkable since such principles are formulated and compared in a discipline, <u>decision theory</u> (briefly discussed in Chapter 3.3), that belongs to psychology and economics, and so <u>prima facie</u> has absolutely nothing to do with physics. So the Everettians' idea is remarkably inventive. The idea is, first, to appeal to such principles of rationality as applied to a person's degrees of belief in the various outcomes of various quantum measurements; and then, to go on to <u>prove</u> that these principles imply that the person's degrees of belief must be given by the Born rule. So let me end this Chapter with some details about this.

In decision theory, there is a tradition of proving what are called '<u>representation theorems</u>'. They are so-called because they show that under certain conditions a rational person's behaviour reveals that their degrees of belief can be <u>represented</u> as numerical probabilities. That is: their degrees of belief must conform to the usual rules about probabilities, viz. that the probabilities of all the envisaged alternatives must add up to 1, and that the probability of either Alternative A or Alternative B, where A and B are incompatible (cannot both be true) is the sum of the individual probabilities. Thus there are theorems to the following effect. Imagine a person whose preferences for gambles (encoding certain degrees of belief and certain desires) conform to a certain set of axioms: axioms that seem rationally compelling, i.e. compulsory for a rational person. The axioms say, for example, that a person who prefers A to B and B to C must also prefer A to C, and that a person would not enter a bet (or a collection of bets) that is guaranteed, whatever the outcomes, to yield a loss. Then the person must have degrees of belief that conform to the rules of probability. In other words, their degrees of belief are represented by a probability distribution.

We do not need further details, technical or even philosophical, about such theorems. But let us note that these theorems do not dictate a unique probability distribution over the various alternatives. This is of course as one would expect. Imagine two people are offered bets on horses in a race. So which bets they are willing to take, and which bets they decline, reveal their degrees of belief in the alternative propositions about which horse wins. Even if the two people are rational in the sense of the axioms listed in the representation theorems, we surely do not expect the two people to accept bets at exactly the same odds. For they can have legitimate differences of opinion. In short: we of course do not expect rationality to dictate specific degrees of belief in arbitrary propositions. (We touched on this in Chapter 2.5's discussion of Hume and inductive logic.)

But the recent Everettian theorems secure precisely this uniqueness, about the specific scenario of a person making gambles on the outcomes of various quantum measurements. Besides: the probability distribution that is uniquely dictated by the axioms about the person---which, as in the tradition of decision theory, seem to encode merely the compulsory requirement that they are rational---is indeed quantum theory's Born-rule probability distribution over the various outcomes.

A bit more precisely: the theorems show that a person who: (i) is an Everettian and is about to observe a sequence of quantum measurements, and also (ii) knows the initial state of the quantum system to be measured, and (iii) is forced to gamble on which outcomes she will see (using the Everettian sense of 'splitting', to interpret the phrase 'she will see'), and (iv) whose gambles are subject to certain rationality axioms---must apportion her degrees of belief (as shown by her betting behaviour) in accordance with the Born-rule as applied to the initial state that she knows (as assumed in (ii)).

To sum up this Section. I began with an objection to the Everettian. It said that they could not answer the quantitative problem of probability, since they seemed obliged to endorse the naïve counting measure: which dooms them to disagreeing with the empirically successful Born-rule probabilities. But the Everettian can and should appeal to decoherence so as to reject the naïve counting measure. And furthermore, Everettians have proved remarkable representation theorems, analogous to those in classical decision theory. But while the classical theorems allow a wide variety of subjective probability distributions, the Everettian theorems dictate that if a rational Everettian knows the relevant quantum state, her degrees of belief about measurement outcomes must be the orthodox Born-rule probabilities. So the Everettians can say that their framework not only accommodates, but even implies, the Born-rule probabilities. Remarkable indeed!

Chapter 4: Notes and Further Reading
As for Chapters 2 and 3, there is a dauntingly large literature. And as in those Chapters, I recommend:
(i) the internet encyclopedias and archives, and the accessible books, listed in items (1) to (3) of the Notes for Chapter 1; all of which cover the problems of interpreting quantum theory, and in particular the Everettian interpretation.
(ii) reading the original masterpieces, some of which are indeed very readable.

I will divide my more specific suggestions in four groups, following the sequence of topics in the Chapter : (1) The measurement problem, and the interpretations of quantum theory mentioned in this Chapter's Section 5; (2) other interpretations (again, I will emphasise how they propose to solve the measurement problem); (3) the Everettian interpretation, in general terms (cf. Sections 5 and 6); (4) the treatments of decoherence (cf. Sections 7 to 10) and of probability (cf. Sections 11 and 12).

(1): For quantum theory and its interpretation, in general terms, I suggest four groups, labelled (A) to (D): of which the last is about historical aspects.

(1.A): For the various interpretations of quantum theory prompted by the measurement problem, pride of place must go to the two essays, by Bell and by Schroedinger, that I lent on in the Chapter's exposition. Namely:
    (i) J. S. Bell, 'Six Possible Worlds of Quantum Mechanics'; which I used in Section 5's taxonomy of interpretations. This is most easily found in the journal, Foundations of Physics, for 1992. It is also reprinted in Bell's collection of articles, Speakable and Unspeakable in Quantum Mechanics (Cambridge University Press 1987; revised edition 2004). Bell's collection also includes his most famous papers, about quantum non-locality (Bell's inequalities): a topic which this Chapter has set aside. This book is available at:
https://www.cambridge.org/core/books/speakable-and-unspeakable-in-quantum-mechanics/E0D032E7E7EDEF4E4AD09F458F2D9DB7
    (ii) E. Schroedinger, 'The Present Situation in Quantum Mechanics', published in 1935 in German, but conveniently translated into English by J. Trimmer, in Proceedings of the American Philosophical Society, volume 124 (1980), pp. 323-338. This is available via J-Stor at:
https://www.jstor.org/stable/986572
    (ii) is also reprinted in J. Wheeler and W. Zurek (eds.), Quantum Theory and Measurement, (Princeton University Press 1983). This anthology of 49 papers reprints many important papers. For example, it includes the two most famous of Bell's non-locality papers. It includes Bohm's 1952 papers re-discovering de Broglie's pilot-wave theory. It also includes seminal papers by Bohr (especially in his debate with Einstein) and by Wigner; and---most relevant to us---it includes Everett's seminal paper from 1957: details in (2) below. This anthology is available at:
https://www.degruyter.com/document/doi/10.1515/9781400854554/html
    Another splendid overview of interpretations, aimed at philosophers, is:
    (iii) D. Lewis, 'How many lives has Schroedinger's cat?', Australasian Journal of Philosophy volume 82 (2004), p. 3-22; available at: https://doi.org/10.1080/713659799

(1.B) There are many fine books that explain quantum theory and the options for its interpretation, for newcomers, especially philosophers. Most of these taxonomize the interpretations in much the same way as my Section 5 (following Bell). Two, whose level makes them natural successors to this Chapter's exposition, are:
    (i) Peter Lewis, Quantum Ontology (Oxford University Press, 2016); which is available at:
https://global.oup.com/academic/product/quantum-ontology-9780190469818?q=Peter%20Lewis%2C%20Quantum%20Mechanics&lang=en&cc=gb
    (ii) By an expert about Everett, but which equally treats other interpretations:
J. Barrett, Conceptual Foundations of Quantum Mechanics (Oxford University Press, 2019); which is available at:
https://global.oup.com/academic/product/the-conceptual-foundations-of-quantum-mechanics-9780198844693?q=Conceptual%20Foundations%20of%20Quantum%20Mechanics&lang=en&cc=gb

(1.C) There are countless more advanced books that address the measurement problem, but presuppose some knowledge of quantum mechanics and facility with mathematics. Here are three books that expound approaches mentioned in my Section 5. They are by eminent leading researchers.

The first two are famous expositions of the pilot-wave theory. The third is P. Pearle's recent exposition of the programme (usually called 'the dynamical reduction programme') that seeks to solve the measurement problem by modifying the Schroedinger equation: a programme to which Pearle is a pre-eminent contributor. Regrettably, my Section 5 gave even less details about this programme than about the pilot-wave theory, simply because it is hard to state briefly how it obtains the right kind of wave-function collapse. But the programme is flourishing; and has been worked on by many eminent physicists, such as Diosi, Ghirardi, Gisin and Penrose, as well as by Pearle.

(i): D. Bohm and B. Hiley, The Undivided Universe, (Routledge 1993); available at: https://www.taylorfrancis.com/books/mono/10.4324/9780203980385/undivided-universe-david-bohm-basil-hiley

(ii) P. Holland, The Quantum Theory of Motion, (Cambridge University Press, 1993); and available since 2010 at:
https://www.cambridge.org/core/books/quantum-theory-of-motion/EF981BAE6222AE87171908E8DB74AF98

(iii): P. Pearle, Introduction to Dynamical Wave Function Collapse (Oxford University Press 2024); it is available at:
https://global.oup.com/academic/product/introduction-to-dynamical-wave-function-collapse-9780198901372?q=Philip%20Pearle&lang=en&cc=gb

(1.D) The history of the interpretations of quantum theory is of course fascinating. I pick just five important items.

(i) Einstein and Bohr had a famous debate from the mid-1920s onwards. For an introduction, see Bohr's essay in the extraordinary anthology of essays in honour of Einstein, Albert Einstein: Philosopher-Scientist, volume 7 in the series Library of Living Philosophers, edited by P.A. Schilpp, Open Court 1949. Bohr's essay is also reprinted in J. Wheeler and W. Zurek (eds.), Quantum Theory and Measurement, mentioned above.

(ii) The Solvay conference in 1927 was epoch-making. Its Proceedings were re-published, with a magisterial editorial introduction, in: G. Bacciagaluppi and A. Valentini (eds.), Quantum Theory at the Crossroads, (Cambridge University Press, 2009). It is available at:
https://www.cambridge.org/core/books/quantum-theory-at-the-crossroads/0F8A6712D61351E330A4D52C7EC8CC2C

(iii) A superb survey of the role of philosophical ideas in the historical development of physics, which I already recommended in Chapter 2, and which also covers the quantum theory, is: J. Cushing, Philosophical Concepts in Physics (Cambridge University Press, 1998; online 2012). Available at: https://www.cambridge.org/core/books/philosophical-concepts-in-physics/F285F13FE71F225BD8BE01F754F8C2E5

(iv) A definitive biography of Bell is: A. Whitaker, John Stewart Bell and Twentieth Century Physics (Oxford University Press, 2020). It is available at:
https://global.oup.com/academic/product/john-stewart-bell-and-twentieth-century-physics-9780198861263?q=A.%20Whitaker%2C%20Bell&lang=en&cc=gb

(v) A recent anthology of commissioned articles about the history is: The Oxford Handbook of the History of Quantum Interpretations, (Oxford University Press, 2022), edited by O. Freire, G. Bacciagaluppi, O. Darrigol et al. It is available at:

https://global.oup.com/academic/product/the-oxford-handbook-of-the-history-of-quantum-interpretations-9780198844495?q=Oxford%20Handbook%20of%20Interpretations%20of%20Quantum%20Theory&lang=en&cc=gb

(2) There are also several important interpretations (including distinctive approaches to the measurement problem) that this Chapter did not mention at all: they are not what my Section 5 called 'the usual suspects'. I will give four examples, under three labels (A) to (C). Then I will end by stressing that physicists recognize the measurement problem as an outstanding problem.

(2.A) My first example is very philosophical. It is the pragmatist interpretation developed by R. Healey, in a series of papers, and in his recent book. Here the label 'pragmatism' is, unlike Bell's usage reported in Section 5, a deliberate echo of the American philosophical movement launched by Peirce, James and Dewey. And being thus philosophical, Healey's interpretation is a cousin of Bohr's complementarity, rather than Section 5's 'practical instrumentalism'.

    Thus Healey's interpretation is one of a handful that reject one or more presuppositions of the way my Sections 4 and 5 (and most discussions) set up the measurement problem. Namely, presuppositions of a philosophically realist stripe; in particular, that the quantum state purports to describe physical reality in the straightforward manner of the logics and semantics expounded in Chapter 3 Section 2f. Being myself a philosophical realist, and an enthusiast about those logics and semantics, I disagree with Healey. (As I do with the rest of this handful of interpretations, for example one called 'quantum Bayesianism'.) But I applaud the imagination and precision with which he has developed the view. So without going into details, I recommend (i) his book and (ii) a laudatory-but-dissenting review of it by a realist, David Wallace: who is also a leading Everettian (cf. (4) below).

    (i) The book is: R. Healey, The Quantum Revolution in Philosophy, (Oxford University Press, 2019); it is available at:
https://global.oup.com/academic/product/the-quantum-revolution-in-philosophy-9780198844679?q=Healey%2C%20The%20quantum%20revolution%20in%20philosophy&lang=en&cc=gb#

    (ii) Wallace's review is in the journal Analysis, volume 80 (2020), pp. 381-388. It is available at: https://academic.oup.com/analysis/article-abstract/80/2/381/5819198?redirectedFrom=PDF&casa_token=IZglh61Zp5AAAAAA:NY5ESD4WQUbapRH5wqa7xgiyRGthpKzblkaSI-nVmdtunq-prgtrp0Ds5bvapoFDk3grTroMmBYbGA

(2.B) My second example is the modal interpretation. Or rather: 'modal interpretations', since it comes in many versions.

    Their common idea is: (i) to say, just like the pilot-wave theory does, that the Schoedinger equation is always correct; and (ii) to assign to a quantum system definite values of quantities, additional to the eigenvalues of its state's eigen-quantities---but not just by postulating ab initio a preferred quantity that is to always have a definite value, in the manner of the pilot-wave theory's postulation of particle position. So there are various precise proposals about which additional quantity is to be definite, and which of its possible values the system is to actually possess (and how these change over time). In these proposals, a common idea is that which quantity is definite, and which value it has, should depend (as a matter of interpretative postulate, not causally) on the quantum state. And in some versions of the modal interpretation, this dependence is not just on the state of the system itself; but also on the state of a larger system of which it is a component.

Of course, this is not the place for details. But here are: two expert and craftsman-like monographs and an anthology of important essays, all from the 1990s; and research articles that are, so far as I know, the most recent expert proposal.

(i) J.Bub, Interpreting the Quantum World (Cambridge University Press, 1997).

(ii) D. Dieks and P. Vermaas (eds.), The Modal Interpretation of Quantum Mechanics (Kluwer Academic 1998).

(iii) P. Vermaas, A Philosopher's Understanding of Quantum Mechanics, (CambridgeUniversity Press, 2000) available at: https://www.cambridge.org/core/books/philosophers-understanding-of-quantum-mechanics/179FE190C668C06A217C160DAB8FD2BC.

(iv) J. Barandes and D. Kagan, The Minimal Modal Interpretation of Quantum Theory; arxiv: https://arxiv.org/abs/1405.6755 ; there is a summary at: https://arxiv.org/abs/1405.6754; and there is a successor article at: https://arxiv.org/abs/1807.07136, and at Foundations of Physics, volume 50 (2020), pp. 1189-1218.

(2.C) My third and fourth examples are technical, and hardly discussed in the philosophical literature about quantum theory. They are due to two physicists, N. Landsman and A. Kent respectively. I choose them because they have the great merits of (i) invoking powerful ideas (and associated technicalities) from other parts of physics to break out of the 'usual suspects' taxonomy; and (ii) being ripe for further development. Thus I think philosophers should take more notice of them . . .

Again, this is not the place for details. Suffice it to say that, broadly speaking, they both contrive to retain the two claims (labelled (1) and (2) in Section 5) whose combination gave the measurement problem: the claims that the Schroedinger equation is always correct, and that quantum systems only have as definite values of quantities, the eigenvalues. And they achieve this "with only one world", i.e. not by having an Everettian splitting, with the consequent assignment of eigenvalues only relative to a world. So they both propose a striking one-world reconciliation of Section 5's two claims, (1) and (2).

The differences between the two proposals lie in the ideas invoked to secure the reconciliation. Landsman invokes ideas about perturbation theory, and spontaneous symmetry breaking. Kent invokes the idea of a final boundary condition. (Kent also aims for an explicit compatibility with relativity theory---a desideratum for any solution to the measurement problem that, for brevity, I have set aside.)

The main references are, respectively:

N. Landsman, The Foundations of Quantum Theory, Springer 2017, Chapter 11 (building on ideas in Chapters 7, 8 and 10). The entire book is an exact and complete survey of the field, packed with information. It is available as Open Access at: https://link.springer.com/book/10.1007/978-3-319-51777-3

A. Kent, 'Solution to the Lorentzian quantum reality problem', Physical Review A, volume 90, 012107 (2014). Available at arxiv: 1311.0249.

Kent has several other articles developing this solution. Further details are in my exposition and assessment (especially as regards quantum non-locality)---which is aimed mostly at philosophers:

J. Butterfield, 'Peaceful Coexistence: Examining Kent's Relativistic Solution to the Quantum Measurement Problem', in Reality and Measurement in Algebraic Quantum Theory (Proceedings of the 2015 Nagoya Winter Workshop), ed. M. Ozawa et al. (Springer Proceedings Mathematics and Statistics, 261: 2018); pp. 277-314. Available at: https://doi.org/10.1007/978-981-13-2487-1_11 http://arxiv.org/abs/1710.07844; http://philsci-archive.pitt.edu/14040

Finally, under (2): I stress that quantum theory's interpretative problems, especially the measurement problem, are accepted by physicists as outstanding problems. This is illustrated by two accessible essays by A. Leggett and S. Weinberg, both very distinguished winners of the Nobel Prize for Physics. Both essays stress the limitation of decoherence noted at the end of my Section 7: in short, that it does not secure a definite outcome at the end of measurements. Both essays are also sympathetic, overall, to the search for a "cousin" of the Schroedinger equation that secures a definite outcome; i.e. to what Bell called the unromantic version of the strategy that the Schroedinger equation 'is not right' (Section 5). (Nor are these two essays "one-off": both authors wrote about their misgivings in several articles.)

A. Leggett, 'Probing quantum mechanics towards the everyday world: where do we stand?', Physica Scripta, volume T102 (2002), p. 69-73.

S. Weinberg, 'The Trouble with Quantum Mechanics' The New York Review of Books 19 January 2017, pp. 51-53. It is available at:
https://goldphysics.unm.edu/phys521/features/WeinbergTroubleWithQMe.pdf

(3): For the Everettian interpretation, I postpone the topics of decoherence and probability until (4) below. Here I confine myself to two topics. First: the basic sources; and second, the idea that the definition of a 'world' (or of 'branch' or 'splitting') should be given in mentalistic terms, i.e. should be allowed to invoke concepts alien to physics such as 'appearance' and 'experience'. The Chapter's discussion gave this idea, often called the 'many minds' version of the Everettian interpretation, short shrift. (I only mentioned it obliquely towards the end of Section 6 as part of my criticism that, in the early Everettian literature, the idea of splitting was too programmatic.) Indeed, I am not sympathetic; but of course, there are good advocacies of it.

(3.A) As to the basic sources, pride of place must go to Everett's original paper:

H. Everett, 'Relative-state formulation of quantum mechanics', Reviews of Modern Physics, volume 29 (1957), pp. 454-462. It is reprinted in various places, including the large anthology mentioned in (1) above: J. Wheeler and W. Zurek (eds.), Quantum Theory and Measurement, (Princeton University Press 1983).

As I mentioned at the start of Section 6, not only are there different versions of the Everettian interpretation, but also there is controversy about how Everett himself understood it. My discussion did not go into these issues, but focussed on what I believe to be the broad version that has been dominant since the 1990s; (which invokes decoherence, macroscopic objects being patterns etc.). For details about Everett's own views, one should consult:

(i) H. Everett, The Everett Interpretation of Quantum Mechanics: Collected Works 1955-1980 with Commentary, edited by J Barrett and P. Byrne, Princeton University Press 2012. It is available at: https://www.jstor.org/stable/j.ctt7t2jf

(ii) The biography of Everett by P. Byrne: The Many Worlds of Hugh Everett III (Oxford University Press, 2012), which is available at:
https://global.oup.com/academic/product/the-many-worlds-of-hugh-everett-iii-9780199659241?q=Byrne%20The%20Worlds%20of%20Hugh%20Everett%20III&lang=en&cc=gb

Another basic source is the popular book by D. Deutsch, a leading Everettian who pioneered the decision theory approach to deducing the Born-rule probabilities (my Section 12, and (4) below). It is:

D. Deutsch, The Fabric of Reality, Penguin Books 1997.

(3.B)  As to the 'many minds' version of the Everettian interpretation, the first articulation of it (so far as I know) was by H.D. Zeh; (a physicist who made profound contributions to the physics of decoherence, cf. (4) below). This was in his paper:

H.D. Zeh, 'On the interpretation of measurement in quantum theory', Foundations of Physics, volume 1 (1970), pp. 69-76.

A book-length advocacy of the interpretation is: M. Lockwood, Mind, Brain and the Quantum (Oxford, Blackwell, 1989).

Some of my own views about this interpretation are in a journal's "symposium" about Lockwood's work. (The paper includes (a) recommending work by M. Donald, and (b) linking the discussion to the topic of probability, as in (4.B) below.) It is at:

J. Butterfield, 'Whither the Minds?', British Journal for the Philosophy of Science, volume 47 (1996), pp. 200-221.

(4): I turn to references about, first, decoherence (Sections 7 to 10), and then probability (Sections 11 and 12) in the Everettian interpretation.

I said in Section 6 that since about 1990, the Everettian literature has appealed to decoherence. Here I should add that in fact the physics of decoherence was much clarified already in the 1970s and 1980s (cf. (4.A) below). But it was only after 1990, especially in papers by S. Saunders, that there was a clear philosophical statement that what decoherence theorists usually called 'branches' (also: 'histories') were a perfect fit for the Everettian's proposed worlds. Two of Saunders' papers on this topic are:

(i) S. Saunders, 'Decoherence, relative states, and evolutionary adaptation', Foundations of Physics, volume 23 (1993), pp. 1553-1585.

(ii) S. Saunders, 'Time, quantum mechanics, and decoherence', Synthese, volume 102 (1995), pp. 235-266.

As to the literature after 1995, pride of place (for both decoherence and probability) goes to a monograph and anthology from about fifteen years ago. They sum up very well Everettian developments about decoherence and probability, from about 1995 to 2010: developments that were spear-headed by D. Deutsch, S. Saunders, and D. Wallace. The two books cover both advocacy (both books) and assessment (the anthology). They are:

D. Wallace, The Emergent Multiverse: Quantum Theory According to the Everett Interpretation, Oxford University Press 2012.

S. Saunders, J. Barrett, A. Kent and D. Wallace (Eds), Many Worlds? Everett, Quantum Theory and Reality, Oxford University Press 2010.

The anthology includes at least a dozen important papers. Among them, there are some of advocacy (e.g. by D. Deutsch, H. Greaves and W. Myrvold, D. Papineau, S. Saunders, D. Wallace) and some of assessment (e.g. by D. Albert, A. Kent, A. Valentini). There are also two technical papers, which I list in (4.A) below. Since these two books, both advocacy and assessment have continued vigorously in the research article literature, including many papers by these authors. I give some references below.

For further references, I will follow the order of the Chapter, discussing (A) decoherence and then (B) probability. For each topic, I first cite technical details, stressing positive results; and then philosophical assessment, stressing misgivings.

(4.A) For the technical physics of decoherence, two outstanding books are:

(i) E. Joos, H.D. Zeh, C. Kiefer et al., Decoherence and the Appearance of a Classical World in Quantum Theory, Springer 2003 (second edition); available at:
https://link.springer.com/book/10.1007/978-3-662-05328-7

(ii) M. Schlosshauer, <u>Decoherence and the Quantum-to-Classical Transition</u>, Springer 2007; available at: https://link.springer.com/book/10.1007/978-3-540-35775-9

Another two expository technical papers, which are readable by philosophers (and which also address the subtleties about the system-environment split that I raised at the ends of Sections 7 and 8), are in the <u>Many Worlds?</u> anthology which I cited above, i.e. in S. Saunders J. Barrett, A. Kent and D. Wallace (Eds). They are:

(iii) J. Hartle, Quasiclassical realms; a revised version is available at: https://arxiv.org/abs/0806.3776

(iv) J. Halliwell, Macroscopic Superpositions, Decoherent Histories, and the Emergence of Hydrodynamic Behaviour; available at: https://arxiv.org/abs/0903.1802

The main theme of philosophical discussion of decoherence in the Everettian context has been the idea that macroscopic objects are patterns. Apart from Wallace's monograph, <u>The Emergent Multiverse</u>, and Saunders' and Wallace's Introduction and Chapter 1 in the <u>Many Worlds?</u> anthology (both cited above), I also recommend: (i) Wallace's first article advocating the idea; and (ii) the most recent (so far as I know) critical assessment, by R. Mulder. They are as follows:

(i) D. Wallace, 'Everett and Structure', <u>Studies in the History and Philosophy of Modern Physics</u>, volume 34 (2003), pp. 86–105; available at: https://www.sciencedirect.com/science/article/pii/S1355219802000850?casa_token=wpZJQ5ckKMwAAAAA:2V62sb_C84WpayehAZYpKAlvcXJBoQehozFfEZfx8nJ2OnYMzaxQsG4XZQRlgjtalNsEldUIKQ

(ii) R. Mulder, 'The classical stance: Dennett's criterion in Wallacian quantum mechanics', <u>Studies in the History and Philosophy of Science</u>, volume 107 (2024), pp. 11-24; available at: <u>https://www.sciencedirect.com/science/article/pii/S0039368124001055</u>.

Another philosophical theme about decoherence in the Everettian context is whether the worlds arising from decoherence can accommodate contingency about which macroscopic objects exist. This theme was Section 8's closing issue, i.e. misgiving, about the sketch definition of 'world'. We will return to this theme in Chapter 6.3's discussion of A. Wilson's proposal to understand the philosophical multiverse of Chapter 3 in terms of the Everettian multiverse; (i.e. in philosophical jargon: to reduce the former to the latter). A recent critical assessment of the proposal is:

(iii) J. Harding, 'Everettian Quantum Mechanics and the Metaphysics of Modality', <u>The British Journal for the Philosophy of Science</u>, volume 72, number 4, December 2021.

(4.B) About Everettian probability, I will give references for three topics: (a) the decision-theory approach which, as I reported in Section 12, is now dominant; (b) a recent approach by S. Saunders; and (c) controversies about interpreting tiny probabilities or amplitudes

(a): The ground-breaking article that first proposed to derive Born-rule probabilities using axioms from decision theory (Section 12) is:

(i) D. Deutsch, 'Quantum Theory of Probability and Decisions', <u>Proceedings of the Royal Society</u> 1999; and available at: <u>https://royalsocietypublishing.org/doi/10.1098/rspa.1999.0443</u> and: <u>https://arxiv.org/abs/quant-ph/9906015</u>.

The idea was much developed by D. Wallace in a series of papers; including Chapter 8 in the <u>Many Worlds?</u> anthology (2010) which I cited above, and his own 2012 book, <u>The Emergent Multiverse</u>. Many further theorems and justifications, and assessments of these, have been given since then.

One issue that the Chapter did not broach is how an Everettian can make sense of quantum theory getting confirmed by experimental statistics, in view of the obvious worry that in many worlds in the multiverse, experiments will yield statistics far from the Born-rule probabilities. A ground-breaking article addressing this is:

(ii) H. Greaves and W. Myrvold, 'Everett and evidence'; Chapter 9 in the 2010 anthology, i.e. (ii) above, edited by S. Saunders, J. Barrett, A. Kent, and D. Wallace.

(b): Recently, S. Saunders has developed a very different approach to deriving the Born-rule in Everettian quantum theory. This approach does not appeal to decoherence or decision theory. But it connects closely to Section 11's topic of probability in classical statistical mechanics, to Section 12's first topic of branch-counting, and to frequentism in the philosophy of probability. Two of his papers about this approach are:

(iii) S. Saunders, 'Finite frequentism explains quantum probability', <u>British Journal for the Philosophy of Science</u> 2024; available at: https://doi.org/10.1086/731544 ; and available at https://arxiv.org/abs/2404.12954v3

(iv) S. Saunders, 'Physical probability and locality in no-collapse quantum theory', forthcoming in <u>Journal of Physics</u>: Proceedings of the DICE 2024 conference; available at https://arxiv.org/abs/2505.06983

(c): The Chapter neglected controversies about the interpretation of amplitudes (and so probabilities) that are small, maybe minuscule. Typically, a wave-function assigns many possible configurations a small amplitude; (these "parts" of the wave-function are called 'tails', like the left and right tails of a bell-shaped probability distribution). But interpretation usually focusses on the wave-function's "peaks". This is especially true of this Chapter's Everettian, who took macroscopic objects to <u>be</u> the peaks of the wave-function (or the cluster of classical configurations under such a peak. This prompts an objection to the Everettian's idea (Section 12 and the references just cited) of deducing the probabilistic interpretation of the squares of amplitudes, from the betting behaviour of an Everettian experimenter who bets on macroscopic outcomes (relative to worlds, of course) of quantum measurements. Namely: how can the Everettian justify ignoring the tails of the wave-function? Surely not by their having a small probability, since the Everettian ignores them <u>en route</u> to their argument justifying the probability interpretation. So how?

In his book, Wallace addresses this objection (2012, pp. 253-254). But not to everyone's satisfaction. In particular, Dawid and Thebault dissented in a 2015 article:

(i) R. Dawid and K. Thebault, 'Many worlds: decoherent or incoherent?' <u>Synthese</u> volume192 (2015), pp. 1539-1580; and available at: https://philsci-archive.pitt.edu/9542/.

Since then, the debate has continued, and in some respects changed. The latest article, which also reviews the literature, is:

(ii) R. Dawid and K. Thebault, 'Decoherence and probability', available at: https://philsci-archive.pitt.edu/23991/ and at: https://arxiv.org/abs/2410.01317

# Chapter 5: All the worlds from the primordial bubbles

This Chapter discusses our third multiverse. It was proposed by some cosmologists from the early 1980s onwards, on the basis of their theories about the Big Bang origin of the cosmos. Like the Everettian multiverse in the last Chapter, it is agreed even by its proponents to be very speculative, and hard to confirm. This common feature is unsurprising. As we discussed in Chapter 1, there might well be, within a branch of physics, theoretical (or more generally: conceptual) reasons for a proposal that is hard to confirm---and whose assessment thus calls on conceptual, even philosophical, arguments.

But there is also a dissimilarity from the Everettian multiverse. For that multiverse arose as a solution to a problem about the general structure of our supremely successful quantum theory. Recall that although we concentrated for simplicity on elementary quantum theory, e.g. the quantum replacement of classical point-particles, all our more advanced quantum theories, including our well-established quantum field theories of electrons, quarks etc. equally face the measurement problem (Chapter 4.3, 4.4). And the reasons in favour of the Everettian solution to that problem apply equally to them; (as do, broadly speaking, the reasons against the Everettian solution).

On the other hand, the cosmological multiverse arises by combining two speculations that are very specific. One is a mechanism called 'inflation', which is speculated to have operated for a tiny fraction of a second in the very early history of the universe, i.e. very soon after the Big Bang. The idea is that for this tiny fraction of a second, the cosmos expanded vastly; and besides, at an accelerating rate. Cosmologists' original rationale, in the early 1980s, for proposing inflation was not that it solved a problem about the general structure of a physical theory, quantum or otherwise. Rather, it answered pressing 'Why?' questions. It promised to explain some facts that were otherwise puzzling, even mysterious. These questions were about the value of a physical quantity "having to be just-so", i.e. having to take a value that was constrained to many decimal places, if our cosmological theory was to adequately describe empirical observations. This just-so-ness is usually called 'fine-tuning'. (I shall focus on two such questions, called the 'flatness problem' and the 'horizon problem'.)

These 'Why?' questions, i.e. 'Why is the value fine-tuned?', prompt more general philosophical questions: 'What counts as an explanation?'; and 'How could science confirm a multiverse proposal?' These will be this Chapter's main philosophical questions, which we pursue from Section 5 onwards.

The other speculation, the second ingredient of the cosmological multiverse, is string theory; (which also began to be developed in the early 1980s). It is a quantum theory that unifies general relativity's successful account of gravitation with quantum field theory's successful account of nature's other fundamental forces. Section 4 will give a few more details about it. But the main point for us is that it combines with inflation so as to give a multiverse. For this, the key contribution from string theory is that: (i) it has many vacuum states (also called 'ground states', or for short, 'vacua'); (ii) these states differ from each other in the values of what we usually call 'constants of nature'; and (iii) each vacuum state has an associated set of states that share with the vacuum state its values for these constants. (String theorists realized this only in about 2000: they originally hoped that the theory would have a unique vacuum state.)

Obviously, a full understanding of (i) (ii) and (iii) requires the demanding technicalities of string theory. But---fortunately, for a philosophical book---we can get by with a sketch. Sections 3 and 4 will give more details. But for the moment, the following three points will suffice.

(i): Beware: 'vacuum' does not mean 'nothing'. Nor does it mean 'no physical system', i..e the sheer absence of the physical system that the theory is intended to describe. Instead, it means 'state of lowest energy'. Hence the synonym 'ground state', with states of higher energy being called 'excited' or 'above the ground state'. By and large, physical systems tend to lose energy and thereby to evolve over time towards their vacuum states. Many physical systems have a unique vacuum state: but by no means all do---a glass has many vacuum states.

(ii): 'Constant of nature' means a value of a physical quantity that, so far as we have measured it, does not vary across the entire observable universe. (So this is unlike Chapter 4's examples of a system's values of position, momentum etc., which obviously vary.) Here are three examples: the amount of electric charge on an electron, the ratio of the strengths of electromagnetic and gravitational forces, and the speed of light. And there are many others: such as the masses of elementary particles such as electrons and quarks, the strengths of the forces between them. And as we shall see, the quantities whose fine-tuning gives the flatness and horizon problems are also examples. But string theory is, so to speak, liberal or unopinionated: in the sense that its different vacuum states have different values for these "constants".

(iii): For each vacuum state, its associated set of states is a set of higher energy, excited, states that share the vacuum state's values for these "constants". (But I shall often drop the double "scare" quotes that are meant to signal that the value can vary across the string theory's state-space; and so I will just write the word 'constant'. Another common jargon avoiding the connotations of 'constant' is to say 'parameter' or 'cosmological parameter'.) We can think of this set as a "tower" of states, standing above the vacuum.

So to sum up (i) to (iii):--- The striking fact about the many vacua of string theory, and their associated towers of excited states, is that the values of the constants listed in (ii) vary from one vacuum and its tower, to another.

One natural (perhaps the most natural) first response to this fact is that string theory should simply restrict itself to the tower of states whose values match the values which we actually measure. That is: it should explore the theoretical features, and experimental consequences, of states in that tower: one of which would then be (according to string theory) the actual state of the cosmos. Thus one might say: it is no demerit of string theory that it could describe various non-actual values of the constants---but also no merit. After all, the same is true of our other theories. For example, classical electromagnetic theory could describe electromagnetism with various non-actual values of the charge on an electron.

But it turns out that combining string theory with inflation, i.e. the very early, very brief but also very rapid expansion of the cosmos, yields a mechanism that makes for a multiverse. Roughly speaking, the expansion makes a quantum state in one tower evolve over time so as to have a component in (an amplitude for) other towers. Besides, this happens in a runaway fashion, called 'eternal inflation'; so that states in very many towers, and so very many combinations of possible values of the "constants", are equally allowed.

It is evident in the light of Chapter 4 that here, we face questions about the interpretation of the quantum state. Thus suppose one takes a non-zero (or at least large enough) amplitude in the quantum state to correspond to something real. (Note that this does not to commit one to being an Everettian in any of the senses discussed in Chapter 4. For they all maintain that the Schroedinger equation is, in Bell's phrase, 'always right'; whereas this supposition is entirely compatible with believing that wave-function collapse is a real dynamical process.) And suppose also that one accepts the cosmologist's quantum state, with amplitudes for states in many towers, as correct. Then one must conclude that each of many combinations of values of the "constants" are real.

This is the cosmological multiverse: whose different universes are described by states in different towers, differing in their constants of nature. These universes are often called 'bubbles' (or 'domains'): hence this Chapter's title.

So much by way of introducing inflation and string theory, and how they combine to give a cosmological multiverse. From a philosophical perspective, it is remarkable how they both raise the philosophical theme of explanation. For I said above that inflation was originally motivated by 'Why?' questions about fine-tuning. And we have just seen that string theory is a framework with constants of nature---including the quantities whose fine-tuning gives the flatness and horizon problems---varying across the multiverse. This confluence obviously prompts an ambitious project: to somehow explain the values of all these constants---i.e. the values that we in this universe actually measure---by invoking some appropriate features of how the values vary across the multiverse. Hence this Chapter's taking explanation and confirmation as its main philosophical topics.

With this background in place, I can now explain the plan of the Chapter. There will be four stages. The first stage (just Section 1) clarifies the relation between the Everettian and cosmological multiverses. This will develop the discussion above, about the cosmological multiverse involving not just inflation but also string theory and its many vacuum states. But again, we will be able to proceed with hardly a mention of the advanced physics involved.

Then in the second stage, I introduce modern cosmology and its multiverse. Section 2 will summarize what cosmology had established by about 1980. Section 3 will do two jobs. First, I report the puzzling facts about fine-tuning that remained unexplained (i.e. the flatness and horizon problems), and how inflation could explain them. Second, I sketch how inflation led to a multiverse, though without details of string theory. Then Section 4 supplies some details about string theory.

The third and fourth stages (Sections 5 onwards) are philosophical: they address explanation and confirmation.

In the third stage (Sections 5 to 7), Section 5 first reviews some of the philosophical literature about explanation. Then it formulates two overall strategies we could adopt in order to explain the values we actually measure. (The distinction between the strategies will not depend on details of the physics.) The first strategy is the obvious one: to argue that the value we measure is in some precise sense generic, or typical, of the values across the multiverse; and is thereby to be expected. But there is a second, less obvious, strategy. It invokes what are called 'selection effects', or (a better-known jargon) 'the anthropic principle', to argue that the value we measure is likely to be observed---even if it is not generic or typical across the whole multiverse, simply because most "regions" of the multiverse have no observers. I will label these two strategies, 'strategy (Gen)' and 'strategy (Obs)': with the labels standing for 'generic' and 'observation' respectively.

The rest of the third stage (the next two Sections) develop the strategy (Gen): that is, explaining a fact by showing that it is generic or typical. Section 6 describes the strategy's successes and its positive features. Section 7 describes its difficulties, including in cosmology.

The fourth stage (Sections 8 and 9) is about the strategy (Obs): explaining a fact by showing it is probable (or probable enough) that it be observed---even if, observation apart, it is improbable (or not probable enough). Section 8 explains the strategy with examples from outside cosmology, indeed from outside physics. It also introduces the jargon of 'selection effects'. Section 9 discusses the strategy within cosmology. It introduces the jargon 'the anthropic principle'. And as an example, it summarises the anthropic explanation of the value of a "constant" (also called 'parameter') which I have not mentioned so far: namely, the cosmological constant.

Finally, Section 10 concludes. Here, I recommend a framework for confirming a theory of the multiverse that incorporates ideas from both the strategies: the idea of being generic, and the idea of being probable (or probable enough) to be observed.

Chapter 5.1: Comparing the Everettian and cosmological multiverses
The mere phrase 'the cosmological multiverse' suggests there should be connections with the Everettian multiverse proposal. For as we saw, the Everettian proposes a quantum state of the cosmos, written with the capital Greek letter 'Psi', i.e. written as Ψ. (Recall Chapter 4.6's decision to use this phrase, not the more usual 'quantum state of the universe'.) So the job of this Section is to describe these connections. I shall of course have to set aside many details of advanced physics, especially string theory: but this omission will, I think, be justified by its not affecting our philosophical questions about explanation and confirmation. We can spell out the connection sin four comments, (1) to (4).

(1): When one first meets the phrase 'the cosmological multiverse', and bears in mind that modern cosmologists of course accept quantum theory, one naturally expects that: (i) some cosmologists will endorse the Everett interpretation; and even (ii) the cosmological multiverse will turn out to be an elaboration of the Everettian one.

The first of these expectations, (i), is indeed true. Among cosmologists interested in interpretative questions about quantum theory---and of course the cosmological multiverse raises such questions---the Everettian interpretation is popular.

But on the other hand, the second expectation (ii) is true only in a liberal, i.e. logically weak, sense of 'elaboration'. Yes, a cosmologist may well accept that there is a quantum state of the cosmos, and may also argue that it describes the cosmological multiverse they advocate. But the universes (or worlds) in this multiverse are very different from the Everettian universes (or worlds) defined by the well-established, continual, ubiquitous and rapid process of decoherence applying to macrosystems, that we discussed in Chapter 4.7. For (as I said in this Chapter's Preamble) the cosmologists' many universes are produced by a specific mechanism, '<u>inflation</u>', that is speculated to have operated for a tiny fraction of a second very soon after the Big Bang. Agreed: in many such universes, there will indeed be (later, long after the universe starts) macrosystems such as dust-particles, or even pointers and cats: macroscopic objects which can decohere. And for these, the Everettian can then claim that the last Chapter's account, with <u>its</u> Ψ, applies.

So the upshot is that if we accept the cosmological multiverse based on the idea of inflation, then the quantum state of the cosmos---in the sense we envisaged in the last Chapter, i.e. Ψ with components, i.e. summands, describing various decoherent worlds: cf. Chapter 4.8--- is at best a description of a <u>single</u> universe within the cosmological multiverse.

Besides, I say 'at best' not just because inflation occurs vastly earlier than any macroscopic objects exist. Also, as I said in the Preamble: the cosmological multiverse is based on string theory (combined with inflation). So the cosmologist's 'quantum state of the cosmos' will be a state in string theory, not a state in a well-confirmed quantum theory. In particular, it is not a state in our well-confirmed theory of electrons, quarks etc., which is nowadays called 'the standard model'. This distinction matters because there is a "large gap" between string theory states and those in established quantum field theories. Indeed, this is a notorious fact about string theory. Namely, it is very hard to get out of string theory any empirical predictions; or even the recovery of specific theoretical postulates of confirmed quantum field theories. So the upshot announced in the last paragraph assumes in effect that this large gap has been bridged. It assumes that we can get by deduction (or by some approximation to deduction) from the official "stringy" state of the cosmos to the states of our confirmed quantum field theories.

So the cosmologist we envisaged two paragraphs above---who accepts that there is a quantum state of the cosmos, and says it describes the cosmological multiverse they advocate--- will mean by 'quantum state of the cosmos' something yet more dizzying than the last Chapter's (the Everettian's) $\Psi$: for two reasons.

First, it means a string theory state, and one needs to bridge the gap to quantum field theory. Second, as announced in the Preamble: it means a state that encompasses various alternative values of (what we call!) the constants of nature. Here, my 'encompasses' is deliberately vague. For the way that the theory (this state) describes eternal inflation's runaway process of the different universes (with their different constants of nature) coming to exist, is complicated and controversial; (not least because of the challenge of the first reason).

But for our purposes, it will suffice to boldly ignore the gap between string theory and quantum field theory, and to construe 'encompasses' as 'has amplitudes for'. That is: it will suffice to take our cosmologist's 'quantum state of the cosmos' to be a sum of "Everettian-Chapter-4 states", i.e. a sum (superposition) of 'quantum states of the cosmos', using this last phrase in the sense of the last Chapter's Everettian. So we take it to be a sum of different Everettian-Chapter-4 states $\Psi_i$, where the suffix 'i' (i = 1, 2, ...) is a label on the summands, i.e. the different universes---which in general disagree with each other about the values of (what we call) constants of nature.

(2): Here, I should give a clarification. For the sake of a clearer exposition, I have in the last six paragraphs taken decoherence as a process that occurs to macrosystems due to their interaction with an environment, e.g. dust-particles immersed in air: just as I did in Chapter 4. As just explained, this gives a crisp contrast between Chapter 4's ideas and this Chapter's new idea, the mechanism of inflation. But cosmologists <u>also</u> apply the idea of decoherence, in various detailed ways, to the cosmos as a whole---including at very early times.

One way this is done is to take all the material in the cosmos to be the system (i.e. the analogue of the dust-particle), and spacetime to be its environment (i.e. the analogue of the air). Here, 'material' will include not just matter e.g. atoms, electrons, quarks, but also radiation e.g. electromagnetic radiation. And in order to give content to the idea of this material <u>interacting</u> with spacetime, one needs states of spacetime that respond to the states of matter, so that one describes spacetime with general relativity, according to which spacetime is indeed responsive in this way. (But as mentioned in Chapter 4.8, the system-environment split is sometimes made in a less "obvious" way, e.g. taking long-wavelength modes to be the system, which is then decohered by short-wavelength modes.)

In this kind of way, even without macrosystems such as dust-particles, decoherence can occur. So cosmologists will talk of all the cosmos' material degrees of freedom being in an improper mixture (cf. the end of Chapter 4.7) of states that are definite for some appropriate quantities. So in short: the contrast between Chapter 4's Everettian multiverse and the cosmological multiverse is not as crisp as I first suggested. That is, the contrast is not as crisp as: decoherence for macrosystems, without variation in constants of nature vs. string theory and inflation, with varying constants.

Nevertheless, in the rest of this Chapter, we can safely take this contrast to be valid. That is: we can think of the cosmologist's 'quantum state of the cosmos' as a sum of different Everettian-Chapter-4 states $\Psi_i$: and these states in general disagree about the values of constants of nature such as the charge of the electron. This picture of the relation between the two multiverse proposals will set us in good stead for our philosophical questions about explanation and confirmation.

Of course, these are not just speculative, but also imprecise and dizzying, ideas. Just as I remarked (in Chapter 4.6) that no Everettian has the faintest idea how to write down in detail the Everettian quantum state of the cosmos $\Psi$, no quantum cosmologist can now write down in detail <u>their</u> quantum state of the cosmos. For as this Chapter will report: we do not know---and we may never know---the underlying physics of inflation. So a sketch-definition of 'universe' for the cosmological multiverse, on analogy with the Everettian's sketch definition of 'world' (Chapter 4.8) is far beyond current knowledge.

(3): But agreed: you can't keep a good idea down. I should also report that some quantum cosmologists, sympathetic to the Everett interpretation, have proposed mathematical formulas for the quantum state of the cosmos at very early times. Of course, the details vary from one author or research programme to another. Thus some of these formulas are independent of whether there was a very brief period of inflation; while some incorporate such a period. Some are string-theoretic, some are not. Some are independent of decoherence at very early times, e.g. of the material degrees of freedom; while some incorporate it. One such formula, proposed by Hartle and Hawking in 1983 (and independently of all three topics above: inflation, string theory and decoherence), is called 'the no-boundary proposal', or 'no boundary condition'. It has been much studied and developed since then: and indeed, it has been related to inflation, string theory and decoherence.

This reflects the more general fact that most quantum cosmologists recognize the relevance of interpretative questions, and related methods and ideas like decoherence, to their scientific work. This relevance is shown by the point above: that cosmologists' models often incorporate decoherence at very early times, long before there were macrosystems like dust-particles.

But for this book's theme of the multiverse, the most vivid example of this relevance will be the point I made in the Preamble (and will develop in Section 3 below): that if (i) one takes a non-zero (or at least large enough) amplitude in the quantum state to correspond to something real, and (ii) one accepts the cosmologist's quantum state, with amplitudes for states in many towers with differing constants, as correct: then one must conclude that each of many combinations of values of the constants are real.

As noted in the Preamble, (i) does not require being Everettian. In particular, it does not require the Schroedinger equation being, in Bell's phrase, 'always right'. And nor does combining (i) with (ii). For there might be a model of wave-function collapse (a revision of the Schroedinger equation of the kind discussed in Chapter 4.5) which describes the formation of a bubble in the expanding inflationary cosmos; and which can also describe the formation of many bubbles---all, thanks to (i), equally real---and thus, a multiverse. But Everettian or not, the relevance of interpretative questions to the cosmological multiverse is vivid. One might well say that taking all the universes ('bubbles') in the cosmological multiverse to be equally real involves assuming some solution, on a cosmic scale, of the quantum measurement problem.

(4): Nor is this vivid example the only place where modern theoretical cosmology meets the measurement problem on a cosmic scale. Even without going back in time as far as the putative period of inflation, and without postulating a multiverse: modern cosmology describes early states of the universe, e.g. a minute, or a year, or ten thousand years after the Big Bang, in terms of quantum theory---and so the measurement problem arises.

Indeed, it arises in connection with something so basic and vivid to us as the existence of stars, planets and galaxies. For as I will explain in the next two Sections, there is a weak electromagnetic signal throughout space, called the 'cosmic microwave background' (CMB) radiation, that dates from about 380,000 years after the Big Bang---and which we can directly observe. We observe this CMB radiation to be very smooth and uniform: it looks the same in all directions. But it has tiny 'wrinkles': which are really variations in the quantum amplitudes for various densities of mass in regions of space. This means that a peak among these wrinkles is a "seed" of a clump of matter becoming gradually localized, under the gravitational attraction of its component parts, in one region of space rather than another. Such a clump, once localized, can grow, pulling yet more matter in by its gravity, and eventually produce, for example, a galaxy.

But note that here, the word 'seed' is a metaphor, that hides the problematic issue of the collapse of the wave-function. For whereas a real seed is an actually existing object that grows into an actually existing plant, this peak of quantum amplitude is only a higher (square root of a) probability for the event of clumping to happen here, rather than there. (Unfortunately, the metaphor is entrenched in textbooks as well as popular expositions. Only the better textbooks admit that this transition, from peaks and troughs of quantum amplitude to a classical, slightly uneven, distribution of mass-density across space, is problematic---since it is a cosmic version of the "collapse of the wave-function", which all agree is problematic.)

So much by way of sketching connections between the Everettian and cosmological multiverses. Or to put it more generally and precisely: so much for the connections between (i) the interpretation of quantum theory, especially its measurement problem, and (ii) quantum-to-classical transitions in the very early cosmos ('primordial bubbles') and in the not-so-early cosmos ('wrinkles' in the CMB).

Obviously, these connections are important, indeed fundamental. But this Chapter will not go in to further details about them, for two reasons: one negative and one positive. The negative reason is that most cosmologists, even quantum cosmologists, believe that these cosmological aspects of the measurement problem (or more neutrally: of the interpretation of quantum states), are not yet precisely enough formulated to be addressed as a problem within physics. In short: we do not know enough, and the time is not yet ripe. The positive reason is that (as we shall see) even if we restrict ourselves to a "classical outlook", there is so much to explore, in both the physics and the philosophy of modern cosmology. At least: there is certainly enough for this Chapter.

Chapter 5.2: A golden age of cosmology
We live in a golden age of cosmology. It began in the twentieth century, especially in its second half. I will describe it in three stages: covering the first half of the twentieth century in (1), and the subsequent history in (2) and (3).

(1): Already in the first half of the century, there were four momentous developments in cosmology: two observational and two theoretical.

First, we discovered that the nebulae, that appeared in telescopes like cloudy smudges rather than point-like stars, were really other galaxies of vast numbers of stars, like our own Milky Way. So the cosmos turned out to be vastly larger than had been envisaged. Second, we discovered that the cosmos is expanding. More precisely: any two galaxies are receding from each other, i.e. the distance between them is increasing. (The speed of recession is approximately proportional to the distance between them.) But this is not an expansion of matter into a pre-existing empty space, like an explosion of a firework or a bomb. Rather, the space itself is expanding. Agreed, that is impossible to visualize. We are bound to think of an ambient or embedding space relative to which the expansion occurs; and it was a struggle for physicists to accept this idea.

In this acceptance, the third development was crucial: namely, Einstein's discovery of general relativity, and its application to the whole cosmos. As mentioned in Chapter 2, general relativity is a relativistic theory of gravitation. According to it, gravitational influence propagates across space at the speed of light; not instantaneously, as in Newton's theory. And like Einstein's earlier theory of special relativity, it unifies space and time into a single entity, spacetime: which, being four-dimensional, is again unvisualizable. (Physicists' acceptance of these unvisualizable ideas, of expanding space or of spacetime, was helped by the rise of pure mathematics, reviewed at the end of Chapter 2: the increasing formalization of mathematics included liberating geometry from visual intuition.)

Einstein himself, immediately after formulating general relativity, applied the theory to the cosmos as a whole. In terms of our jargon of systems and their state-spaces (cf. Chapter 3.3): he took the cosmos as his system, and he described it as a spacetime, extending not just arbitrarily far in all directions in space, but also throughout the past and future. (This is reminiscent of our description of the cosmos, "our world", in Chapter 1.1.) So he aimed to find a solution to the equations of general relativity that described the whole of spacetime: not of course in its myriad details, but in the broadest possible terms. In particular, matter was treated as smoothed out uniformly across space, although of course it is in fact concentrated in great clumps. (For the galaxies are clumps; and within them, the stars are clumps.) By the mid-1920s, solutions of general relativity describing an expanding cosmos (whose matter is smoothed out uniformly across space) had been found, and in the following years they were investigated and elaborated. In some of these solutions, the expansion began from a very hot, very dense state: which came to be called 'the primeval fireball'.

The fourth development was the rise of astrophysics: i.e. the physics of stars. Quantum theory, discovered in the 1920s, was applied to describe in detail, not just how stars shine by burning helium, but also much else: the different types of stars, how other chemical elements are formed in stars (called '<u>nucleosynthesis</u>', since the stars synthesize i.e. make the nuclei of, elements), how and why some stars explode and others implode.

By the late 1940s, the third and fourth developments had been combined. For the ideas and results of astrophysics were applied to describe nucleosynthesis in the conjectured primeval fireball. This led to detailed predictions about the cosmic abundances of light elements like hydrogen, helium and lithium; and to the prediction of a pervasive but very faint electromagnetic radiation with a characteristic wavelength, that was a remnant of the fireball.

(2): Thus matters stood in about 1960. It was the following years that really ushered in the golden age. Again, they were several momentous developments, both observational and theoretical. I shall pick out three, that were all well underway by 1980. Developments after 1980, including the proposed multiverse, will be treated in the next Section.

Foremost among observations was the discovery (by accident, in 1964) of the predicted remnant radiation. It is called the 'cosmic microwave background' ('CMB') radiation. It was immediately recognized as confirming general relativity's expanding cosmological solutions. In just a few years, almost all cosmologists accepted that the cosmos originated in a primeval fireball about fourteen billion years ago. (The fireball was soon renamed '<u>the Big Bang</u>': a name that had originally been suggested by sceptics, as a derisory label.) Besides, in the following decades, the CMB has proven to be an extraordinarily rich source of information about the early cosmos.

The second main observational development between 1960 and 1980 was the invention and deployment of several new sorts of telescope that enabled astronomers and cosmologists to study types of electromagnetic radiation other than visible light. For radio waves (with wavelengths much longer than visible light), there was ground-based radio astronomy, which had been pioneered in the 1940s. For microwave and infra-red radiation (i.e. wavelengths a bit longer than visible light) and X-rays (shorter than visible light), one needed to get above the Earth's atmosphere. For these wavelengths, dozens of satellite missions have yielded a profusion of data, both for astrophysics and cosmology: for example, data about the CMB and the cosmic abundances of elements.

The third development was theoretical. After 1960, there was a renaissance in the study of general relativity, both as regards its mathematics and its applications. Here, one highpoint was a cluster of theorems saying that among the solutions of general relativity (i.e. the spacetimes that are possible according to the theory), singularities are generic, i.e. typical. The idea of a singularity is a breakdown in the smooth structure of spacetime; and the theorems showed that such a breakdown must occur under certain conditions. These conditions included when a star whose mass is above a certain limit, having burnt all its fuel, implodes under its own weight. This is called the 'gravitational collapse' of the star; and it leads to a black hole, in which the singularity lies. But more relevant to us: among these conditions were conditions that were understood to prevail in the early cosmos. This gave a new perspective on the simple expanding cosmological solutions of general relativity that had been recognized, already in the 1920s, as having an initial singularity: i.e. the original "point where the fireball began" (though not a point in spacetime itself). Namely: this initial singularity came to be regarded, not as an artefact due to the solution's admittedly very idealized treatment (especially its smoothing out the matter), but as a robust feature of the solution that might well be physically real.

The result of these three developments was that by the mid-1970s, cosmologists had agreed on a model of the history of the cosmos, with an initial singularity about fourteen billion years ago, followed by a hot primeval fireball that cooled and expanded. It was called 'the standard model'. (This is not to be confused with its namesake, the standard model in high-energy physics. That describes the physics of electrons, photons, neutrinos, quarks i.e. the constituents of protons and neutrons, and also of unstable particles. It also was formulated in the mid-1970s.)

In the last fifty years, this standard model of cosmology has stood up amazingly well; (as has its namesake in high-energy physics). Agreed: in addition to elaborating ideas and methods that were already formulated in the 1970s, two major new ideas have had to be added so as to accommodate observations. One such idea is that much, indeed the majority, of the mass of the cosmos is of an as-yet unknown form; this is called 'dark matter'. Another is that although one would expect the expansion of the cosmos to slow down i.e. to decelerate (since the stars and galaxies, having mass, pull on each other gravitationally), the expansion is in fact accelerating. This is called 'dark energy'.

(Incidentally, these two ideas have prompted the standard model to be re-named as the 'Λ-CDM model'. Here, 'CDM' means 'cold dark matter'; and Λ, i.e. the Greek capital Lambda, represents dark energy. It is the cosmological constant, to which we will return in Sections 7 and 9.)

But for the most part, these two ideas, dark matter and dark energy, need not concern us, for two reasons. First: their bearing on our topic, the multiverse, is slight. Most proposals about the nature of dark matter and dark energy do not give reasons for, or against, a multiverse. In effect, they are compatible with inflation's producing a multiverse, but do not especially support it.

Second: there is good theoretical reason to think that whatever the detailed physics of dark matter and dark energy turn out to be, it will not overturn the main outlines of what the standard model claims to have established.

This is well illustrated by two of the standard model's "grand narratives" of the history of the cosmos: the thermal history of the cosmos (i.e. its density, temperature, pressure etc. at successive stages); and the history of the synthesis of elements, both in the primeval fireball and later in the stars. Indeed: it is a striking testimony to how well confirmed this standard model now is, that for both these narratives, the detailed story given in a technical exposition written today matches closely the detailed story in expositions written some fifty years ago.

So I will end this review of our fortunate golden age in cosmology by taking as an example, the thermal history of the cosmos. This will set the scene for the next Section's description of the puzzling features that prompted cosmologists to postulate an even earlier, very brief period, of accelerating expansion: inflation.

(3): I will give just three "snapshots" of what the temperature, density and relative size of the universe was, at the following times: (1) a millionth of a second after the Big Bang, (2) a hundredth of a second after it, and (3) ten million million seconds, i.e. about 380,000 years, after it. Note that we are now about a hundred thousand million million seconds, i.e. about fourteen billion years, after the Big Bang.

Before I give the numbers, let me adopt the <u>exponent</u> (or <u>index</u>) <u>notation</u>, using a superscript to indicate the number of noughts. So one hundred is $10^2$; and a million is $10^6$. Similarly, we use negative exponents to represent reciprocals, i.e. 1 divided by a larger number. So one hundredth is $10^{-2}$; and a millionth is $10^{-6}$.

This notation prompts another important point. It will be helpful (though I admit, it is difficult) to think <u>logarithmically, not arithmetically</u>: to think, for example, that since the present time is about $10^{17}$ seconds after the Big Bang, the time $t = 10^{-17}$ seconds before the Big Bang is <u>as much before t = 1 second, as we are after it</u>.

Though this sounds blatantly wrong, the rationale for it is that a great deal of physics is a matter of <u>scales</u>. That is: if you change the situation you wish to describe by a factor of about 10, you may well need a very different description: and this is even more likely if you change by a factor of about 100. This trend holds whether the quantity whose value you change is time, or distance, or energy or temperature. So when cosmologists puzzle over what was the state of the universe, at say $t = 10^{-6}$ seconds, or how physical processes changed as a result of the cooling between, say, $t = 10^{-6}$ and $t = 10^{-2}$ seconds, we should not accuse them of straining at gnats, i.e. of foolishly concentrating on events that are so transient that they cannot be very important for the physics. For, agreed: the universe was changing unbelievably rapidly (arithmetically speaking!); but the relevant processes change---and so our description must change---in crucial ways, depending logarithmically on the earlier time.

So here are the three snapshots.

(1): $t = 10^{-6}$ seconds after the Big Bang:--- The temperature was about $10^{13}$ degrees Centigrade. This is when protons and neutrons, i.e. the constituents of atomic nuclei, form: for at yet higher temperatures, they "melt" into their own yet-smaller constituents, quarks. The size of the observable universe relative to its size today was $10^{-12}$, and the mass density was about $10^{17}$ grams per cubic centimetre.

(2): $t = 10^{-2}$ seconds after the Big Bang:--- The temperature was about $10^{11}$ degrees Centigrade. Atomic nuclei form: i.e. at higher temperatures, they "melt" into their constituent protons and neutrons. The size of the observable universe relative to its size today is $10^{-11}$, and the mass density was about $10^9$ grams per cubic centimetre.

(3): $t = 10^{13}$ seconds (i.e. about 380,000 years) after the Big Bang:--- This is when atoms formed: by free electrons combining with nuclei, so as to form electrically neutral atoms of the familiar kind. The universe thereby became for the first time transparent to electromagnetic radiation. So for cosmology, this is a crucially important time. For it means that our direct observations of electromagnetic radiation cannot go back any earlier than this time. (But amazingly, we <u>do</u> observe this time: the CMB, the remnant radiation from the Big Bang, dates from exactly this time.) It is known as the '<u>recombination time</u>': though all agree that 'combination' would be a much better name, since the electrons and nuclei were not stably combined at any earlier time. The temperature was about 3000 degrees Centigrade. (By way of comparison, the temperature at the surface of the Sun is about 6000 degrees.) The size of the observable universe relative to its size today was $10^{-3}$, and the mass density was $10^{-21}$ grams per cubic centimetre.

I said that these claims about the universe's thermal history were now established. But I agree that when presented with these stupendous figures---so high for temperature and density, so tiny for time and size---one of course asks: 'Is all this really established as fact?'

I think the answer is Yes. Of course, the evidence is technical and varied---and I cannot go in to details. But I note that physicists' description of even my earliest snapshot, i.e. the description of protons and neutrons "melting" into quarks by the standard model of high energy physics, has been confirmed by terrestrial experiments. Indeed, I could have chosen an earlier snapshot. For it is common nowadays to take the boundary between known and speculative physics to be at about $10^{-11}$ seconds after the Big Bang. But agreed: there is a spectrum of caution and confidence (as we discussed in Chapter 1.4), and one could reasonably be more cautious, even taking e.g. one second as the start of what one calls 'established'.

Chapter 5.3: Inflation … eternally

So much by way of celebrating our golden age of cosmology, and its standard model as formulated in the mid-1970s and developed since then by e.g. the admission of dark matter. In this Section, I first describe, in (1), two puzzling features of the model, which were recognized by 1980, and which prompted the idea of inflation. I introduce inflation in (2), and its conjectured mechanism in (3). In (4), I describe how this mechanism yields a multiverse.

(1): The two puzzling features are called the '<u>flatness problem</u>', and the '<u>horizon problem</u>'. (I set aside a third puzzling feature, called 'the monopole problem': not just for brevity, but also because inflation's treatment of it is similar.) For both of them, the problem is not one of empirical adequacy. That is: the problem is not that the standard model from the mid-1970s gets some observational prediction wrong. The problem is that according to the model, an empirically measured number amounts to a coincidence so enormous that, as the saying goes: 'it cries out for explanation'.

So first, the flatness problem. The expanding solutions of general relativity fall into three classes:

> (i): those in which the matter is on average dense enough that gravitation eventually overcomes the expansion so that there is a contraction and ultimately a "Big Crunch"; this is called a 'closed universe';
> 
> (ii): those in which the average density is low enough that expansion goes on forever at some non-zero rate (with of course lower densities making for a higher final rate): this is called an 'open universe'; and
> 
> (iii): those "between" (i) and (ii) in that the average density (a) is low enough that gravitation cannot overcome the expansion, but also (b) is high enough that the final rate of expansion is zero. This is called a 'flat universe', since the instantaneous geometry of space, across the whole universe, gets ever closer to being Euclidean---so 'eventually-flat universe' would be a more accurate name.

The density in (iii) is special. Not only is it the boundary between the regimes (i) and (ii). Also, once a solution has that density it will have it forever. It is called the '<u>critical density</u>'.

These ideas are often put in terms of the ratio between the universe's actual density (of course, as usual: taking the matter as smoothed out over all space) and the critical density. So this number is a pure number. For it is defined by dividing one density by another; and so it has no units. It is written as the Greek letter 'Omega', i.e. $\Omega$.

So here is the enormous coincidence. In fact, we have measured $\Omega$ to be now close to 1; and indeed to have been close to 1 at all times later than about one second after the Big Bang. (This means that ever since that time, the universe has been almost flat: its spatial geometry has been almost Euclidean.) But in these solutions of general relativity, any difference of Ω from 1 in the early universe is very rapidly amplified. For example: if at one second after the Big Bang, Ω = 1.08, then already at ten seconds Ω = 2; and thereafter Ω keeps increasing exponentially. And on the other hand: if at one second after the Big Bang, Ω = 0.92, then already at ten seconds Ω = 0.5; and thereafter Ω keeps decreasing exponentially. In short: in these solutions, Ω = 1 represents an equilibrium---but a very unstable equilibrium. In particular, for Ω to be about 1 today requires that it be stunningly close to this privileged value soon after the Big Bang. For example, at one second after the Big Bang it has to differ from 1 by at most $10^{-16}$. This is about the ratio between the width of a human hair (viz. a tenth of a millimeter) and the average distance between Earth and Mars (viz. 225 million kilometres).

Indeed, this is an enormous coincidence, crying out for explanation.

Second, the horizon problem. It has a similar structure. Namely: although the standard model of the late 1970s is empirically adequate, it requires a feature of the CMB to be "just so" to an extreme degree. Indeed: to a degree so extreme that it is implausible to treat it as a brute fact, without explanation.

The problem arises from the fact that the CMB, dating from 380,000 years after the Big Bang, is <u>very</u> uniform across the sky. Its wavelength, amplitude etc. is almost the same in whatever direction you point your telescope: its wrinkles are minuscule. More precisely: their proportional size is $10^{-5}$. That is like having, on the surface of a pool of water one meter deep, a wave which is only a hundredth of a millimetre high.

One naturally asks why it should be so uniform. And this question becomes all the more urgent in the context of the standard model of the late 1970s. For it says that for two directions in the sky with a sufficient angle between them---about one degree or more (the visual width of the moon, or more)---the two emission-events of the CMB that lie along those directions (about 13 billion years ago) have no <u>common causal past</u>. This means: no yet-earlier event could affect both of the two emission-events, by influences travelling to each of them at most as fast as light. That is, there is no event that could influence them both via causal, i.e. no-faster-than-light, processes. Hence the phrase 'no common causal past'. Relativity theory has some other helpful jargon for this. Given any event, the set of events to its past that <u>could</u> affect it by an influence travelling at most as fast as light, is called the event's <u>past light-cone</u>. So the point is: the standard model says that the two past light-cones of the two CMB emission-events do not overlap.

This makes our question urgent. For this means there could not have been any kind of interaction between events in the past of the first emission-event and events in the past of the second emission-event. But since the emission events are so strongly correlated---their quantitative properties differ by at most the tiny factor $10^{-5}$---one would naturally expect some such interaction. For think of how we explain various systems with uniform properties throughout their extent; (such states are called '<u>homogeneous</u>'). For example: a cup of tea with milk throughout it, or an iron bar with its temperature equal along its length. We explain these by a past process of interaction. Namely, the system started in a non-uniform (heterogeneous) state: then the milk spread through the tea, the heat spread along the bar. (A process that ends in such an equilibrium state is called '<u>equilibration</u>'.) But here, the standard model forbids such a process of achieving uniformity by an earlier interaction. For it says that no causal process of any kind could affect the two signals of CMB coming to us from these two directions in the sky.

In short: the standard model tells us to accept these signals' strong correlation as a brute fact, which is encoded in the state of the universe's matter and radiation at times earlier than the recombination time. That is hard to accept.

Besides, it is all the harder to accept when one calculates that the angle between directions sufficient to imply (according to the standard model) no common causal past, is very small. It is only about one degree—the visual width of the moon. For this angle being so small means that, according to the standard model, there are a stupendous number of patches of the sky whose CMB radiations are strongly correlated with each other (to within the tiny factor $10^{-5}$), even though there was no interactions in their pasts. That is surely incredible.

So the flatness and horizon problems have a common structure. They each take a certain feature ($\Omega$, and the smoothness of the CMB, respectively) to be <u>just so</u>. That is: the feature has a value specified to many decimal places (also called: 'many significant figures'), without the standard model giving any account of why the feature is so exactly specified. This "just-so-ness" is called '<u>fine-tuning</u>'. (Later in this Chapter, this phrase will get a more specific meaning in the context of <u>selection effects</u>.)

(2): Enter the idea of <u>inflation</u>. It turns out that if we change the standard model by "inserting" into it a very brief and very early epoch of rapid, indeed accelerating, expansion, then we can solve both problems.

The basic idea of both solutions is quite simple. It turns out that whatever the value of $\Omega$ at the onset of the inflationary epoch, $\Omega$ will be driven close to 1 by the end of the epoch, and will remain close to 1 for a very long time thereafter---including until now. And recalling that $\Omega$ being one is a matter of a flat Euclidean spatial geometry, we can see the simple idea behind this calculation: an expansion of a highly curved surface makes a local patch flatter. Think of blowing up a balloon; or how the fact that the earth is large makes our local patch of it seem flat.

The situation is similar for the horizon problem. A suitable inflationary epoch changes the spacetime geometry in just the right way. Namely, it implies that the past light-cones of all emission events of the CMB---even for points on opposite sides of the sky---do in fact overlap. So with inflation, the cosmos' very early spacetime geometry allows for a suitable process of equilibration that made the CMB's properties so uniform.

The details of these solutions also work out well, in the sense that when one calculates how much inflation, and when, is needed so as to solve these two problems, one gets approximately the same results, despite the two problems being so different. Namely, the solutions are quantitatively correct, if we postulate:

    (a) the inflationary epoch ends at about $10^{-34}$ seconds (which corresponds to a temperature of $10^{28}$ degrees Centigrade);

    (b) the inflationary expansion is exponential, and started at, for example, $5 \times 10^{-35}$ seconds with a characteristic expansion time (i.e. the time in which the radius of the universe is multiplied by about 3) of $10^{-36}$ seconds.

Taking (a) and (b) implies that in the course of the inflationary epoch, the size of universe expanded by a factor of about $10^{22}$.

Agreed, these are dizzying figures; and the epoch is proposed to occur at times and energies very far beyond those we have confirmed in experiments or observed. So we are undoubtedly in the realm of extreme speculation; and accordingly, caution is in order. It would certainly be reasonable to give low credence to the idea of inflation; and therefore, to the details in the rest of this Section and the next. (But if so, the philosophical discussion of explanation from Section 5.5 onwards would still stand.)

(3): Having solved the flatness and horizon problems, i.e. avoided two fine-tunings by conjecturing a process of expansion, one naturally asks: 'What caused this expansion: what is its mechanism?' For one might suspect that unless we can cite a plausible cause, we should conclude that it is just a coincidence that the same quantitative details about the expansion solve both problems. In answer to this question, the advocate of inflation has, as the saying goes: good news and bad news.

The good news is that a mechanism has been formulated. Indeed: there are many proposed mechanisms which, needless to say, remain conjectural. Most of them involve postulating a new physical field (called '<u>the inflaton field</u>', and written φ) which evolves i.e. changes over time according to a postulated potential energy function, written V(φ). And fortunately, from such a field and potential one can deduce some characteristic features of the CMB: namely, characteristic probabilities for the amplitudes and frequencies of the slight wrinkles (unevennesses) in the CMB. These features have been observed by a sequence of increasingly refined instruments, mostly on satellites: (the famous acronyms/names are 'COBE', 'WMAP' and 'Planck'). So nowadays, these confirmed predictions are regarded as more important evidence that there was a brief epoch of expansion, than the epoch's solving the flatness and horizon problems.

But there is also bad news. The data we now have, and maybe all the data we will ever have, leave wide open which of the many possible mechanisms---which sort of field φ, and which potential V(φ)---actually occurred. There are two aspects to this .The data leaves open the formal mathematics, e.g. what is the function V(φ). And it leaves open what the physical nature of φ is: it is widely believed <u>not</u> to be one of the known fields.

So far in this Section, we have reviewed two problems that were solved by the idea of an inflationary epoch, and broached the question of what mechanism led to that epoch. Now we are ready for the punch-line: that is, the punch-line for someone interested in the multiverse.

(4): In explaining this punchline, I will expand on the summary I gave in this Chapter's Preamble and Section 1, but postpone details about string theory till the next Section.

It turns out that many models of the inflaton field and its potential involve a <u>branching structure</u> in which, during the epoch, countless spacetime regions branch off and then expand to yield other universes. Here, 'branch off and then expand' means that the region stops its accelerating expansion, and expands only slowly, like our observable universe does. (Note that I said 'many models'; so again, caution is in order.) As a result, the whole structure gives a multiverse, whose component slowly-expanding universes cannot now directly observe (nor otherwise interact with) each other, since they are causally connected only through their common origin during the inflationary epoch. (Incidentally, a pair of neighbouring points on the inflating background space also separate from each other so rapidly as to lose causal contact: so such pairs also "soon cannot see each other".)

Besides, in a universe that branches off---called a '<u>bubble</u>' or '<u>domain</u>' or '<u>pocket universe</u>'---yet another universe can branch off; and also from that one, there can be a branching … and so on. In short: bubbles (domains) spawn more bubbles, endlessly. This idea of open-ended, maybe infinite, branching towards the future is called '<u>eternal inflation</u>'.

Of course, all this is a quantum process. So the quantum state of the cosmos (in the cosmologist' sense, explained in Section 1) contains components corresponding to (amplitudes for) the different bubbles. And one is committed to the bubbles being real, if one accepts that to a non-zero (or at least large enough) component/amplitude there corresponds something real. (This acceptance was (i) in Section 1: and as I said there, it does not imply an Everettian view, since it is compatible with dynamical models of wave-function collapse.)

There are two main types of model that yield eternal inflation, labelled '<u>false-vacuum</u>' and '<u>slow-roll</u>'. In both types of model, the inflationary expansion coming to an end, i.e. the beginning of a slow expansion, is a matter of the inflaton field evolving, i.e. changing over time, to its state of lowest energy. And as discussed in the Preamble and Section 1: such states are called '<u>vacuum states</u>', or for short, '<u>vacua</u>'; or '<u>ground states</u>'. (So again: 'vacuum' does not mean 'nothing' or 'no physical system'.)

By and large, physical systems tend to lose energy and to evolve to their vacuum states. So also here. Thus eternal inflation is a matter of the prevention of the inflaton field's tendency to get into its vacuum state (which state would render the expansion slow). The system being prevented in this way is usually called its 'being <u>frustrated</u>'. Thus a region where this frustration does <u>not</u> occur, is where the new bubble branches off. And since the region where there <u>is</u> frustration continues to expand exponentially, there is very soon a vastly larger region in tiny patches of which new branching will occur. Hence: eternal inflation.

Chapter 5.4: Glimpsing the landscape of string theory
So much by way of introducing the idea of eternal inflation. In this Section, I turn to string theory, building on the Preamble and Section 1: but again, necessarily omitting a lot of advanced physics. I will confine myself to just three topics. First: in (1), I will state the initial idea of string theory. Then in (2) I will develop the idea that it has many vacuum states (each with an associated "tower" of excited states) that differ about the constants of nature. Then in (3), the dauntingly large number of vacuum states will prompt us to face philosophical questions about explanation and confirmation---which will dominate the rest of the Chapter.

(1): String theory is a speculative attempt, that began in the mid-1980s, to unify general relativity's successful account of gravitation with quantum field theory's successful account of nature's other fundamental forces. (These are: the electromagnetic forces between charged particles, and two other forces between sub-atomic particles such as electrons, neutrinos and quarks, which are called the 'weak' and 'strong' forces.) The sense in which string theory aims to unify these forces is like the unification of electric and magnetic forces that Maxwell achieved (Chapter 4.1). Roughly speaking, the four apparently diverse forces are to be revealed as aspects of a single force.
     We can glimpse why this is hard to achieve by looking at string theory's initial key idea. Namely, it "does for a string, what elementary quantum theory did for a point-particle". (Hence its name.) Thus recall from Chapter 4.2 that elementary quantum theory replaced the state of a classical point-particle—in effect, its single actual position---by an entire function on all possible such positions, mapping each position to a "square root" of a probability (an amplitude) to be found there, if measured. Now, a point-particle is extensionless, and can be thought of as zero-dimensional; while an infinitely thin line, a mathematical curve, is one-dimensional. So just as we can think of classical point-particles as idealizations of tiny spheres, we can think of an infinitely thin line as an idealization of a thin filament---a string, though without the spiral threads. Such a string has, of course, no single position. Each of its constituent points has a position, and the configuration of the string as a whole is the infinite set of those configurations: which we might call a 'placement' of the string. So the quantum replacement of the state of this classical string is a function on all possible placements, mapping each placement to an amplitude … No wonder that advanced physics is needed.

(2): But for this book's purposes, all we need is the upshot: that string theory predicts the system, i.e. the set of all the quantum strings, has very many <u>different possible vacuum states</u>. (This was realized in about 2000: until then, string theorists had hoped there was only one vacuum state.)
     Note that here, the jargon can be confusing. For a state that is not the overall lowest-energy state (the state with energy lower than <u>all</u> others), but is only a local minimum with an energy lower than all its near neighbours in the state space, is often called a '<u>false vacuum</u>'. (As I mentioned at the end of the last Section: this jargon is used in cosmology, as the name of one main types of eternal inflation.) But it is 'false' only in the sense that the minimum is <u>local</u>. So this jargon is rather like calling a valley in a mountain range a 'false valley', just because it is higher above sea-level than the lowest valley in the entire mountain range. But this analogy with

valleys and peaks has also prompted a more helpful jargon. In string theory, the whole state space, i.e. the set of states of the quantum strings (varying in energy, with some lower and some higher), is called 'the landscape'.

(Incidentally: biology makes an analogous use of 'landscape'. In the theory of natural selection, the attributes of an organism, or of a population of organisms, make it more or less fit: where fitness is, roughly, a matter of having more offspring who live long enough to reproduce. Over time, natural selection, "the survival of the fittest", increases the proportion of fitter organisms. So over time, the population (descendants of the original organisms) gets a higher fitness "score". So the population "climbs to a peak in the fitness landscape". So high fitness is analogous to high energy: the biology-physics difference is that in biology fitness increases over the generations, while in physics an individual system tends to a lower energy---to a valley, not a peak. From the philosophical perspective of Chapter 3, the interesting point here is of course that this is another vivid illustration of our science being up to its neck in modality: almost all positions in the fitness landscape are not inhabited by a real organism or population---it is a realm of possibilities.)

Returning to physics: each of string theory's vacuum states has a tower of associated higher energy states. The idea goes back to Chapter 4.3's comment that in quantum field theory, particles are really energetic excitations of fields. There, the tower was called Fock space. It is a sum of infinitely many subspaces. Namely: first, the space containing the given zero-particle i.e. lowest-energy state; then, the space of one-particle states; then, the space of two-particle states, then the space of three-particle states, and so on. But the idea of particles as excitations is more general than this simple sum-of-subspaces structure of Fock space, and it carries over to string theory.

In string theory, all the various higher states got by exciting a given vacuum state will share with it the values of physical parameters, like the electric charge on an electron, the ratio of strengths between the electromagnetic and gravitational forces, or the speed of light. (We tend to call the charge on an electron a 'physical parameter', not a 'physical quantity', since we think of the value of a quantity, e.g. the position or energy of a system, varying across the states within a single tower, while the value of a parameter is the same for all states in a tower.)

But for different vacua, these values will vary. This is so even for parameters that we take to be constants of nature: i.e. constant in value across the observable universe, like the examples just listed. (This uniformity, this "geographical" unity of the observable universe, is itself very remarkable. A priori, there could well be regions of the universe in which the charge on the electrons, or the relative strengths of the forces, or the speed of light-signals, is different from what we measure hereabouts.)

(3): This variation prompts an ambitious but alluring project. For we now have a framework so broad that it encompasses scenarios (formally: towers of states above certain vacuum states) that differ from each other in the values of parameters that we usually call 'constants of nature'---though this framework means we are now envisaging that they vary across a wider landscape. So this suggests: let us invoke this framework to answer the obvious big 'Why?' question about such a parameter, namely: 'Why does it have the value that it does?' That is: let us try to find features in the framework that in some sense favour, and thereby explain, the value.

But this project runs up against a major problem: a problem that suggests the project will stumble, unless one appeals to some philosophically contentious ideas. For it turns out that there is a dauntingly large number of vacuum states. And this problem confronts not only string theory taken alone, irrespective of cosmology and inflation. It also confronts inflationary cosmology.

Thus a recent estimate of the number of string theory vacua is $10^{500}$. This is enormously larger than all the numbers in more established branches of physics. For example, the number of elementary particles in our universe, i.e. setting aside the cosmological multiverse, is estimated to be about $10^{100}$. So the number of string theory vacua is larger by a factor---not of 400, but of--- $10^{400}$. Similarly, in the cosmological multiverse: estimates of the number of bubble universes give vast numbers.

Obviously, to explore this set of states, this landscape---i.e. to understand it in quantitative detail, classifying valleys and peaks---is forever beyond human, or even superhuman, ability.

In the face of this impossibility, the project envisaged above, of explaining the values of the parameters, seems to stumble. For as I said in this Chapter's Preamble, the obvious overall strategy for getting such an explanation would be to argue that the value we measure is in some precise sense <u>generic</u>, or <u>typical</u>, of the values across the multiverse; and is thereby to be expected. One aims to explain the actual value we see by showing that is typical, and to be expected. But how can we do that, without understanding the set of states (the landscape, the towers above the vacua) in quantitative detail?

The rest of this Chapter will discuss suggestions for how to do this---albeit contentious ones. Section 5 sets the scene by discussing explanation in general, and formulating: first, the obvious strategy above, which I will label 'strategy (Gen)' (for 'generic'); and another, 'strategy (Obs)' (for 'observation') which invokes selection effects. Subsequent Sections will treat these strategies, in order.

Chapter 5.5: *Angst* about explanation:
Let us for a moment take a step back from the details of physics, and ask: how does one explain any fact? How does one answer any 'Why?' question?

There is a large philosophical literature about explanation, with rival accounts of what an explanation is, and what role explanations fulfil in the enterprise of science. But for our purposes, these accounts' agreements matter more than their disagreements; so this Section will for the most part summarize the agreements. But unfortunately, these agreements will not settle the questions raised at the end of the last Section. So those questions will have to wait for the next two Sections.

The accounts agree that in everyday life what counts as a correct or appropriate answer to a 'Why?' question obviously depends strongly on what the enquirer (and no doubt, also the respondent) knows, what their interests are, etc. These accounts also agree that such contextual and pragmatic features also apply to scientific explanation.

They also agree that in both everyday life and science, there is a spectrum of requirements one can impose on the answer, along the lines of: whether it must be believed by the respondent, or must be true, or even must be known to be true. Again, it is a contextual and pragmatic matter which requirement lying on this spectrum we should impose on a would-be explanation, in order for it to count as a genuine explanation.

These accounts of explanation also agree on some helpful jargon, from Latin. The fact to be explained (or the proposition expressing the fact) is called the '<u>explanandum</u>', and what does the explaining (or the proposition expressing what does it) is called the '<u>explanans</u>'.

More substantively, and relevant to us: they also agree that what facts count as <u>needing</u> explanation is a contingent, and often historically determined, matter: no less in science than in everyday life.

This is not just the obvious points that explanation must come to an end somewhere, and that where the respondent's chain of explanations terminates depends on their state of knowledge, which is a contingent and historically determined matter. (And as every five-year-old who persistently asks 'Why?' learns: where the chain of explanations terminates can depend on the parent's inventiveness, or patience.)

There is also the more interesting point that acceptance of a scientific theory (or more loosely: of a research tradition or framework) can influence, even determine, which sorts of fact are taken to need explanation, and which do not. This point merits two examples.

A standard example of this from the history of physics is Kepler's endeavour (in his *Mysterium Cosmographicum* of 1596) to explain the relative sizes of the planets' orbits, and the number of known planets (viz. 6), by interpolating the five Platonic solids between the orbits. (The known planets were: Mercury, Venus, Earth, Mars, Jupiter and Saturn.) Thus Kepler believed that such a major structural feature of our solar system should have a systematic explanation. But nowadays, we accept, not just that there are more planets---Uranus was discovered in 1781, and Neptune in 1846---but that the sizes of the orbits are "merely" accidents of the history of the solar system. They no doubt have a very complicated causal explanation---if only we could know it. The explanation would involve the various radii (i.e. distances from the sun) at which the planets were first formed, how they interacted gravitationally etc. But these are (at least for the most part) matters of sheer happenstance, about how the solar system happens to have evolved. We do not expect the number of planets, or their orbits' sizes, to have any general or systematic explanation. (Needless to say, this is not to disparage Kepler's endeavour. Given his overall world-view, that there are six planets orbiting the Sun is a main, even pre-eminent, fact about the solar system about which it is very natural to ask 'why six?'.)

Nor is it only particular facts that can come to be seen as not needing a general or systematic explanation. Very general patterns of behaviour can also fall out of the purview of explanation. (I say 'patterns of behaviour' to set aside controversy about laws of nature (cf. Chapter 3.6); but as we will see, the pattern might well be called a 'law' of a given theory.) A standard example of this is the idea of "<u>natural motion</u>".

In the long history, since the ancient Greeks' geometry and astronomy, of the precise quantitative description of motion, 'natural motion' is an inevitably vague term. But the rough idea is: motion that needs no explanation, since the body is "moving without interference". Thus in Aristotelian cosmology, the natural motion for the element earth (one of four: the others being air, fire and water) was downward---towards the Earth. But by the mid-seventeenth century, the mechanical philosophers (cf. Chapter 2.2) maintained that natural motion was motion in a straight line at constant speed---it was other motion that was "forced". For example, a body's accelerating towards the Earth was due to the Earth's gravitational force; and a block's slowing as it slid down an inclined plane was due to friction with the plane. This came to be called '<u>the principle of inertia</u>'. It was given its first clear formulation by Descartes; and later, it was Newton's First Law of Motion---that a body subject to no force at all moves in a straight line at constant (maybe zero) speed.

Thus for both the mechanical philosophers and Newton, the motion of a projectile in a straight line and at constant speed (neglecting gravity and air resistance) needs, in a sense, no explanation. Agreed: one can ask what causes set this projectile moving, i.e. what launched it. And agreed: the motion, once underway, is an instance of the principle of inertia; and so it can be deduced from the principle. But the motion, once underway, needs no explanation in the sense that no <u>causes</u> need to be cited. It is enough that the motion instantiates, and can be deduced from, the principle of inertia. (Here, my 'it is enough' deliberately echoes Hume's and Newton's lowering our sights about the rationalist understanding of nature, discussed in Chapter 2.5 and 2.6).

So much by way of examples. The final issue on which these philosophical accounts of explanation also agree is the core idea of explanation: namely, that a <u>successful explanation shows that the explanandum was to be expected</u>.

Here, I choose the words 'successful' 'shows' and 'to be expected' deliberately. Thus the first two words signal how my formulation of the agreement between these accounts deliberately steers clear of some controversies that---irrelevantly for this book's purposes--- dominate the philosophical literature; as follows.

I say 'successful', in order to signal flexibility about pragmatic factors such as the enquirer's interests, and about whether the explanans must be true, or even known to be true.

I say 'shows', in order to signal flexibility about whether: (i) there must be an outright deduction of the (proposition expressing the) explanandum from the propositions comprising the explanans (which would be a strong sense of 'show'); or (ii) it is sufficient to render the explanandum probable (usually in the sense of having a high enough probability, conditional on the explanans).

(Of course, there are other controversies I have not touched on. For example: must an explanation of an individual event or fact (as against a general proposition) cite the causes of the event or fact? And: is explanation fundamentally contrastive, i.e. about answering 'Why A rather than B?' not just 'Why A?')

On the other hand, my third deliberately chosen phrase, 'to be expected', signals a return to the questions at the end of the last Section. I deliberately choose a phrase that is ambiguous; and ambiguous in a way crucial to our concern with multiverse proposals' endeavour to explain the values of parameters such as constants of nature. For the phrase can be understood as referring to either one of two different strategies for explaining some fact; in particular, explaining some apparent fine-tuning of a parameter's value. These strategies are:

> (Gen): showing that the fact: either is deducible (from the explanans); or is generic or typical, i.e. roughly speaking, one of the alternatives that have high enough probability; or

> (Obs): showing that although the fact is not deducible, and is not even generic or typical: it is likely (or has high enough probability) to be observed. (As I mentioned in this Chapter's Preamble: this strategy invokes 'selection effects', or (a better-known jargon) 'the anthropic principle'.)

The next two Sections explore strategy (Gen). The subsequent Sections explore strategy (Obs).

Chapter 5.6: Expected because generic

In this Section, I discuss strategy (Gen) in general terms, without considering the multiverse---though with examples from physics. I will first, in (1), cast inflation's answer to the flatness and horizon problems---problems of fine-tuning, in a single universe---as an example of strategy (Gen). Then in (2), I mention some other examples. This will prompt a more general statement of what fine-tuning amounts to. Then in (3), I will give a bit more detail about three ways in which one can make precise the idea that the value of a parameter is to be expected, because it is generic. I put them under the labels: 'topology', 'effective field theory', and 'probability'. This Section will emphasize the first two of these. But probability will be a large topic for us, also in connection with strategy (Obs): so although I will introduce it here, the details will be postponed to subsequent Sections.

So the overall shape of this Section will be to start with fine-tuning as a problem, and to end with three approaches to answering the problem by saying that the parameter's value is in fact generic. Thus the tone of this Section's assessment of strategy (Gen) will be positive. With the examples and approaches considered here, the strategy has successes. But in the next Section, the difficulties that the strategy faces will move to centre-stage.

(1): First, let us recall the flatness and horizon problems from Section 3. In both, a certain feature ($\Omega$, and the smoothness of the CMB, respectively) had to be just so. That is: the feature has a value specified to many decimal places, without our cosmological model (i.e. the model that was standard in the 1970s) giving any account of why the value is so tightly constrained---in short, fine-tuning. As we saw, inflation solved these problems by changing the theoretical context so substantially that the required values could arise through an (admittedly, conjectural) dynamical process, from generic initial states. For a suitable inflationary epoch drives $\Omega$ to become close to 1 by the end of the epoch; and thereafter, the standard cosmological model's (non-inflationary) dynamics makes $\Omega$ remain close to 1 for a very long time, including until now. Similarly, a suitable inflationary epoch makes the past light-cones of all emission events of the CMB---even for points on opposite sides of the sky---overlap.

We can now construe this discussion in terms of the last Section's ideas about a successful explanation showing that the explanandum is to be expected, either by deduction, or by getting a high enough probability, from the explanans. And since it is generic initial states (i.e. states before the inflationary epoch) that lead to $\Omega$ being close to 1, and to the past light-cones overlapping, the explanation is insensitive to what exactly is the initial state. Such an explanation (or deduction, or calculation of high probability) is often called 'robust' or 'stable' or 'resilient'. In short: here, the phrase 'to be expected', in 'showing the explanandum is to be expected', has the straightforward sense (Gen) at the end of the last Section.

These examples of inflation solving the flatness and horizon problems also illustrate other ideas from the last Section, such as the ideas that:
- (i) explanations inevitably come to an end somewhere; and
- (ii) the theoretical context moulds one's judgments about what is generic or probable (or probable enough to count as being explained), and what is not.

For after all, one can ask: (i) what explains (and-or what caused) the pre-inflation state, no matter how generic or probable one accepts it to be; and (ii) what justifies one's judgment about what is a generic or probable enough pre-inflation state. (I will shortly return to this topic of making precise the idea of a state of the whole cosmos being generic or probable.)

But although one can raise these questions, inflation's account of why $\Omega$ is close to 1, namely as a robust feature of a dynamical mechanism, is generally agreed to be a successful explanation---even though the dynamics is very conjectural. In short, it is a successful example of strategy (Gen).

(2): Besides, there are several other examples in physics where a fine-tuned value gets explained as generic by a suitable change in the theoretical context. One such case was the explanation in the early 1970s of some fine-tuning in sub-atomic physics by postulating a new kind of particle (viz. a charmed quark): which was later empirically confirmed.

Indeed, such examples fall in to a wider category: of explaining a value (not necessarily fine-tuned) of a parameter, by a suitable change in the theoretical context---but not necessarily by showing the value to be generic or typical, e.g. by having high or at least moderate probability.

A famous case of this wider category---with the merit that it is simple enough to describe---is Maxwell's explanation of the speed of light. When Maxwell formulated his unified theory of electricity and magnetism (mentioned in Chapter 4.1) he found that some solutions to his equations described waves of the electric and magnetic fields that his theory postulated. That is: the theory described oscillating patterns of (values of) these fields (vectors located at points in our familiar 3-dimensional physical space). These patterns propagated across space at a speed that is a simple function of two fundamental constants (called 'permittivity' and 'permeability'), that are mentioned in the theories of electricity and magnetism, and whose values were known. When Maxwell calculated this simple function from the two known constants, the answer turned out to be the speed that had already been measured as the speed of light (viz. 300,000 kilometres per second). Maxwell then inferred that light is waves of the electric and magnetic fields.

This is often, and rightly, celebrated as a reduction of one field of physics, the theory of light i.e optics, to another, the theory of electromagnetism (as we now call it). (Here, 'reduction' is meant in the sense of Chapter 3.1: viz. deriving the doctrine of one theory from that of another by augmenting the latter with suitable definitions.) But once light is indeed identified as being such waves, Maxwell's calculation can also be taken as an explanation of the speed of light. In effect the explanation is: 'these waves of the electric and magnetic fields must travel at this simple function of the permittivity and permeability constants; and given the actual values of those constants, the speed must therefore be 300,000 kilometres per second---as is observed.'

So much by way of examples of successfully explaining the value of a parameter. I turn now to formulating a bit more generally what fine-tuning really amounts to: what is the problematic 'just-so-ness' of a parameter's value. The idea will be that the parameter should not be a function of other parameters, that depends very sensitively on those other parameters' values.

To take perhaps the simplest example: the value of a parameter should not be an arithmetical difference of two other physically significant numbers that are nearly equal, but are both vastly larger in magnitude than the parameter itself.

Thus imagine a theoretical framework in which the chosen parameter, which I call p, is a millionth: $p = 10^{-6}$. And imagine this is "because" (i.e. because, according to the given framework) p is the difference of two other numbers, q and r, that themselves have some physically significant interpretations, and that are nearly equal but are both vastly larger than p. For examples, they might have values $10^6 + 10^{-6}$ and $10^6$. That is: $q = 10^6 + 10^{-6}$ and $r = 10^6$ and $p = q - r = 10^{-6}$.

So the imagined framework makes the value of the parameter p fine-tuned. It is extremely sensitive to the exact values of these other numbers q and r: in my example, sensitive to their thirteenth digit. Had q and r been slightly different (in terms of proportions of their actual values, e.g. in their last digit), then p's value would have been vastly different (proportionately) from its actual value. In short: the framework, with its equation $p = q - r$, gives us an unsatisfactorily fragile derivation of p's value, not a robust explanation of it.

(Of course, since the value of a parameter usually depends on a human choice of units, these numbers $10^{-6}$ etc. should be dimensionless. That is, they should be pure numbers without a physical unit involved. For example, they could be a ratio of two masses, or of two densities, or of two electric charges, or of strengths of two forces.)

This example illustrated the idea of sensitivity with an arithmetical difference being tiny, and so liable to be vastly (proportionately) changed by a change in the numbers whose difference is being taken. But as I said, fine-tuning need not be a matter of an arithmetical difference. The function involved, whose values depend very sensitively on its arguments, could be a function very different (in particular, more complicated) than addition. All such cases of fine-tuning prompt strategy (Gen): the value of a parameter should be shown to be generic or typical, in some precise sense that is defined by an appropriate theoretical framework. (Of course, this framework is often not the one in which the parameter's value is first observed or known. Recall for example how $\Omega$ was measured to be close to 1, before the framework of inflation was suggested.)

But can we make precise this idea of being generic, in a more general way? That is: can we do so without invoking case-studies, with their case-specific functions like arithmetical difference, and case-specific theoretical frameworks that show the value to be generic? ('Being generic' is sometimes called 'genericity': a word so ugly that I avoid it.)

(3): In my opinion, mathematics and physics provides three overall approaches to doing so. I suggest the labels: topology, effective field theory and probability. I will discuss the first two, which are comparatively specific to mathematics and physics. Then I will briefly discuss probability: which of course extends far beyond mathematics and physics, and which will also occupy us in subsequent Sections.

Topology is a major branch of pure mathematics that focusses on the idea of a continuous transformation: which means, roughly speaking, a transformation that preserves the nearness relations holding between the objects being transformed. Here, being near need not be a matter of a numerical distance. It can be a qualitative relation, and it can come in degrees. Thus topology has jargon like 'closeness', 'neighbourhood' etc.; and a set of objects endowed with such nearness relations is called a 'topological space'. Thus for a transformation T that shifts objects a and b respectively to T(a) and T(b) (in the same or a different topological space), we say that T is continuous if: whenever a and b are near, so are T(a) and T(b). On the other hand, T is discontinuous if there are objects a and b that "get pulled apart" by T, i.e. are such that T(a) and T(b) are not near.

In this way, a discontinuous transformation can express the idea of 'sensitive dependence on the inputs' (here: a and b), even without invoking number-valued functions. Or more precisely: it expresses one version of this idea, without invoking number-valued functions. Applying ideas like these, mathematicians have made precise the idea that an object a in a certain set of objects {a,b,…} is generic, in the sense that it is like (in appropriate respects) the other objects in the space that are near it.

Mathematicians have even defined topological spaces whose objects, i.e. elements a, b, … of the set, are possible physical systems, each taken as subject to certain forces. (Since specifying a system and the forces on it prompts the traditional format of a physics problem, viz. 'For a given initial condition, how will this system change over time?', such spaces are often called 'spaces of problems'.) In such cases, the elements of the space i.e. the physical systems are usually described by a mathematical function such as a potential energy function which encodes the forces on the system. So nearness of the elements of the space is a matter of the systems having potential energy functions that are "nearly the same", according to some criterion for the approximate equality of functions.

I will not go in to these ideas in more detail. But I cannot resist noting that: (i) the jargon of the subject includes the alluring phrases, 'catastrophe theory' and 'structural stability'; (ii) since the elements of such a space are possible physical systems, very few of which are actual, we again see that physics is up to its neck in modality (cf. Chapter 3.3).

I call the second approach to making precise the idea of being generic, or typical, the effective field theory approach. It is like the topological approach, in two ways. It eschews probabilities; (which will be the third approach). And it describes a set of complicated entities with such mathematical precision, both about each entity and about their mutual relations, that the set deserves the name 'mathematical space'. (As discussed in Chapters 3.3 and 4.2, mathematicians call a set that is endowed with various structures, especially structures inspired by geometric or visual intuition, a 'space', even though its elements have nothing to do with points or regions of physical space.)

But there is also a contrast with the topological approach. There, the entities were physical systems, each taken as subject to certain forces; (as noted: often called 'problems'). But in the effective field theory approach, the entities, the elements in the mathematical space, are physical theories, where each theory is identified by, roughly speaking, the set of parameters that occur in its specification of the forces on the systems the theory describes. A bit more precisely: the forces are encoded in a special function: the Lagrangian. Or in some formulations, they are encoded in a function that is a mathematical "cousin" of the Lagrangian, viz. the Hamiltonian. Both the Lagrangian and the Hamiltonian functions have as arguments states in the system's state-space; so they are functions on state-space; and their value is a certain difference, or a certain sum, of different kinds of energies of the state.

(Of course, only a few Lagrangians (Hamiltonians) will be instantiated by an actual physical system. So most of the elements in this space of theories are not actualized. So as in my comment (ii) at the end of the topological approach: we are up to our necks in modality.)

The Lagrangian (or Hamiltonian) contains parameters, especially those whose value specifies how strong a force is (called '<u>coupling constants</u>'). The list of these parameters' values specifies the theory. That is, it specifies the element in the postulated space of theories.

The point of postulating this space of theories---and the link to expressing our topic of being generic---lies in the key idea that the parameters (including, despite their name: the coupling constants) are <u>not</u> really constant. For at high energies, they take different values than at low energies.

So as you mentally traverse a curve in the space of theories, from higher energies to successively lower energies, you can consider the functional dependence of a parameter's values, on various other parameters. In the mathematics, traversing such a curve is a matter, in effect, of summarizing the influence of the physical phenomena at higher energies (the physics described by points, i.e. theories, on the curve which we have already traversed) on the lower-energy physical phenomena that are described by the point i.e. theory you are currently at---i.e. the theory you are currently considering. A bit more precisely: 'influence' here means 'mathematical implications' not 'causal effects'; and 'summarizing the influence' is a matter of averaging the numerical values implied by the higher-energy phenomena. Or in yet more technical jargon: it is a matter of integrating out higher-energy modes of the system.

Traversing such a curve is called 'following the <u>renormalization group flow</u>'. For the way that a parameter's value changes as we consider lower energies is described by a mathematical structure called 'the renormalization group'. And the approach is called '<u>effective field theory</u>' because in physics, '<u>effective</u>' means, not 'efficacious' i.e. 'having a big or strong or intended effect', but: 'approximate in a useful way'. So in physics, an 'effective theory' is a theory which is believed, and often known, to <u>not</u> be completely correct; but which is correct to a sufficient approximation that it is useful.

So the overall idea here is that although we do not know, and may never know, the correct theory of physics at the higher energies that our experiments cannot probe, we can hope to formulate an effective theory of physics at the lower energies that our experiments <u>can</u> probe. (The reason for saying 'effective field theory' not just 'effective theory', is that these ideas were developed (in the period 1965-1975), mainly in the context of quantum field theory, and its topic of renormalization. In these developments, Ken Wilson (1936-2013) was a leading light.)

Besides, traversing a curve from high to low energies, in the above way, amounts to deducing which low-energy theory (low-energy point on the curve) is implied by the high-energy theory (point) at which the curve began. More precisely, the clause 'is implied by….' means: is implied by (i) the high-energy theory, taken together with (ii) the chosen way of summarizing the influence of higher-energy phenomena---the way that the curve defines.

So the question arises: do the curves along which one mentally proceeds, from two different (though close) points i.e. theories at some high energy, towards lower energies, diverge or converge?

If they diverge, that means that small differences in one or other of the parameters of the high-energy theory from which one started will imply large differences in one or more parameters of the low-energy theory which one arrives at. That is: divergence means the low-energy physics, i.e. the physics we can now observe, is extremely sensitive to the values of at least one parameter describing high-energy physics, i.e. the physics we cannot now, and might never, probe with our experiments. So divergence is bad news. For it means fine-tuning of one or more parameters of the low-energy theory; and we might never be able to probe the high-energy physics on which the parameter depends.

On the other hand, convergence of the curves would mean that the values of parameters describing low-energy physics are <u>robust</u> to variations in high-energy physics. They stay approximately the same, when we envisage different high-energy theories, even substantially different ones. This is good news, in that we can hope to argue, even without knowing the correct high-energy theory, that the value of a parameter at low energies should be (approximately) thus-and-so. In short: we can hope to argue that what we see is <u>generic</u>. For whatever the unknown high-energy physics, we would see approximately this value.

Finally, I turn to the third approach to making precise the idea of being generic, or typical. Of the three, it is by far the oldest: one might even call it 'venerable'. The idea is to appeal to <u>probability</u>. There should be a probability distribution over the possible values of the parameter, and the actual value should not have too low a probability.

This connects of course with statistical inference: in both everyday life and science, far beyond physics. There, it is standard practice to say that if a probability distribution for some variable is hypothesized, then an observation that the value of the variable lies 'in the tail of the distribution'---often called: 'has a low likelihood', i.e. a low probability, conditional on the hypothesis that the distribution is correct---disconfirms the hypothesis that the distribution is the correct one. That is: it disconfirms the hypothesis that the distribution truly governs the variable.

This scheme for understanding typicality seems to me, and most interested parties---be they scientists or philosophers---sensible, perhaps even mandatory, as part of scientific method. Agreed: questions remain about:

(a) how far under the tail of the distribution---how much of an outlier---an observation can be without it disconfirming the hypothesis, i.e. without it being deemed to be atypical;

(b) how in general we should understand 'confirm' and 'disconfirm', e.g. whether in Bayesian terms or in traditional (Neyman-Pearson) terms; and relatedly:

(c) whether the probability distribution is subjective or objective; and more generally:

(d) what probability really means (Chapter 4.11, 4.12); and, after Hume's critique of his predecessors (Chapter 2.4, 2.5), what is the philosophical justification for induction.

But these questions are obviously not specific to physics, let alone our more specific topic of the cosmological multiverse. So I will not pursue them in general terms. But this is not to suggest that they are easy, or irrelevant to our topic. We will see them crop up several times in what follows.

Chapter 5.7: Difficulties about being generic
The previous Section described strategy (Gen), i.e. explaining a parameter by showing it to be generic; and some of its successes. In this Section, I report some of the difficulties it faces. I will begin in (1) with some examples, and then move in (2) and (3) to general issues; (3) will be focussed on cosmology. These difficulties will prompt us, in the following Sections, to consider the other strategy, (Obs).

(1): First, we should note that in recent decades in physics, strategy (Gen)'s "track-record" has been mixed. The previous Section noted some successes: especially inflation's explaining $\Omega$ and the homogeneity of the CMB; and the charmed quark. But by no means every apparently fine-tuned parameter has been explained along the lines of strategy (Gen). I will report two such recalcitrant examples. Both examples fall under what the last Section called the 'effective field theory approach'. So they are examples of dismaying fine-tuning. Small differences in one or more parameters of a high-energy theory imply large differences in one or more parameters of the low-energy theory whose predictions we can observe. As in the previous Section, one example is from cosmology, the other is from high-energy physics. As we will see in later Sections, both examples have prompted some physicists, in particular cosmologists, to shift to strategy (Obs).

The first example is the <u>cosmological constant</u>. Introduced by Einstein as a possible emendation of his field equations for general relativity, and written as Λ, (i.e. the capital Greek letter, lambda), this constant represents a repulsive force between two masses. So in a cosmological context, Λ amounts to a cause or tendency for the universe to expand; so it opposes the gravitational force that tends to make matter clump together. And the eventual destiny of a universe that, like ours, is in fact expanding---whether to expand forever, or to come to a stop and re-contract---will be determined by the balance between Λ and gravitation. The current evidence is that the universe's expansion is accelerating: this means that we measure Λ to be positive.

However, we have no good explanation, following strategy (Gen), of the value of Λ: or even of its approximate value. Worse, theoretical estimates of the value of Λ (using the framework of quantum field theory) are wildly wrong. For unless we assume a lot of fine-tuning, the estimates are wrong by very many factors of ten. In some estimates, the error is a factor of $10^{120}$; (that is, 120 factors of ten, called '120 orders of magnitude'.) This discrepancy is called 'the cosmological constant problem'. It is of course agreed to be a major problem for physics. Regardless of one's philosophical views about explanation, and in particular about strategy (Gen): it suggests a very basic conflict between quantum theory and general relativity. (But as mentioned, we will see that strategy (Obs) fares better in dealing with it.)

The second example is the mass of the <u>Higgs boson</u>. (This particle, first postulated in the mid 1960s, was discovered in 2012 at the particle accelerator at CERN, Geneva.) This is another dismaying example of fine-tuning. Thus recall the scenario in the middle of the last Section: a parameter p is defined as a tiny arithmetical difference of two other numbers q and r, each of which is vastly larger than p and also has an appropriate physical interpretation. This scenario implies that the value of p is very sensitive to (changes enormously with) changes in q and r. My toy example was: $q = 10^6 + 10^{-6}$ and $r = 10^6$; so that $p = q - r = 10^{-6}$.

The mass of the Higgs boson is just such a parameter p. The exact value of the exponent depends on which higher value of energy one envisages the (unknown) higher-energy theory being valid; in other words, on how far beyond energies which we can observe one expects new physics to "kick in". The higher this value of energy, the larger is the exponent. For example, if one sets it very high, at the Planck energy---an energy at which the need to reconcile quantum field theory with general relativity becomes acute---then the exponent is 34. That is: the Higgs mass is a difference of two numbers that, written in decimal notation, match in their first thirty-four digits, and then differ in the thirty-fifth digit. Indeed, dismaying.

Again, strategy (Gen) stumbles here. For the most popular version of strategy (Gen) for this case is to appeal to an idea that extends the standard model of high-energy physics that (as mentioned in Section 2) was consolidated in the mid 1970s. And this version of strategy (Gen) turns out to have a fine-tuning problem of its own.

The idea being appealed to here is called '<u>supersymmetry</u>'. We do not need details about it, but can just note the following. Supersymmetry comes in various versions. The "good news" is that some versions imply that the observed mass of the Higgs boson is generic in strategy (Gen)'s sense: the observed mass lies in a range where it is expected to lie. But the trouble is that in order to have this implication, these versions also imply that that there are other particles with a mass similar to the Higgs: particles that have not been observed. These particles, predicted by supersymmetry, are supersymmetric partners of known particles, and are called '<u>superpartners</u>'. But as I say, no such particles have been observed: not even with masses far from that of the Higgs.

We can put the problem a bit more precisely. The only way that supersymmetry can avoid the embarrassing implication that there are superpartners with a mass similar to the Higgs, is to postulate higher masses for the superpartners: so high that our particle accelerators---more generally, our experiments---cannot detect them. But to postulate this requires . . . fine-tuning the masses of the superpartners.

(A note about jargon: in physics, especially high-energy physics, '<u>naturalness</u>' is used to mean, in effect, the opposite of 'fine-tuning'. So in this jargon: the hope was that supersymmetry would show the Higgs mass to be natural. But the problem is that if all masses are natural, then the masses of the superpartners should be similar to that of the Higgs.)

(2): So much by way of reporting examples where strategy (Gen) stumbles. I now move to a general statement of the difficulties strategy (Gen) faces.

As I see matters, it is clearest to distinguish:

(a): a group of difficulties each of which is not specific to cosmology, but arises from the variety, and often, the context-dependence or subjectivity, of the considerations that determine whether something counts as generic, and-or as not needing explanation, so that it can provide, or at least contribute to, an <u>explanans</u>: these are difficulties that we have already seen at various places in our discussion;

(b): difficulties that are specific to cosmology, i.e. that arise from the fact that the system we are concerned with is the entire cosmos: these are difficulties we have touched on, but not focussed on.

I will treat (a) here; and (b) in (3) below, which will lead in to the next Section.

As to (a), here are three main ways in which we have seen variety, or context-dependence or subjectivity, about what counts as generic, and-or as not needing explanation. I present them in the order in which they came up in the previous two Sections.

First: In both Sections, we discussed how judgments about what is generic or not needing explanation are often moulded by the context of enquiry, and by pragmatic or even subjective factors. For example, recall Kepler's effort to explain why there were six (known) planets in terms of Platonic solids (cf. the start of Section 5 above); and the judgment that an initial state ('initial condition' in the jargon of physics) is generic at least in the sense that it provides an <u>explanans</u> for a later state (cf. (i) and (ii) at the start of Section 6 above).

Second: In what I called the topological and the effective field theory approaches to making 'generic' precise, there is again variety and context-dependence of judgments. This may seem surprising, since these approaches' definitions of notions like a topology or a renormalization group and its flow (on a space of physical systems, or problems, or theories) are, after all, mathematical.

But of course, being mathematical does not imply being unique. In general, a set can have many different topologies defined on it; and similarly, for defining renormalization flows on a set of Lagrangians (or Hamiltonians). Agreed: of the many mathematically consistent definitions, only some will be natural or significant, from the point of view of physics. But in general, the notion of 'physically natural or significant' is too vague and-or ambiguous to pick out a unique definition. So there will be a choice to be made, depending on context or aims.

Third: On the probabilistic approach to making 'generic' precise, there are similar difficulties. For a set can have many different probability distributions (in a more mathematical jargon: probability measures) defined on it. So the obvious question arises: what justifies the one being used? And even after accepting for whatever reason one distribution as correct, or at least as correct for one's purposes: there are the questions that I listed as (a) and (b) at the end of the previous Section (Section 6). Namely: how far under the tail of the distribution---how much of an outlier---must an observation, for example of the value of a variable, be in order to count as not generic, as atypical? (And so to count as disconfirming the hypothesis that the distribution is correct.) And even after accepting an answer to this, there is the more general question of what framework of statistical inference (Bayesian? Neyman-Pearson?) we should adopt. That is: how should we infer from observations to hypotheses about what is the correct probability distribution?

(3): So much for my group (a) of difficulties. I turn to my second group, (b): difficulties that arise from the fact that the system we are concerned with is the entire cosmos. So we return to focus on the main question we first formulated at the end of Section 4 above. Namely: Suppose we are given a cosmological multiverse of many bubble universes (domains), across which the value of a fundamental physical parameter (normally called a 'constant of nature'!), such as the cosmological constant or the electric charge on an electron, varies. Then the question is: How can we implement our explanatory strategy (Gen), so as to explain the value as "to be expected"?

As I see matters, there are two main points to make about this question.

The first is the obvious point that in this cosmological setting, the difficulties in group (a) are aggravated. For the cosmological multiverse, the theoretical context is so speculative that rigorous definitions are hardly to be had. So the difficulties are not just about having to appeal to context, pragmatic factors etc., so as to single out your preferred precise definition of being generic. Also, the daunting complexity of the relevant state-space (recall the $10^{500}$ vacua at the end of Section 4 above) makes it hard to give rigorous definitions of topologies, or renormalization flows, or probability measures.

We can express this point as a further comment on the problem which we first admitted back at the beginning of Chapter 4's discussion of the Everettian multiverse (Chapter 4.6). Namely: no Everettian knows how to write down the state of the cosmos---the symbol $\Psi$, with its honorific use of a capital letter, is a promissory note. We also admitted in this Chapter's Section 1 that eternal inflation aggravated this problem in several ways; although in order to keep the discussion as simple as possible, I proposed that we take the envisaged quantum state of the cosmological multiverse to be a sum or superposition, over the countless bubbles (domains), of each of their Everettian states $\Psi$ in the sense of Chapter 4 (especially Chapter 4.8's sketch definition of 'world'). That is: the multiverse's state is a sum of states $\Psi_i$, where the label 'i' on the summands labels the different universes. And recall from the end of this Chapter's Section 4, that there might be dauntingly many states in the sum, dauntingly many values of the label 'i': in the context of string theory, $10^{500}$!

So much by way of gathering previous discussions' threads about our not knowing how to write down the state of the cosmos. The present point is the further comment that not only are we unable to write down the state. Also, we cannot rigorously define such notions as the state being generic, or the appropriate probability distribution on states.

The second main point is about probability, and more specifically about <u>confirmation</u>: a topic which will be developed in the following Sections. Suppose that despite the difficulties above, we could define various probability distributions on the states of the multiverse, and make sense of the idea that one of them is correct. Still, we would face the question: 'How can we gather evidence about which one is correct?'

The problem is obvious. We presumably cannot get empirical data about bubble universes other than the one we are in: for the spatiotemporal connection between "our bubble universe" and any others is through the inflationary epoch, which for us is long gone. (And even apart from inaccessibility: its extreme conditions, for example of temperature, put it so far beyond established physics that we could hardly expect to get interpretable data from it.) But without such data, it is very unclear how we could gather evidence about which probability distribution is correct. After all: our understanding of the phrase 'correct probability' derives from cases where there is a set of actual systems or events (tosses of a coin or coins, rolls of a die or dice, etc.) that are, or are believed to be, suitably similar. We then estimate probabilities by counting the proportions, the relative frequencies, with which certain features (heads or tails, scores on a die) occur. Agreed, there is debate about how best to make these estimates from observed frequencies: a debate addressed by theories of statistical inference---recall questions (a) to (d), at the end of Section 6. But all parties agree that there are deep connections between probabilities and frequencies. So if we do not know any frequencies, how can we guess probabilities?

Chapter 5.8: Biased sampling: Eddington's net
In the last two Sections, I discussed the successes, and then the difficulties, of what (at the end of Section 5) I labelled 'strategy (Gen)' for explaining a fact: namely by deducing it, or showing it is generic, given some appropriate framework or explanans. In that discussion, the sort of fact to be explained has been, since the end of Section 4, facts about the value of a cosmological parameter, or a constant of nature---though we now envisage that such "constants of nature" may vary from one bubble universe to another.

So now I turn to what I labelled 'strategy (Obs)': explaining a fact, that one admits may well not be deducible or generic, by showing that it is likely (or at least has high enough probability) to be observed. In this Section, I discuss this strategy, and how it differs from strategy (Gen), in everyday terms, regardless of cosmology---to which the next Section will return.

The distinction between these two strategies lies in the fact that what is most probable to occur is not necessarily what is most probable to be observed. That is, we need to distinguish: (a) having high, or high enough, probability (or frequency) in a total population of cases; and (b) having high, or high enough, probability (or frequency) in the sub-population that we observe. The distinction between (a) and (b) arises from something familiar from very elementary applications of probability theory: biased sampling.

For example: when you take a sample from a set, say ten adults from a population of 10,000, in order to estimate the average height, your sample might be biased, in the sense that the frequency of the attributes of interest (here, being such-and-such metres tall) within the sample is different from its frequency in the total population of 10,000. Agreed: some difference in frequencies is to be expected. Almost always, the sample frequency does not exactly equal the population frequency (or population probability)---this is called 'stochastic variation'. And what counts as a large enough difference to earn the label 'bias' is a matter of how big a difference counts as significant for one's purposes---and so is partly a matter of judgment. For example, the sample might be biased, with large heights more frequent (as a proportion) than in the total population, simply because you chose the ten people from your local basketball club. And whether your consequent over-estimate of the population's average height is large enough to matter will depend on your purposes. For example, a five-centimetre over-estimate would matter if you planned to sell shirts to the population, but not if you planned to sell them lottery tickets.

So far, so obvious. But our interest lies in cases where the sample is biased, not as a matter of coincidence (as might well occur in the example of the basketball club), but as a systematic effect of the method of observation, or data-gathering. This is called a 'selection effect'; (or: 'effect of observational selection').

A famous example occurred in the 1936 US Presidential election. The incumbent Democratic President, Roosevelt, beat his Republican challenger, Landon, by a large margin. But one magazine had predicted that Landon would win, on the grounds that it posted questionnaires to ten million subscribers---of whom about two million responded, mostly favouring Landon. But this was a selection effect. The subscribers were disproportionately

Republican, compared with the nation at large; and the subscribers with the interest to send back a response were even more disproportionately Republican.

There is also a famous and vivid metaphor for selection effects, invented by the British physicist Arthur Eddington (1882-1944). In his book, *The Philosophy of Physical Science* (1938), he wrote:

'Let us suppose that an ichthyologist is exploring the life of the ocean. He casts a net into the water and brings up a fishy assortment. Surveying his catch, he proceeds in the usual manner of a scientist to systematise what it reveals. He arrives at two generalisations: (1) No sea creature is less than two inches long. (2) All sea creatures have gills. These are both true of his catch, and he assumes tentatively that they will remain true however often he repeats it.'

To sum up: Fishermen whose net has a mesh of, say, two inches, and who therefore observe that all the fish <u>in their catch</u> are longer than two inches, should not infer that all the fish <u>in the lake</u> are also longer than two inches. (Incidentally, Eddington intended his metaphor to teach a different and more contentious moral than just 'Beware selection effects'. Namely, a moral about the relation between physics and philosophy---which we will return to in the Notes at the end of this Chapter and in Chapter 6.)

So far, we have thought of selection effects as a bug: as a hindrance to making good estimates of the probability or frequency of an attribute in the total population from which we take our sample. That is right: they are a hindrance, especially if we do not know the details of how our sample is biased. If we do know those details, we can try to "build in" the details to the procedure by which we make an estimate, so as to compensate for the bias.

How to do this is a topic in the statistical theory of estimation. The rough idea is of course to conditionalize one's probabilities on a description of the sampling process. Similarly, if we know only some details, or have only probabilistic information about how the sampling is or might be biased: we "build in" what we know about the process. In short, this is familiar ground in the practice of statistical inference: specifically, in the theory of estimation. There may be practical difficulties about learning the details of the sampling process, and debate within statistical theory about how best to "build in" those details to the procedure for making an estimate. But there is no general or philosophical problem about the fact that we need to allow for these details.

But there is also another way to think of a selection effect. Namely, as <u>explaining</u> the frequency of an attribute that we observe (the height of a human, or their political views, or the length of a fish), despite that frequency being different from those for the total population. It is of course this perspective that is encapsulated by the strategy (Obs).

Again, I think there is no general or philosophical problem about this strategy, although there may well be difficulties about the details of the sampling process. As above, these could include, first, practical difficulties about learning the details. For example, does the fishermen's two-inch mesh really prevent any fish longer than two inches, if such there be, from getting caught? And there could be theoretical difficulties about how the calculation of an estimate should allow for such details. For example, how should my estimate of average height allow for my having sampled heights from a basketball club? Being told to conditionalize on the proposition 'All the people in my sample play basketball' gives very little guidance, if I do not know any details about how much playing basketball favours the tall.

I do not mean to downplay these difficulties, whether in physics or in other sciences. In all sciences, the observational process is indeed liable to be biased, i.e. the value of the variable we wish to observe may be correlated with the process of observation; and it can be a hard and complicated matter to recognize this, and to understand it in enough detail so as to compensate for the bias. Just think of the care that goes in to calibrating scientific instruments. But the point is that this is familiar ground in the practice of science and statistics: there is no a general or philosophical problem hereabouts.

Or rather: there is no such problem, outside cosmology. But when we consider cosmology, there may be such problems---as I discuss in the next Section.

Chapter 5.9: Selection effects in cosmology: the anthropic principle and the cosmological constant

So we return to the main question that we first formulated at the end of Section 4. Namely: Given a cosmological multiverse of many bubble universes (domains), across which the value of a fundamental physical parameter, such as the cosmological constant or the electric charge on an electron, varies: How can we explain the value that we measure? And relatedly, since one confirms a scientific theory by its predicting---and one hopes: explaining---results of measurements and observations: how, if at all, can we confirm (or disconfirm) a theory postulating such a multiverse? I begin in (1) with general remarks; and then in (2) turn to specifics.

(1): In general terms, our trouble about addressing these two questions is that it is not enough to say that a cosmological theory will assign differing probabilities to various values of such a parameter, of which each bubble universe exhibits one value; and that this enables us to assess the theory by ordinary statistical inference---along the lines that if the observed value is too much of an outlier, i.e. in the tail of the probability distribution, then we should conclude that the theory is disconfirmed.

Indeed, it is not enough for two reasons. We spelt out the first reason at the end of Section 7. Namely: we measure and observe only our own bubble universe, our own "cosmic parish". So for a parameter that describes an entire such universe, like a "constant of nature", we only get one number---we cannot count frequencies. That is a miserably meagre basis on which to judge which probability distribution is correct. Indeed, it is too meagre even for estimating the average value of the attribute in question: imagine trying to estimate the average height of a population of 10,000 by measuring the height of just one person. And for the cosmological multiverse, we expect the number of bubble universes to be vastly larger than 10,000.

The second reason is, of course, selection effects: which prompt our explanatory strategy (Obs), viz. that we explain a value of a parameter by its being probable (or at least has probable enough) to be observed. For in the context of measuring fundamental parameters in a multiverse, biased sampling threatens to be a significant problem. The problem is not just that, as in the first reason, each bubble universe exhibiting only one value means the sample size is so small as to be useless, thanks to stochastic variation. Also, established theories in both physics and chemistry show that many of the parameters at issue, such as the cosmological constant and the charge on the electron, are indeed correlated with what the last Section called the 'process of observation'. That is, they are correlated with facts that underpin humans' being able to measure the parameter: such as the fact that life on earth depends on a suitable abundance of carbon and oxygen, or that stars exist, and have planets orbiting them, for times long enough for the complex carbon chemistry of life to evolve on a planet.

With this mention of how the process of observation involves humans (or their complex carbon chemistry), we thus arrive at last at the well-known phrase: 'the anthropic principle'. For some fifty years, it has been the topic of heated debate in both cosmology and philosophy. (The phrase was suggested in 1973 by Brandon Carter (1942 - ) a theoretical astrophysicist. People also talk of 'anthropic reasoning'.)

(2): The general idea of these correlations is that modern astronomy and cosmology (as reported in Section 2) has shown our universe to be very unified: not just in what I called the 'geographical' sense that a parameter, such as the charge on the electron, takes the same value throughout our universe (Section 4); but also in the sense that what happens to its smaller parts, such as a star or galaxy, depends on truly global, i.e. universe-wide, features.

      A good example is the parameter I mentioned first in this Chapter: the density parameter $\Omega$, which is the ratio of the universe's density to the critical density that would make the final rate of expansion zero. (Cf. Sections 2 and 3. But here we are concerned, not with the speculative inflationary phase perhaps explaining $\Omega$'s fine-tuning, but with $\Omega$'s value much later on, and so within established cosmology: say after the recombination time 380,000 years after the Big Bang.) If the early universe were very dense, i.e. $\Omega$ was much greater than 1, the universe would have re-collapsed in far less than thirteen billion years, so that life would not have had time to evolve on planets; while if $\Omega$ was much less than 1, no stars, and therefore no planets, would have formed.

      Another example, which I will return to later in this Section, is the <u>cosmological constant</u> $\Lambda$ (introduced at the start of Section 7). Since it represents a universal expansion, it having a much larger value than it actually does is like $\Omega$ being much less than 1. Namely: a much larger value would have made the universe's expansion too fast for stars to form.

      A third example is the other parameter I mentioned, the charge on the electron: more precisely, the ratio of the strengths of electromagnetic to gravitational forces (which is greater, the greater the charge on the electron). If this had been much smaller than it is, gravitation would have been comparatively stronger, and the universe would have re-collapsed in far less than thirteen billion years, so that life would not have had time to evolve on planets.

      These correlations are often "tight", in the sense that they are not probabilistic. They are not a matter of the probability of one proposition, the parameter's value, being altered by conditionalizing on another proposition, such as the proposition that there is abundant carbon and oxygen. They are a matter of one value, or range of values, being mathematically dependent on, i.e. a function of, another value or range of values. So these mathematical dependences can be, and often are, summed up in what Chapter 3.7 called a 'counterfactual conditional', along the lines: '<u>If the parameter had taken a different (or different enough) value than its actual one, then there would be no observations---at least, no observations by humans (understood as having such-and-such carbon chemistry)</u>.' Witness the examples above.

      Besides, the correlations are in several cases 'tight' in a distinct numerical sense. Namely, only a very narrow range of values of the parameter, such as a few percentage points around its actual value, is compatible with a fact like there being abundant carbon and oxygen. Hence this is also called '<u>fine-tuning</u>'.

      Agreed, this is not stupendous fine-tuning to many decimal places, such as we saw in the flatness and horizon problems (Section 3) and in the problem of the Higgs mass (Section 7). (The fine-tuning of $\Omega$ in the flatness problem was $10^{-16}$, i.e. a hundred-million-millionth of one per cent.) But each of these stupendous fine-tunings occurred within a single theoretical framework; while here, many such frameworks are in play. For the physical and chemical facts and processes that link global features like cosmic expansion, or the comparative strengths of fundamental forces, to local features like the existence of heavier elements such as carbon, or the existence of life on rocky planets orbiting stars, are very diverse. They range from nucleosynthesis in the early universe, and in stars, through planet-formation and the chemistry of water, to the evolution of life. So it is indeed very striking that our established theories of these diverse facts and processes, when conjoined together, provide a "patchwork description" of the facts and processes that---despite its diverse ingredients---implies these numerical quantitative links, constrained to within a few percentage points.

But this is not to suggest that it is straightforward to spell out these implications: that it is straightforward to quantify the correlation. As I said at the end of the last Section, in connection with calibrating scientific instruments: even within a single scientific theory, compensating for the fact that the observational process is biased can be a hard and complicated matter. All the more so, when there are several theories or frameworks in play, and when the parameter in question is correlated, via various different mechanisms, with various different aspects of our making observations.

For example, there are many different necessary conditions, each scientifically describable, of our observing the charge on the electron. The observer is alive; and life requires---one may well argue---complex carbon chemistry. Carbon requires stellar nucleosynthesis. And the complex chemistry of life requires---one may argue---that a planet orbit a star at a suitable distance (neither too hot nor too cold, like Goldilocks' porridge), and for a long enough time, so that life can evolve. All these correlations, and the mechanisms underpinning them, and these mechanisms' mutual relations, are very hard to disentangle. And this is so, even if we somehow settle on some exact definition of 'observation' or 'life'. Besides, this is so even for a single parameter such the charge on the electron, let alone all the physical parameters of interest.

But despite the complexities just mentioned, some examples <u>are</u> comparatively straightforward to calculate. So I end this Section with a bit more detail about one such example. Namely, Weinberg's renowned explanation of the value of the <u>cosmological constant</u> as an observation selection effect. Recall from the start of Section 7 that the cosmological constant $\Lambda$ represents a repulsive force between two masses. So in a cosmological context, $\Lambda$ amounts to a tendency for the universe to expand; so it opposes the gravitational force that tends to make matter clump together.

Weinberg recognized that the requirement that life evolves in an expanding universe of the type considered in the standard model of cosmology is correlated with the value of the cosmological constant, through a single, and comparatively simple, mechanism. Thus he wrote:

> … in a continually expanding universe, the cosmological constant (unlike charges, masses etc.) can affect the evolution of life in only one way. Without undue anthropocentrism, it seems safe to assume that in order for any sort of life to arise in an initially homogeneous and isotropic universe, it is necessary for sufficiently large gravitationally bound systems to form first . . . However, once a sufficiently large gravitationally bound system has formed, a cosmological constant would have no further effect on its dynamics, or on the eventual evolution of life.

So the idea is that the evolution of life constrains the cosmological constant in a simple way, because we can think of (a positive value of) the constant as a long-range repulsive ('anti-gravity') force. Thus one assumes that (i) life can only exist on planets, and (ii) life takes a long time, say billions of years, to evolve. Since (i) requires that matter has the chance to clump together under gravity so as to form planets, the initial expansion cannot be too powerful. That is: (i) implies that there is an upper bound on the cosmological constant (i.e. a number that it must be less than). On the other hand, (ii) means that the universe must last long enough for life to evolve. So gravity cannot be so powerful (the initial expansion cannot be so weak, $\Lambda$ cannot be so small or negative) that gravity overcomes the initial expansion in a Big Crunch that happens so early that life does not have time enough to evolve. Thus (ii) implies that there is also a lower bound on the cosmological constant.

Indeed, the calculation along these lines in 1997, by Weinberg and co-authors, amounted to showing that the observed value of $\Lambda$ (tentative in 1997, but confirmed a year later) fell "safely" between this lower and upper bound. So it is natural to see this calculation as explaining the observed value of $\Lambda$ as an observation selection effect.

Chapter 5.10: Confirming a theory of the multiverse

The last Section's report of the fine-tuned correlations, between cosmological parameters and facts about observers, focussed on the one universe we are in. But now let us consider these correlations in the context of the cosmological multiverse.

I shall first state---without repeating all the difficulties presented in Sections 7 to 9---the basic predicament that besets observers in a single bubble universe, trying to confirm a cosmological theory. Then I shall sketch a scheme for overcoming this predicament: a scheme which I find clarifying. In (1), I report the scheme's general ideas; in (2), I report how it incorporates ideas from both the strategies (Gen) and (Obs). But I admit of course that the scheme is fiendishly difficult to apply, i.e. calculate with, except in simple toy-models of such theories.

The basic predicament is that when we observe our universe, we are like Eddington's fishermen. Our observations of a physical parameter (e.g. the cosmological constant) are like measurements of the length of fish in the catch. And so we should not infer that in unobserved bubble universes---in domains other than ours---the parameter takes, or is likely to take, a value close to what we observe. For our established theories describe how the parameter's value is correlated with whether the domain has observers in it. So if in some other bubble, those theories are true, or approximately true, and there is no observer there, the value would be different. And if in this other bubble, those theories are badly wrong, i.e. not even approximately true, then anyway---all bets are off about the parameter's value.

(1): The scheme I favour was first proposed by Srednicki and Hartle (about fifteen years ago). It has of course been developed since then: with proposals by such authors as Aguirre, Azhar, Hertog and Tegmark, and by Hartle and Srednicki themselves. But for simplicity and brevity, I will sketch a very simplified version: (for references, see the Notes at the end of this Chapter).

The scheme aims to incorporate appropriately the ideas from both strategies, (Gen) and (Obs): the idea of being generic, and the idea of being probable (or probable enough) to be observed. And it does this in a Bayesian way.

In a bit more detail, this means: the scheme prescribes probabilities for data D (say, the value of a cosmological parameter such as $\Omega$ or $\Lambda$) to be observed, conditional on the cosmological theory T, and other propositions, for example propositions encoding selection effects. (Probabilities like these, i.e. probabilities of data or evidence conditional on a theory or hypothesis, are often called 'likelihoods'.) One then uses Bayes' theorem to calculate the probability of the conjunction of T with the other propositions, conditional on the data D.

Here, we need not go into any detail about Bayes' theorem. For us it is sufficient that the theorem provides a way to calculate from the conditional probability $P(A/B)$ of a proposition A conditional on a probability of B, the "opposite" conditional probability, $P(B/A)$. (The calculation uses the values of other probabilities, additional to $P(A/B)$.) The basic idea for how to use the theorem so as to make inferences (called 'Bayesianism' or 'Bayesian statistical inference') is then: we take B to be some hypothesis or theory we wish to assess (confirm or disconfirm), while A is some proposition reporting evidence such as an experimental result.

Thus we imagine that the hypothesis B prescribes a value of $P(A/B)$: the probability, assuming B is true, of the evidence A. $P(A/B)$ might be high; or it might be low, with A "lying under the tails" of the probability distribution $P( /B)$ prescribed by B. Then the "magic" of Bayes' theorem is that we can then calculate $P(B/A)$. And then the Bayesian statistician tells us that if we in fact learn the evidence or result A---and so set our credence (subjective probability) in A equal to 1---we should adjust our credence in the hypothesis B, from our initial $P(B)$, to be equal to $P(B/A)$. (So $P(B/A)$ is often called the 'posterior probability' of B.)

Also, the Bayesian says that A confirms B provided that this prescription makes our credence go up, i.e. provided that P(B/A) is larger than P(B). This Bayesian account of how evidence confirms or disconfirms hypotheses has many merits. In many cases, it gives the intuitively right verdicts about confirmation. In particular, observing some evidence A that "lies under the tails" of the probability distribution P( /B) prescribed by B will, generally speaking, disconfirm B.

Returning to cosmology: the idea of Srednicki and Hartle's scheme will thus be that after receiving the data D (about say, the value of $\Omega$ or $\Lambda$) we should apply Bayes' theorem to the probability of D conditional on the conjunction of the theory T and propositions about selection effects etc, so as to set our credence in the conjunction equal to the posterior probability.

(2): To convey how ideas from both my strategies, (Gen) and (Obs), get incorporated in the scheme, I will begin by summarizing the problems one faces in extracting predictions from cosmological theories of the kinds currently envisaged (including inflation). This will lead in to details about Srednicki and Hartle's proposal, which they call a 'framework'.

I summarize the problems under three headings. These headings will echo the difficulties I presented in Sections 7 to 9, about such matters as: (i) the definition of a probability function on a very large space of possibilities, (ii) how to specify the "fact about life or observation" on which we need to conditionalize so as to accommodate selection effects, etc. (My three headings are also a simplification, or amalgamation, of an analysis by Aguirre. He lists seven problems, or headings, rather than my three; again, the Notes for this Chapter give references.)

I call the headings 'Measure', 'Conditionalization', and 'Typicality'. They are as follows.

(1): Measure. As discussed in Sections 6 and 7: What are the elements of the set (called in probability theory: 'the sample space') on which the probability distribution (called: 'measure') is to be defined? Should they be domains, i.e. bubble universes, even though these vary greatly in volume? Or should the elements be spacetime regions of equal volume? Or some other option? And once a sample space is defined: which measure on it should we adopt?

(2) Conditionalization. As discussed in Sections 8 and 9: we need to allow for selection effects. But how exactly should we characterize our observational situation? How detailed should the proposition describing it, on which we will conditionalize, be?

(3): Typicality. As discussed in Sections 6 and 7, there are various problems about how to make precise the idea of a fact (in particular, our explanandum) being generic, or typical. In particular: How much "under the tails" of a probability measure can our observation turn out to be, without our then inferring that the theory is disconfirmed?

This trio of headings leads directly to Srednicki and Hartle's proposed Bayesian scheme for discussing the confirmation of cosmological theories. Srednicki and Hartle define a 'framework' as a conjunction of:--- i): A cosmological theory T (though often cosmologists will say 'model'). This is taken as solving the Measure problem (1) above. So we write a probability P( /T); where the argument-place i.e. the gap will be filled by a proposition about the value of a physical parameter, e.g. the cosmological constant;

ii): A 'selection proposition' that describes our observational situation: which is called a 'conditionalization scheme', and labelled as C. So we conditionalize on C as well as on T, and we consider: P( /T, C): where we expect the argument-place to be filled by a proposition D about "our seeing" some specific data;

iii): A probability distribution, denoted by the Greek letter $\xi$ (pronounced 'xi'), and called the 'xerographic distribution' by Srednicki and Hartle. This is defined on those domains, i.e. bubble universes, that have a non-zero measure according to P( /T, C). $\xi$ encodes typicality assumptions: as Srednicki and Hartle discuss, it need not be a flat distribution .

So the idea is that i) to iii) jointly implement our envisaged solutions to the problems listed under (1) to (3) above. The upshot is that we are to consider: P(  /T, C, ξ). Srednicki and Hartle call P(D / T, C, ξ ) the '<u>first-person</u>' likelihood of seeing data D. It is a probability "to be observed", rather than a probability "to be", thanks to its conditioning on C and on ξ: i.e. its encoding our observational situation and also the typicality assumption we are making.

Srednicki and Hartle then propose a Bayesian framework to compute degrees of confirmation of the framework, i.e. the conjunction of T, C, and ξ. That is, they use Bayes' theorem to calculate: P(T, C, ξ/ D). They (and other authors, such as Azhar) go on to give examples of the framework in action. They show, for various 'toy' cosmological models/theories T (e.g. with finitely many bubble universes, so as to assume the Measure problem has been solved), how various conditionalization schemes C, and typicality assumptions ξ, fare in the light of various data D.

To sum up: I suggest that this Srednicki-Hartle scheme of frameworks is a clear and convincing scheme for handling both selection effects and assumptions of about being generic (typicality). It gives both of them an appropriate role in the endeavour of confirming a theory postulating a cosmological multiverse.

Chapter 5: Notes and Further Reading
As for Chapters 2, 3 and 4, there is a dauntingly large literature. And as in those Chapters, I recommend:
(i) The internet encyclopedias and archives, and the accessible books, listed in items (1) to (2) of the Notes for Chapter 1; all of which cover the cosmological multiverse.
(ii) The seminal works, some of which are indeed very readable.

I will first, in (1), give more details about some items under these headings, (i) and (ii). Within (1), I will begin with more accessible, even popular, items, and then turn to academic (though still readable) books. Then in (2), I will list some research articles, in an order corresponding to the Chapter's sequence of Sections.

(1): First, as regards internet encyclopedias: I recommend three entries in the <u>Stanford Encyclopedia of Philosophy</u>. They are about: fine-tuning (by S. Friederich), the philosophy of cosmology (by C. Smeenk and G. Ellis), and cosmology and theology (by H. Halvorson and H. Kragh). They are at:
https://plato.stanford.edu/entries/fine-tuning/
https://plato.stanford.edu/entries/cosmology/#Mult
<u>https://plato.stanford.edu/entries/cosmology-theology/#5.2</u>

As regards seminal works, I firstly recommend three books. They are all very accessible; the first two are by physicists, the third is by a philosopher.

First: J. Barrow and F. Tipler, <u>The Anthropic Cosmological Principle</u> (Oxford University Press, 1986). This is undeniably the <u>locus classicus</u> for the topics of this Chapter. Though it came out very soon after inflation was proposed, it does cover inflation. I especially recommend Chapters 4 to 8, of which Chapters 6 and 7 are focussed on inflation and quantum theoretic aspects.

Second: B. Carr (ed.), Universe or Multiverse? (Cambridge University Press 2007). This book collects articles arising from four conferences (from 2001 to 2005), all focussed on the topics of this Chapter. Almost all the articles do not require knowledge of advanced physics (not even within cosmology). Most of the authors are prominent researchers in cosmology. Some of them advocate the cosmological multiverse, and endorse anthropic explanations of parameters such as the cosmological constant (e.g. Linde, Rees, Susskind, Tegmark and Weinberg); while some criticize both these tenets (e.g. Ellis and Smolin). So this collection is an invaluable resource; and in my list of research articles, (2) below, I will cite (for more specific reasons) the articles by Aguirre, Vilenkin and Weinberg.

Third: S. Friederich, Multiverse Theories: a philosophical perspective (Cambridge University Press, 2021) is an excellent recent philosophical assessment of the cosmological multiverse. The material mostly closely related to this Chapter is about how we could confirm a multiverse theory (Friederich's Chapters 7 to 9). But his Chapter 10 also briefly (though sceptically) discusses our other two multiverse proposals, i.e. the logical and quantum multiverses of our Chapters 3 and 4. The book is available at Cambridge University Press Core, at: https://www.cambridge.org/core/books/multiverse-theories/68CE18BE78DE31550C67855107A57942

There are also many excellent popular books about all the topics, or some of the topics, of this Chapter. Here are four popular books about all the topics, ordered by how sharply they focus on them, i.e. this Chapter's concerns, to the exclusion of other themes.

First, there is Max Tegmark's The Mathematical Universe (2014). It gives a detailed, though popular, advocacy of the cosmological multiverse. As mentioned in (2) of the Notes for Chapter 1, I stand by my criticisms of this book: both in the review which I cited there, and in the long item (1) in the Notes for Chapter 3, which criticized Tegmark's Pythagoreanism. But no matter: these criticisms do not undermine his advocacy of the cosmological multiverse.

Second, there are two books by M. Rees: Before the Beginning: our universe and others (Simon and Schuster, 1997); and Just Six Numbers: the Deep Forces that Shape the Universe, (Weidenfeld and Nicholson 1999). Both books also give a lot of detail about astrophysics and cosmology, apart from the multiverse: for example, about the early universe, the cosmic background radiation, stars, galaxies and black holes---thus filling out my Section 2's review of the current golden age cosmology.

Finally, there is J. Barrow, The Book of Universes (Bodley Head, 2011). As the title hints, this gives a lot of detail about the "other" universes (spacetimes) that are admitted as possible solutions by general relativity (i.e. by Einstein's theory of gravity) but are not part of the standard Big-Bang cosmological model. But its Chapters 9 and 10 discuss inflation and the multiverse.

Here are four other popular science books, each giving a detailed account of one topic of this Chapter. I list them in the chronological order in which their topic took centre-stage in cosmology; which mostly corresponds to the topics' order in this Chapter's Sections.

First, there is S. Singh, The Big Bang: the origin of the universe (Fourth Estate, 2004). This focusses on how the standard Big-Bang cosmological model was confirmed by about 1980, largely as a result of the discovery of the CMB; (cf. Section 2).

Second: A. Guth, The Inflationary Universe: the quest for a new theory of cosmic origins, (Penguin, 1998). This masterpiece of popular science is by one of the inventors of inflationary cosmology (and so also an advocate of the multiverse); (cf. Section 3).

Third: Brian Greene, The Elegant Universe: superstrings, hidden dimensions and the quest for the ultimate theory (Random House 1999). This is an excellent exposition of string theory, though without emphasis on the "landscape" of many vacuum states; (cf. Section 4).

Fourth: T. Hertog, <u>On the Origin of Time</u> (Penguin 2023), focusses on quantum cosmology, including times even earlier than the putative inflationary epoch; and therefore on the no boundary proposal for the initial state of the cosmos, mentioned in Section 1.

This last topic, the initial state of the cosmos, prompts me to issue a warning.

Recall (from this Chapter's preamble, and Section 1) that in physics, 'vacuum' does not mean 'nothing', or 'no physical system'. It means 'state of lowest energy' (hence the synonym 'ground state'). There is a widespread tendency in popular physics books, even by physicists, to forget this; and thereby to suggest that by postulating a vacuum state as the initial state of the cosmos, we can explain the creation of the cosmos---out of nothing! In this widespread but pernicious mistake, the phrase 'vacuum fluctuation' gets abused similarly to 'vacuum' state.

So far as I know, the most egregious example of this mistake is the book by L. Krauss, <u>The Universe from Nothing</u> (Free Press, 2011)---which I therefore do <u>not</u> recommend.

But I do recommend two antidotes to this sort of error. First: the devastating review of Krauss by D. Albert, in the New York Times (25 March 2012), and available at: https://www.nytimes.com/2012/03/25/books/review/a-universe-from-nothing-by-lawrence-m-krauss.html
Second: J. Weatherall, <u>Void: the strange physics of nothing</u> (Yale University Press, 2016), is a magisterial, but very readable, account of physics' changing conception of vacuum (void, empty space) from the time of Descartes and Newton, through the rise of classical field theories such as electromagnetism and of quantum field theories, till today---including the landscape of countless string vacua, and thus the cosmological multiverse.

I turn to recommending a few academic books. They are: (1.A) cosmology textbooks; (1.B) philosophy of cosmology books; (1.C) histories of twentieth-century cosmology.

(1.A): Here are three cosmology textbooks, in roughly ascending order of difficulty.

M. Longair, <u>Our Evolving Universe</u> (Cambridge University Press, 1996). Its final Chapter, 'The Origin of the Universe', is a very readable "immediate successor" to this Chapter.

M. Rowan-Robinson, Cosmology (Oxford University Press, 2004: fourth edition). Its Epilogue, 'Twenty controversies in cosmology', covers inflation. It is available at: https://academic.oup.com/book/52969?login=false

A. Liddle, <u>An Introduction to Modern Cosmology</u> (Wiley, 2015; third edition). This is slightly more theoretical than Rowan-Robinson's book: and it also covers inflation.

An expository article for the celebratory 'Einstein's legacy' issue of <u>Science</u> magazine in 2005, by A. Guth (an inventor of inflationary cosmology) and D. Kaiser, is: Inflationary cosmology: exploring the uv from the smallest to the largest scales; which is available at: https://arxiv.org/abs/astro-ph/0502328; and at
https://www.science.org/doi/10.1126/science.1107483

(1.B): Here are four books on the philosophy of cosmology. The first is a collection of invited articles. The second is a magisterial monograph covering many topics in the philosophy, or foundations, of general relativity as well as inflationary cosmology (its Chapter 5). The third is Friederich's recent monograph devoted to the multiverse; which I already recommended as 'seminal' above, especially for its discussion how we could confirm a multiverse theory (i.e. the topic of my Section 5 onwards, especially Section 10). The fourth is a collection of invited articles discussing in detail fine-tuning of the conditions for complex chemistry and life, as well as for planet formation etc. (i.e. the topic of my Section 9).

K. Chamcham, J. Silk, J. Barrow and S. Saunders (eds.), The Philosophy of Cosmology Cambridge University Press, 2017). This has many good articles. I will cite those by Smeenk, Hartle and Hertog in my list of research articles, below. The book is available at: https://www.cambridge.org/core/books/philosophy-of-cosmology/2E9F97DDF98A672256D35B46C3F574B4

J. Earman, Bangs, Crunches, Whimpers and Shrieks: singularities and acausalities in relativistic spacetimes, (Oxford University Press, 1995). The book is at Oxford Scholarship Online at: https://academic.oup.com/book/49463?login=false. It is also available, with almost all Earman's other work in philosophy of physics, at: https://sites.pitt.edu/~jearman/

S. Friederich, Multiverse Theories: a philosophical perspective (Cambridge University Press, 2021). The book is available at Cambridge University Press Core, at: https://www.cambridge.org/core/books/multiverse-theories/68CE18BE78DE31550C67855107A57942

D. Sloan, R. Batista, M. Hicks, and R. Davies (eds), Fine-Tuning in the Physical Universe, (Cambridge University Press, 2020). Available at: https://www.cambridge.org/core/books/finetuning-in-the-physical-universe/DAAE3182CBC72F012EFF589E67178F1C

(1.C): Here are three histories of twentieth-century cosmology. The first book narrates the establishment of the standard Big-Bang model of cosmology. (The stress is on its defeat of main rival, the steady-state theory---which my Section 2 did not mention.) The second book is more technical: it also covers the entire twentieth century and covers astrophysics as much as cosmology. Its Chapter 16 focusses on this Chapter's topics. The third book is a recent collection of articles.

H. Kragh, Cosmology and Controversy: the historical development of two theories of the universe (Princeton University Press, 1996).

M. Longair, The Cosmic Century: a history of astrophysics and cosmology (Cambridge University Press, 2006).

H. Kragh and M. Longair (eds.), The Oxford Handbook of the History of Modern Cosmology (Oxford University Press 2019), available at: https://academic.oup.com/edited-volume/34295.

(2): I now list some research articles, grouped by letters A, B, C etc. Their order mostly corresponds to this Chapter's sequence of Sections. The first three groups, (2.A) to (2.C), relate to the Preamble and Section 1. Groups (2.D) and (2.E) are about the difficulties of confirmation---in philosophical jargon: the under-determination of theory by data---for cosmology in general, and for inflationary cosmology in particular. So these correspond to Sections 2, 3 and 5 to 7. Groups (2.F) and (2.G) are about observation selection effects, in general and in cosmology; and correspond to Sections 7 to 9. Finally, (2.H) gives more details about the Srednicki-Hartle proposal of frameworks for confirming multiverse theories (Section 10).

(2.A): The relation of the Everettian and cosmological multiverses was a topic in this Chapter's Preamble and Section 1. Again, there is a large literature.

As a place to begin, I recommend the articles by Carter (the inventor of the phrase 'anthropic principle') and Mukhanov in the collection edited by Carr, Universe or Multiverse?, listed at the start of (1) above. (Both authors advocate an Everettian approach.)

An example of a research article about this relation---in fact, advocating that the two multiverses are the same---is: R. Bousso and L. Susskind, Multiverse interpretation of quantum mechanics, <u>Physical Review D 85</u> (2012); available at: arxiv: 1105.3796

(2.B): In Section 1 (and more briefly in the Preamble and Section 3), I mentioned how cosmology confronts the quantum measurement problem on a cosmic scale. In particular, there was the idea that a peak in (the amplitude of) the wrinkles in the CMB is a seed for later gravitational clumping, and thereby for the later existence of galaxies and stars. I warned that 'seed' is a metaphor, since the transition is from a quantum amplitude to a classical event of aggregation---a 'collapse of the wave-function'.

Recall also my warning, just before (1.A) above, of a widespread mistaken tendency to slur over this transition, especially with the buzz-words of 'vacuum state', and 'vacuum fluctuation'.

In the light of this, I recommend some research tying the dynamical reduction programme---i.e. the effort to suitably modify the Schroedinger equation: cf. the citation of Pearle in (iii) of the Notes to Chapter 4---to cosmology. In this endeavour, work by Sudarsky and co-authors has been prominent. So here are three such articles.

A. Perez, H. Sahlmann and D. Sudarsky, On the quantum origin of cosmic structure, <u>Classical and Quantum Gravity</u> (2006), and arxiv: general relativity-qc/0508100.

J. Berjon, E. Okon and D. Sudarsky, Critical review of prevailing explanations for the emergence of classicality in cosmology, <u>Physical Review D</u>, (2021); arxiv: 2009.09999

R. Lechuga and D. Sudarsky, Eternal inflation and collapse theories, <u>Journal of Cosmology and Astroparticle Physics</u> (2024); arxiv: 2308.01383.

(2.C) My Section 1 also mentioned quantum cosmology's efforts to formulate an initial quantum state of the cosmos, including Hartle and Hawking's no boundary proposal of 1983. This is a central topic of the popular book <u>On the Origin of Time</u> by T. Hertog, cited just before (1:A) above. In a large literature, a very fine recent research article is:

J. Halliwell, J Hartle and T. Hertog, What is the no-boundary wave function of the universe?, <u>Physical Review D</u> (2019); arxiv: 1812.01760.

(2.D) The stupendous achievements of modern cosmology, reviewed in Section 2, naturally prompt the question: 'Can all these details about the physics in places and times so very distance from us-now really be established?' The worry, in philosophical jargon, is the under-determination of theory by data; and <u>prima facie</u>, it seems this must be a big problem for cosmology.

About this, I will here cite three survey articles (two by me). (2.E) cites more specialized articles, especially about the problems of confirming inflation.

J. Butterfield, On Under-determination in Cosmology, <u>Studies in the History and Philosophy of Modern Physics</u>, <u>46</u> (2014), pp 57-69; At: <u>arxiv.org/abs/1406.4747</u> <u>https://philsci-archive.pitt.edu/9866/</u>; doi:<u>10.1016/j.shpsb.2013.06.003.</u>

F. Azhar and J. Butterfield, Scientific Realism and Primordial Cosmology'; available at <u>arxiv.org/abs/1606.04071</u>; <u>https://philsci-archive.pitt.edu/12192/</u>

C. Smeenk, Philosophical aspects of cosmology, in Kragh and Longair (eds.), <u>The Oxford Handbook of the History of Modern Cosmology</u> (Oxford University Press 2019), cited in (1.C) above. The article is available at:
https://doi.org/10.1093/oxfordhb/9780198817666.013.13

(2.E): Some articles about the specific difficulties of confirming inflation (and doubts about its providing explanations), in chronological order:

J. Earman, <u>Bangs, Crunches, Whimpers and Shrieks: singularities and acausalities in relativistic spacetimes</u>, (Oxford University Press, 1995); Chapter 5. Cited in (1.B) above.

C. Smeenk, Predictability crisis in early universe cosmology, <u>Studies in History and Philosophy of Modern Physics</u> 46 (2014), pp. 122-133.

C. McCoy, Does Inflation Solve the Hot Big Bang Model's Fine-Tuning Problems?, <u>Studies in History and Philosophy of Modern Physics</u> 51 (2015), pp. 23–36. doi: 10.1016/j.shpsb.2015.06.002.

C. Smeenk, Testing inflation, in K. Chamcham et al (eds.), cited in (1.B) above. The article is available at https://www.cambridge.org/core/books/philosophy-of-cosmology/testing-inflation/4095B5F8D7991E344D203CCBE0B369C8

A. Koberinski and C. Smeenk, Establishing a theory of inflationary cosmology, <u>British Journal for the Philosophy of Science</u> 2024; https://doi.org/10.1086/733886

F. Azhar and N. Linnemann, Rethinking the anthropic principle, in <u>Philosophy of Science</u> 2025; doi:10.1017/psa.2024.41 ;and available at: https://philsci-archive.pitt.edu/23934/

Finally, here are two articles which also cover the problems of confirming theories of dark matter and of dark energy (mentioned in Section 2).

M. Longair and C. Smeenk, Inflation, dark matter and dark energy, in H. Kragh and M. Longair (eds.), cited in (1.B) above. The article is at https://doi.org/10.1093/oxfordhb/9780198817666.013.11

P. Ferreira, W. Wolf and J. Read, The Spectre of Underdetermination in Modern Cosmology; available at: https://arxiv.org/abs/2501.06095; and at: https://philsci-archive.pitt.edu/24537/

(2.F): About observation selection effects in general, here are two items. The first is about supersymmetry: whose confirmational difficulties were a topic in Section 7. It is a Bayesian analysis; and so far as I know, the most recent one.

R. Dawid and J. Wells, A Bayesian Model of Credence in Low Energy Supersymmetry (2024), available at: https://philsci-archive.pitt.edu/24172/

The second item is a fun historical point. It returns us to Eddington's famous metaphor of the net, in his 1938 book, <u>The Philosophy of Physical Science</u>. As I quoted it in Section 8:

'Let us suppose that an ichthyologist is exploring the life of the ocean. He casts a net into the water and brings up a fishy assortment. Surveying his catch, he proceeds in the usual manner of a scientist to systematise what it reveals. He arrives at two generalisations: (1) No sea creature is less than two inches long. (2) All sea creatures have gills. These are both true of his catch, and he assumes tentatively that they will remain true however often he repeats it.'

Although that passage is famous, philosophers should also take notice of---and take encouragement from!---what Eddington goes on to say just afterwards---which is almost never quoted. For Eddington takes the net to stand, not just for our means of observation in the specific science "ichthyology" (so that naively, we might infer that all fishes are longer than two inches), but also for our scientific method as a whole. Thus Eddington's moral is not just the obvious one I stressed in Section 8, viz. 'conditionalize your credence on your means of observation'; but also that we should allow for types of knowledge inaccessible to the scientific method. This open-mindedness is bound to be welcome to a philosopher . . .  Thus Eddington writes:

'In applying this analogy, the catch stands for the body of knowledge which constitutes physical science, and the net for the sensory and intellectual equipment which we use in obtaining it. The casting of the net corresponds to observation; for knowledge which has not been or could not be obtained by observation is not admitted into physical science. An onlooker may object that the first generalisation is wrong. 'There are plenty of sea-creatures under two inches long, only your net is not adapted to catch them.' The icthyologist dismisses this objection contemptuously. 'Anything uncatchable by my net is ipso facto outside the scope of icthyological knowledge. In short, 'what my net can't catch isn't fish.' Or — to translate the analogy — 'If you are not simply guessing, you are claiming a knowledge of the physical universe discovered in some other way than by the methods of physical science, and admittedly unverifiable by such methods. You are a metaphysician. Bah!''

(2.G): About observation selection effects in cosmology (Section 9), I begin with two survey articles. The first is philosophical and sceptical about inferring from fine-tuning to a multiverse; the second is scientific (focussed on fine-tuning of conditions for stars, planets and life) and more optimistic about such inferences.

N. Landsman, The fine-tuning argument, in N. Landsman and E. van Wolde (eds.), <u>The Challenge of Chance</u>, Springer 2016; available at: https://library.oapen.org/bitstream/handle/20.500.12657/27974/1/1002025.pdf#page=115 and downloadable from the publications page of Landsman's site, https://www.math.ru.nl/~landsman/eprints.html

M. Livio and M. Rees, Fine-tuning, complexity, and life in the multiverse; in D. Sloan et al (eds.) Fine-Tuning in the Physical Universe, cited in (1.B) above. This article is also available at arxiv: 1801.06944.

I turn to the cosmological constant $\Lambda$, and anthropic explanations of it (as in the second half of Section 9). A superb historico-philosophical overview of the cosmological constant is:

J. Earman, Lambda: the constant that refuses to die, <u>Archive for History of the Exact Sciences</u> 2001; it is available, with almost all Earman's other outstanding work in philosophy of physics, at: https://sites.pitt.edu/~jearman/

Weinberg's (and his co-authors') anthropic explanation of the value of $\Lambda$ is discussed in two articles in the collection, B. Carr (ed.), <u>Universe or Multiverse?</u> (Cambridge University Press 2007), which I recommended at the start of (1) above.

The first article is by S. Weinberg himself: 'Living in the multiverse', which is also available at <u>https://arxiv.org/abs/hep-th/0511037</u>. (My quotation from Weinberg in Section 9 is from his 1987 paper in <u>Physical Review Letters.)</u>

The second article (more detailed than Weinberg's) is: A. Vilenkin, Anthropic predictions: the case of the cosmological constant. It is also available at https://arxiv.org/abs/astro-ph/0407586

(2.H): Here are some details about Srednicki and Hartle's proposals for how to confirm multiverse theories. I outlined this in Section 10. My three headings, 'Measure', 'Conditionalization' and 'Typicality', for the problems that such confirmation faces were an amalgamation of the seven problems listed by A. Aguirre in his excellent analysis: 'Making predictions in a multiverse: conundrums, dangers, coincidences', in the collection, B. Carr (ed.), <u>Universe or Multiverse?</u> which I recommended at the start of (1) above.

Srednicki and Hartle's first two papers are:

J. Hartle and M. Srednicki, Are we typical? Physical Review D 75: 123523 (2007). arXiv:0704.2630 [hep-th].

# Chapter 6: Multiverses compared---and combined?

In the last three Chapters, we have surveyed three proposed multiverses: from logic and philosophy, from quantum physics, and from cosmology. Each dizzies the mind. Each has powerful arguments in its favour: in the latter two cases, arguments that are in part empirical. But each is very controversial, since there are rival accounts of the phenomena it treats. (For the first multiverse, these are logico-linguistic phenomena like our commitment to modal language. For the second and third multiverse, they are physical phenomena.) And each multiverse proposal, for all its allure, throws up major problems as well as solving some: conceptual, indeed philosophical, problems. I have emphasized several such problems, and we have seen how these problems connect the different conceptions of a multiverse. I especially emphasized three problems, prompted by the three multiverses in succession: What is a possibility? What is objective probability? And what is an explanation? These are difficult open problems in philosophy.

So I will <u>not</u> conclude the book by firmly endorsing, or firmly rejecting, even one of these multiverse proposals. This demurral is unsurprising given my previous admonitions that each of us must decide for ourselves what we can, or cannot, honestly believe, and how ambitious or modest is our intellectual temperament (cf. Chapters 1.3 and 1.4). But there are three loose ends that I should tie up.

First, I briefly announced in Chapter 1.2 what I myself believe about the three proposals. Now that we have seen the detail of Chapters 3 to 5, I should say a bit more about my position (Section 1).

Second, I said in Chapter 1.6 that all three proposals can make a good case that the different universes are isolated, i.e. unable to communicate with one another. So again: now that we have seen the detail of Chapters 3 to 5, I should say a bit more about this; and about the more general topic of getting empirical evidence for a multiverse. I do this in Section 2.

Thirdly, I briefly announced that the Everettian interpretation of quantum theory prompts a proposal for what a possible world actually is. Namely: it is a branch of the Everettian multiverse, and so represented by a summand in the Everettian quantum state of the cosmos. (I mentioned this both in Chapter 3.10 (about Lewis' modal realism), and in Chapter 4 at the end of Section 8 (A sketch definition of 'world') and in Note (4.A).) So in Section 3, I will discuss this proposal in more detail.

Finally, I will end with two salutary quotations (Section 4).

Chapter 6.1: What I believe
I briefly announced in Chapter 1.2 what I myself believe about the three multiverse proposals. I said that I believe in the philosophical multiverse, but not the Everettian one, and I am undecided about the cosmological multiverse. As I put it: my verdicts are: 'Yes, No and Maybe'. Now I give some more details: again, with the qualification 'for what it is worth', since each of us must decide for ourselves what we can, or cannot, honestly believe (cf. Chapters 1.3 and 1.4). I will again proceed in order, treating the philosophical multiverse in (1), and then the physical multiverses in (2) to (4).

(1): I believe in the philosophical multiverse. For recall from Chapter 3.3 that both in everyday thought and talk, and in technical science, we are up to our necks in modality. This commitment was further illustrated by my discussions of state spaces in physics, both classical (Chapter 3.3 and 3.8) and quantum (Chapter 4.2 and 4.3). And recall the benefits of explicitly accepting a realm of possibilities (of which the maximally specific possibilities are the possible worlds). Thus Chapter 3.4 to 3.8 paraded these benefits, for understanding not only everyday thought and talk, but also technical science. (Besides, 'understanding' might here be construed ambitiously, as providing a conceptual analysis in the sort of sense discussed in Chapter 3.1 and 3.2.)

But the benefits of this philosopher's paradise (Lewis' phrase, echoing Hilbert's homage to Cantor, cf. Chapter 3.4) do not require Lewis' own conception of it, i.e. his modal realist version of the philosophical multiverse (Chapter 3, Preamble and Section 3.10). Nor did Lewis believe the paradise required his modal realism. Rather, he advocated it as being, on balance, the best conception. Recall the problems and disadvantages of other conceptions that I reviewed in Chapter 3's long Section 9: in particular, Section 9.C, and its objections (i) to (iii) against the widespread view that the actual world is "concrete", while all the others are "abstract".
On the other hand, I admitted that despite the strength of Lewis' arguments, I simply do not---cannot---believe his modal realist account. So as in Chapter 3 and the Preamble above: I leave the nature of possibilities, in particular possible worlds (and as I said: similar notions like proposition) as an open, and very difficult, problem in philosophy.

(2): I turn to the two multiverse proposals from physics. The first, and obvious, point to make is that they have in common a contrast with the philosophical multiverse: a contrast that will be a recurrent theme in all the Sections of this Chapter. For since these proposals are prompted by physical theories that have been formulated to describe, explain and predict empirical phenomena, they have, in contrast to the philosophical multiverse, both an apparent advantage and an apparent disadvantage---as follows.

First: they apparently could be supported by the specific empirical data that confirms the underlying physical theory; (more especially, data that confirm those of its claims which prompt, or are conceptually closest to, the multiverse proposal). Agreed: the philosophical multiverse is supported (on some or other account of it) by the great raft of all our commitments, everyday and technical, to possibilities, so as to understand such topics as semantics, counterfactuals and determinism (and the others listed in Chapter 3.5 to 3.8). But it is not supported by any data (even about what we say with modal language) with the kind of specificity enjoyed by a physical theory's evidence.

Second: on the other hand, any physical theory is a human construction, moulded by the scientific community's conceptual framework and the available evidence at a certain stage in enquiry. Therefore any physical theory is fallible, and all too likely to be superseded later on. The evidence that favoured the theory may later on be better accounted for---described and explained---by a successor theory. So here lies, apparently, an advantage of the philosophical multiverse. For whatever the vicissitudes of physical theorizing, or more generally empirical enquiry, turn out to be: we can be sure that they will not overturn our great raft of commitments to possibilities (on some or other account).

(Of course: the vicissitudes of empirical enquiry might prompt us to give up specific modally involved claims, such as determinism, or notions like that of laws of nature. But that does nothing to dispose of the realm of possibilities that Chapter 3 argued to be indispensable for formulating those claims and notions. Besides, we saw that this realm is indispensable for formulating claims and notions that are more everyday, and less technical (such as reference and counterfactuals, Chapters 3.5 to 3.7): and so surely less prone to be overturned by empirical enquiry.)

Of course, we have seen this contrast before. We first saw it in general terms, in Chapter 1's Preamble and Section 1. Besides, Chapter 5 considered the idea of empirically confirming, not just the inflationary epoch, but also the ensuing multiverse proposal (Chapter 5.10). And on the other side, my emphasizing that the inflationary epoch is still a speculation signalled that after all, the theory of inflation might in the future be superseded---and its multiverse thereby fall by the wayside. Nevertheless, it is worth stating this contrast explicitly here. For not only will it be a theme throughout this Chapter. Also, although it was <u>not</u> developed in Chapter 4, it will play a specific role in Section 3's discussion of Everett.

So much by way of stating this contrast between the philosophical multiverse and the two multiverse proposals from physics. Let me now spell out a little how it applies to these two multiverse proposals. The results will be ambivalent and tentative; and they will lead to my stating that I cannot believe the Everettian one, but might be persuaded about the cosmological multiverse. Again, I will treat first the Everettian, then the cosmological, multiverse.

(3): Chapter 4 emphasized that Everettians have yet to establish their interpretation of quantum theory. But it did not discuss how they could try to find evidence in its favour. I think it is clear that the broad strategy must have two components.

First, they need to show that even macroscopic systems, if strictly isolated, obey the Schroedinger equation, not some cousin equation as advocated by the dynamical reduction programme (Chapter 4.5). This is in essence a challenge of experiment, not of theory. And it is a very daunting challenge since even a few atoms in a vacuum chamber around the system, or even the photons i.e. quantum particles of light in the CMB (cosmic background radiation: cf. Chapter 5.2), are enough of an environment to decohere a macroscopic system. And this will mean that the system's state (an improper mixture) prescribes probabilities for all possible measurements that match, or are very hard to discern from, the state predicted by a cousin equation advocated by the dynamical reduction programme (Chapter 4.).

Second, Everettians need to refute the other no-collapse interpretations of quantum theory, which are also "one-world" theories, i.e. without a multiverse (as is the dynamical reduction programme). Chapter 4.5 described the best known example, the pilot-wave theory; and item (2) of the Notes to Chapter 4 described others, such as the modal interpretation. This is likely to be a challenge of theory, at least as much as experiment: namely to show defects, perhaps with an experimental signature, of those other interpretations.

So much by way of the first half of the above contrast: a physical multiverse's (here, the Everettian's) apparent advantage in being able to garner empirical support. I turn to the second half of the contrast: the apparent disadvantage that the physical theory in question (here, quantum theory) might be superseded, and its multiverse thereby fall by the wayside.

For quantum theory, we are in this regard in a very ambivalent position. The theory is stupendously well confirmed: it describes countless phenomena with stunning accuracy (i.e. to many decimal places), many of them phenomena that we can prove to be inexplicable by classical physics. But of course, this does not mean that it will never be superseded. And there (at least) two broad grounds for believing that it will be.

The first is very familiar from Chapter 4. Quantum theory's glorious successes are in the last analysis a matter of its predicted probabilities matching statistics gathered in experiments---which says nothing about the definiteness of outcomes in a single run of the experiment. In other words: the theory faces the grave embarrassment of the measurement problem. Though opinions about how to best solve it vary, many expect that only some very basic changes in the theory can do so---and there is no reason to expect such changes to suggest some analogue of the Everettian multiverse.

The second ground for expecting quantum theory to be superseded has <u>not</u> (hitherto) been a topic of this book; though it is a staple of the physics, and popular physics, literature. I mean the deep tensions between the concepts and detailed formalism of quantum theory and the concepts and formalism of general relativity (Einstein's theory of gravitation). For decades, these tensions have prompted efforts to formulate a new theory reconciling quantum theory and general relativity, dubbed 'quantum gravity'. These efforts have led to many deep insights, and all sorts of proposed frameworks, and even theories; but none has won acceptance. And with one exception, these frameworks and theories do <u>not</u> suggest some analogue of the Everettian multiverse. (The exception is of course string theory: which I discussed in Chapter 5, and will again here treat as an aspect of the cosmological multiverse.)

For me, the upshot of this discussion is that, as it happens, I cannot believe in the Everettian multiverse. For firstly: assuming that quantum theory (i.e. the Schroedinger equation, with no collapse of the wave function) is exactly true, I am nevertheless sufficiently sympathetic to solutions to the measurement problem other than Everett's; (cf. items (1) and (2) of the Notes to Chapter 4). And secondly: assuming that quantum theory will be superseded one day, e.g. so as to reconcile it with gravitation, the successor theory may well not suggest some analogue of the Everettian multiverse . . . But again: I say 'as it happens' so as to signal that this is my fallible verdict. Each of us must make our own judgment about the evidence, both conceptual and empirical.

(4): I turn to the cosmological multiverse. Again, I will lead up to my overall position by considering both the apparent advantage of a physical multiverse (here, the cosmological multiverse) that it can garner empirical support; and its apparent disadvantage that the cosmological theory (here, inflation and string theory) might be superseded, and its multiverse thereby fall by the wayside.

Indeed, Chapter 5 discussed how cosmologists could try to find evidence for their theory, while recognizing that it is fiendishly difficult (Chapter 5.10). This represented a contrast with Chapter 4. But it is an unsurprising contrast, since the Everettian view is usually called---and I called it—an <u>interpretation</u> of quantum theory. After all, it has no postulates or mechanisms additional to those of orthodox quantum theory; (a parsimony that Everettians often advocate as an advantage of their view). On the other hand, inflationary cosmology is called a <u>theory</u>, albeit a speculative one, rather than an interpretation, precisely because it has postulates and mechanisms additional to those in the Big-Bang theory of the 1970s. (It is called a theory: (a) despite our having no agreed choice for what is the inflaton field, or for its properties, e.g. the potential function; and (b) independently of whether one invokes string theory to give a mechanism of the inflationary expansion.)

Broadly speaking, this contrast makes me more persuaded on the cosmological multiverse than the Everettian one. Hence my joke in Chapter 1.2 that for the cosmological multiverse, I say---with the film producer Sam Goldwyn---'a definite Maybe'.

Agreed, there are good reasons not to be so sanguine: some of which are reasons not to discriminate in one's credence between the Everettian and cosmological multiverses. Here are four. (1): Although postulates and mechanisms may earn inflationary theory the honorific label of being 'scientific', rather than 'interpretative' or 'philosophical', that by no means implies that there is any evidence in its favour. (2): Indeed, there is so far no evidence in favour of a specific model of inflation as against others. (3): Besides, inflationary theory might be superseded by a successor theory that "does away with" its multiverse; similarly to how, as I said above, quantum theory might be superseded, doing away with the Everettian multiverse. (Indeed, as one would guess: respectable rivals to inflationary cosmology without any multiverse have already been developed.) (4): We saw in Chapter 5.1 and 5.4 (and items (2.A) to (2.C) of Chapter 5's Notes) the subtlety of the relations between the Everettian and cosmological multiverses, thanks not least to string theory's central role in the latter. Such subtleties undermine this Section's contrast with the Everettian multiverse, i.e. my comparatively negative verdict about it, above.

So you of course ask me: why do I maintain this contrast? Here I must confess. I think the reason lies, not in some objective superiority of the quantity or quality of the arguments and evidence, conceptual or empirical, for the cosmological multiverse compared with the Everettian one, but in two more instinctive points. The first is about intellectual history, the second about causal connections. So the first is less technical, and less internal to the discussions in this book. Indeed the second will return us to the subtleties of the relations between the Everettian and cosmological multiverses

First: over the more than five hundred years since Copernicus, we have learned various lessons that humanity is not central, or more generally special, in the universe. First, the Earth was displaced from the centre. Then the solar system was discovered to lie in a "suburb" of the Milky Way: which turns out to be a disc, one hundred thousand light-years across, of about one hundred billion stars. Then the nebulae (so-called since their images in telescopes were cloudy) were discovered to be yet other galaxies (with on average a hundred million stars). And there are so many: nowadays, the number of galaxies in the observable universe is estimated to be between two hundred billion and two trillion. In the light of these successive belittlements, it can seem a small step to accept that beyond the observable universe, there are other similar "vast expanses".

Second: there is a contrast between the Everettian multiverse, as described in Chapter 4, and the cosmological multiverse---about causal connections. The former involves countless realities each of which is vast and intricate, yet inhabiting the very same spacetime as the apparent reality we know, being so to speak overlaid on it and not interacting with it. That seems very hard to believe; or even to get one's mind around as a proposition, with a view to assessing its truth. On the other hand, the cosmological multiverse involves countless realities, each now vast and intricate, that <u>do</u> have causal relations to us, albeit very distant and convoluted relations. For they and the observable universe which we now see were "spawned together" in an unimaginably small and hot regime. This means that causal relations can in principle be traced by, so to speak, following a cosmic itinerary: starting from "our end", one first traces back to the inflationary epoch, and then goes forwards into (one or another) bubble universe. That branching web of causally connected regions seems easier to believe in than the Everettians' myriad co-present realities.

But I admit: these are both only instinctive points. Besides, the second as I stated it uses Chapter 4's account of the Everettian multiverse, in particular its sketch definition of 'world' (Chapter 4.8): which assumed implicitly a common background spacetime "shared" by the different worlds or branches. But we saw in Chapter 5 that there are ways to combine the Everettian and cosmological multiverse, especially by invoking the landscape of string theory (Chapter 5.1, 5.4 and items (2.A) to (2.C) of Chapter 5's Notes). Such combinations may well undermine my second instinctive point above, about a contrast of causal connections.

So much by way of stating---for what it is worth---my own verdicts on the three multiverse proposals. But this last topic, causal connections, leads us back to my comment in Chapter 1.6, that all three proposals can make a good case that their different universes are isolated, i.e. causally disconnected. With the detail of Chapters 3 to 5 in hand, we can say a bit more about this. I now do this in Section 2.

Chapter 6.2: Why don't we see the other universes?
The verb 'see' is of course metaphorical. Our topic is not just vision, but other observable traces (effects or signs) of the other universes; which is of course part of the more general topic of getting empirical evidence for their existence, i.e. for a multiverse.

In this Section, I will emphasize the Everettian and philosophical multiverses. I discuss them respectively, in (1) and (2) below. My reason for this emphasis is that for the cosmological proposal, the topic of causal connections leads to details of advanced physics that are beyond this book's scope; while on the other hand, for the Everettian and philosophical multiverses, a non-technical discussion is possible.

So first, I will deal very briefly with the cosmological multiverse. For it, the causal connections between different universes were mentioned at the end of the last Section. Namely: one can in principle---very much in principle!---trace back in time within one universe to the inflationary epoch, and then forwards in time to another. Such a causal link is extremely tenuous. But what matters about such links is of course, not the second half of my description, the "forwards in time to another universe", but the idea that physical events within the different universes' common past, i.e. within the inflationary epoch, might leave some kind of observable trace in some (our?) universe that is a sign of the existence of the others. Indeed, there are proposals for how this could be. But they lie outside our scope; (they involve such notions as primordial gravitational waves, and subtle imprints in the structure of the cosmic background radiation (CMB)). I only note that such traces would of course greatly aid the effort to confirm specific multiverse theories: an effort which we discussed in Chapter 5.10 in terms of statistical inferences from measurements of cosmological parameters.

(1): For the Everettian multiverse, the question 'Why don't we see the other universes?' can be answered with some details at the expository level we adopted in Chapter 4. The relevant ideas are in Chapter 4.7, about decoherence, and in Chapter 4.8, about using decoherence to define the Everettian's universes (worlds, branches).

The first point we need from Chapter 4.7 is that the difference between a superposition (superposed quantum state) and a mixture (mixed state) is encoded in numerical differences between the probability distributions that the two states prescribe. These differences are called '<u>interference terms</u>'. Here, the word 'interference' doesn't connote disturbance, but comes from the physics of waves. For when two peaks of two waves e.g. water waves meet to form a yet higher peak, we say the waves interfere constructively; similarly, when two troughs meet to make a yet deeper trough; and when a peak meets a trough so that they form a level surface, we say the waves interfere destructively.

What Chapter 4.7 did not say---it was in a hurry to expound decoherence---is that there is a paradigm experimental set-up that exhibits these ideas; and this set-up helps answer the question about not seeing other universes.

Indeed, there is both a classical and a quantum set-up, both called '<u>the double-slit experiment</u>'. The classical version may well be familiar from school physics. The quantum version is famous: it began as a thought-experiment, or teaching device, to illustrate that quantum wave-function indeed behaves like a wave, but it has also been realized in the laboratory.

The classical version is a shallow pool of water, across which lies a barrier with two small slits. On one side of the barrier a wave-machine generates a "train" of waves moving towards the barrier, each wave parallel to the barrier. As a result, on the other "downstream" side of the barrier, from each slit there flows a train of circular waves. These two trains spread out, meeting each other---and interfering. If we put a screen parallel to the barrier on this downstream side, at a suitable distance, we can see the interference pattern, of doubly high peaks and doubly low troughs.

The quantum version again has a barrier with two slits, on one side of which an "electron gun" fires a train of electrons, each with the same definite (or nearly definite) momentum, straight ahead towards the barrier. (Although the apparatus of course exists in three-dimensional space, we can arrange to make a better analogy with the two-dimensional classical pool of water, with the slits being indeed not holes, but narrow slits extending in the third dimension.)

The definite momentum means that the wave function of each electron is like a train of waves all parallel to the barrier. As a result, on the downstream side of the barrier, there emerges from each slit a circularly symmetric wave-function, i.e. a train of semi-circular waves. The two trains interfere, and we can see the interference pattern on a suitably placed screen. Indeed, even if we arrange for the electron gun to emit electrons intermittently so that at any one time there is at most one electron in the set-up, there is an interference pattern.

(Each individual electron makes a dot, a scintillation, on the screen, which is like a TV screen; so the pattern builds up gradually, showing bands (lying along the third dimension) that each consist of many closely-spaced dots, alternating with bands without any dots. Of course, the transition from the spatially extended wave-function to the localized dot is the notorious collapse of the wave-function, which is at the heart of Chapter 4's measurement problem.)

This interference pattern indicates that the state of each electron as it passes through the barrier, is a superposition---namely, of passing through each slit---so that the wave-function shortly afterwards is the sum of the two circularly symmetric wave-functions each centred at a slit.

For if each electron definitely passed through just one slit, one would expect the screen to show, not an interference pattern with its many alternating bands, but just two "humps" (clusters of dots): one with its peak i.e. most-closely spaced dots directly behind one of the slits, and the other with its peak directly behind the other slit. Such a two-hump pattern corresponds, not to a superposition, but to a mixture of going through one slit and going through the other. (And if it is a 50-50 i.e. equi-weighted mixture, we expect an experiment with many runs i.e. the gun firing many electrons one after another, to produce two humps with approximately equal numbers of dots.)

So much by way of expounding the double-slit experiment. I can now say how it helps answer the question about not seeing other universes, i.e. our not detecting other Everettian worlds or branches. We only need to recall Chapter 4.7's main theme, viz. decoherence. Recall that as a result of the rapid and ubiquitous process of decoherence, the states of macroscopic objects are (very close to) mixtures of states that are definite for quantities like position that we intuitively want to be definite so as to solve the measurement problem. Applying this to the double-slit experiment: decoherence effects (for example, collisions with air molecules downstream of the barrier, which would amount to a probe system monitoring through which slit the electron passes) would produce a mixture, i.e. the two hump pattern. Putting it the other way around: the double-slit's interference pattern is realized in the laboratory by making sure, by clever engineering, that decoherence does not smear the pattern into two humps. For example, the chamber needs to be a nearly perfect vacuum, with almost no air molecules.

But we also saw that Everettians define their universes (worlds, branches) in terms of the definite-valued states that are the components of the mixture obtained from decoherence (cf. Chapter 4.8). The upshot is that for all the countless macroscopic objects and set-ups that are <u>not</u> cleverly engineered within a quantum physics laboratory to be shielded from decoherence, the interference terms characteristic of a quantum superposition are strongly suppressed, so that within a single universe (world, branch) there is no observable evidence of other universes. Putting it, again, the other way around: one can think of the double-slit's interference pattern as revealing, to the perspective of the "mini-universe" that is defined by being definitely at one slit, the existence the other "mini-universe" defined by being definitely at the other slit.

(2): I turn to the philosophical multiverse. For this proposal, this Section's question 'Why don't we see the other universes?' seems at first confused. For one's first thought is: whatever we eventually conclude about the exact nature of possible worlds in Chapter 3's sense---about which Chapter 3 ended in anxious agnosticism---they will surely <u>not</u> be the sort of entity that can be seen i.e. observed, one from another. More generally, they will surely not be the sort of entity that has causal relations from one to another; or from one part (event, state of affairs) within one world to another part of another world.

Broadly speaking, this misgiving---one might even say: accusation---is surely right. But it is nevertheless worth pressing the question. There are two aspects to this. First, the question prompts one to ask what is causation and how does it relate to possibility. Second, for Chapter 3's great advocate of possible worlds, David Lewis, the question <u>is</u> germane; and he had a full and interesting reply to it. I will address these two aspects in turn.

I will treat the first aspect more briefly. For to say more would need a thorough discussion of what is causation. But we can surely agree that causation is a relation between localized matters of fact. Here, my phrase 'localized matter of fact' is intended to be neutral between various more specific conceptions that philosophers have advocated as being the relata of causation. (These conceptions are often given everyday words, like 'event' or 'state of affairs', as labels; but used thus, the everyday word becomes a technical term of art.) And once we accept that the relata of causation, causes and effects, are localized matters of fact, then two reasons to <u>deny</u> that there can be causation from one possible world to another present themselves as plausible. I will prefer the second reason.

First, one might invoke the distinction between "concrete" and "abstract", and then maintain that: (i) actual matters of fact (events, states of affairs) are concrete, while non-actual ones are abstract (in line with Chapter 3.9:C's suggestion that a possible world is a sentence, or a set of sentences, or something similar); and (ii) that causation requires its relata to be concrete. From these tenets it obviously follows that there is causation only within the actual world. But I think this line of thought stumbles. For first: if there is indeed no causation at non-actual worlds, how are we to understand our many (surely true) statements about possible causes and effects, such as 'this short-circuit could have caused a fire'? This line of thought apparently vetoes understanding such statements using possible worlds. Second, and more fundamentally: the distinction between concrete and abstract is not in good order---as I urged, following Lewis, in Chapter 3.9:C (first objection) and Chapter 3.10 (a).

Second: one might say that causation requires its relata to be spatiotemporally related to one another, i.e. to be at some spatial distance and temporal interval (both possibly zero) from one another. Notice that this requirement is vastly weaker than requiring these relata, a cause and an effect, to be contiguous (adjacent) in space and time, as demanded by the principle of contact action (cf. Chapter 2.2). In particular, this requirement makes no prohibition against action-at-a-distance as in Newton's theory of gravity (Chapter 2.3). It also admits, as one surely should, causation in non-actual worlds (more precisely: in those worlds that have a space and a time): unlike the last paragraph's line of thought. Besides, it secures the desired answer to our question, i.e. the verdict that there is no causation between possible worlds, provided no two parts of two different worlds are spatiotemporally related to each other … which for the philosophical multiverse, though not of course for the Everettian or cosmological one, seems plausible.

So much by way of general discussion of causation and how it relates to possibility. I turn to what I called the second aspect: how Chapter 3's great advocate of possible worlds, David Lewis, answered this Section's question, i.e. argued that there is no causation between possible worlds. His answer is interesting, for three reasons: of which, the first two relate to the last two paragraphs.

First: after he meticulously formulates several disambiguations of the concrete/abstract distinction (a distinction which, as we have discussed, he diagnoses as multiply ambiguous), he admits that on most disambiguations, worlds are indeed, according to him, concrete. So (following the first line of thought above) this conclusion would seem to exacerbate the threat of causation between worlds.

Second: Lewis argues, quite independently of the topic of causation, for the proviso at the end of the general discussion above: that no two parts of two different worlds are spatiotemporally related to each other. (His argument, in short, is that overall the best definition or conceptual analysis of what it is for two objects to be in the same possible world is precisely that they are spatiotemporally related.) So that meshes well with my sympathy to the second line of thought above, that causation requires its relata to be spatiotemporally related to one another.

Third: Lewis has, independently of all his views about possible worlds, a theory of causation. And this provides fuel for a proof that according to this theory, there is no causation between worlds. I will not go into details about this proof. For us, it suffices to say that his theory analyses causation in terms of counterfactual conditionals about the cause and effect, along the lines of 'If the cause had not occurred, the effect would not have occurred'. Here, he understands the counterfactual conditional in terms of similarity between worlds, i.e. along the lines of the logical theories invented by Stalnaker and Lewis himself (and expounded in Chapter 3.7). But in any case, his proof that there cannot be causation between possible worlds would go through, i.e. remain valid, if one instead adopted various other semantics for counterfactual conditionals.

Furthermore, it is interesting for this Chapter's purposes, viz. comparing the different multiverse proposals, that the above three reasons come together---as Lewis realizes very well. For he ends his discussion of my second and third reason with a passage in which he explicitly says that---while he has proven that there is no causation (nor any spatiotemporal relations) between objects (or localized matters of facts: events, states of affairs) in different possible worlds in his logical, i.e. Chapter 3's sense---he is very open to causation between what an Everettian or cosmologist might call two different worlds (in our jargon: different universes), each a part of what for Lewis is a single possible world. Agreed, he does not cite Everett or the ideas of inflation; rather he mentions science fiction. But the intent is clear.

Besides, he says all this vividly, indeed wittily, in his main book about possible worlds. So let me end by quoting the passage . After this Section's back-and-forth of reasons for and against various claims, a tiring ping pong of dialectic, it is a relief to read:

But if you would like to see a world where Napoleon conquered all, don't give up hope. Maybe ours is one of those big worlds with many world-like parts, spatiotemporally related in some peculiar way. Then you might get your wish, near enough, by means of a special telescope or a special spaceship that operates entirely within our single world. You won't see the world-like part where Napoleon himself is, of course; you're there already, and he didn't conquer all. But I presume you'd be content with a world-like part where the conqueror was an excellent counterpart of Napoleon. I would be the last to denounce decent science fiction as philosophically unsound. No; tales of viewing or visiting 'other worlds' are perfectly consistent. They come true at countless possible worlds. It's just that the 'other worlds' that are viewed or visited never can be what I call 'other worlds'. (On the Plurality of Worlds, end of Section 1.6).

Chapter 6.3: One reality to rule them all?
Recall that Chapter 3 left unresolved the hard question what exactly a possible world is. But as I mentioned there (Chapter 3.10) and reported in Chapter 4.8 (and in Chapter 4's Note (4.A)): the Everettian interpretation prompts a proposed answer. Namely: a possible world is a branch of the Everettian multiverse---and so is represented by a summand in the Everettian quantum state of the cosmos. In this Section, I briefly discuss this proposal. (I will confine myself to Chapter 4's account the Everettian multiverse, setting aside the subtleties arising from cosmology which we saw in Chapter 5.1 and 5.4.)

My discussion is brief, not least because a recent book by A. Wilson develops and defends the proposal: (details in this Chapter's Notes). First, in (1) I will state the proposal. Then in (2) I will discuss how it overturns the view, usual among philosophers, of the relations between logic and physics; and make a specific objection.

(1): The proposal begins by adopting Chapter 4's Everettian multiverse. This involves not just the "one-liner" idea (Chapter 4.6) that to each possible outcome of a quantum measurement, there corresponds a branch (or world or universe---and maybe many such); but also the appeal to decoherence to more precisely define the branches (Chapter 4.7 to 4.9); and also the appeal to indexicality and decision theory (Chapter 4.11 and 4.12) to make sense of probability, and in particular to justify the Born-rule probability assignment. (In this Section, it will be clearer to talk of Everettian branches, rather than worlds (or universes): for then I can reserve 'world' for use in the phrase 'possible world', and thus signal the connotations of Chapter 3's concern with modality.)

So there is what Chapter 4 called 'the quantum state of the cosmos' (usually called in the literature 'the universal quantum state'). It has its countless decoherent branches. Countless of those <u>are</u> (not merely correspond to---for the Everettian) macroscopic realms of the sort we imagine or mention in our modal thought and language. Suppose for example that we make a counterfactual supposition, either in everyday thought and talk or in technical science. To take Chapter 3.9's example: we suppose that Butterfield is in Rome in August 2024, by saying 'If Butterfield were in Rome in August 2024, then . . .'. Or we suppose that kangaroos have no tails, or that planet Earth does not exist.

Now the proposal is plain as day. Namely: these branches or realms <u>are</u> the possible worlds that give, along the lines of Chapter 3, the semantics, the truth-conditions, of our modal thought and talk. For example, let us consider counterfactual conditionals, and adopt the Lewis-Stalnaker semantics for them (Chapter 3.7). And let us consider 'If Butterfield were in Rome in August 2024, then he would be on holiday', as said by me or you (in the actual branch containing us both). The proposal is that this is true (at the actual branch) if the branches that are most similar to the actual one while making true that Butterfield is in Rome in August 2024, also make true that he is on holiday there and then.

(Of course, the Everettian branches of the sort we imagine or mention in our modal thought and language are just a subset of all the branches; probably a small subset in some precise sense of 'small'. That is: there are countless other decoherent branches that correspond, <u>not</u> to anything we imagine, but to much weirder possibilities for which we have no appropriate words and concepts.)

And similarly for other aspects or topics in our modal thought and talk. Thus the proposal is that: (i) we should transcribe logical and semantic accounts of various phenomena, conceptual and linguistic (including in technical science), from the sort of framework expounded in Chapter 3 to that of Chapter 4, replacing 'possible world' by 'Everettian branch'; and (ii) by doing this, we get an account of the phenomena that is, not just tenable, but superior to others--- that is, superior provided we accept the truth of the Everettian interpretation of quantum theory.

Evidently, this proposal is bold. It brings in to contact, and comparison, with each other two detailed frameworks (and their literatures) which were developed to answer very different questions from each other. Besides, it claims that this comparison succeeds, in the sense that the two sides mesh. More precisely: there are on the logical side (cf. Chapter 3) accounts that, once transcribed, fit the Everettian interpretation of quantum theory well.

I will not go into great detail. But for this book's topics, I should report that as it turns out, several of these accounts which, once transcribed, fit Everett well are Lewisian accounts. So—such is Lewis' influence---they are familiar to modal logicians and philosophers of modality. Here are two examples.

(1): Lewis' account of 'actual' as indexical (in the sense of Chapter 4.11, used to explain Everettian probability) fits well, once transcribed. Thus Lewis says that 'actual' is indexical, referring to the world of the speaker/thinker: in just the same way that 'now' refers to the time of the speaker's/thinker's words or thought, and 'here' refers to the place of the speaker's/thinker's words or thought. Once transcribed, this becomes: 'actual' refers to the branch of the speaker/thinker---which fits well the Everettian's treatment (Chapter 4.11) of an agent's uncertainty about the outcome of an imminent quantum measurement.

(2): Lewis' account of determinism and indeterminism fits well, once transcribed. Recall from the last part of Chapter 3.8 that the broad idea of determinism is determination (i.e. supervenience) of the sequence of the system's future states, by the system's present state taken together with the sequence of all its past states. (A stronger formulation says: determination by the present state alone. But here, little will turn on this variation of the broad idea.) Recall also that Lewis accepts the idea of a law of nature (Chapter 3.6), and therefore the idea of the conjunction of the laws of nature at a given possible world, say <u>w</u>: what we might call 'the theory of <u>w</u>'. (Here we recall Chapter 1.4's spectrum of confidence through to caution about contentious concepts; so Lewis is confident.) With these ideas in hand, one then has two possible formulations of the claim that the theory of a given possible world w is a deterministic theory; (and correlatively, that it is an indeterministic theory). Lewis argues in favour of the first.

First, one can say: the theory of <u>w</u> is deterministic if and only if among all the possible worlds that share with <u>w</u> their laws of nature (i.e. their 'theory of the world'): if two such worlds utterly match each other at each time up to a given time, then they also match at each and every later time. Note that this formulation allows for many such matching pairs of worlds. For it requires only that for worlds with the same theory of the world as <u>w</u>, their utterly matching on an initial segment of history up to a given time implies their also utterly matching at all later times. Correlatively: indeterminism is a matter of there being at least one pair of worlds that utterly match up to a time, but do not match at some later time (maybe many later times). Lewis proposes a jargon for indeterminism in this sense. He calls it '<u>divergence</u> of worlds'.

On the other hand, there is another possible formulation, which is clearest to state for the idea of indeterminism. Namely, that a <u>single</u> world (among those with the same theory of the world as w) at some time <u>splits</u>, in the sense of itself having two or more sequences of later states. So in the simple case of a single splitting in two, one can picture such a world, with time going up the page, as a 'Y'. There is a single "thread" until a time, and thereafter two threads. Thus Lewis' proposed jargon for this meaning of indeterminism is '<u>splitting</u> of worlds'.

In Chapter 4's discussion of Everettian branching, I did not distinguish these two meanings: there was plenty else to discuss . . . But it is clear that however exactly the Everettian defines branches (called 'worlds' in Chapter 4.8), the distinction of meanings carries over: which prompts the question whether an Everettian should be a "diverger" or a "splitter". I will not go into this. Suffice it to say that there are good reasons to be a "diverger". And furthermore, some of those reasons are what I have called the transcriptions of Lewis' reasons to be a diverger within his possible worlds framework.

So to summarize the proposal: it compares two multiverses that are apparently very different, having been developed to answer very different questions. And it makes a detailed case that the bold identification of possible worlds with Everettian branches is correct. The evidence for this is the fact that several correspondences of ideas that are implied or suggested by this identification, e.g. correspondences about indexicality and about indeterminism, work out very well.

(2): So much by way of summary: I turn to assessing the proposal. I will entirely set aside misgivings about the Everettian interpretation of quantum theory. These were discussed in Chapter 4, and do not need to be repeated. And I will focus on the proposal's main idea, the identification of possible worlds with Everettian branches, not on details such as my examples of indexicality and indeterminism. I will start with generalities, and then end with a specific objection.

The first thing to say is that this is an utterly <u>naturalistic</u> account of modality. The entire realm of modality----all the possibilities imagined in our counterfactual suppositions, and the countless others that we never imagine---is to be incorporated into the physical cosmos as nowadays described by quantum theory. There could hardly be a more radical, indeed breath-taking, "take-over bid" of the subject-matter of logic and semantics, the traditional preserve of logicians and philosophers, by another discipline. Not by physics itself of course, but by another band of philosophers, viz. the philosophical interpreters of quantum theory.

It is worth setting this take-over bid in the wider context of the usual view of the connections between this book's three proposed multiverses. On this view, the logical and semantical investigations that prompt Chapter 3's 'philosophers' paradise' of possible worlds are---if not wholly <u>a priori</u>, then at least---independent of the contingent discoveries of physics, or other empirical sciences. So whatever our answer to Chapter 3's anxious question about the nature of possible worlds eventually turns out to be: the framework of possible worlds, the 'philosophers' paradise', provides a vast reality, an empire, in which physics and other sciences take their place as, so to speak, a province. They investigate an aspect of reality: namely, the contingent details of how the actual cosmos works. But they are not "the whole story". And while physics may discover a multiverse, whether Everettian or cosmological or both: that physical multiverse is <u>all within</u> the one actual possible world (i.e. cosmos), in the sense of Chapter 3's framework.

This usual view is undoubtedly the one that would be endorsed by the vast majority of logicians and philosophers After all, it accords with the traditional, indeed centuries-old, idea that logic and philosophy investigate features of reality that are very general, independent of the details of observation and experiment, and maybe even <u>a priori</u>; whereas the sciences investigate specific and presumably contingent features, via detailed observation and experiment.

Obviously, this proposal's take-over bid completely denies this view. And speaking for myself, I confess that for that reason, I resist it. That is: even if I assume the Everettian interpretation of quantum theory, I still believe sufficiently in the autonomy, generality and maybe even <u>a priori</u> status of logic, and in particular of logical investigations of modality, to <u>not</u> incorporate all the possible worlds as "mere" branches in the quantum cosmos. I would rather suffer anxiety from my not knowing how to answer Chapter 3's question about the nature of possible worlds, than have quantum physics take over logic.

Of course, this confession accords with the Humeanism I admitted in various passages of Chapters 1 to 3 (especially Chapter 2.5, 2.6 and 3.6): both my acceptance that the results of science, in particular the laws of nature, are contingent; and more generally a low-key or modest estimate of the kind of understanding that human enquiry can secure.

It is also worth noticing a consequence for this proposal of the two facts (discussed in Section 1 above) that: (i) the Everettian interpretation has not been established, and might one day be agreed to be wrong, with another interpretation being endorsed, and (ii) quantum theory (as a physical theory, however interpreted) may one day be superseded. These facts imply that this proposal's official semantics for our modal talk, e.g. counterfactual conditionals, is a hostage to fortune. If (i) the Everettian interpretation is one day discarded, or (ii) quantum theory is superseded by a successor theory with no multiverse that could supply the truth-conditions of our modal talk, then the proposal's advocate will have to concede that their semantics has been refuted by the progress of physics---and that a new semantics must be formulated. (For they could hardly say that their semantics remains right, i.e. that we have for all these centuries been talking modally about the strictly fictional multiverse of a twentieth-century physical theory that has just been refuted.)

This consequence, that semantics could be refuted by physics, is worth noticing since it makes vivid how opposed the proposal is to the usual view of the relations between philosophy and physics. But agreed: I do not intend this consequence as an objection. Indeed, the proposal's advocate is likely to be so steeped in naturalism as to take this consequence in their stride.

Finally, here is a specific objection to the proposal. The idea is that the proposal conflicts with the contingency, not just of the laws of quantum theory (in particular the Schroedinger equation), but also of two features much more specific and particular than the theory's laws: namely, what is the physical state of the quantum cosmos, and what are the forces operating within it. (These forces are encoded in a term in the Schroedinger equation, the Hamiltonian.) Of course a Humean like me will want to take the laws of quantum theory as contingent (Chapters 2.5 and 3.6). But it is not only Humeans like me who endorse the contingency of the other two features, viz. what is the state and what forces are operating. For non-Humeans, including those who say that the laws of nature are necessary, can and almost always do agree that both the state of a system, and the forces acting on it (causing changes, determining its future evolution), are contingent: they could have been otherwise.

      Thus the objection is plain as day. The proposal implies that "a lot more" is necessary than is usual believed to be necessary. Not only does it imply that the laws of quantum theory are necessary. Non-Humeans about laws, who accept quantum theory (as they should!), might well accept that. But the proposal also implies that the quantum state of the cosmos (usually called: the universal quantum state) could <u>not</u> have been other than it is. For according to the proposal, the state defines, via its set of decoherent branches, what is possible. And here, 'what is possible' means---<u>not</u> what is nomologically possible in the restricted-modality sense explained at the end of Chapter 3.6---but what is possible <u>tout court</u>. Such necessity is hard to believe. Similarly, the proposal implies that the forces operating within the quantum cosmos (encoded in the Hamiltonian that, as I stressed in Chapters 4 and 5, nobody knows how to write down!) could <u>not</u> have been other than they are. That is also hard to believe.

      There is also an ancillary problem. This necessity of the quantum state, and of the operative forces, also threatens to cause trouble for the second issue raised at the end of Chapter 4.8. I mean the Everettian's need to allow that the factorization of the cosmos' state-space, into the factor (component) state-spaces for the various macrosystems, depends on what the overall state is. In Chapter 4.8, we saw that this allowance promised to give the Everettian the wherewithal to secure the fact that there surely could have been different macrosystems than there in fact are. This fact seems a "non-negotiable" fact of our modal thought and talk. After all, we say things like 'I might have had more children'. But if the quantum state of the cosmos is necessary---given "once and for all" in the most absolute sense, since it defines the entire realm of possibilities---then the above strategy for securing the fact that there could have been different macrosystems stumbles badly.

Let me sum up this Section. I have stated and assessed a proposal to reduce (in Chapter 3.1's sense) the philosophical multiverse to the Everettian one; (I restricted myself to Chapter 4's non-cosmological Everettian). I admired the boldness of this proposal. But I cannot believe it. My reasons were not just my belief in the autonomy and generality of logic compared with physics. Also more specifically: being a Humean, I see the results of science, even results so glorious as those of quantum theory, as both contingent and fallible---so that this proposal gives us an <u>embarrass de richesse</u> of necessities.

<u>Chapter 6.4: Envoi</u>
One theme has been so prominent throughout this book as to merit a closing quotation. I mean the theme that I derived from Bacon, Locke and especially Hume: that we should not be beguiled by words, and we should be modest about the kind of understanding of nature that human enquiry---even the glories of modern physics, with their breathtaking quantitative precision---can secure. (See the discussions in Chapters 1.5 and 2.5, 2.6; and item (1) in Chapter 2's Notes.)

In that masterpiece, <u>An Enquiry concerning Human Understanding</u>, Hume writes near the end of Part I of Section IV (which is entitled 'Sceptical doubts concerning the operations of the understanding'):

Hence we may discover the reason, why no philosopher, who is rational and modest, has ever pretended to assign the ultimate cause of any natural operation, or to show distinctly the action of that power, which produces any single effect in the universe. It is confessed, that the utmost effort of human reason is, to reduce the principles, productive of natural phenomena, to a greater simplicity, and to resolve the many particular effects into a few general causes, by means of reasonings from analogy, experience, and observation. But as to the causes of these general causes, we should in vain attempt their discovery; nor shall we ever be able to satisfy ourselves, by any particular explication of them. These ultimate springs and principles are totally shut up from human curiosity and enquiry. Elasticity, gravity, cohesion of parts, communication of motion by impulse; these are probably the ultimate causes and principles which we shall ever discover in nature; and we may esteem ourselves sufficiently happy, if, by accurate enquiry and reasoning, we can trace up the particular phenomena to, or near to, these general principles. The most perfect philosophy of the natural kind only staves off our ignorance a little longer: As perhaps the most perfect philosophy of the moral or metaphysical kind serves only to discover larger portions of it. Thus the observation of human blindness and weakness is the result of all philosophy, and meets us, at every turn, in spite of our endeavours to elude or avoid it.

Wise words. But they imply that deciding what to believe about the various multiverse proposals is hard . . .

So after this book's relentless, indeed humourless, dialectical weighing-up of the arguments for and against, it is amusing to see that---if you can forgive the ambiguity of the word 'world', meaning either 'planet' or 'cosmos'---Alexander the (so-called) Great was similarly daunted. Thus Plutarch, in Section 4 of his '<u>On the Tranquillity of Mind</u>', writes:

Alexander wept when he heard from Anaxarchus that there were an infinite number of worlds; and his friends asking him if any accident had befallen him, he returns this answer: 'Do you not think it a matter of lamentation that when is such a vast multitude of them, we have not yet conquered one?'

Indeed, conquering the multiverse is hard.

<u>Chapter 6: Notes and further reading</u>
I shall give a few suggestions for each of Sections 2 to 4 of this Chapter. (Section 1, on what I myself believe, referred a good deal to Sections in previous Chapters; so it needs no further suggestions.) As perhaps befits a final Chapter, some of these suggestions will return us to references given in the Notes to previous Chapters.

(1): For Section 2, about why we cannot "see" or have causal contact with other universes, I have two suggestions: both of them recommended in previous Chapters.

First, the double-slit experiment is used by John Bell in the opening pages of his pedagogic exposition of the interpretations of quantum theory, to explain the measurement problem: specifically, the collapse of the wave-function to produce the dots (scintillations) forming the interference pattern on the screen. That is: J. S. Bell, 'Six Possible Worlds of Quantum Mechanics'. As I said in the Notes to Chapter 4: this is most easily found in the journal, Foundations of Physics, for 1992. It is also reprinted in Bell's collection, Speakable and Unspeakable in Quantum Mechanics (Cambridge University Press 1987; revised edition 2004): which is available at:
https://www.cambridge.org/core/books/speakable-and-unspeakable-in-quantum-mechanics/E0D032E7E7EDEF4E4AD09F458F2D9DB7

Second, Lewis' masterpiece book-length defence of his modal realism, On the Plurality of Worlds (Backwell, 1986), has two Sections that bear directly on the question why we cannot "see" or have causal contact with other universes.

The first is Section 1.6, entitled 'Isolation'. (My Section 2 quoted its closing passage.) Its first two-thirds argues that any two objects (across all the worlds) are spatiotemporally related iff they are in the same world (in Lewis' jargon: iff they are worldmates). And the last third of Section 1.6 argues that there can be no causation a la Lewis between worlds.

The second relevant Section is Lewis' Section 1.7, entitled 'Concreteness'. This shows in detail that the concrete/abstract distinction is not nearly as clear as many philosophers presume, since it is multiply ambiguous. This is a topic that bears not only on the understanding of causation; but also (as we have seen in Chapter 3.9C, and (1) of the Notes to Chapter 3) on (i) the nature of possible worlds and (ii) M. Tegmark's advocacy of a "Pythagoreanist" mathematical multiverse.

(2): For Section 3, about the proposal to identify the philosopher's possible worlds with Everettian branches, the main source is A. Wilson's book-length advocacy of this proposal:

A. Wilson, The Nature of Contingency (Oxford University Press 2020); available at: https://global.oup.com/academic/product/the-nature-of-contingency-9780198846215?q=The%20Nature%20of%20Contingency&lang=en&cc=gb

The book was preceded by several articles. Here are three. In the first two, Wilson argues in favour of interpreting the indeterminism of branching in terms of diverging branches/worlds rather than splitting branches/worlds. (Cf topic (2) in Section 3's exposition of Wilson's proposal.) In the third, he argues that his proposal accommodates, and indeed supports, the decision-theoretic approach to Everettian probability that I expounded in Chapter 4.11 and 4.12.

A. Wilson, 'Macroscopic ontology in Everettian quantum mechanics', The Philosophical Quarterly volume 61 (2011), pp. 363-382.

A. Wilson, 'Everettian quantum mechanics without branching time', Synthese volume 188 (2012), pp. 67–84; DOI 10.1007/s11229-011-0048-9

A. Wilson, 'Objective probability in Everettian quantum mechanics', British Journal of Philosophy of Science volume 64 (2013), pp. 709–737.

As to the literature's responses to the proposal, I recommend:

(i) the reviews of the book in several philosophy journals, such as Mind (Oxford University Press), and Notre Dame Philosophical Reviews, which is open access and online only at: https://ndpr.nd.edu/about/

(ii) a critical assessment, and suggested improvement, of the proposal by: J. Harding, 'Everettian Quantum Mechanics and the Metaphysics of Modality', The British Journal for the Philosophy of Science, volume 72, number 4, December 2021.

(3): For Section 4, and thus for the close of the book: it seems appropriate to celebrate the synergy between physics and philosophy---what used to be called 'Natural Philosophy' (as mentioned in Chapter 2.1). I will do this by first recalling what I said in Chapter 5 about what

Eddington really intended with his famous metaphor of the fishing net: namely, to underline the synergy, the mutual relevance, of physics (or more generally, science) and philosophy. Then I will close by recommending two articles by Carlo Rovelli.

In the Notes to Chapter 5 (item (2.F)), I quoted the whole passage containing Eddington's famous metaphor of the fishing net (in his 1938 book, <u>The Philosophy of Physical Science)</u>. The net is usually taken to stand just for our means of observation in the specific science at hand. In the fishing example, this means the biology of fish, which Eddington calls 'ichthyology'; so that using a net with a two-inch mesh, we might naively infer that all fishes are longer than two inches.

    But as I explained: if one reads a little beyond the frequently-quoted metaphor, one realizes Eddington takes the net to stand, not just for our means of observation, but also for our scientific method as a whole. Thus Eddington's moral is not just the obvious one I stressed in Chapter 5 Section 8, viz. 'conditionalize your credence on your means of observation'; but also that we should allow for types of knowledge inaccessible to the scientific method. This open-mindedness is bound to be welcome to a philosopher. We should take notice of---and take encouragement from---it.

    So here again is a portion of the later part of the passage, a portion that makes clear Eddington's intent. He writes:

'An onlooker [i.e. a philosopher savvy about such matters as observation selection effects] may object to [the icthyologist's] generalisation [that all fish are longer than two inches]. 'There are plenty of sea-creatures under two inches long, only your net is not adapted to catch them.' The icthyologist dismisses this objection contemptuously. 'Anything uncatchable by my net is ipso facto outside the scope of icthyological knowledge. In short, 'what my net can't catch isn't fish.' Or — to translate the analogy — 'If you are not simply guessing, you are claiming a knowledge of the physical universe discovered in some other way than by the methods of physical science, and admittedly unverifiable by such methods. You are a metaphysician. Bah!'

Finally, here are two articles by Carlo Rovelli, a theoretical physicist with a great gift for philosophy, and for work at the interface between physics and philosophy. They are not about the multiverse: indeed, I think Rovelli would be sceptical of all three of this book's proposals. But each of them is a manifesto for synergy between physics and philosophy, and so very much in the spirit of this book; and an eloquent passage in the second merits being quoted.

    The first declares this in its title: which also gives a salutary message to us philosophers to not be introspective, but to be open to what physics can offer us. It is from 2018:---

    C. Rovelli, 'Physics Needs Philosophy. Philosophy Needs Physics', <u>Foundations of Physics</u> volume 48 (2018) pp. 481–491; <u>https://doi.org/10.1007/s10701-018-0167-y</u>; and on arxiv at: 1805.10602.

    The second article is from 1997. It is: 'Halfway through the woods', in <u>The Cosmos of Science: Essays of Exploration</u>, (Pittsburgh University Press) edited by J. Earman and J.Norton.

    The title alludes to the opening lines of Dante's Divine Comedy. Rovelli's idea is that like Dante, modern physics is in the middle of a journey, and the path ahead is very unclear. For the quantum and relativity revolutions are by no means settled---there is much still to do. He draws an analogy with the one hundred and fifty years from Copernicus' heliocentrism to Newton's mechanics and theory of gravity (cf. Chapter 2.3). In this analogy, the quantum and relativity revolutions are like natural philosophy after Copernicus; and like natural philosophers between Copernicus and Newton---figures like Galileo and Descartes---we ourselves are halfway through the woods, searching for a synthesis of the insights from quantum theory and relativity. And---happily for us philosophers---he sees this predicament as an opportunity for philosophers. He writes:

General relativity and quantum mechanics are discoveries as extraordinary as the Copernican discovery. I believe they are, like Kepler's ellipses and Descartes's principle of inertia, fragments of a future science. I think that it is time to take them seriously, to try to understand what we have actually learned about the world by discovering relativity and quantum theory, and to find the fruitful questions. Maybe the Newtonian age has been an accident and we will never again reach a synthesis. If so, a major project of natural philosophy has failed. But if a new synthesis is to be reached, I believe that philosophical thinking will be once more one of its ingredients. . . . As a physicist involved in this effort, I wish the philosophers who are interested in the scientific description of the world would not confine themselves to commenting and polishing the present fragmentary physical theories, but would take the risk of trying to look <u>ahead.</u>

And after from this call to arms, he also gives us an uplifting conclusion:

We are not close to the end of physics, nor to the final theory of everything. We are very much in the dark. We left the sunny grasses of Cartesian-Newtonian physics and are traveling through the woods, armed with everything we have learned and with our weak intuition, always wishing we were smarter. It would be a discouraging state of confusion, and we would feel lost, if it weren't that the trip is wonderful and the landscape so breathtaking.